\documentclass[11pt,letterpaper]{db}    

\usepackage{color}
\usepackage{wasysym}

  \usepackage[dvips]{graphicx} 

\usepackage{rotating}

\newcommand{\tab}{\hspace*{2em}}

\makeatletter   
\renewcommand{\labelitemi}{$\m@th\circ$}
\makeatother
\renewcommand{\thefigure}{\thesection.\arabic{figure}}
\renewcommand{\thetable}{\thesection.\arabic{table}}

\let\Otemize =\itemize
\let\Onumerate =\enumerate
\let\Oescription =\description
\def\Nospacing{\itemsep=0pt\topsep=0pt\partopsep=0pt\parskip=0pt\parsep=0pt}

\begin{document}

\setlength{\baselineskip}{2.6ex}
\pagestyle{empty}
\begin{figure}
\parbox{0.35\linewidth}{
\includegraphics*[height=3.5cm, angle=-90]{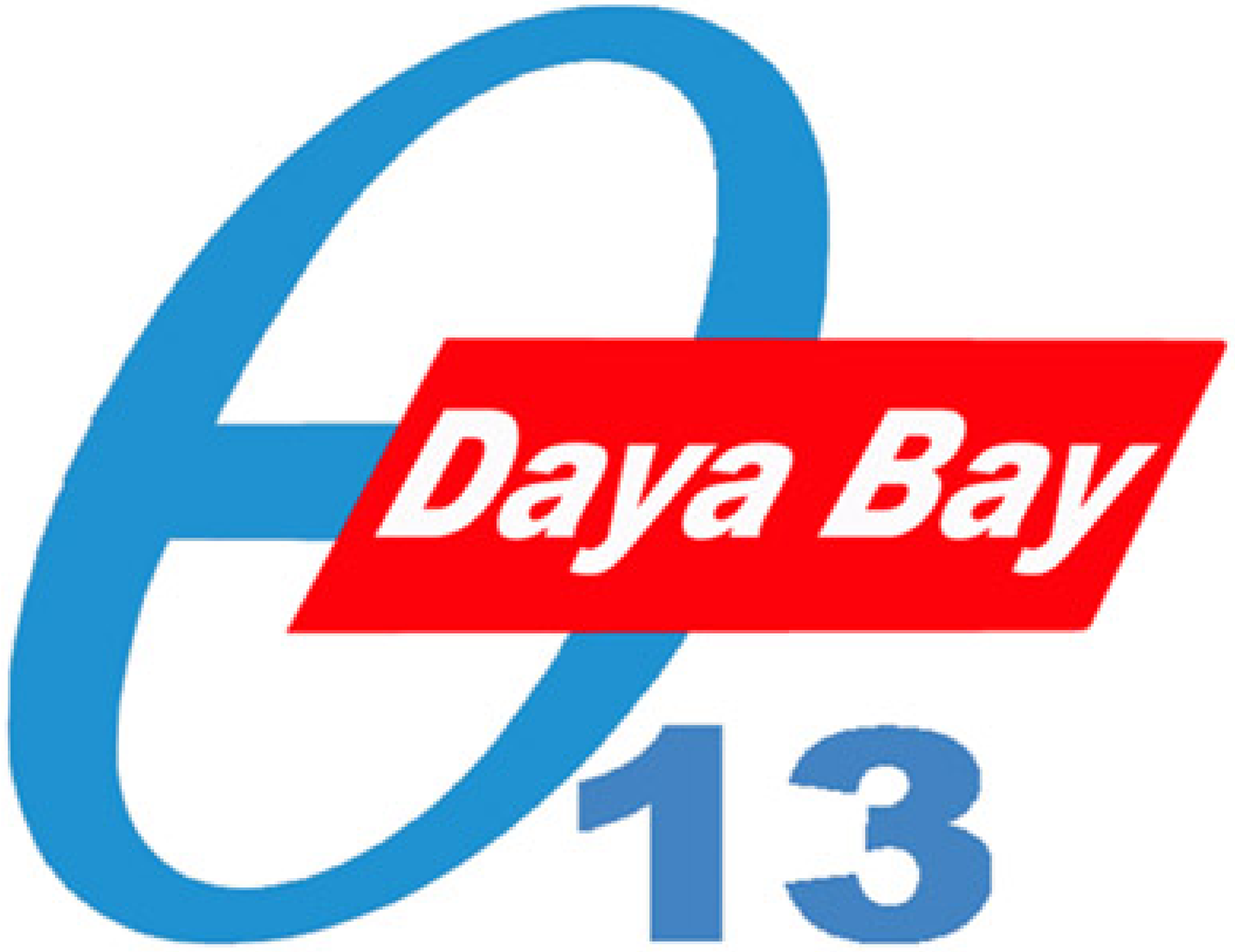}}
\parbox{0.29\linewidth}
{\hspace*{\fill}}
\parbox{0.35\linewidth}{
\includegraphics*[height=3.5cm, angle=-90]{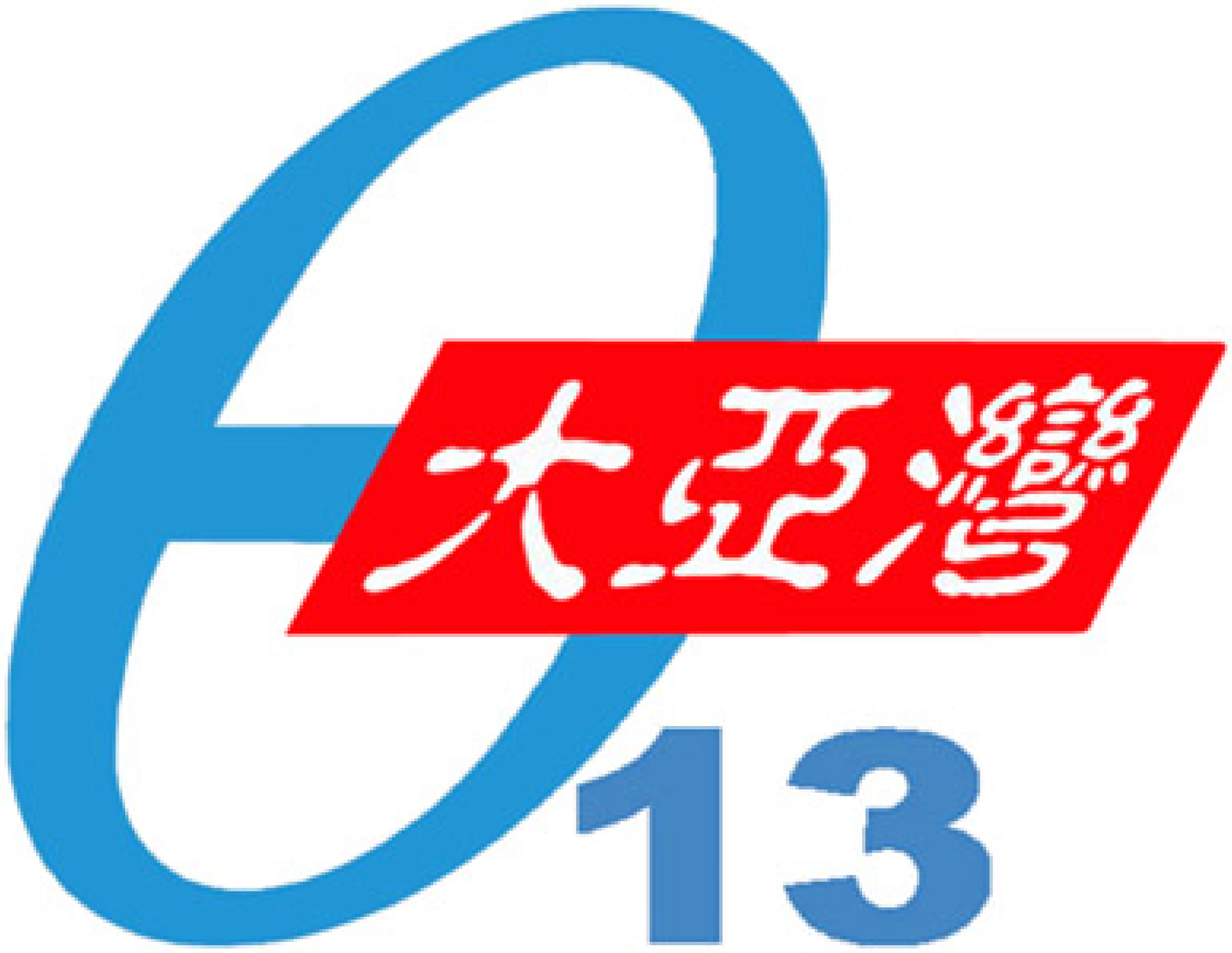}}
\end{figure}

\vspace*{0.5cm}
\LARGE
\centerline{Daya Bay}
\bigskip
\Huge
\centerline{Proposal}
\bigskip
\LARGE
\centerline{December 1, 2006}
\bigskip
\begin{center}
A Precision Measurement of the Neutrino Mixing Angle $\theta_{13}$
Using Reactor Antineutrinos At Daya Bay
\end{center}
\normalsize
\vfill
\begin{figure}[h]
\begin{center}
\includegraphics*[height=0.9\linewidth, angle=-90]{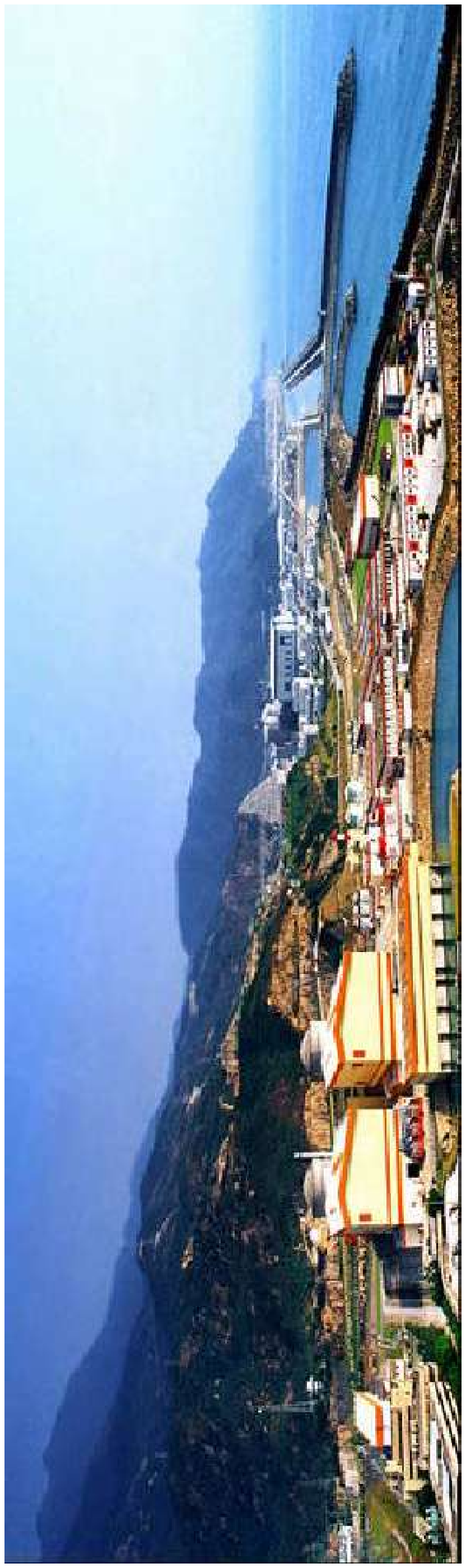}
\end{center}
\end{figure}

\newpage
\centerline{\bf \large Daya Bay Collaboration}

\begin{description}

\item[Beijing Normal University] \tab \newline
Xinheng Guo, Naiyan Wang, Rong Wang

\item[Brookhaven National Laboratory] \tab \newline
Mary Bishai, Milind Diwan, Jim Frank, Richard L. Hahn, Kelvin Li, Laurence Littenberg, David Jaffe, Steve Kettell, Nathaniel Tagg, Brett Viren, Yuping Williamson, Minfang Yeh 

\item[California Institute of Technology] \tab \newline
Christopher Jillings, Jianglai Liu, Christopher Mauger, Robert McKeown

\item[Charles Unviersity] \tab \newline
Zdenek Dolezal, Rupert Leitner, Viktor Pec, Vit Vorobel

\item[Chengdu University of Technology] \tab \newline
Liangquan Ge, Haijing Jiang,  Wanchang Lai, Yanchang Lin

\item[China Institute of Atomic Energy] \tab \newline
Long Hou, Xichao Ruan, Zhaohui Wang, Biao Xin, Zuying Zhou

\item[Chinese University of Hong Kong,] \tab \newline
Ming-Chung Chu, Joseph Hor, Kin Keung Kwan, Antony Luk

\item[Illinois Institute of Technology] \tab \newline
Christopher White

\item[Institute of High Energy Physics] \tab \newline
Jun Cao, Hesheng Chen, Mingjun Chen, Jinyu Fu, Mengyun Guan, Jin Li,
Xiaonan Li, Jinchang Liu, Haoqi Lu, Yusheng Lu, Xinhua Ma, Yuqian Ma,
Xiangchen Meng, Huayi Sheng, Yaxuan Sun, Ruiguang Wang, Yifang Wang,
Zheng Wang, Zhimin Wang, Liangjian Wen, Zhizhong Xing, Changgen Yang,
Zhiguo Yao, Liang Zhan, Jiawen Zhang, Zhiyong Zhang, Yubing Zhao,
Weili Zhong, Kejun Zhu, Honglin Zhuang

\item[Iowa State University] \tab \newline
Kerry Whisnant, Bing-Lin Young

\item[Joint Institute for Nuclear Research] \tab \newline
Yuri A. Gornushkin, Dmitri Naumov, Igor Nemchenok, Alexander Olshevski

\item[Kurchatov Institute] \tab \newline
Vladimir N. Vyrodov

\item[ Lawrence Berkeley National Laboratory and University of California at Berkeley] \tab \newline
Bill Edwards, Kelly Jordan, Dawei Liu, Kam-Biu Luk, Craig Tull

\item[Nanjing University] \tab \newline
Shenjian Chen, Tingyang Chen, Guobin Gong, Ming Qi

\item[Nankai University] \tab \newline
Shengpeng Jiang, Xuqian Li, Ye Xu

\item[National Chiao-Tung University] \tab \newline
Feng-Shiuh Lee, Guey-Lin Lin, Yung-Shun Yeh

\item[National Taiwan University] \tab \newline
Yee B. Hsiung

\item[National United University] \tab \newline
Chung-Hsiang Wang

\item[Princeton University] \tab \newline
Changguo Lu, Kirk T. McDonald

\item[Rensselaer Polytechnic Institute] \tab \newline
John Cummings, Johnny Goett, Jim Napolitano, Paul Stoler

\item[Shenzhen Univeristy] \tab \newline
Yu Chen, Hanben Niu, Lihong Niu

\item[Sun Yat-Sen (Zhongshan) University] \tab \newline
Zhibing Li

\item[Tsinghua University] \tab \newline
Shaomin Chen, Hui Gong, Guanghua Gong, Li Liang, Beibei Shao,
Qiong Su, Tao Xue, Ming Zhong

\item[University of California at Los Angeles] \tab \newline
Vahe Ghazikhanian, Huan Z. Huang, Charles A. Whitten, Stephan Trentalange

\item[University of Hong Kong,] \tab \newline
K.S. Cheng, Talent T.N. Kwok, Maggie K.P. Lee, John K.C. Leung, Jason C.S. Pun, Raymond H.M. Tsang, Heymans H.C. Wong

\item[University of Houston] \tab \newline
Michael Ispiryan, Kwong Lau, Logan Lebanowski, Bill Mayes, Lawrence Pinsky, Guanghua Xu

\item[University of Illinois at Urbana-Champaign] \tab \newline
S. Ryland Ely, Wah-Kai Ngai, Jen-Chieh Peng

\item[University of Science and Technology of China] \tab \newline
Qi An, Yi Jiang, Hao Liang, Shubin Liu, Wengan Ma, Xiaolian Wang, Jian Wu, Ziping Zhang, Yongzhao Zhou

\item[University of Wisconsin] \tab \newline
A. Baha Balantekin, Karsten M. Heeger, Thomas S. Wise

\item[Virginia Polytechnic Institute and State University] \tab \newline
Jonathan Link, Leo Piilonen

\end{description}

\vfill
Preprint numbers:\\
BNL-77369-2006-IR \\
LBNL-62137 \\
TUHEP-EX-06-003

\newpage
\pagestyle{headings}
\renewcommand{\thepage}{\Roman{page}}
\setcounter{page}{1}
\section*{Executive Summary}
\addcontentsline{toc}{section}{Executive Summary}

This document describes the design of the Daya Bay reactor neutrino
experiment.  Recent discoveries in neutrino physics have shown that
the Standard Model of particle physics is incomplete. The observation
of neutrino oscillations has unequivocally demonstrated that the
masses of neutrinos are nonzero. The smallness of the neutrino masses
($<$2~eV) and the two surprisingly large mixing angles measured have
thus far provided important clues and constraints to extensions of the
Standard Model.

The third mixing angle, $\theta_{13}$, is small and has not yet been
determined; the current experimental bound is $\sin^22\theta_{13} <$
0.17 at 90\% confidence level (from Chooz) for $\Delta m^2_{31} = 2.5
\times 10^{-3}$ eV$^2$. It is important to measure this angle to
provide further insight on how to extend the Standard Model.  A
precision measurement of $\sin^22\theta_{13}$ using nuclear reactors
has been recommended by the 2004 APS Multi-divisional Study on the
Future of Neutrino Physics as well as a recent Neutrino Scientific
Assessment Group (NuSAG) report.

We propose to perform a precision measurement of this mixing angle by
searching for the disappearance of electron antineutrinos from the
nuclear reactor complex in Daya Bay, China.  A reactor-based
determination of $\sin^22\theta_{13}$ will be vital in resolving the
neutrino-mass hierarchy and future measurements of $CP$ violation in
the lepton sector because this technique cleanly separates
$\theta_{13}$ from $CP$ violation and effects of neutrino propagation
in the earth.  A reactor-based determination of $\sin^22\theta_{13}$
will provide important, complementary information to that from
long-baseline, accelerator-based experiments.  The goal of the Daya
Bay experiment is to reach a sensitivity of 0.01 or better in
$\sin^22\theta_{13}$ at 90\% confidence level.

\vspace{1em}
{\bf The Daya Bay Experiment}
\vspace{1em}

The Day Bay nuclear power complex is one of the most prolific sources
of antineutrinos in the world. Currently with two pairs of reactor
cores (Daya Bay and Ling Ao), separated by about 1.1~km, the complex
generates 11.6~GW of thermal power; this will increase to 17.4~GW by
early 2011 when a third pair of reactor cores (Ling Ao II) is put into
operation and Daya Bay will be among the five most powerful reactor
complexes in the world. The site is located adjacent to mountainous
terrain, ideal for siting underground detector laboratories that are
well shielded from cosmogenic backgrounds. This site offers an
exceptional opportunity for a reactor neutrino experiment optimized to
perform a precision determination of $\sin^22\theta_{13}$ through a
measurement of the relative rates and energy spectrum of reactor
antineutrinos at different baselines.  In addition, this project
offers a unique and unprecedented opportunity for scientific
collaboration involving China, the U.S., and other countries.

\begin{figure}[!htb]
\begin{center}
\includegraphics[height=10cm]{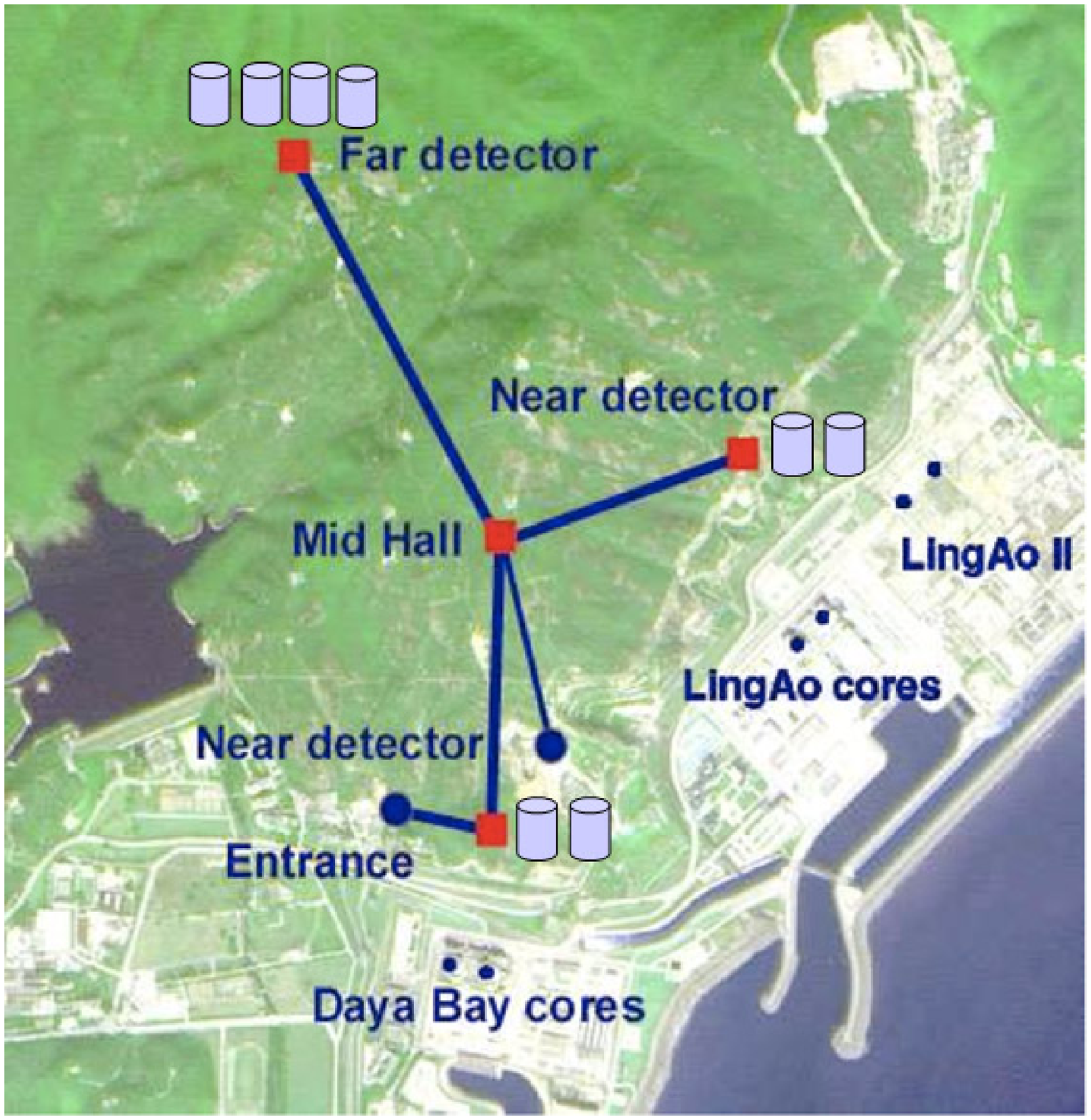}
\caption{Default configuration of the Daya Bay experiment, optimized
for best sensitivity in sin$^22\theta_{13}$. Four detector modules are
deployed at the far site and two each at each of the near sites.}
\label{fig:deployment}
\end{center}
\end{figure}

The basic experimental layout of Daya Bay consists of three
underground experimental halls, one far and two near, linked by
horizontal tunnels. Figure~\ref{fig:deployment} shows the detector
module deployment at these sites.
Eight identical cylindrical detectors, each consisting of three nested
cylindrical zones contained within a stainless steel tank, will be
deployed to detect antineutrinos via the inverse beta-decay reaction.
To maximize the experimental sensitivity four detectors are deployed
in the far hall at the first oscillation maximum.  The rate and energy
distribution of the antineutrinos from the reactors are monitored with
two detectors in each near hall at relatively short baselines from
their respective reactor cores, reducing the systematic uncertainty in
$\sin^2 2\theta_{13}$ due to uncertainties in the reactor power levels
to about 0.1\%.  This configuration significantly improves the
statistical precision over previous experiments (0.2\% in three years
of running) and enables cross-calibration to verify that the detectors
are identical.  Each detector will have 20 metric tons of 0.1\%
Gd-doped liquid scintillator in the inner-most, antineutrino target
zone. A second zone, separated from the target and outer buffer zones
by transparent acrylic vessels, will be filled with undoped liquid
scintillator for capturing gamma rays that escape from the target
thereby improving the antineutrino detection efficiency.  A total of
224 photomultiplier tubes are arranged along the circumference of the
stainless steel tank in the outer-most zone, which contains mineral
oil to attenuate gamma rays from trace radioactivity in the
photomultiplier tube glass and nearby materials including the outer
tank. The detector dimensions are summarized in Table~\ref{tab:dimen}.
\begin{table}[htdp]
\begin{center}
\begin{tabular}{|l||c|c|c|} \hline
Dimensions           & Inner Acrylic & Outer Acrylic	& Stainless Steel \\ \hline\hline
Diameter (mm)        & 3200	     & 4100	        & 5000	          \\ \hline
Height (mm)	     & 3200	     & 4100	        & 5000	          \\ \hline
Wall thickness (mm)  &   10	     &   15		&   10		  \\ \hline
Vessel Weight (ton)  &  0.6	     &  1.4		&   20		  \\ \hline	
Liquid Weight (ton)  &$\sim$20	     & $\sim$20		& $\sim$40	  \\ \hline
\end{tabular}
\caption{Summary of antineutrino detector properties. The liquid weights 
are for the mass of liquid contained only within that zone.}
\label{tab:dimen}
\end{center}
\end{table}

With reflective surfaces at the top and bottom of the detector
the energy resolution of the detector is about 12\% at 1 MeV.

The mountainous terrain provides sufficient overburden to suppress
cosmic muon induced backgrounds to less than 1\% of the antineutrino
signal.  The detectors in each experimental hall are shielded by 2.5~m
of water from radioactivity and spallation neutrons in the surrounding
rock.  The detector halls include a muon detector system, consisting
of a tracker on top of the water pool and water Cherenkov counters in
the water shield, for tagging the residual cosmic muons.

With this experimental setup, the signal and background rates at the
Daya Bay near hall, Ling Ao near hall and the far hall are summarized
in Table~\ref{tab:rates}.
\begin{table}[htdp]
\begin{center}
\begin{tabular}{|l||c|c|c|} \hline
		& Daya Bay Near	& Ling Ao Near		& Far Hall	     \\ \hline\hline
Baseline (m)	& 363		& 481 from Ling Ao	& 1985 from Daya Bay \\
		&		& 526 from Ling Ao II	& 1615 from Ling Ao's \\ \hline
Overburden (m)  &  98		& 112			&  350		     \\ \hline
Radioactivity (Hz)  &  $<$50    &  $<$50                &    $<$50           \\ \hline
Muon rate (Hz)  &  36           &  22                   &    1.2             \\ \hline
Antineutrino Signal (events/day)& 930	& 760		& 90                 \\ \hline
Accidental Background/Signal (\%)  &$<$0.2 & $<$0.2	& $<$0.1             \\ \hline
Fast neutron Background/Signal (\%)& 0.1 &    0.1	&    0.1	     \\ \hline
$^8$He+$^9$Li Background/Signal (\%)& 0.3 &    0.2     	&    0.2	     \\ \hline	
\end{tabular}
\caption{Summary of signal and background rates for each detector
module at the different experimental sites.}
\label{tab:rates}
\end{center}
\end{table}

Careful construction, filling, calibration and monitoring of the
detectors will reduce detector-related systematic uncertainties to a
level comparable to or below the statistical uncertainty.
Table~\ref{tab:syst} is a summary of systematic uncertainties for the
experiment.
\begin{table}[htdp]
\begin{center}
\begin{tabular}{|l||c|} \hline
Source				&	Uncertainty		 \\ \hline\hline
Reactor Power			&	0.087\% (4 cores)	 \\
				&	0.13\% (6 cores)	 \\ \hline
Detector (per module)	        &	0.38\% (baseline)	 \\
				&	0.18\% (goal)		 \\ \hline
Signal Statistics	        &	0.2\%	                 \\ \hline
\end{tabular}
\caption{Summary of uncertainties. The baseline value is
realized through proven experimental methods, whereas the goal value
should be attainable with additional research and development.}
\label{tab:syst}
\end{center}
\end{table}

The horizontal tunnels connecting the detector halls will facilitate
cross-calibration and offer the possibility of swapping the detectors
to further reduce systematic uncertainties.

Civil construction is scheduled to begin in the spring of
2007. Deployment of the first pair of the detectors in one of the near
halls will start in February 2009. Data taking using the baseline
configuration of two near halls and the far hall will begin in June
2010. With three years of running and the estimated signal and
background rates as well as systematic uncertainties, the sensitivity
of Daya Bay for $\sin^22\theta_{13}$ is 0.008 or better, relatively
independent of the value of $\Delta m^2_{31}$ within its currently
allowed range.

\newpage
\renewcommand{\thepage}{\roman{page}}
\setcounter{page}{1}
\tableofcontents
\newpage
\listoffigures
\newpage
\listoftables
\newpage
\renewcommand{\thepage}{\arabic{page}}
\setcounter{page}{1}
\renewcommand{\thesection}{\arabic{section}}
\setcounter{figure}{0}
\setcounter{table}{0}
\setcounter{footnote}{0}

\section{Physics}
\label{sec:physics}

Neutrino oscillations are an ideal tool for probing neutrino mass and
other fundamental properties of neutrinos.  This intriguing phenomenon
depends on two neutrino mass differences and three mixing angles.  The
neutrino mass differences and two of the mixing angles have been measured
with reasonable precision.  The goal of the Daya Bay reactor
antineutrino experiment is to determine the last unknown neutrino
mixing angle $\theta_{13}$ with a sensitivity of 0.01 or better in
sin$^22\theta_{13}$, an order of magnitude better than the current
limit.  This section provides an overview of neutrino oscillation, the
key features of reactor antineutrino experiments, and a summary of
the Daya Bay experiment.

\subsection{Neutrino Oscillations}
\label{ssec:physics_overview}

The last decade has seen a tremendous advance in our understanding
of the neutrino sector~\cite{BMW}. There is now robust evidence
for neutrino flavor conversion from solar, atmospheric, reactor
and accelerator experiments,
using a wide variety of detector technologies.
The only consistent explanation for these results is that
neutrinos have mass and that the mass eigenstates are not the same
as the flavor eigenstates (neutrino mixing).  Neutrino
oscillations depend only on mass-squared differences and neutrino
mixing angles.  The scale of the mass-squared difference probed by an
experiment depends on the ratio $L/E$, where $L$ is the baseline
distance (source to detector) and $E$ is the neutrino energy.
Solar and long-baseline reactor experiments are sensitive to a
small mass-squared difference, while atmospheric, short-baseline
reactor and long-baseline accelerator experiments are sensitive to
a larger one. To date only disappearance
experiments have convincingly indicated the existence of neutrino
oscillations.

The SNO experiment~\cite{SNO} utilizes heavy water to measure 
high-energy $^8$B solar neutrinos via charged current (CC), 
neutral current (NC) and elastic scattering (ES) reactions. 
The CC reaction is sensitive
only to electron neutrinos whereas the NC reaction is sensitive to
the total active solar neutrino flux ($\nu_e$, $\nu_\mu$ and
$\nu_\tau$). Elastic scattering has both CC and NC components and
therefore serves as a consistency check.  The neutrino flux
indicated by the CC data is about one-third of that given by the
NC data, and the NC data also agrees with the standard solar model
prediction for the $^8$B neutrino flux.  Since only $\nu_{\rm
e}$'s are produced in the sun, the SNO data can only be explained
by flavor transmutation $\nu_e \to \nu_\mu$ and/or $\nu_\tau$.
Super-Kamiokande has also measured the ES flux for
the $^8$B neutrinos~\cite{Super-K-sol} with a water Cherenkov counter 
and their data agree with the SNO results.

Radiochemical experiments can also measure lower-energy solar
neutrinos, in addition to $^8$B neutrinos. The Homestake
experiment~\cite{Homestake} is sensitive to $^7$Be and pep neutrinos
using neutrino capture on $^{37}$Cl. The SAGE, GALLEX and GNO
experiments~\cite{Ga} are sensitive to all sources of solar neutrinos,
including the dominant pp neutrinos, using neutrino capture on
$^{71}$Ga. A global fit to all solar neutrino data yields a unique
region in the oscillation parameter space, known as the Large Mixing Angle
(LMA) solution.

Using a liquid scintillator detector, the KamLAND
experiment~\cite{KamLAND} measured a deficit of electron antineutrinos from
reactors ($L/E$ sensitive to the mass-squared difference
indicated by the solar neutrino data) 
consistent with neutrino oscillations.  Furthermore, KamLAND has also
observed a spectral distortion that can only be explained by neutrino
oscillations. The oscillation parameters indicated by KamLAND agree
with the LMA solution.  Since they were done in completely different
environments, the combination of solar neutrino and KamLAND data rules
out exotic explanations such as nonstandard neutrino interactions or
neutrino magnetic moment~\cite{BMW}.

The atmospheric-neutrino induced $\mu$-like events of Super-Kamiokande
show a depletion at long flight-path compared to the theoretical
predictions without oscillations, while the $e$-like events agree with the
non-oscillation expectation~\cite{Super-K-atm}. The detailed
energy and zenith angle distributions for both electron and muon
events agree with the oscillation predictions if the dominant
oscillation channel is $\nu_\mu \to \nu_\tau$.  More recently, the
long-baseline accelerator experiments K2K~\cite{K2K} and
MINOS~\cite{MINOS}, 
have measured $\nu_\mu$ survival that is
consistent with the atmospheric neutrino data. The mass-squared
difference indicated by the atmospheric neutrino data is about 30
times larger than that obtained from the fits to solar data. The
existence of two independent mass-squared difference scales means
that the three neutrinos have different masses.

The Chooz~\cite{Chooz} and Palo Verde~\cite{PaloVerde} experiments, which
measured the survival probability of reactor electron antineutrinos at an
$L/E$ sensitive to the mass-squared difference indicated by the
atmospheric neutrino data, found no evidence for oscillations,
consistent with the lack of $\nu_e$ involvement in the atmospheric
neutrino oscillations. However, $\nu_e$ oscillations for this
mass-squared difference are still allowed at roughly the 10\%
level or less.

There exists another set of neutrino oscillation data from the
LSND short-baseline accelerator experiment~\cite{LSND}, which
found evidence of the oscillation $\bar{\nu}_\mu\rightarrow
\bar{\nu}_{\rm e}$.  A large region allowed by the LSND data has
been ruled out by the KARMEN experiment~\cite{KARMEN} and
astrophysical measurements~\cite{astro}.  The remaining allowed region
is currently being tested by the MiniBooNE
experiment~\cite{MiniBooNE}. If confirmed, the LSND signal would
require the existence of new physics beyond the standard
three-neutrino oscillation scenario.

\subsection{Neutrino Mixing}
\label{sssec:physics_mixing}

The phenomenology of neutrinos is described by a mass matrix.  For $N$
flavors, the neutrino mass matrix consists of $N$ mass eigenvalues,
$N(N-1)/2$ mixing angles, $N(N-1)/2$ $CP$ phases for Majorana
neutrinos or $(N-1)(N-2)/2$ $CP$ phases for Dirac neutrinos.  The
mixing phenomenon is caused by the misalignment of the flavor
eigenstates and the mass eigenstates which are related by a mixing
matrix.  The mass matrix which is commonly expressed in the flavor
base is diagonalized using the mixing matrix.  For three flavors, the
mixing matrix, usually called the
Maki-Nakagawa-Sakata-Pontecorvo~\cite{MNSP} mixing matrix, is defined
to transform the mass eigenstates ($\nu_1$, $\nu_2$, $\nu_3$) to the
flavor eigenstates ($\nu_{\rm e}$, $\nu_\mu$, $\nu_\tau$) and can be
parameterized as
\begin{eqnarray}
 U_{\rm MNSP}
 & = &\left(\begin{array}{ccc}
      1 & 0 & 0 \\ 0 & C_{23} & S_{23} \\ 0 & -S_{23} & C_{23}
      \end{array} \right)
 \left(\begin{array}{ccc}
      C_{13} & 0 & \hat{S}^*_{13} \\ 0 & 1 & 0 \\
      -\hat{S}_{13} & 0 & C_{13}
      \end{array} \right)
 \left(\begin{array}{ccc}
      C_{12} & S_{12} & 0 \\ -S_{12} & C_{12} & 0 \\ 0 & 0 & 1
      \end{array} \right)
 \left(\begin{array}{ccc}
    e^{i\phi_1}  &  &    \\
    &  e^{i\phi_2}  &    \\
    &               & 1    \end{array} \right) \nonumber \\
&=&
  \left( \begin{array}{ccc}
   C_{12}C_{13} & C_{13}S_{12} & \hat{S}^*_{13}  \\
   -S_{12}C_{23} - C_{12}\hat{S}_{13}S_{23}  &
         C_{12}C_{23} - S_{12}\hat{S}_{13}S_{23} & C_{13}S_{23} \\
    S_{12}S_{23} - C_{12}\hat{S}_{13}C_{23}  &
        -C_{12}S_{23} - S_{12}\hat{S}_{13}C_{23} & C_{13}C_{23}
\end{array} \right)
\left( \begin{array}{ccc}
    e^{i\phi_1}  &  &    \\
    &  e^{i\phi_2}  &    \\
    &               & 1      \end{array} \right)
\label{MMatrix}
\end{eqnarray}
where $C_{\rm jk}=\cos\theta_{\rm jk}$, $S_{\rm jk}=\sin\theta_{\rm
jk}$, $\hat{S}_{13}= e^{i\delta_{CP}}
\sin\theta_{13}$. The ranges of the mixing angles and the phases
are: $0 \leq \theta_{jk}\leq \pi/2$, $0 \le \delta_{CP},
\phi_1, \phi_2 \leq 2\pi$. The neutrino oscillation phenomenology
is independent of the Majorana phases $\phi_1$ and $\phi_2$, which
affect only {\it neutrinoless} double beta-decay experiments.

For three flavors, neutrino oscillations are completely described by
six parameters: three mixing angles $\theta_{12}$, $\theta_{13}$,
$\theta_{23}$, two independent mass-squared differences,
$\Delta{m}^2_{21}\equiv m^2_2-m^2_1$, $\Delta{m}^2_{32}\equiv
m^2_3-m^2_2$, and one $CP$ phase angle $\delta_{CP}$ (note that
$\Delta{m}^2_{31}\equiv m^2_3-m^2_1 =
\Delta{m}^2_{32}+\Delta{m}^2_{21}$).  An extensive discussion of
theoretical effects of massive neutrinos and neutrino mixings can be
found in~\cite{APStheory}.

\subsubsection{Current Knowledge of Mixing Parameters}
\label{sssec:physics_current}

Various solar, atmospheric, reactor, and accelerator neutrino
experimental data have been analyzed to determine the mixing
parameters separately and in global fits. In the three-flavor
framework there is a general agreement on solar and atmospheric
parameters. In particular for global fits in the $2\sigma$ range,
the solar parameters $\Delta{m}^2_{21}$ and $\sin^2\theta_{12}$ have been determined to
9\% and 18\%, respectively; the
atmospheric parameters $|\Delta{m}^2_{32}|$ and $\sin^2\theta_{23}$
have been determined to 26\% and 41\%, respectively. Due to the absence of a signal,
the global fits on $\theta_{13}$ result in upper bounds which vary
significantly from one fit to another. The sixth parameter, 
the $CP$ phase angle $\delta_{CP}$, is inaccessible to the
present and near future oscillation experiments.

We quote here the result of a recent global fit with $2\sigma$ (95\% C.L.) ranges~\cite{Fogli}:
\begin{eqnarray}
 \Delta{m}^2_{21} &=& 7.92(1.00 \pm 0.09)\times 10^{-5}~{\rm
 eV^2} \hspace{3ex}
 \sin^2\theta_{12} ~=~ 0.314(1.00 ^{+0.18}_{-0.15}) \label{SNOsalt} \\
 |\Delta{m}^2_{32}| &=& 2.4(1.00 ^{+0.21}_{- 0.26})\times
     10^{-3}~{\rm eV^2} \hspace{7ex}
 \sin^2\theta_{23} ~=~ 0.44(1.00 ^{+0.41} _{-0.22}) \label{eq:atmos}\\
 \sin^2\theta_{13} &=& (0.9  ^{+2.3} _{-0.9})\times 10^{-2} \label{eq:theta13}
\end{eqnarray}

\begin{figure}[!htb]
\begin{center}
\includegraphics[height=7cm,angle=0]{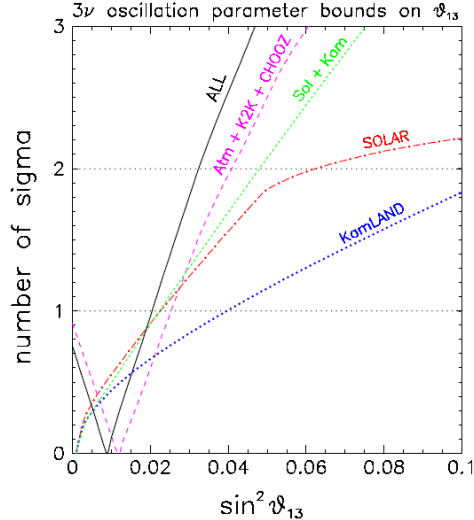}
\caption[Global fits to $\sin^2\theta_{13}$.]{Global fits to $\sin^2\theta_{13}$, taken from~\cite{Fogli}.
\label{fig:theta-13}}
\end{center}
\end{figure}

A collection of fits of $\sin^2\theta_{13}$ with different inputs as
given in~\cite{Fogli} is reproduced in Fig.~\ref{fig:theta-13}.
Note that fits involving solar or atmospheric data separately have
$\theta_{13}=0$ coinciding with the minima of the chi-square.
However global analyses taking into account both solar and
atmospheric effects show $\chi^2$ minima at a non-vanishing value
of $\theta_{13}$. 
Another very recent global fit~\cite{Schwetz} with different inputs finds
allowed ranges for the oscillation parameters that
overlap significantly with the above results even at $1\sigma$ (68\% C.L.).
The latest MINOS neutrino oscillation results~\cite{MINOS}
significantly overlap those in the global fit~\cite{Fogli}.
All these signify the convergence to a set of accepted values of
neutrino oscillation parameters $\Delta{m}^2_{21}$,
$|\Delta{m}^2_{32}|$, $\sin^2\theta_{12}$, and
$\sin^2\theta_{23}$. 

At 95\% C.L., the upper bound of $\theta_{13}$ extracted from
Eq.~\ref{eq:theta13} is about 10$^\circ$.  This corresponds to a value of
$\sin^22\theta_{13}$ of 0.12, which should be compared to the upper
limit of 0.17 at 90\% C.L. obtained by Chooz (see
Section~\ref{sssec:physics_past}).  We can conclude that, unlike $\theta_{12}$ and
$\theta_{23}$, the mixing angle $\theta_{13}$ is relatively small.

At present the three parameters that are not determined by the solar,
atmospheric, and KamLAND data are $\theta_{13}$, the sign of
$\Delta m^2_{32}$ which fixes the hierarchy of neutrino masses,
and the Dirac $CP$ phase $\delta_{CP}$.

\subsection{Significance of the Mixing Angle $\theta_{13}$}
\label{ssec:physics_theta13}

As one of the six neutrino mass parameters measurable in neutrino
oscillations, $\theta_{13}$ is important in its own right and for
further studies of neutrino oscillations. In addition, $\theta_{13}$
is important in theoretical model building of the neutrino mass
matrix, which can serve as a guide to the theoretical understanding of
physics beyond the standard model. Therefore, on all considerations,
it is highly desirable to significantly improve our knowledge on
$\theta_{13}$ in the near future.  The February 28, 2006 report of the
Neutrino Scientific Assessment Group (NuSAG)~\cite{nusag}, which
advises the DOE Offices of Nuclear Physics and High Energy Physics and
the National Science Foundation, and the APS multi-divisional study's
report on neutrino physics, {\it the Neutrino Matrix}~\cite{aps}, both
recommend with high priority a reactor antineutrino experiment to
measure sin$^22\theta_{13}$ at the level of 0.01.

\subsubsection{Impact on the experimental program}

The next generation of neutrino oscillation experiments has several
important goals to achieve: to measure more precisely the mixing
angles and mass-squared differences, to probe the matter effect, to
determine the hierarchy of neutrino masses, and very importantly to
determine the Dirac $CP$ phase.  The mixing matrix element which
provides the information on the $CP$ phase angle $\delta_{CP}$ appears
always in the combination $U_{\rm e 3}=\sin\theta_{13}
e^{-i\delta_{CP}}$. If $\theta_{13}$ is zero then it is not possible
to probe leptonic $CP$ violation in oscillation experiments.  Given
the known mixing angles $\theta_{12}$ and $\theta_{23}$ which are both
sizable, we thus need to know the value of $\theta_{13}$ to a
sufficient precision in order to design the future generation of
experiments to measure $\delta_{CP}$. The matter effect, which can be
used to determine the mass hierarchy, also depends on the size of
$\theta_{13}$. If $\theta_{13} > 0.01$, then the design of future
oscillation experiments is relatively straightforward~\cite{E-WhitePaper}.
However, for smaller $\theta_{13}$
new experimental techniques and accelerator technologies are likely required to carry out the
same sets of measurements.

\subsubsection {Impact on theoretical development}

The observation of neutrino oscillation has far reaching theoretical
implications. To date, it is the only evidence of physics beyond the
standard model in particle physics.  The pattern of the neutrino
mixing parameters revealed so far is strikingly different from that of
quarks. This has already put significant constraints and guidance
for constructing models involving new physics.  Driven by the value of
$\theta_{13}$, studies of the neutrino mass matrix have reached some
interesting general conclusions.

In general, if $\theta_{13}$
is not too small i.e., close to the current upper 
limit of $\sin^22\theta_{13}\approx 0.1$ and $\theta_{23}\neq
{\pi\over 4}$, the neutrino mass matrix does not have to have any
special symmetry features, sometimes referred to as anarchy
models, and the specific values of the mixing angles can be
understood as a numerical accident.

However, if $\theta_{13}$ is much smaller than the current
limit, special symmetries of the neutrino mass matrix will
be required. As a concrete example, the study of Mohapatra~\cite{Mohapatra}
shows that for $\theta_{13} < {\Delta{m}^2_{\rm sol}
\over\Delta{m}^2_{\rm atm}}\approx 0.03$ a $\mu$-$\tau$ 
lepton-flavor-exchange symmetry is required. It disfavors a quark-lepton
unification type theory based on $SU_c(4)$ or $SO(10)$ models.

For a larger value of $\theta_{13}$, it leaves open the question
of quark-lepton unification.

\subsection{Complementarity of Reactor-based and Accelerator-based Neutrino Oscillation
            Experiments}
\label{sec:synergy}

Long-baseline accelerator experiments with intense $\nu_\mu$ beams and
very large detectors, in addition to improving the measurements of
$|\Delta m^2_{32}|$ and $\theta_{23}$ via the study of $\nu_\mu$ survival, will
also be able to search for $\nu_e$ appearance due to $\nu_\mu \to \nu_e$
oscillations. A
measurement of both $\nu_\mu \to \nu_e$ and $\bar\nu_\mu \to
\bar\nu_e$ oscillations allows one to measure $\theta_{13}$, test
for $CP$ violation in the lepton sector, and determine the hierarchy 
of the neutrino masses, provided that $\theta_{13}$
is large enough. However, there are potentially three two-fold
parameter degeneracies leading to the following ambiguities~\cite{BMW,bmw}:

\begin{enumerate}
\item the $\delta_{CP} - \theta_{13}$ ambiguity,
\item  the ambiguity in the sign of $\Delta m^2_{32}$  and
\item  the $\theta_{23}$ ambiguity,
which occurs because only $\sin^22\theta_{23}$, {\it not $\theta_{23}$},
is measured in $\nu_\mu$ survival.
\end{enumerate}
The degeneracies can all be present
simultaneously, leading to as much as an eight-fold ambiguity in
the determination of $\theta_{13}$ and $\delta_{CP}$. Another
problem is that Earth-matter effects can induce fake $CP$
violation, which must be taken into account in any determination
of $\theta_{13}$ and $\delta_{CP}$.  One advantage of matter
effects is that they may be able to distinguish between the two possible
mass hierarchies.

There are experimental strategies that can overcome some of these
problems.  For example, by combining the results of two
long-baseline experiments at different baselines, the sign of
$\Delta m^2_{32}$ could be determined if $\theta_{13}$ is large
enough~\cite{hierarchy}. By sitting near the peak of the leading
oscillation with a narrow-band beam, $\theta_{13}$ can be removed
from the $\delta_{CP} - \theta_{13}$ ambiguity~\cite{peak}. However,
neither of these approaches resolves the $\theta_{23}$ ambiguity, and
$\theta_{13}$ may not be uniquely determined.

\begin{figure}[!htb]
\begin{center}
\includegraphics*[viewport=30 150 570 670, height=8cm]{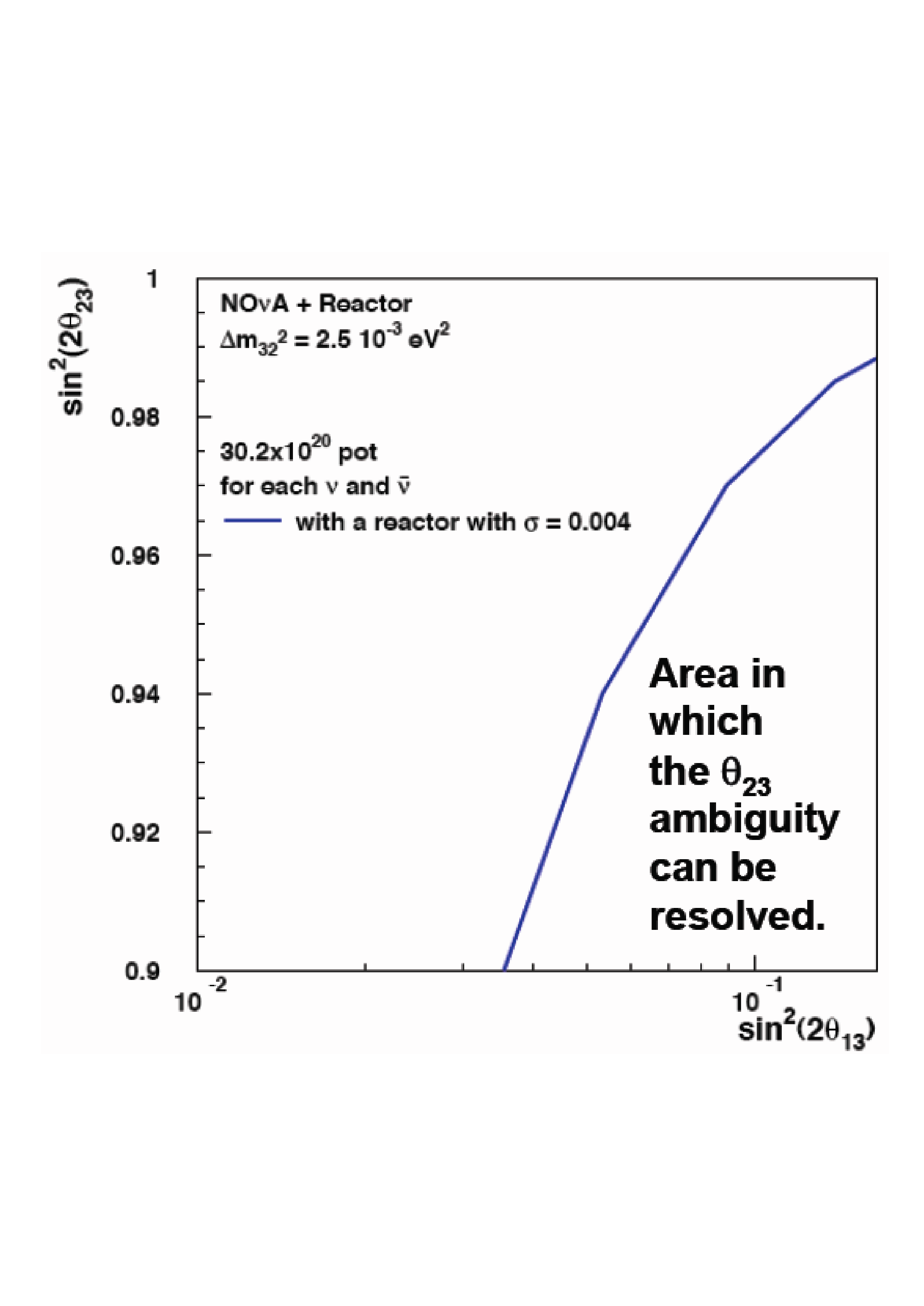}
\caption[Resolving ambiguity in $\theta_{23}$ with $\sin^22\theta_{13}$
determined by reactor experiments. Not only does the reactor
experiment provide a precise measurement of $\sin^22\theta_{13}$, but
it provides a precise measurement of $\theta_{23}$ by resolving an
ambiguity in the interpretation of the accelerator data.]{Resolving
ambiguity in $\theta_{23}$ with $\sin^22\theta_{13}$ determined by
reactor experiments.  Not only does the reactor experiment provide a
precise measurement of $\sin^22\theta_{13}$, but it provides a precise
measurement of $\theta_{23}$ by resolving an ambiguity in the
interpretation of the accelerator data. The blue line is the 95\%
C.L. curve averaged over the two mass-hierarchy solutions and possible
values of $\delta_{CP}$.}
\label{fig:synergy}
\end{center}
\end{figure}

The $\bar\nu_e$ survival probability for reactor antineutrinos
at short baseline depends only on $\theta_{13}$ and $\Delta m^2_{31}$, and is
independent of $\delta_{CP}$ and insensitive to $\theta_{12}$ and
$\Delta m_{21}^2$. Furthermore, matter effects are
negligible due to the short distance. Therefore, a short-baseline
reactor antineutrino experiment is an ideal method for measuring
$\theta_{13}$ with no degeneracy problem.  If $\theta_{13}$ can be
unambiguously determined by a reactor antineutrino experiment, then 
the $\delta_{CP} - \theta_{13}$ ambiguity is resolved and long-baseline
accelerator experiments can measure $\delta_{CP}$ and
determine the sign of $\Delta m^2_{32}$~\cite{synergy}.
Figure~\ref{fig:synergy} is an illustration of the synergy between
reactor experiments and the future very long-baseline accelerator
experiment, NOvA.
For $\sin^22\theta_{23} > 0.87$ at 68\% C.L.~\cite{MINOS}, using both muon 
neutrino and antineutrino beams, NOvA cannot distinguish the value of $\theta_{23}$.  
Yet, for $\Delta m^2_{32} = 2.5 \times 10^-3$ eV$^2$ and $\sin^22\theta_{13} > 0.035$, 
a reactor antineutrino experiment with an error of 0.004 can help single out the 
correct value for $\theta_{23}$.

\subsection{Reactor Antineutrino Experiments}
\label{ssec:physics_reactors}

Nuclear reactors have played crucial roles in experimental
neutrino physics.  Most prominently, the very first observation of
the neutrino was made at the Savannah River Nuclear Reactor in 1956 by
Reines and Cowan~\cite{Reines}, 26 years after the neutrino was first
proposed.  Recently, again using nuclear reactors, KamLAND observed
disappearance of reactor antineutrinos at long baseline and distortion
in the energy spectrum, strengthening the evidence of neutrino
oscillation.  Furthermore, as discussed in Section~\ref{sec:synergy},
reactor-based antineutrino experiments have the potential of uniquely
determining $\theta_{13}$ at a low cost and in a timely fashion.
  
In this section we summarize the important features of nuclear
reactors which are crucial to reactor-based antineutrino experiments.

\subsubsection{Energy Spectrum and Flux of Reactor Antineutrinos}
\label{sssec:physics_flux}

A nuclear power plant derives its power from the fission of uranium
and plutonium isotopes (mostly $^{235}$U and $^{239}$Pu) which are
embedded in the fuel rods in the reactor core. The fission produces
daughters, many of which beta decay because they are neutron-rich.
Each fission on average releases approximately 200~MeV of energy and
six antineutrinos.  The majority of the antineutrinos have very low
energies; about 75\% are below 1.8~MeV, the threshold of the inverse
beta-decay reaction that will be discussed in
Section~\ref{sssec:physics_ibd}. A typical reactor with 3~GW of
thermal power (3~GW$_{\rm th}$) emits $6\times 10^{20}$ antineutrinos
per second with antineutrino energies up to 8~MeV.

Many reactor antineutrino experiments to date have been carried out at
pressurized water reactors (PWRs).  The antineutrino flux and energy
spectrum of a PWR depend on several factors: the total thermal power
of the reactor, the fraction of each fissile isotopes in the fuel, the
fission rate of each fissile isotope, and the energy spectrum of
antineutrinos of the individual fissile isotopes.

The antineutrino yield is directly proportional to the thermal power that is
estimated by measuring the temperature, pressure and the
flow rate of the cooling water.
The reactor thermal power is measured
continuously by the power plant with a typical precision of
about 1\%.

\begin{figure}[!htb]
\begin{center}
\includegraphics[height=0.33\textwidth]{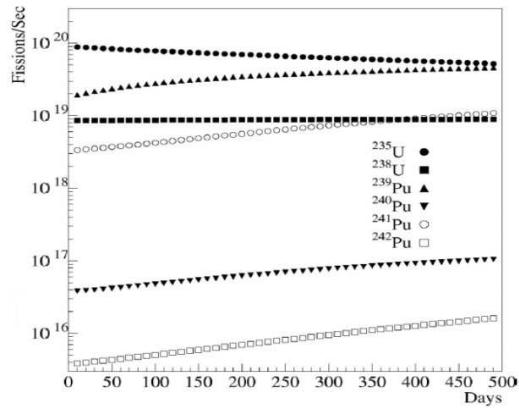}
\caption[Fission rate of reactor isotopes as a function of time.]{Fission rate of reactor isotopes as a function of time from
a Monte Carlo simulation~\cite{3miller}.} \label{fig:fisrate}
\end{center}
\end{figure}

Fissile materials are continuously consumed while new fissile isotopes
are produced from other fissionable isotopes in the fuel (mainly $^{238}$U) by
fast neutrons. Since the antineutrino energy spectra are slightly
different for the four main isotopes, $^{235}$U, $^{238}$U, $^{239}$Pu, and
$^{241}$Pu, the knowledge on the fission composition and its evolution
over time are therefore critical to the determination of the antineutrino
flux and energy spectrum. From the average thermal power and the
effective energy released per fission~\cite{3efission}, the average
number of fissions per second of each isotope can be calculated as a
function of time. Figure~\ref{fig:fisrate} shows the results of a
computer simulation of the Palo Verde reactor cores~\cite{3miller}.

It is common for a nuclear power plant to replace some of the fuel
rods in the core periodically as the fuel is used up. Typically, a
core will have 1/3 of its fuel changed every 18 months. At the
beginning of each refueling cycle, 69\% of the fissions are from
$^{235}$U, 21\% from $^{239}$Pu, 7\% from $^{238}$U, and 3\% from
$^{241}$Pu. During operation the fissile isotopes $^{239}$Pu and
$^{241}$Pu are produced continuously from $^{238}$U. Toward the end of the
fuel cycle, the fission rates from $^{235}$U and $^{239}$Pu are about
equal.
The average (``standard'') fuel composition is 58\% of
$^{235}$U, 30\% of $^{239}$Pu, 7\% of $^{238}$U, and 5\%
$^{241}$Pu~\cite{Kopeikin01}.

\begin{figure}[!htb]
\begin{minipage}[t]{0.45\textwidth}
\includegraphics[width=0.85\textwidth]{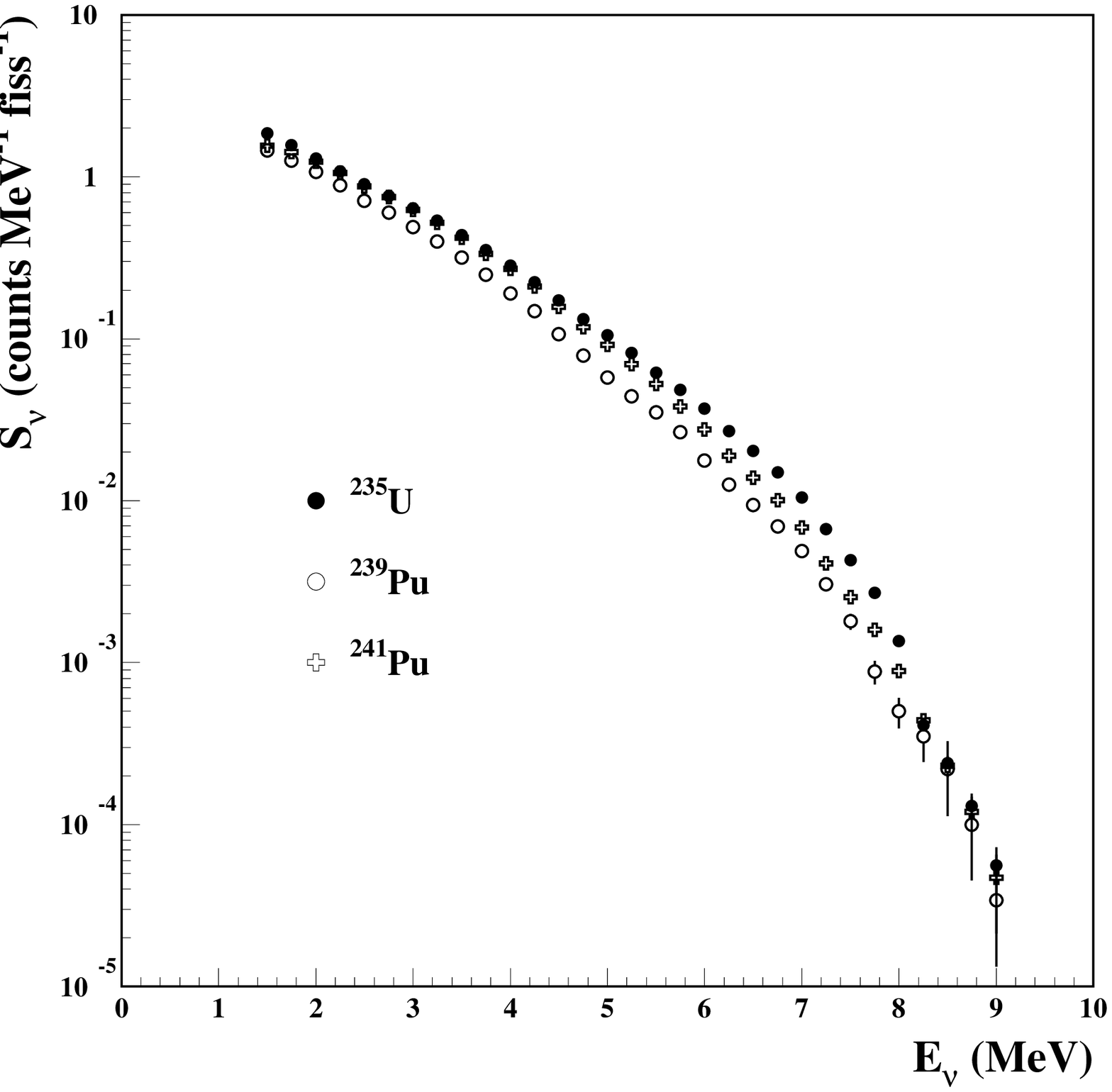}
\caption[Yield of antineutrinos per fission for the several isotopes.]{Yield of antineutrinos per fission for the several isotopes.
These are determined by converting the corresponding measured
$\beta$ spectra~\protect\cite{3illbeta}.} \label{fig:illyield}
\end{minipage}
\hfill
\begin{minipage}[t]{0.45\textwidth}
\includegraphics[width=0.85\textwidth]{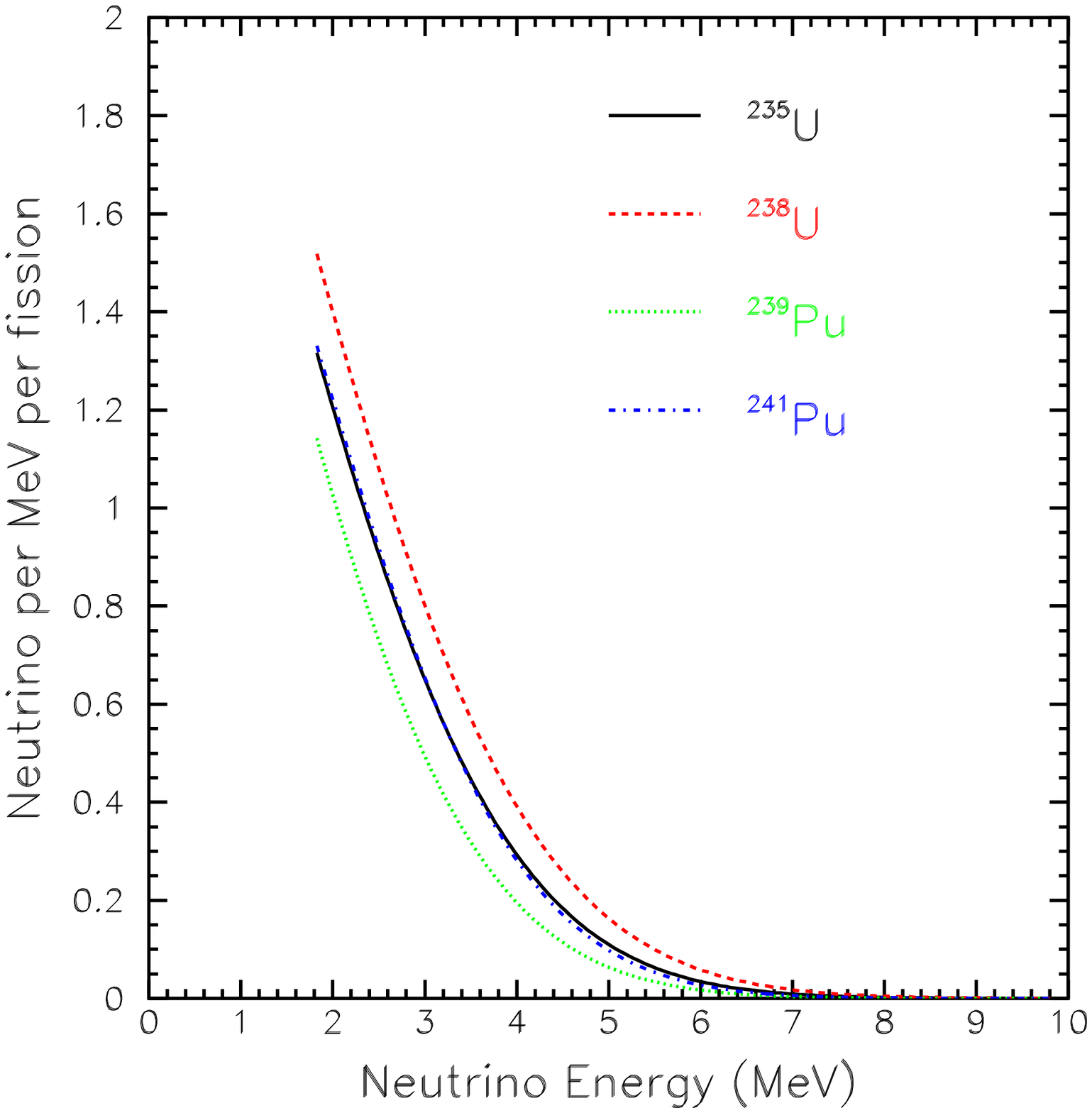}
\caption[Antineutrino energy spectrum for four isotopes.]{Antineutrino energy spectrum for four isotopes following
 the parameterization of Vogel and Engel~\protect\cite{3vogel}.}
\label{fig:nyth}
\end{minipage}
\end{figure}

In general, the composite antineutrino energy
spectrum is a function of the time-dependent contributions of the
various fissile isotopes to the fission process.
The Bugey 3 experiment compared three different models of the
antineutrino spectrum with its measurement~\cite{Bugey}. Good agreement was
observed with the model that made use of the $\bar\nu_e$ spectra derived from the
$\beta$ spectra~\cite{3illbeta} measured at the Institute Laue-Langevin (ILL).
The ILL derived spectra for isotopes
$^{235}$U, $^{239}$Pu, and $^{241}$Pu are shown in
Fig.~\ref{fig:illyield}.
However, there is no data for $^{238}$U; only the theoretical
prediction is used. The possible discrepancy between the predicted and
the real spectra should not lead to significant errors since the
contribution from $^{238}$U is never higher than 8\%.
The overall normalization uncertainty of the ILL
measured spectra is 1.9\%. A global shape uncertainty is also
introduced by the conversion procedure.

A widely used three-parameter parameterization of the antineutrino
spectrum for the four main isotopes, as shown in Fig.~\ref{fig:nyth},
can be found in~\cite{3vogel}.

\subsubsection{Inverse Beta-Decay Reaction}
\label{sssec:physics_ibd}

\begin{figure}[!htb]
 \begin{center}
 \includegraphics[height=7cm]{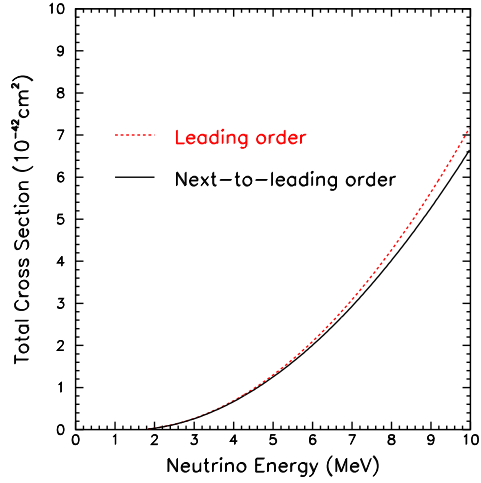}
 \caption{Total cross section for inverse beta-decay
calculated in leading order and next-to-leading order.}
 \label{fig:nloxsec}
 \end{center}
\end{figure}

The reaction employed to detect the $\bar \nu_e$ from a reactor is
the inverse beta-decay $\bar \nu_e + p \to e^+ + n$. The total
cross section of this reaction, neglecting terms of order
$E_\nu/M$,
where $E_\nu$ is the energy of the antineutrino and $M$ is the nucleon mass,
is
\begin{equation}
\sigma^{(0)}_{tot} = \sigma_0 (f^2 + 3g^2)(E^{(0)}_e p^{(0)}_e /1
{\rm MeV^2})
\end{equation}
\noindent where $E^{(0)}_e = E_\nu - (M_n - M_p)$ is the positron
energy when neutron recoil energy is neglected, and $p^{(0)}_e$ is
the positron momentum. The weak coupling constants are $f=1$ and
$g=1.26$, and $\sigma_0$ is related to the Fermi coupling constant
$G_F$, the Cabibbo angle $\theta_C$, and an energy-independent
inner radiative correction. The inverse beta-decay process has a threshold
energy in the laboratory frame $E_\nu=[(m_{\rm n}+m_{\rm
e})^2-m^2_{\rm p}]/2m_{\rm p}$ = 1.806~MeV. The leading-order
expression for the total cross section is
\begin{equation}
\sigma^{(0)}_{tot} = 0.0952 \times 10^{-42} {\rm cm}^2 (E^{(0)}_e
p^{(0)}_e/1 {\rm MeV^2})
\end{equation}
Vogel and Beacom~\cite{3vogel99} have recently extended the
calculation of the total cross section and
angular distribution to order $1/M$ for the inverse beta-decay reaction. Figure~\ref{fig:nloxsec}
shows the comparison of the total cross sections obtained in the
leading order and the next-to-leading order calculations.
Noticeable differences are present for high antineutrino energies. We
adopt the order $1/M$ formulae for describing the inverse beta-decay
reaction.  The calculated cross section can be related to the neutron
lifetime, whose uncertainty is only 0.2\%.

\begin{figure}[!hbt]
\begin{minipage}[t]{0.45\textwidth}
\includegraphics[width=0.85\textwidth]{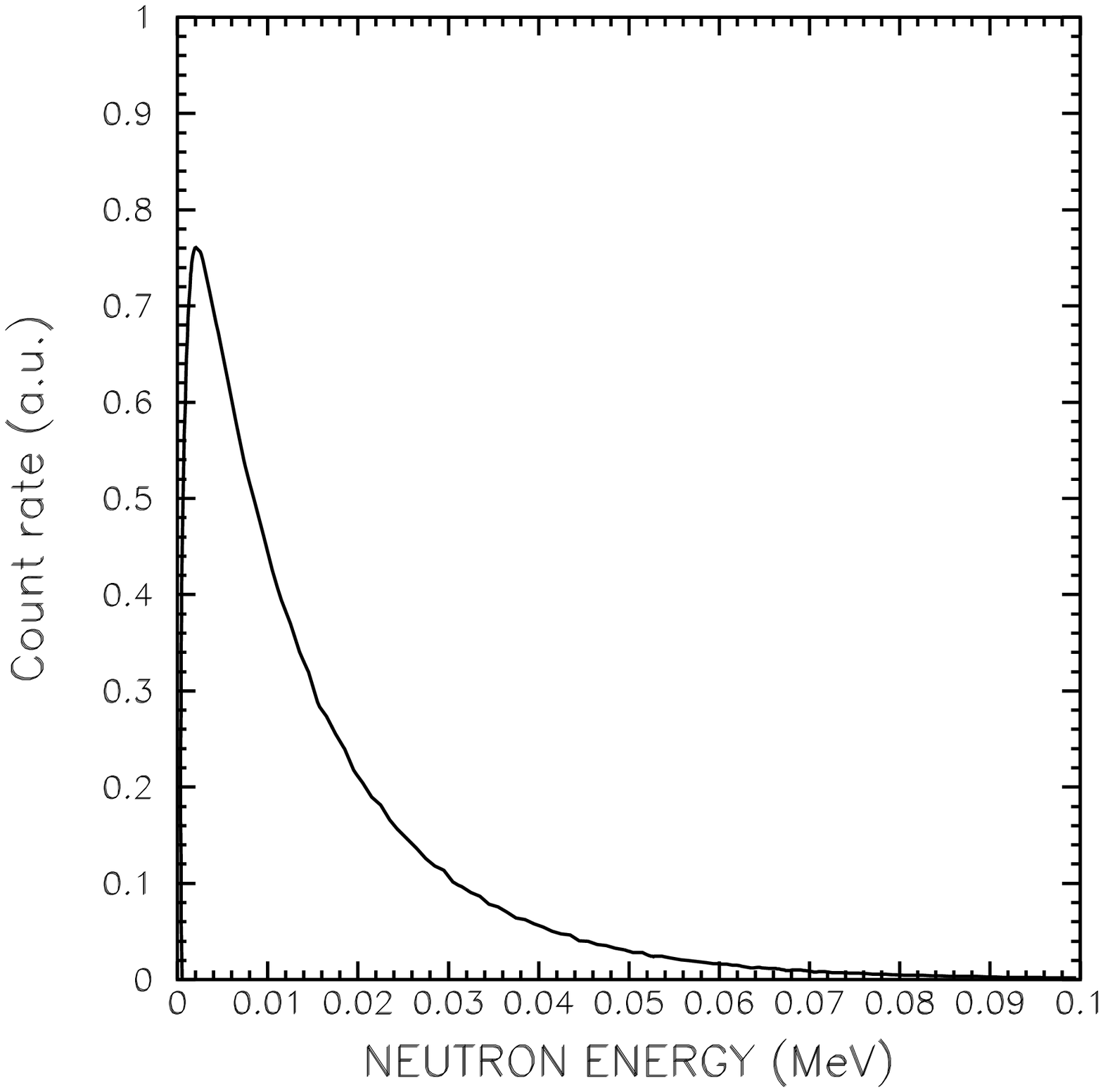}
\caption{Recoil neutron energy spectrum from inverse beta-decay
weighted by the antineutrino energy spectrum.} \label{fig:erecoil}
\end{minipage}
\hfill
\begin{minipage}[t]{0.45\textwidth}
\includegraphics[width=0.85\textwidth]{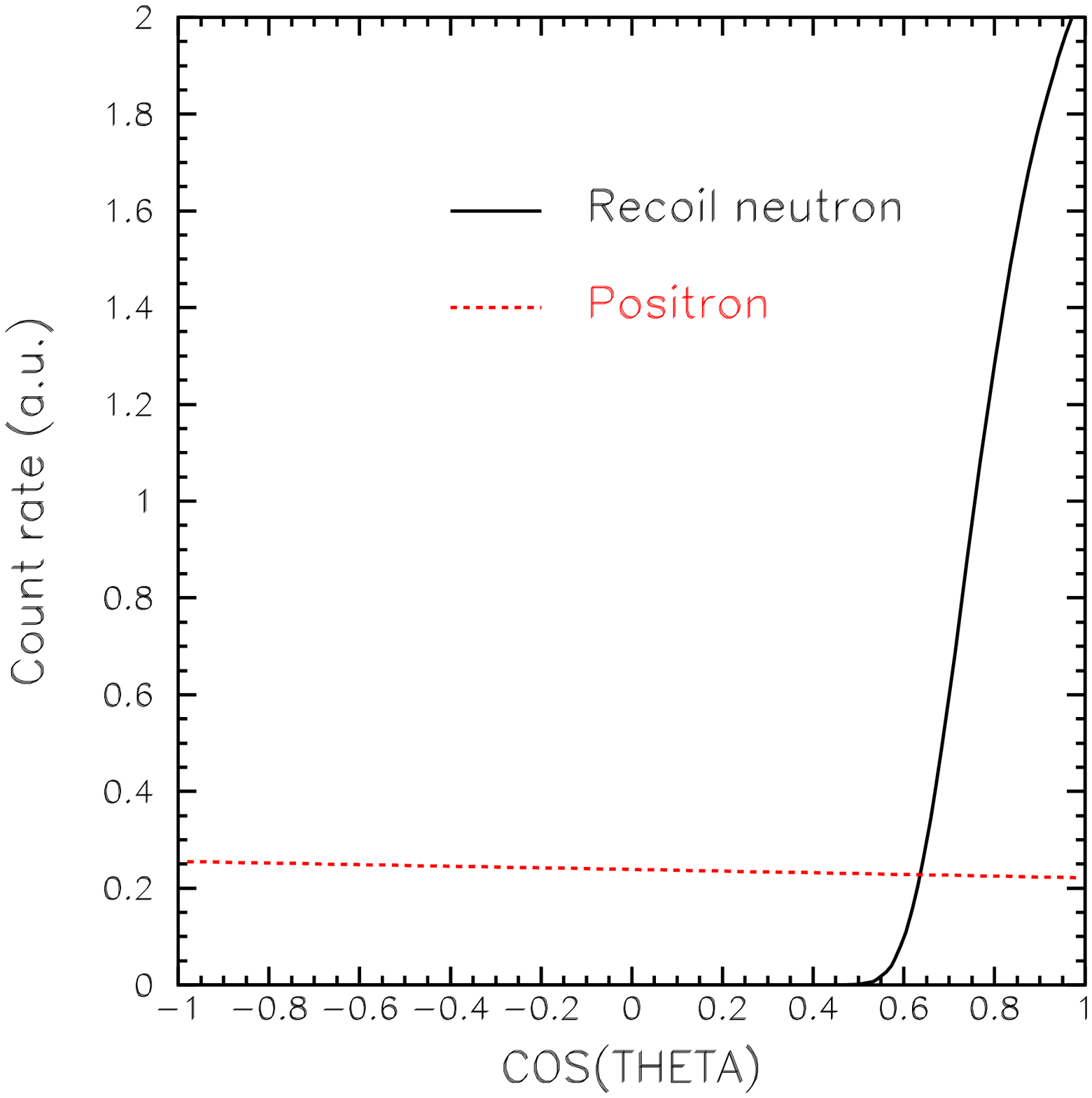}
\caption{Angular distributions of positrons and recoil neutrons
from inverse beta-decay in the laboratory frame.} \label{fig:angle}
\end{minipage}
\end{figure}

The expected recoil neutron energy spectrum, weighted by the
antineutrino energy spectrum and the $\bar \nu_e + p \to e^+ + n$
cross section, is shown in Fig.~\ref{fig:erecoil}.
Due to the low antineutrino energy relative to the mass of the
nucleon, the recoil neutron has low kinetic energy. While the positron
angular distribution is slightly backward peaked in the laboratory
frame, the angular distribution of the neutrons is strongly forward
peaked, as shown in Fig.~\ref{fig:angle}.

\subsubsection{Observed Antineutrino Rate and Spectrum at Short Distance}
\label{sssec:physics_predicted}

\begin{figure}
 \begin{center}
 \includegraphics[height=7cm]{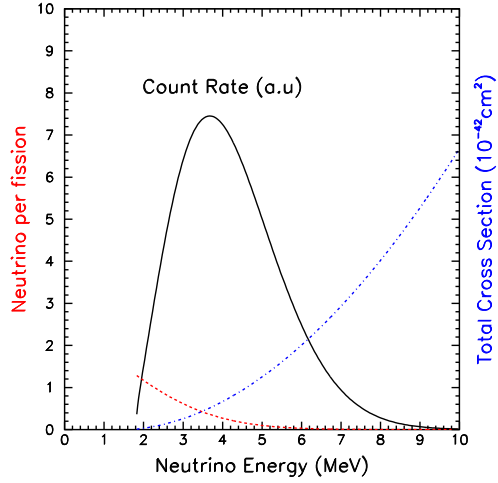}
 \caption{Antineutrino energy spectrum,
total inverse beta-decay cross section, and count rate as a
function of antineutrino energy.}
 \label{fig:bxsec}
 \end{center}
\end{figure}

The observed antineutrino spectrum in a liquid scintillator detector,
which is rich in free protons in the form of hydrogen, is a product of
the reactor antineutrino spectrum and the cross section of inverse
beta-decay.  Figure~\ref{fig:bxsec} shows the differential
antineutrino energy spectrum, the total cross section of the inverse
beta-decay reaction, and the expected count rate as a function of the
antineutrino energy. The differential energy distribution is the sum
of the antineutrino spectra of all the radio-isotopes in the fuel. It
is thus sensitive to the variation of thermal power and composition of
the nuclear fuel.

By integrating over the energy of the antineutrino, the number of
events can be determined.  With one-ton\footnote{Throughout this 
document we will use the term ton to refer to a metric ton of 1000~kg.}  
of liquid scintillator, a typical rate is about 100 antineutrinos per day 
per GW$_{\rm th}$ at 100~m from the reactor.  The highest count rate 
occurs at $E_\nu \sim 4$~MeV when there is no oscillation.

\subsubsection{Reactor Antineutrino Disappearance Experiments}
\label{sssec:physics_reactor_mixing}

In a reactor-based antineutrino experiment the measured quantity is
the survival probability for $\bar{\nu}_{\rm
e}\rightarrow\bar{\nu}_{\rm e}$ at a baseline of the order of hundreds
of meters to about a couple hundred kilometers with the
$\bar{\nu}_{\rm e}$ energy from about 1.8~MeV to 8~MeV.  The matter
effect is totally negligible and so the vacuum formula for the
survival probability is valid. In the notation of Eq.~\ref{MMatrix},
this probability has a simple expression
\begin{eqnarray}
P_{\rm sur}&=&1-C^4_{13}\sin^22\theta_{12}\sin^2\Delta_{21}
     -C^2_{12}\sin^22\theta_{13}\sin^2\Delta_{31}
     -S^2_{12}\sin^22\theta_{13}\sin^2\Delta_{32}
\label{Probab}
\end{eqnarray}

where
\begin{eqnarray}
\Delta_{\rm jk} &\equiv& 1.267 \Delta{m}^2_{\rm jk}({\rm eV^2})
                \times 10^3 {L ({\rm km})\over E ({\rm MeV})}  \label{eq:phase}\\
\Delta{m}^2_{\rm jk} &\equiv& m^2_{\rm j}-m^2_{\rm k} \nonumber
\end{eqnarray}
$L$ is the baseline in km, $E$ the antineutrino energy in MeV, and
$m_{\rm j}$ the $j$-th antineutrino mass in eV. The $\nu_{\rm
e}\rightarrow\nu_{\rm e}$ survival probability is also given by
Eq.~\ref{Probab} when $CPT$ is not violated.  Eq.~\ref{Probab} is
independent of the $CP$ phase angle $\delta_{CP}$ and the mixing angle
$\theta_{23}$.

To obtain the value of $\theta_{13}$, the depletion of $\bar{\nu}_{\rm
e}$ has to be extracted from the experimental $\bar\nu_{\rm e}$
disappearance probability, 
\begin{eqnarray}
P_{\rm dis} &\equiv& 1-P_{\rm sur} \nonumber \\
      &=& C^4_{13}\sin^22\theta_{12}\sin^2\Delta_{21}
          + C^2_{12}\sin^22\theta_{13}\sin^2\Delta_{31}
          + S^2_{12}\sin^22\theta_{13}\sin^2\Delta_{32} \,
\label{Pdis}
\end{eqnarray}
Since $\theta_{13}$ is known to be less than 10$^{\circ}$,
we define the term that is insensitive to $\theta_{13}$ as
\begin{equation}
P_{12} =  C^4_{13}\sin^22\theta_{12}\sin^2\Delta_{21} \approx  \sin^22\theta_{12}\sin^2\Delta_{21} \,
\label{P12}
\end{equation}
Then the part of the disappearance probability directly related to
$\theta_{13}$ is given by
\begin{eqnarray}
P_{13} &\equiv& P_{\rm dis}-P_{12}  \nonumber \\
      &=&           + C^2_{12}\sin^22\theta_{13}\sin^2\Delta_{31}
          + S^2_{12}\sin^22\theta_{13}\sin^2\Delta_{32}
\label{Pnet}
\end{eqnarray}

\begin{figure}[!htb]
\begin{center}
\includegraphics*[viewport=0 0 460 350, height=7cm]{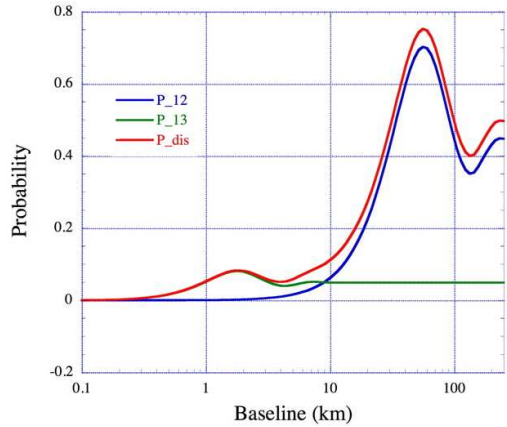}
\caption[Reactor antineutrino disappearance probability as a function of distance from the source.]{Reactor antineutrino disappearance probability as a function of distance from the source.
The values of the mixing parameters are given in
Eq.~\ref{eq:oscpar}. $P_{12}$ is the slowly rising blue
curve. $P_{13}$ is the green curve that has a maximum near 2 km.  The
total disappearance probability $P_{\rm dis}$ is the red curve.}
\label{fig:probs}
\end{center}
\end{figure}

The above discussion shows that in order to obtain $\theta_{13}$ we
have to subtract the $\theta_{13}$-insensitive contribution $P_{12}$
from the experimental measurement of $P_{\rm dis}$. To see their
individual effect, we plot $P_{13}$ in Fig.~\ref{fig:probs}
together with $P_{\rm dis}$ and $P_{12}$ as a function of the baseline
from 100~m to 250~km. The antineutrino energy is integrated from 1.8~MeV to
8~MeV.  We also take $\sin^22\theta_{13}=0.10$, which will be used for
illustration in most of the discussions in this section. The other
parameters are taken to be
\begin{equation}
\theta_{12} = 34^\circ,~~~\Delta{m}^2_{21} = 7.9 \times 10^{-5}
{\rm eV}^2,~~~\Delta{m}^2_{31} = 2.5 \times 10^{-3} {\rm eV}^2
\label{eq:oscpar}
\end{equation}

The behavior of the curves in Fig.~\ref{fig:probs} are quite clear
from their definitions, Eqs.~(\ref{Pdis}), (\ref{P12}), and
(\ref{Pnet}).  Below a couple kilometers $P_{12}$ is very small, and
$P_{13}$ and $P_{\rm dis}$ track each other well.  This suggests that
the measurement can be best performed at the first oscillation maximum
of $P_{13}({\rm max})\simeq \sin^22\theta_{13}$. Beyond the first
minimum $P_{13}$ and $P_{\rm dis}$ deviate from each other more and
more as $L$ increases when $P_{12}$ becomes dominant in $P_{\rm dis}$.

When we determine $P_{13}$({\rm max}) from the difference $P_{\rm
dis}-P_{12}$, the uncertainties on $\theta_{12}$ and $\Delta m^2_{21}$
will propagate to $P_{13}$.  It is easy to check that, given the best
fit values in Eq.~\ref{SNOsalt}, when $\sin^22\theta_{13}$ varies from
0.01 to 0.10 the relative size of $P_{12}$ compared to $P_{13}$ is
about 25\% to 2.6\% at the first oscillation maximum. Yet the
uncertainty in determining $\sin^22\theta_{13}$ due to the uncertainty
of $P_{12}$ is always less than 0.005.

\begin{figure}[!htb]
\begin{center}
\includegraphics[height=7cm]{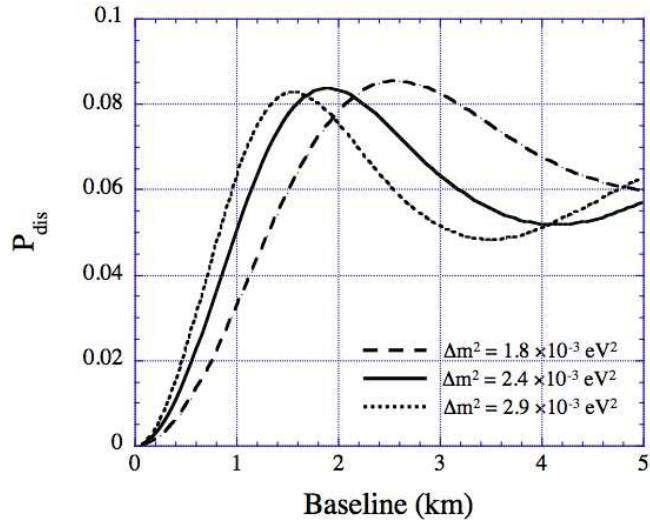}
\caption{Reactor antineutrino disappearance probability due to the mixing 
angle $\theta_{13}$ as a function of the baseline $L_{\rm
far}$ over the allowed 2$\sigma$ range in
$\Delta{m}^2_{32}$.} 
\label{fig:deltam32}
\end{center}
\end{figure}

In Fig.~\ref{fig:deltam32} $P_{13}$ integrated over $E$ from 1.8 to
8~MeV is shown as a function of $L_{\rm far}$, the best possible
distance for the detector, for three values of $\Delta{m}^2_{32}$ that
cover the allowed range of $\Delta{m}^2_{32}$ at 95\% C.L.  as given
in Eq.~\ref{eq:atmos}.
The curves show that $P_{dis}$ is sensitive to $\Delta{m}^2_{32}$.
For $\Delta{m}^2_{32}=(1.8,~2.4,~2.9)\times 10^{-3}$ eV$^2$, the
oscillation maxima correspond to a baseline of 2.5~km, 1.9~km, and
1.5~km, respectively.  From this simple study, placing the detector
between 1.5~km and 2.5~km from the reactor looks to be a good choice.

We conclude from this phenomenological investigation that the
choice of $L_{\rm far}$ be made so that it can cover as large a
range of $\Delta{m}^2_{31}$ as possible.

\subsection{Determining $\theta_{13}$ with Nuclear Reactors}

In this section previous attempts to measure $\sin^22\theta_{13}$
are summarized, and a new method for a high-precision
determination of $\sin^22\theta_{13}$ is presented.

\subsubsection{Past Measurements}
\label{sssec:physics_past}

In the 1990's, two reactor-based antineutrino experiments,
Chooz~\cite{Chooz} and Palo Verde~\cite{PaloVerde}, were carried out
to investigate neutrino oscillation.  Based on $\Delta{m}_{32}^2$ =
$1.5 \times 10^{-2}$ eV$^2$ as reported by Kamiokande~\cite{Kamioka},
the baselines of Chooz and Palo Verde were chosen to be about 1~km.
This distance corresponded to the location of the first oscillation
maximum of $\nu_e \to \nu_{\mu}$ when probed with low-energy reactor
$\bar{\nu}_e$.  Both Chooz and Palo Verde were looking for a deficit
in the $\bar{\nu}_e$ rate at the location of the detector by comparing
the observed rate with the calculated rate assuming no oscillation
occurred. With only one detector, both experiments had to rely on the
operational information from the reactors, in particular, the
composition of the nuclear fuel and the amount of thermal power
generated as a function of time, for calculating the rate of
$\bar{\nu}_e$ produced in the fission processes.

Chooz and Palo Verde utilized Gd-doped liquid scintillator (0.1\% Gd
by weight) to detect the reactor $\bar{\nu}_e$ via the inverse
beta-decay $\bar{\nu}_e + p \to n + e^+$ reaction. The ionization loss
and subsequent annihilation of the positron give rise to a fast signal
obtained with photomultiplier tubes (PMTs).  The energy associated
with this signal is termed the prompt energy, $E_{p}$.  As stated in
Section~\ref{sssec:physics_ibd}, $E_{p}$ is directly related to the
energy of the incident $\bar{\nu}_e$.  After a typical moderation time
of about 30~$\mu$s, the neutron is captured by a Gd
nucleus,\footnote{The cross section of neutron capture by a proton is
0.3~b and 50,000~b on Gd.} emitting several $\gamma$-ray photons with
a total energy of about 8~MeV. This signal is called the delayed
energy, $E_{d}$.  The temporal correlation between the prompt energy
and the delayed energy constitutes a powerful tool for identifying the
$\bar{\nu}_e$ and for suppressing backgrounds.

The value of $\sin^2{2\theta_{13}}$ was determined by comparing the
observed antineutrino rate and energy spectrum at the detector with
the predictions that assumed no oscillation.  The number of detected
antineutrinos $N_{det}$ is given by
\begin{eqnarray}
N_{\rm det}&=&\frac{N_p}{4\pi L^2}\int{\epsilon\sigma P_{\rm sur}S dE}
\label{eq:absolute}
\end{eqnarray}
where $N_p$ is the number of free protons in the target, $L$ is the
distance of the detector from the reactor, $\epsilon$ is the
efficiency of detecting an antineutrino, $\sigma$ is the total cross
section of the inverse beta-decay process, $P_{sur}$ is the survival
probability given in Eq.~\ref{Probab}, and $S$ is the differential
energy distribution of the antineutrino at the reactor shown in
Fig.~\ref{fig:bxsec}.

Since the signal rate is low, it is desirable to conduct reactor-based
antineutrino experiments underground to reduce cosmic-ray induced
backgrounds, such as neutrons and the radioactive isotope $^9$Li.  Gamma
rays originating from natural radioactivity in construction materials
and the surrounding rock are also problematic. The random coincidence
of a $\gamma$ ray interaction in the detector followed by a neutron
capture is a potential source of background.  For Chooz, their
background rate was $1.41 \pm 0.24$ events per day in the 1997 run,
and $2.22 \pm 0.14$ events per day after the trigger was modified in
1998.  The background events were subtracted from $N_{\rm det}$ before
$\sin^2{2\theta_{13}}$ was extracted.

The systematic uncertainties and efficiencies of
Chooz are summarized in Tables~\ref{tab:Choozerr}
\begin{table}[!htb]
\begin{center}
\begin{tabular}{|l||c|} \hline
 parameter  & relative uncertainty (\%) \\ \hline\hline
 reaction cross section      & 1.9  \\
 number of protons           & 0.8  \\
 detection efficiency        & 1.5  \\
 reactor power               & 0.7  \\
 energy released per fission & 0.6  \\ \hline\hline
 combined                    & 2.7  \\ \hline
\end{tabular}
 \caption{Contributions to the overall systematic uncertainty in
the absolute normalization of Chooz.}
 \label{tab:Choozerr}
 \end{center}
\end{table}
and~\ref{tab:Choozeff} respectively.
\begin{table}[!htb]
\begin{center}
\begin{tabular}{|l||c|c|}
 \hline
 selection & $\epsilon$ (\%) & relative error (\%) \\ \hline\hline
 positron energy           & 97.8 & 0.8 \\
 positron-geode distance   & 99.9 & 0.1 \\
 neutron capture           & 84.6 & 1.0 \\
 capture energy containment& 94.6 & 0.4 \\
 neutron-geode distance    & 99.5 & 0.1 \\
 neutron delay             & 93.7 & 0.4  \\
 positron-neutron distance & 98.4 & 0.3  \\
 neutron multiplicity      & 97.4 & 0.5 \\ \hline\hline
 combined                  & 69.8 & 1.5  \\ \hline
\end{tabular}
 \caption{Summary of the antineutrino detection efficiency in Chooz.}
 \label{tab:Choozeff}
\end{center}
\end{table}

\begin{figure}[!htb]
\begin{center}
\includegraphics[height=7cm]{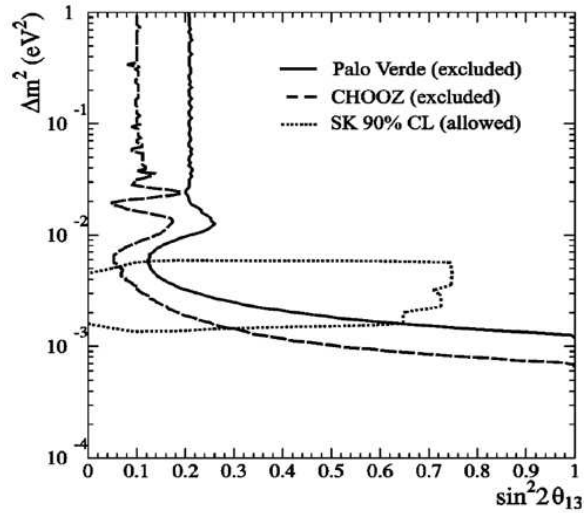}
\caption[Exclusion contours determined by Chooz, Palo Verde along with the
allowed region obtained by Kamiokande.] {Exclusion contours determined by Chooz, 
Palo Verde along with the allowed region obtained by Kamiokande.~\cite{Chooz}}
\label{fig:limits}
\end{center}
\end{figure}

Neither Chooz nor Palo Verde observed any deficit in the
$\bar{\nu}_e$ rate. This null result was used to set a limit on the neutrino
mixing angle $\theta_{13}$, as shown in Fig.~\ref{fig:limits}.
Chooz obtained the best limit of 0.17 in sin$^{2}2\theta_{13}$ for
$\Delta{m}^2_{31} = 2.5 \times 10^{-3}$ eV$^2$ at the 90\%
confidence level.

\subsubsection{Precision Measurement of $\theta_{13}$}
With only one detector at a fixed baseline from a reactor, according
to Eq.~\ref{eq:absolute}, Chooz and Palo Verde had to determine the
absolute antineutrino flux from the reactor, the absolute cross
section of the inverse beta-decay reaction, and the efficiencies of
the detector and event-selection requirements in order to measure
$\sin^22\theta_{13}$.  The prospect for determining
$\sin^22\theta_{13}$ precisely with a single detector that relies on a
thorough understanding of the detector and the reactor is not
promising.  It is a challenge to reduce the systematic uncertainties
of such an absolute measurement to sub-percent level, especially for
reactor-related uncertainties.

Mikaelyan and Sinev pointed out that the systematic uncertainties can
be greatly suppressed or totally eliminated when two detectors
positioned at two different baselines are utilized~\cite{Russian}.
The near detector close to the reactor core is used to establish the
flux and energy spectrum of the antineutrinos. This relaxes the
requirement of knowing the details of the fission process and
operational conditions of the reactor.  In this approach, the value of
$\sin^22\theta_{13}$ can be measured by comparing the antineutrino
flux and energy distribution observed with the far detector to those
of the near detector after scaling with distance squared.  According
to Eq.~\ref{eq:absolute}, the ratio of the number of antineutrino
events with energy between $E$ and $E+dE$ detected at distance $L_{\rm
f}$ to that at a baseline of $L_{\rm n}$ is given by
\begin{eqnarray}
\frac{N_{\rm f}}{N_{\rm n}}&=&
\left (\frac{N_{\rm p,f}}{N_{\rm p,n}}\right )
\left (\frac{L_{\rm n}}{L_{\rm f}}\right)^2
\left (\frac{\epsilon_{\rm f}}{\epsilon_{\rm n}}\right )
\left [\frac{P_{\rm sur}(E,L_{\rm f})}{P_{\rm sur}(E,L_{\rm n})}\right]
\end{eqnarray}

If the detectors are made identical and have the same efficiency, the ratio
depends only on the distances and the survival probabilities. 
By placing the near detector close to the core such that there is no significant
oscillating effect, $\sin^22\theta_{13}$ is approximately given by
\begin{eqnarray}
\sin^22\theta_{13}&\approx&\frac{1}{A(E,L_{\rm f})}\left[1-\left(\frac{N_{\rm f}}{N_{\rm n}}\right)
\left (\frac{L_{\rm f}}{L_{\rm n}}\right)^2\right]
\end{eqnarray}
where $A(E,L_{\rm f})$ = $\sin^2\Delta_{31}$ with $\Delta_{31}$
defined in Eq.~\ref{eq:phase} is the analyzing power when the
contribution of $\theta_{12}$ is small.  Indeed, from this simplified
picture, it is clear that the two-detector scheme is an excellent
approach for pin-pointing the value of $\sin^22\theta_{13}$ precisely.
In practice, we need to extend this idea to handle more complicated
arrangements involving multiple reactors and multiple detectors as in
the case of the Daya Bay experiment.

\subsection{The Daya Bay Reactor Antineutrino Experiment}
\label{sssec:physics_DB}

As discussed in Section~\ref{sssec:physics_reactor_mixing}, probing
$\sin^22\theta_{13}$ with excellent sensitivity will be an important
milestone in advancing neutrino physics.  There are proposals to
explore $\sin^22\theta_{13}$ with sensitivities approaching the level
of 0.01~\cite{E-WhitePaper}.  The objective of the Daya Bay experiment
is to determine $\sin^22\theta_{13}$ with sensitivity of 0.01 or
better.

In order to reach the designed goal, it is important to reduce both
the statistical and systematic uncertainties as well as suppress
backgrounds to comparable levels in the Daya Bay neutrino oscillation
experiment.

\begin{figure}[!htb]
\begin{center}
\includegraphics[width=0.8\textwidth]{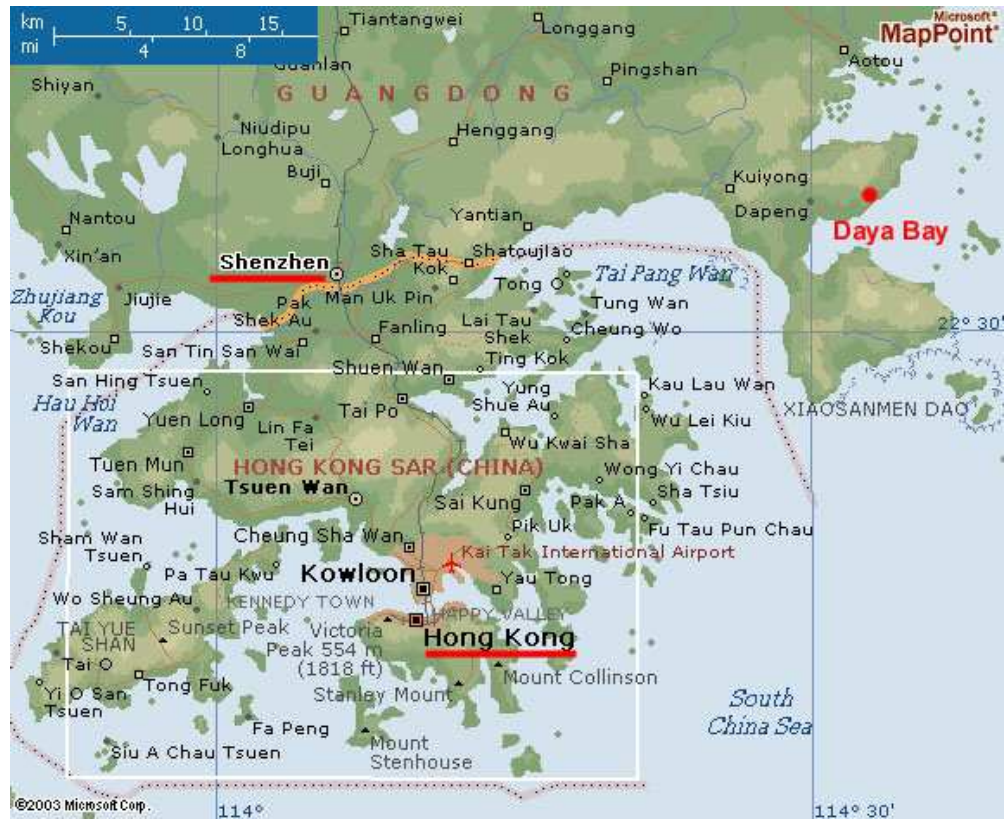}
\caption{Daya Bay and vicinity: The nuclear power complex is
located about 55~km from central Hong Kong.}
\label{fig:DayaVicinity}
\end{center}
\end{figure}

\begin{figure}[!htb]
\begin{center}
\includegraphics[width=0.3\textwidth,angle=-90]{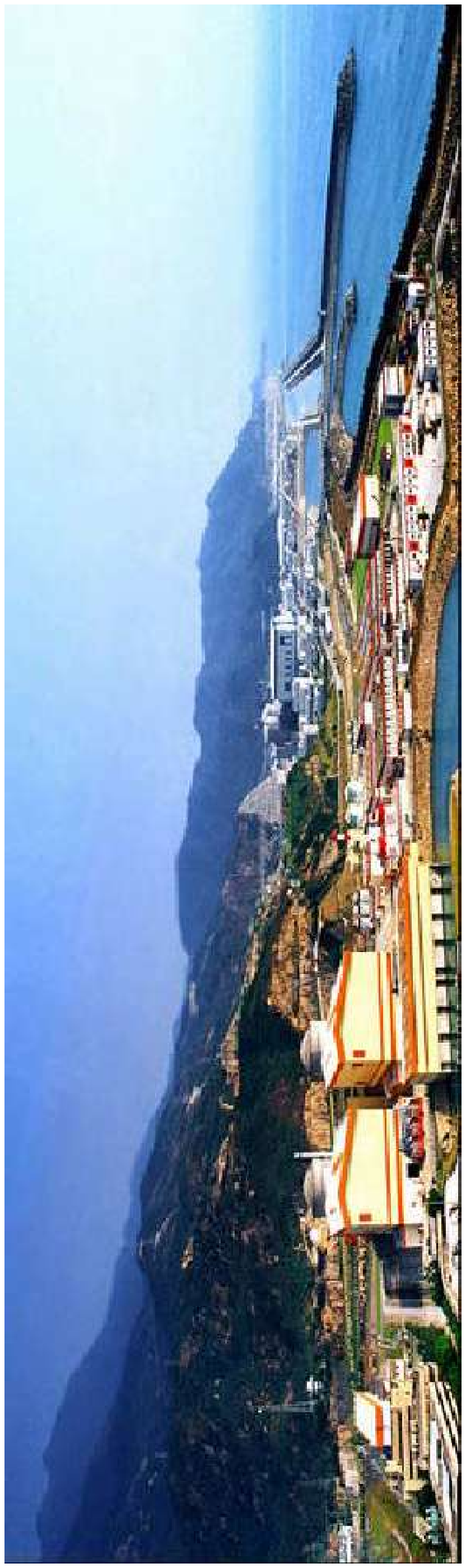}
\caption{The Daya Bay nuclear power complex. The Daya Bay nuclear
power plant is in the foreground. The Ling Ao nuclear power plant is
in the background. The experimental halls 
will be underneath the hills to the left.}
\label{fig:DayaNPP}
\end{center}
\end{figure}

This experiment will be located at the Daya Bay nuclear power complex
in southern China. Its geographic location is shown in
Fig.~\ref{fig:DayaVicinity}.
The experimental site is about 55~km north-east from Victoria Harbor
in Hong Kong.  Figure~\ref{fig:DayaNPP}
is a photograph of the complex. The complex consists of three nuclear
power plants (NPPs): the Daya Bay NPP, the Ling Ao NPP, and the Ling
Ao II NPP. The Ling Ao II NPP is under construction and will be
operational by 2010--2011. Each plant has two identical reactor
cores. Each core generates 2.9~GW$_{\rm th}$ during normal
operation. The Ling Ao cores are about 1.1~km east of the Daya Bay
cores, and about 400~m west of the Ling Ao II cores.  There are
mountain ranges to the north which provide sufficient overburden to
suppress cosmogenic backgrounds in the underground experimental
halls. Within 2~km of the site the elevation of the mountain varies
generally from 185~m to 400~m.

The six cores can be roughly grouped into two clusters, the Daya Bay
cluster of two cores and the Ling Ao cluster of four cores. We plan to
deploy two identical sets of near detectors at distances between 300~m
and 500~m from their respective cluster of cores to monitor the
antineutrino fluxes. Another set of identical detectors, called the
far detectors, will be located approximately 1.5~km north of the two
near detector sets. Since the overburden of the experimental site
increases with distance from the cores, the cosmogenic background
decreases as the signal decreases, hence keeping the
background-to-signal ratio roughly constant.  This is beneficial to
controlling systematic uncertainties. By comparing the antineutrino
fluxes and energy spectra between the near and far detectors, the Daya
Bay experiment will determine $\sin^22\theta_{13}$ to the designed
sensitivity.  Detailed design of the experiment, including baseline
optimization accounting for statistical and systematic uncertainty,
backgrounds and topographical information will be discussed in the
following chapters.

It is possible to instrument a mid detector site between the near and
far sites. The mid detectors along with the near and far detectors can
be used to carry out measurements for systematic studies and for
internal consistency checks. In combination with the near detectors
close to the Daya Bay NPP, they could also be utilized to provide a
quick determination of $\sin^2{2\theta_{13}}$, albeit with reduced
sensitivity, in the early stage of the experiment.

\newpage
\renewcommand{\thesection}{\arabic{section}}
\setcounter{figure}{0}
\setcounter{table}{0}
\setcounter{footnote}{0}

\section{Experimental Design Overview}
\label{sec:over}

To establish the existence of neutrino oscillation due to
$\theta_{13}$, and to determine sin$^{2}2\theta_{13}$ to a precision
of 0.01 or better, at least 50,000 detected events detected at the far
site are needed, and systematic uncertainties in the ratios of
near-to-far detector acceptance, antineutrino flux and background have to
be controlled to a level below 0.5\%, an improvement of almost an
order of magnitude over the previous experiments.  Based on recent
single detector reactor experiments such as Chooz, Palo Verde and
KamLAND, there are three main sources of systematic uncertainty:
reactor-related uncertainty of (2--3)\%, background-related
uncertainty of (1--3)\%, and detector-related uncertainty of
(1--3)\%. Each source of uncertainty can be further classified into
correlated and uncorrelated uncertainties.  Hence a carefully designed
experiment, including the detector design and background control, is
required.  The primary considerations driving the improved performance
are listed below:
\begin{itemize}

\item {\bf identical near and far detectors}
Use of identical detectors at the near and far sites to cancel
reactor-related systematic uncertainties, a technique first proposed
by Mikaelyan et al. for the Kr2Det experiment in
1999~\cite{3mikaelyan}.  The event rate of the near detector will be
used to predict the yield at the far detector. Even in the case of a
multiple reactor complex, reactor-related uncertainties can be
controlled to negligible level by a careful choice of locations of the
near and far sites.

\item {\bf multiple modules}
Employ multiple, identical modules at the near and far sites to cross
check between modules at each location and reduce detector-related
uncorrelated uncertainties. The use of multiple modules in each site
enables internal consistency check to the limit of
statistics. Multiple modules implies smaller detectors which are
easier to move.  In addition, small modules intercept fewer cosmic-ray
muons, resulting in less dead time, less cosmogenic background and
hence smaller systematic uncertainty.  Taking calibration and
monitoring of detectors, redundancy, and cost into account we settle
on a design with two modules for each near site and four modules for
the far site.

\item {\bf three-zone detector module}
Each module is partitioned into three enclosed zones. The innermost
zone, filled with Gd-loaded liquid scintillator, is the antineutrino
target which is surrounded by a zone filled with unloaded liquid
scintillator called the $\gamma$-catcher.  This middle zone is used to
capture the $\gamma$ rays from antineutrino events that leak out of the
target. This arrangement can substantially reduce the systematic
uncertainties related to the target volume and mass, positron energy
threshold, and position cut. The outermost zone, filled with
transparent mineral oil that does not scintillate, shields against
external $\gamma$ rays entering the active scintillator volume.

\item {\bf sufficient overburden and shielding}
Locations of all underground detector halls are optimized to ensure
sufficient overburden to reduce cosmogenic backgrounds to the level
that can be measured with certainty.  The antineutrino detector
modules are enclosed with sufficient passive shielding to attenuate
natural radiation and energetic spallation neutrons from the
surrounding rocks and materials used in the experiment.

\item {\bf multiple muon detectors}
By tagging the incident muons, the associated cosmogenic background
events can be suppressed to a negligible level. This will require the
muon detector to have a high efficiency and that it is known with
small uncertainty.  Monte Carlo study shows that the efficiency of the
muon detector should be better than 99.5\% (with an uncertainty less
than 0.25\%). The muon system is designed to have at least two
detector systems.  One system utilizes the water buffer as a Cherenkov
detector, and another employs muon tracking detectors with decent position
resolution. Each muon detector can easily be constructed with an
efficiency of (90--95)\% such that the overall efficiency of the muon
system will be better than 99.5\%.  In addition, the two muon
detectors can be used to measure the efficiency of each other to a
uncertainty of better than 0.25\%.

\item {\bf movable detectors}
The detector modules are movable, such that swapping of modules between
the near and far sites can be used to cancel detector-related
uncertainties to the extent that they remain unchanged before and
after swapping. The residual uncertainties, being secondary, are
caused by energy scale uncertainties not completely taken out by
calibration, as well as other site-dependent uncertainties.  The goal
is to reduce the systematic uncertainties as much as possible by
careful design and construction of detector modules such that
swapping of detectors is not necessary.  Further discussion of
detector swapping will be given in Chapters~\ref{sec:sys}
and~\ref{sec:ops}.

\end{itemize}

\par
With these improvements, the total detector-related systematic
uncertainty is expected to be $\sim$0.2\% in the near-to-far ratio per
detector site which is comparable to the statistical uncertainty of
$\sim$0.2\% (based on the expected 250,000 events for three years of
running at the far site). Using a global $\chi^2$ analysis (see
Section~\ref{sssec:sys_chi}), incorporating all known systematic and
statistical uncertainties, we find that $\sin^22\theta_{13}$ can be
determined to a sensitivity of better than 0.01 at 90\% confidence
level as discussed in Sec.~\ref{sssec:sys_sensitivity}.

\subsection{Experimental layout}
\label{ssec:exp_layout}

\par
Taking the current value of $\Delta m^2_{31}=2.5\times 10^{-3}$ eV$^2$
(see equation~\ref{eq:oscpar}), the first maximum of the oscillation
associated with $\theta_{13}$ occurs at $\sim$1800~m.  Considerations
based on statistics alone will result in a somewhat shorter baseline,
especially when the statistical uncertainty is larger than or comparable to
the systematic uncertainty. For the Daya Bay experiment, the
overburden influences the optimization since it varies along the
baseline. In addition, a shorter tunnel will decrease the civil
construction cost.

\par
\begin{figure}[!htb]
\begin{center}
\includegraphics[width=0.5\textwidth]{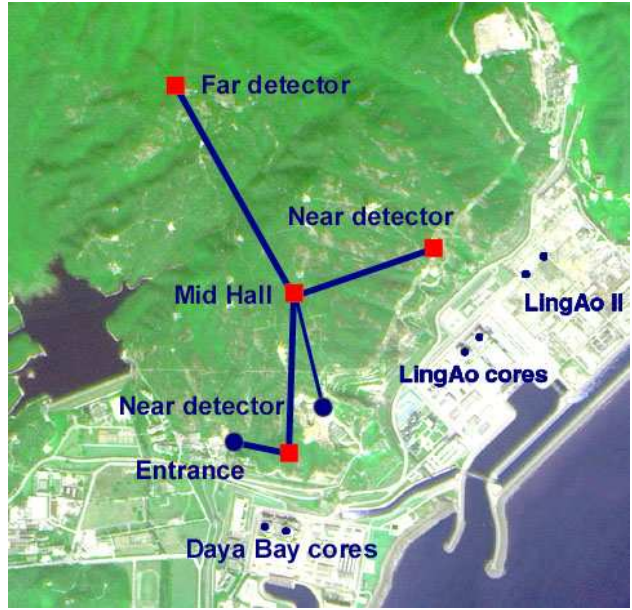}
\caption{Layout of the Daya Bay experiment.}
  \label{fig:eng_bw}
\end{center}
\end{figure}

The Daya Bay experiment will use identical detectors at the near and
far sites to cancel reactor-related systematic uncertainties, as well
as part of the detector-related systematic uncertainties. 
The Daya Bay site
currently has four cores in two groups: the Daya Bay 
NPP and the Ling Ao NPP. The two Ling Ao II cores will 
start to generate
electricity in 2010--2011.  Figure~\ref{fig:eng_bw} shows the
locations of all six cores.
The distance between the two cores in each NPP is about 88~m. Daya Bay
is 1100~m from Ling Ao, and the maximum distance between cores will be
1600~m when Ling Ao II starts operation. The experiment will locate
detectors close to each reactor cluster to monitor the antineutrinos
emitted from their cores as precisely as possible. At least two near
sites are needed, one is primarily for monitoring the Daya Bay cores
and the other primarily for monitoring the Ling Ao---Ling Ao II
cores. The reactor-related systematic uncertainties can not be
cancelled exactly, but can be reduced to a negligible revel, as low as
0.04\% if the overburden is not taken into account. A global
optimization taking all factors into account, especially balancing the
overburden and reactor-related uncertainties, results in a residual
reactor uncertainty of $<$0.1\%

\par
Three major factors are involved in optimizing the locations of the
near sites.  The first one is overburden. The slope of the hills near
the site is around 30 degrees.  Hence, the overburden falls rapidly as
the detector site is moved closer to the cores.  The second concern is
oscillation loss. The oscillation probability is appreciable even at
the near sites. For example, for the near detectors placed
approximately 500~m from the center of gravity of the cores, the
integrated oscillation probability is $0.19\times\sin^22\theta_{13}$
(computed with $\Delta m_{31}^2=2.5\times 10^{-3}$ eV$^2$). The
oscillation contribution of the other pair of cores, which is around
1100~m away, has been included. The third concern is the near-far
cancellation of reactor uncertainties.

\par
\begin{figure}[!htbp]
  \centering
  \includegraphics[height=6cm]{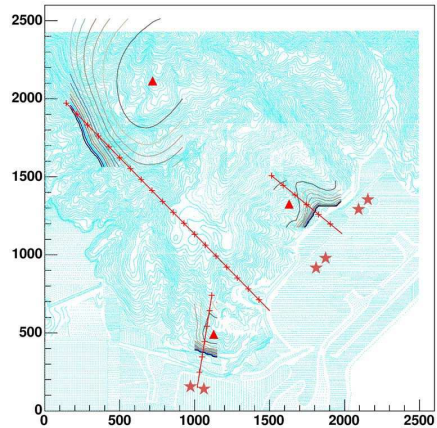}
  \caption{Site optimization using the global $\chi^2$ analysis. The
   optimal sites are labelled with red triangles. The stars show the
   reactors. The black contours show the sensitivity when one site's location is varied
   and the other two are fixed at optimal sites. The red lines with
   tick marks are the perpendicular bisectors of the reactor pairs. The
   mountain contours are also shown on the plot (blue lines).}
  \label{fig:optimal}
\end{figure}

After careful study of many different experimental designs, the best
configuration of the experiment is shown in Fig.~\ref{fig:eng_bw}
together with the tunnel layout.  Based on this configuration, a
global $\chi^2$ fit (see Eq.~\ref{eqn:chispec}) for the best
sensitivity and baseline optimization was performed, taking into
account backgrounds, mountain profile, detector systematics and
residual reactor related uncertainties.  The result is shown in
Fig.~\ref{fig:optimal}.

\par
Ideally each near detector site should be positioned equidistant from
the cores that it monitors so that the uncorrelated reactor
uncertainties are cancelled.  However, taking overburden and
statistics into account while optimizing the experimental sensitivity,
the Daya Bay near detector site is best located 363~m from the center
of the Daya Bay cores.  The overburden at this location is 98~m
(255~m.w.e.).\footnote{The Daya Bay near detector site is about 40~m
east of the perpendicular bisector of the Daya Bay two cores to gain
more overburden.} The Ling Ao near detector hall is optimized to be
481~m from the center of the Ling Ao cores, and 526~m from the center
of the Ling Ao II cores\footnote{The Ling Ao near detector site is
about 50~m west of the perpendicular bisector of the Ling Ao-Ling Ao
II clusters to avoid installing it in a valley which is likely to be
geologically weak, and to gain more overburden.} where the overburden
is 112~m (291~m.w.e).

\par
The far detector site is about 1.5~km north of the two near
sites. Ideally the far site should be equidistant between the Daya Bay
and Ling Ao---Ling Ao II cores; however, the overburden at that
location would be only 200~m (520~m.w.e). At present, the distances
from the far detector to the midpoint of the Daya Bay cores and to the
mid point of the Ling Ao---Ling Ao II cores are 1985~m and 1615~m,
respectively. The overburden is about 350~m (910~m.w.e). A summary of
the distances to each detector is provided in Table~\ref{tab:distances}.
\begin{table}[!htb]
\begin{center}
\begin{tabular}[c]{|l||r|r|r|} \hline
 Sites       &  DYB &  LA  & Far  \\ \hline\hline
DYB cores    &  363 & 1347 & 1985 \\ \hline
LA  cores    &  857 &  481 & 1618 \\ \hline
LA II  cores & 1307 &  526 & 1613 \\ \hline
\end{tabular}
\caption{Distances in meters from each detector site to the centroid of each pair of
reactor cores. \label{tab:distances}}
\end{center}
\end{table}

\par
It is possible to install a mid detector hall between the near and far
sites such that it is 1156~m from the midpoint of the Daya Bay cores
and 873~m from the center of the Ling Ao---Ling Ao II cores. The
overburden at the mid hall is 208~m (540~m.w.e.). This mid hall could
be used for a quick measurement of $\sin^22\theta_{13}$, studies of
systematics and internal consistency checks.

\par
There are three branches for the main tunnel extending from a junction
near the mid hall to the near and far underground detector
halls. There are also access and construction tunnels. The length of
the access tunnel, from the portal to the Daya Bay near site, is
295~m. It has a grade between 8\% and 12\%~\cite{3BINEReport}. A
sloped access tunnel allows the underground facilities to be located
deeper with more overburden. The quoted overburdens are based on a
10\% grade.

\subsection{Detector Design}
\label{ssec:exp_detector}

\begin{figure}[!htb]
\begin{center}
\includegraphics[width=0.7\textwidth]{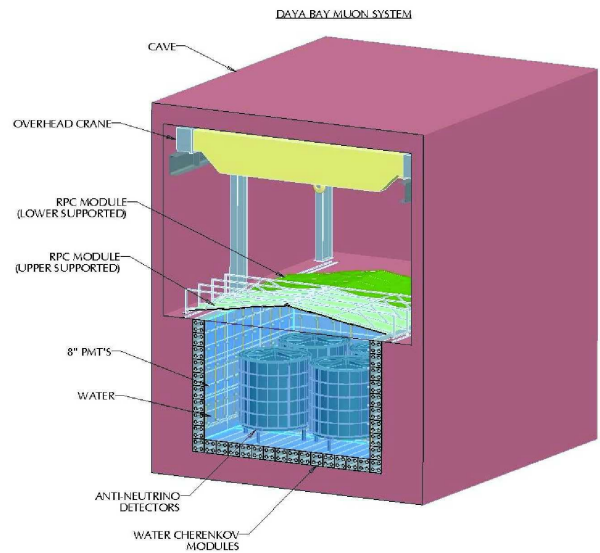}
\caption{Layout of the baseline design of the Daya Bay detector.
Four antineutrino detector modules are shielded by a 1.5~m-thick active
water Cherenkov  buffer. Surrounding this buffer is another 1-meter of water Cherenkov tanks serving as
muon trackers. The muon system is completed with RPCs at the top.}
 \label{fig:detector}
 \end{center}
 \end{figure}

As discussed above, the antineutrino detector employed at the near
(far) site has two (four) modules while the muon detector consists of a
cosmic-ray tracking device and active water buffer. There are
several possible configurations for the water buffer and the muon
tracking detector as discussed in Section~\ref{sec:muon}. The 
baseline design is shown in Fig.~\ref{fig:detector}.

The water buffer in this case is a water pool, equipped with
photomultiplier tubes (PMTs) to serve as a Cherenkov detector.  The
outer region of the water pool is segmented into water tanks made of
reflective PVC sheets with a cross section of 1~m$\times$1~m and a
length of 16~m. Four PMTs at each end of the water tank are installed
to collect Cherenkov photons produced by cosmic muons in the water
tank. The water tank scheme first proposed by
Y.F.~Wang~\cite{3water_tank} for very long baseline neutrino
experiments as a segmented calorimeter is a reasonable choice as a
muon tracking detector for reasons of both cost and technical
feasibility.  Above the pool the muon tracking detector is made of
light-weight resistive-plate chambers (RPCs). RPCs offer good
performance and excellent position resolution for low cost.

The antineutrino detector modules are submerged in the water pool that
shields the modules from ambient radiation and spallation neutrons.
Other possible water shielding configurations will be discussed in
Section~\ref{sec:designs}.

\subsubsection{Antineutrino detector}

Antineutrinos are detected by an organic liquid scintillator (LS) with
high hydrogen content (free protons) 
via the inverse beta-decay reaction:
$$\bar\nu_{e} + p \longrightarrow e^+ + n$$ The prompt positron signal
and delayed neutron-capture signal are combined to define a neutrino
event with timing and energy requirements on both signals.  In LS
neutrons are captured by free protons in the scintillator emitting
2.2~MeV $\gamma$-rays with a capture time of 180~$\mu$s. On the other
hand, when Gadolinium (Gd), with its huge neutron-capture cross
section and subsequent 8~MeV release of $\gamma$-ray energy, is loaded
into LS the much higher $\gamma$ energy cleanly separates the signal from
natural radioactivity, which is mostly below 2.6~MeV, and the shorter
capture time ($\sim$30~$\mu$s) reduces the background from accidental
coincidences.  Both Chooz~\cite{3Chooz} and Palo Verde~\cite{3piepke}
used 0.1\% Gd-loaded LS that yielded a capture time of 28~$\mu$s, about
a factor of seven shorter than in undoped liquid
scintillator. Backgrounds from random coincidences will thus be
reduced by a factor of seven as compared to unloaded LS.

The specifications for the design of the antineutrino detector module follow:
\begin{itemize}
\item
Employ three-zone detector modules partitioned with acrylic tanks as
shown in fig~\ref{fig:wyfbw}. 
\begin{figure}[!htb]
 \begin{center} \includegraphics[width=0.4\textwidth]{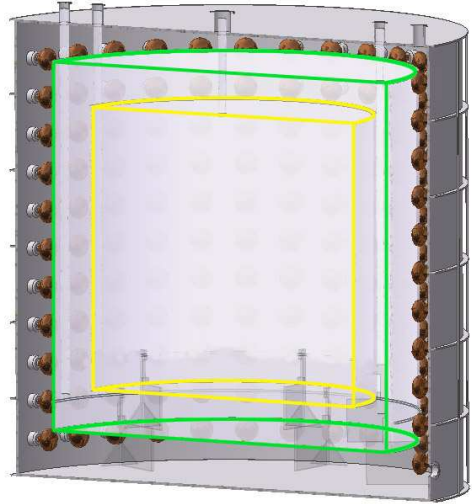}
 \caption{Cross sectional slice of a 3-zone antineutrino detector
 module showing the acrylic vessels holding the Gd-doped liquid
 scintillator at the center (20~T), liquid scintillator between the
 acrylic vessels (20~T) and mineral oil (40~T) in the outer
 region. The PMTs are mounted on the inside walls of the stainless
 steel tank.}  \label{fig:wyfbw} \end{center}
\end{figure}
The target volume is defined by the physical dimensions of the central
region of Gd-loaded liquid scintillator. This target volume is surrounded by an
intermediate region filled with normal LS to catch $\gamma$s escaping
from the central region.  The liquid-scintillator regions are
embedded in a volume of mineral oil to separate the PMTs from the
scintillator and suppress natural radioactivity from the PMT glass and
other external sources.

Four of these modules, each with 20~T fiducial volume, will be deployed
at the far site to obtain sufficient statistics and two modules will be
deployed at each near site, enabling cross calibrations. The matching
of four near and four far detectors allows for analyzing data with matched
near-far pairs.

In this design, the homogeneous target volume is well determined
without a position cut since neutrinos captured in the unloaded
scintillator will not in general satisfy the neutron energy
requirement.  Each vessel will be carefully measured to determine its
volume and each vessel will be filled with the same set of mass-flow
and flow meters to minimize any variation in relative detector volume and
mass. The effect of neutron spill-in and spill-out across the
boundary between the two liquid-scintillator regions will be cancelled
when pairs of identical detector modules are used at the near and far
sites. With the shielding of mineral oil, the singles rate will be
reduced substantially.  The trigger threshold can thus be lowered to
below 1.0~MeV, providing $\sim$100\% detection efficiency for the
prompt positron signal.

 \item The Gd-loaded liquid scintillator, which is the antineutrino
target, should have the same composition and fraction of hydrogen
(free protons) for each pair of detectors (one at a near site and the
other at the far site).  The detectors will be filled in pairs (one
near and one far detector) from a common storage vessel to assure that
the composition is the same. Other detector components such as
unloaded liquid scintillator and PMTs will be characterized and
distributed evenly to a pair of detector modules during assembly to
equalize the properties of the modules.

 \item The energy resolution should be better than 15\% at 1~MeV. Good
energy resolution is desirable for reducing the energy-related
systematic uncertainty on the neutron energy cut.  Good energy
resolution is also important for studying spectral distortion as a
signature of neutrino oscillations.

\item The time resolution should be better than 1~ns
for determining the event time and for studying backgrounds.
\end{itemize}

Detector modules of different shapes, including cubical, cylindrical,
and spherical, have been considered. From the point of view of ease of
construction cubical and cylindrical shapes are particularly
attractive. Monte Carlo simulation shows that modules of cylindrical
shape can provide better energy and position resolutions for the same
number of PMTs. Figure~\ref{fig:wyfbw} shows the structure of a
cylindrical module. The PMTs are arranged along the circumference of
the outer cylinder.  The surfaces at the top and the bottom of the
outer-most cylinder are coated with white reflective paint or other
reflective materials to provide diffuse reflection. Such an
arrangement is feasible since 1) the event vertex is determined only
with the center of gravity of the charge, not relying on the
time-of-flight information,\footnote{Although time information may not
be used in reconstructing the event vertex, it will be used in
background studies. A time resolution of 0.5~ns can be easily realized
in the readout electronics.} 2) the fiducial volume is well defined
with a three-zone structure, thus no accurate vertex information is
required. Details of the antineutrino detector will be discussed in
Chapter~\ref{sec:det}.

\subsubsection{Muon detector}
\label{sssec:exp_muon}

\par
Since most of the backgrounds come from the interactions of
cosmic-ray muons with nearby materials, it is desirable to
have a very efficient active muon detector coupled with a tracker for
tagging the cosmic-ray muons. This will provide a means for
studying and rejecting cosmogenic background events. The three
detector technologies that are being considered are water Cherenkov
counter, resistive plate chamber (RPC), and plastic scintillator strip.
When the water Cherenkov counter is  combined with a tracker, the
muon detection efficiency can be close to 100\%. Furthermore, these two
independent detectors can cross check each other. Their
inefficiencies and the associated uncertainties can be well determined by
cross calibration during data taking. We expect the inefficiency
will be lower than 0.5\% and the uncertainty of the inefficiency
will be lower than 0.25\%.

\par
\begin{figure}[!htb]
 \begin{center}
 \includegraphics[width=0.4\textwidth, angle=-90]{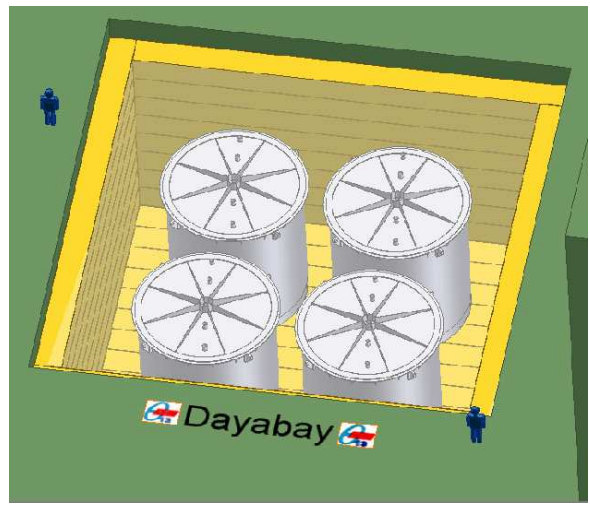}
 \caption{The water pool with four antineutrino detector modules inside.
 water tanks of dimension 1~m$\times$1~m are used as an outer muon tracker.}
 \label{fig:wmodule}
 \end{center}
\end{figure}

Besides being a shield against ambient radiation, the water buffer can also be utilized as a
water Cherenkov counter by installing PMTs in
the water. The water Cherenkov detector is based on proven technology,
and known to be very reliable. With sufficient PMT coverage and
reflective surfaces, the
efficiency of detecting muons should exceed 95\%. The current
baseline design of the water buffer is a water pool, similar to a
swimming pool with a dimensions of 16~m (length) $\times$ 16~m
(width) $\times$ 10~m (height) for the far hall containing four
detector modules, as shown in Fig.~\ref{fig:wmodule}. 
The PMTs of the water Cherenkov counters are mounted facing the inside
of the water pool. This is a simple and proven technology with very
limited safety concerns. The water will effectively shield the
antineutrino detectors from radioactivity in the surrounding rocks and
from Radon, with the attractive features of being simple,
cost-effective and having a relatively short construction time.

\par
The muon tracking detector consists of water tanks and RPCs.  RPCs
are very economical for instrumenting large areas, and simple to
fabricate. The Bakelite based RPC developed by IHEP for the BES-III
detector has a typical efficiency of 95\% and noise rate of
0.1~Hz/cm$^2$ per layer.~\cite{3ZhangJW}. A possible configuration is
to build three layers of RPC, and require two out of three layers hit
within a time window of 20~ns to define a muon event. Such a scheme
has an efficiency above 99\% and noise rate of $<$0.1~Hz/m$^2$.  Although
RPCs are an ideal large area muon detector due to their light
weight, good performance, excellent position resolution and low cost,
it is hard to put them inside water to fully surround the water pool.
The best choice seems to use them only at the top of the water pool.

\par
Water tanks with a dimension of 1~m$\times$1~m and a length of 16~m as
the outer muon tracking detector have a typical position resolution of
about 30~cm. Although not as good as other choices, the resolution is
reasonably good for our needs, in particular with the help of RPCs at
the top. Actually the water tanks are not really sealed
tanks, but reflective PVC sheets assembled on a stainless steel
structure, so that water can flow freely among water pool and water
tanks, and only one water purification system is needed for each
site. Water tanks can be easily installed at the side of the water
pool, but must be cut into sections at the bottom to leave space for
the supporting structure of antineutrino detector modules.  Each tank
will be equipped with four PMTs at each end to collect Cherenkov
photons produced by cosmic-muons. A few more PMTs are needed for
the bottom tanks to take into account optical path obstruction by the
supporting structure of the antineutrino detector modules.  A 13~m
long prototype has been built and tested~\cite{3water_tank}. A
detailed Monte Carlo simulation based on the data from this prototype
shows that the total light collected at each end is sufficient, as will
be discussed in detail in Chapter~\ref{sec:muon}. The technology
employed in this design is mature and the detector is relatively easy
and fast to construct.

\subsection{Alternative Designs of the Water Buffer}
\label{sec:designs}

We have chosen a water pool as the baseline experimental design (see
Fig.~\ref{fig:detector}). The two near detector sites have two
antineutrino detector modules in a rectangular water pool, whereas the
far site has four antineutrino detector modules in a square water
pool. The distance from the outer surface of each antineutrino
detector is at least 2.5~m to the water surface, with 1~m of water
between each antineutrino detector.

\begin{figure}[!htb]
\begin{center}
\includegraphics[width=0.65\textwidth,angle=0]{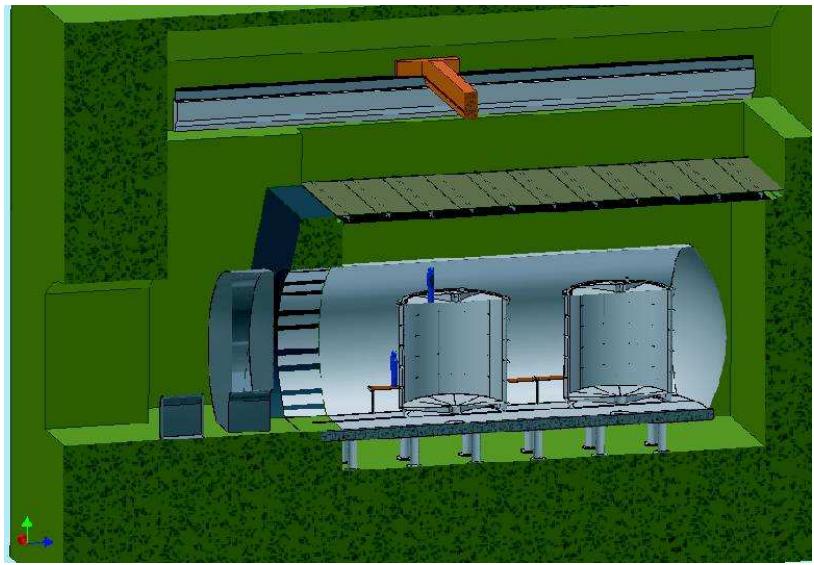}
\caption{Side cutaway view of a near detector site aquarium with two detectors visible.
\label{fig:alternate1}}
\end{center}
\end{figure}

\begin{figure}[!htb]
\begin{center}
\includegraphics[width=0.7\textwidth,angle=0]{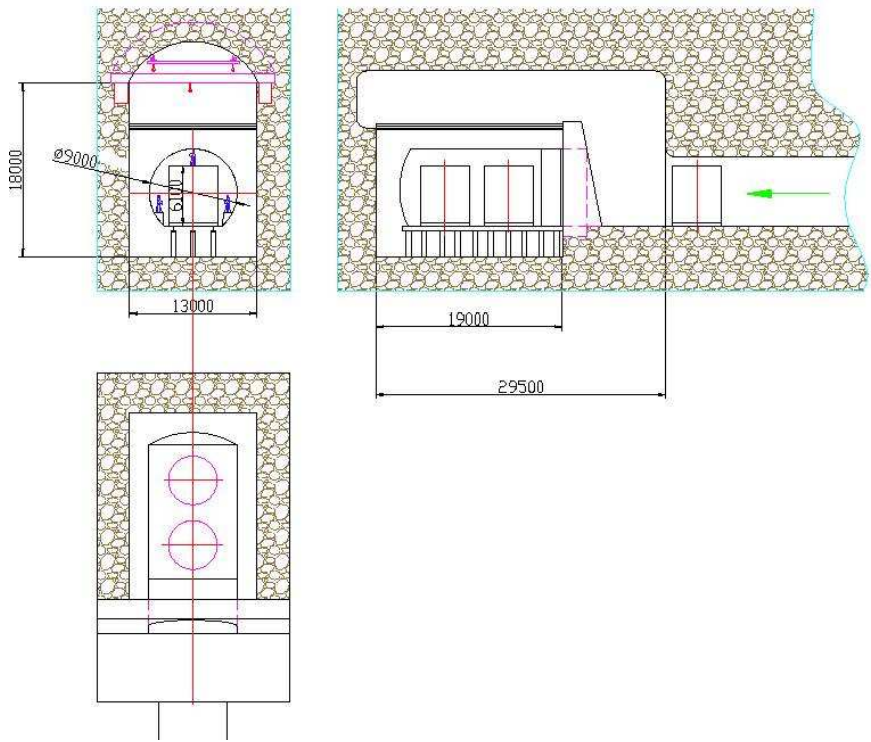}
\caption{End, side and top views of the conceptual design of a near detector site aquarium. All
distances are in mm.
\label{fig:chp4_aquarium_2}}
\end{center}
\end{figure}

Our primary alternative to the baseline design is the ``aquarium''
option. A conceptual design, showing a cut-away side view is provided in
Fig.~\ref{fig:alternate1}. 
Several views are shown in Fig.~\ref{fig:chp4_aquarium_2}.
The primary feature of this aquarium design is that the antineutrino
detector modules do not sit in the water volume, but are rather in
air.  The advantages of this design are ease of access to the
antineutrino detectors, ease of connections to the antineutrino
detectors, simpler movement of the antineutrino detectors, more
flexibility to calibrate the antineutrino detectors and a muon system
that does not need to be partially disassembled or moved when the
antineutrino detectors are moved. The primary disadvantages of this
design include the engineering difficulties of the central tube and
the water dam, safety issues associated with the large volume of water
above the floor level, cost, maintenance of the antineutrino detectors
free of radon and radioactive debris. This is preserved as the primary
option for a ``dry detector'' and serves as our secondary detector
design option.  Other designs that have been considered include:
ship-lock, modified aquarium, water pool with a steel tank, shipping
containers, and water pipes, among others.

The cost drivers that we have identified for the optimization of the
experimental configuration include:
\begin{itemize}
  \item Civil construction
  \item Cranes for the antineutrino detectors
  \item Transporters for the antineutrino detectors
  \item Safety systems in the event of catastrophic failure
  \item Storage volume of purified water
  \item Complexity of seals in water environment
\end{itemize}
The physics performance drivers that we have identified include:
\begin{itemize}
  \item Uniformity of shielding against $\gamma$'s from the rock and cosmic muon induced neutrons
  \item Cost and complexity of purifying the buffer region of radioactive impurities
  \item Amount and activity of steel near the antineutrino detectors (walls and mechanical support structures)
  \item Efficiency of tagging muons and measurement of that inefficiency
\end{itemize}
The primary parameters that we have investigated in the optimization
of the detector design are the thickness of the water buffer,
the optical segmentation of this water Cherenkov detector, the PMT
coverage of this water Cherenkov detector, the size and distribution
of the muon tracker system, the number of PMTs in the antineutrino
detector, the reflectors in the antineutrino detector.
The study is ongoing, but existing work favors the water pool.

\newpage
\renewcommand{\thesection}{\arabic{section}}
\setcounter{figure}{0} \setcounter{table}{0}
\setcounter{footnote}{0}

\section{Sensitivity \& Systematic Uncertainties}
\label{sec:sys}

The control of systematic uncertainties is critical to achieving the
$\sin^22\theta_{13}$ sensitivity goal of this experiment. The most
relevant previous experience is the Chooz experiment~\cite{2chooz}
which obtained $\sin^22\theta_{13}<0.17$ for $\Delta
m^2_{31}=2.5\times10^{-3}$eV$^2$ at 90\% C.L., the best limit to
date, with a systematic uncertainty of 2.7\% and statistical
uncertainty of 2.8\% in the ratio of observed to expected events at
the `far' detector. In order to achieve a $\sin^22\theta_{13}$
sensitivity below 0.01, both the statistical and systematic
uncertainties need to be an order of magnitude smaller than Chooz.
The projected statistical uncertainty for the Daya Bay far detectors
is 0.2\% with three years data taking. In this section we discuss
our strategy for achieving the level of systematic uncertainty
comparable to that of the statistical uncertainty. Achieving this
very ambitious goal will require extreme care and substantial
effort, and can only be realized by incorporating rigid constraints
in the design of the experiment.

\par
There are three main sources of systematic uncertainties: reactor,
background, and detector. Each source of uncertainty can be further
classified into correlated and uncorrelated uncertainties.

\subsection{Reactor Related Uncertainties}
\label{ssec:sys_reactor}

For a reactor with only one core, all uncertainties from the reactor,
correlated or uncorrelated, can be canceled precisely by using one far
detector and one near detector (assuming the distances are
precisely known) and forming the ratio of measured antineutrino
fluxes~\cite{7mikaelyan}. In reality, the Daya Bay nuclear power
complex has four cores in two groups, the Daya Bay NPP and the Ling Ao
NPP.  Another two cores will be installed adjacent to Ling Ao, called
Ling Ao II, which will start to generate electricity in 2010--2011.
Figure~\ref{fig:eng_bw} shows the locations of the Daya Bay cores, Ling
Ao cores, and the future Ling Ao II cores. Superimposed on the figure
are the tunnels and detector sites.  The distance between the two
cores at each NPP is about 88~m. The midpoint of the Daya Bay cores is
1100~m from the midpoint of the Ling Ao cores, and will be 1600~m from
the Ling Ao II cores. For this type of arrangement, with more reactor
cores than near detectors, one must rely upon the measured reactor
power levels in addition to forming ratios of measured antineutrino
fluxes in the detectors.  Thus there is a residual uncertainty in the
extracted oscillation probability associated with the uncertainties in
the knowledge of the reactor power levels. In addition to the reactor
power uncertainties, there are uncertainties related to uncertainties
in the effective locations of the cores relative to the detectors.

\subsubsection{Power Fluctuations}
\label{sssec:sys_power}

Typically, the measured power level for each reactor core will have
a correlated (common to all the reactors) uncertainty of the order of 2\%
and an uncorrelated uncertainty of similar size. Optimistically, we may be
able to achieve uncorrelated uncertainties of 1\%, but we conservatively
assume that each reactor has 2\% uncorrelated uncertainty in the
following. (We note that both Chooz and Palo Verde achieved total
reactor power uncertainties of 0.6--0.7\%. The appropriate value
for the Daya Bay reactors will need to be studied in detail with the
power plant and could hopefully be reduced below 2\% per
core.) If the distances are precisely known, the correlated uncertainties
will cancel in the near/far ratio.

For the geometry of the Daya Bay experiment, we have (effectively) two
near detectors. One near site primarily samples the rate from the two
Daya Bay cores and and the other primarily samples the rate from the
(two or four) Ling Ao cores. The detectors at the far site do not sample
the reactor cores equally, so one needs to consider the weighting of
the data from the near sites relative to the far site. In order to
provide optimal relative weights of the near sites one can utilize
the following combination of ratios in the event rates of the far
and near detectors:
\begin{equation}\label{eqn:nratio}
  \rho = \left[ \alpha \sum_r \frac{\phi_r}{(L_{r}^{DB})^2}
  \, +  \, \sum_r \frac{\phi_r}{(L_{r}^{LA})^2} \right] \left/ \sum_r
  \frac{\phi_r}{(L_{r}^f)^2} \right.
\end{equation}
where $\phi_r$ is the antineutrino flux at unit distance from core
$r$,  $L_{r}^f$ is the distance from reactor $r$ to the far site,
$L_{r}^{DB}$ ($L_{r}^{LA}$) is the distance from reactor $r$ to the
near Daya Bay (Ling Ao) site, and $\alpha$ is a constant chosen to
provide the proper weighting of the near site data and minimize the
sensitivity of $\rho$ to the uncertainties in the relative reactor
power levels. (In Eq.~\ref{eqn:nratio} we have neglected neutrino
oscillations. In the absence of oscillations and given a value of
$\alpha$, the quantity $\rho$ is completely determined by the
geometry. Thus a measurement of $\rho$ that differs from this value
could then be used to determine the oscillation probability that
depends upon $\sin^2 2 \theta_{13}$ with minimal systematic
uncertainty due to the uncorrelated reactor power uncertainties.)

To illustrate the utility of the ratio $\rho$ in
Eq.~\ref{eqn:nratio}, we can consider a slightly simplified geometry
where there are only two cores, each very close to a near detector.
Then the cross-talk in a near detector from the other core can be
neglected and the value of $\alpha= (L^f_{LA}/L^f_{DB})^2$ will
correct the ratio $\rho$ for the fact that the two reactors are not
sampled equally by the far detector. (Here $L^f_{DB}$ and $L^f_{LA}$
are the distances of the far detector from the two reactor cores.)
Then the ratio $\rho$ would be independent of the reactor power
uncertainties.

For the more complex situation as in Fig.~\ref{fig:eng_bw}, the
optimal choice of the weighting factor $\alpha$ is somewhat
different, and can be computed from knowledge of the relative
distances and powers of the reactor cores. One can also determine
$\alpha$ by Monte Carlo simulations that minimize the systematic
uncertainty in $\rho$ due to uncorrelated reactor power
uncertainties. The weighting of near sites using $\alpha$ does
introduce a slight degradation (in our case $<$11\% fractional
increase) in the statistical uncertainty. The correlated
uncertainties of the reactors are common to both the numerator and
denominator of the ratio $\rho$, and therefore will cancel.

Using the detector configuration shown in Fig.~\ref{fig:eng_bw},
with two near sites at $\sim$500~m baselines to sample the reactor
power and a far site at an average baseline of $\sim$1800~m, an
uncorrelated uncertainty of 2\% for each core and optimal choice of
$\alpha$ leads to the estimated reactor power contribution to
$\sigma_\rho$ (i.e., the fractional uncertainty in the ratio $\rho$)
shown in Table~\ref{taberrsite} for the case of four (six) reactor
cores. In Section~\ref{sssec:sys_chi} below, we study the
sensitivity of the Daya Bay experiment to neutrino oscillations and
$\sin^2 2 \theta_{13}$ using a more general $\chi^2$ analysis that
includes all the significant sources of systematic uncertainty. The
optimal weighting of near sites in that analysis is implemented by
allowing all the reactor core powers to vary in the $\chi^2$
minimization associated with the measured rates in the different
detectors.
\begin{table}[!htb]
\begin{center}
\begin{tabular}[c]{|c||c|c|c|c|} \hline
 Number of cores & $\alpha$ &$\sigma_\rho$(power) & $\sigma_\rho$(location) &
 $\sigma_\rho$(total) \\ \hline  \hline
 4 & 0.338 &0.035\% & 0.08\% & 0.087\% \\ \hline
 6 & 0.392 &0.097\% & 0.08\% & 0.126\% \\ \hline
\end{tabular}
\caption{Reactor-related systematic uncertainties for different
reactor configurations. The uncorrelated uncertainty of the power of a single
core is assumed to be 2\%. \label{taberrsite}}
\end{center}
\end{table}

\subsubsection{Location Uncertainties}
\label{sssec:sys_location}

The the center of gravity of the antineutrino source in each core will
be determined to a precision of about 30~cm. We assume that the
location uncertainties are uncorrelated, and so their combined effect
will be reduced by $\sim \sqrt{N_r}$ where $N_r$ is the number of
reactor cores. The resulting fractional uncertainty in the near/far
event ratio is estimated to be 0.08\% for the near baseline of
$\sim$500~m.

\subsubsection{Spent Fuel Uncertainties}
\label{sssec:sys_fuel}
\par
In addition to fission, beta decay of some fission products
can also produce antineutrinos with energy higher than the inverse
beta decay threshold 1.8~MeV. Some of these have long lifetimes, such
as~\cite{kopeikin}
\begin{eqnarray}
 ^{106}{\rm Ru}(T_{1/2}=372\,d)&\rightarrow & ^{106}{\rm Rh}(T_{1/2}=20\, s, E_{\rm
max}=3.54 {\rm MeV}) \nonumber\\
 ^{144}{\rm Ce}(T_{1/2}=285\, d)&\rightarrow & ^{144}{\rm Pr}(T_{1/2}=17\, m, E_{\rm
max}=3.00 {\rm MeV}) \nonumber\\
 ^{90}{\rm Sr}(T_{1/2}=28.6\, y)&\rightarrow& ^{90}{\rm Y}(T_{1/2}=64\, h, E_{\rm
max}=2.28 {\rm MeV})
\end{eqnarray}
These isotopes will accumulate in the core during operations.
Normally a fuel rod will produce power in the core for 2--3 years. The inverse
beta decay rate arising from these fission products will increase
to 0.4--0.6\% of the total event rate. In the 1.8--3.5~MeV range,
the yield will increase to about 4\%. Neutron capture by fission products
will also increase the total rate by 0.2\%~\cite{kopeikin}. 

\par
The Daya Bay and Ling Ao NPPs store their spent fuel in
water pools adjoining the cores. A manipulator moves the
burnt-out fuel rods from the core to the water pool during
refueling. The long lived isotopes mentioned in the previous paragraph
will continue to contribute to the antineutrino flux. 
The spent fuel data, as well as the realtime running
data, will be provided to the Daya Bay Collaboration by the power
plant.

\par
Taking the average of all fuel rods at different life cycles, and
the decay in the spent fuel, these isotopes are estimated to
contribute $<$0.5\% to the event rate (prior to receiving the
detailed reactor data). All of these events are in the low energy
region. Since the spent fuel is stored adjoining to the core, the
uncertainty in the flux will be canceled by the near-far relative
measurement, in the same way as the cancellation of the reactor
uncertainties. The uncertainty associated with the spent fuel is
much smaller than the assumed 2\% uncorrelated uncertainty of
reactor fission, and thus we expect it will have negligible impact
on the $\theta_{13}$ sensitivity.

\subsection{Detector Related Uncertainties}
\label{ssec:sys_detector}

For the detector-related uncertainties, we estimate two values for the
Daya Bay experiment: baseline and goal. The baseline value is what
we expect to be achievable through essentially proven methods with
straightforward improvement in technique and accounting for the fact
that we need to consider only {\it relative} uncertainties between near and
far detectors. The goal value is that which we consider achievable
through improved methods and extra care beyond the level of previous
experiments of this type. The results are summarized in
Table~\ref{tabsyserr} and discussed in the rest of this section.
\begin{table}[!hbt]
\begin{center}
\begin{tabular}[c]{|l|l||c|c|c|c|} \hline
 \multicolumn{2}{|c||}{Source of uncertainty} & Chooz &
 \multicolumn{3}{|c|}{Daya Bay ({\it relative})} \\
  \cline{4-6}
 \multicolumn{2}{|l||}{} & ({\it absolute}) & Baseline & Goal & Goal w/Swapping \\  \hline  \hline
 \# protons & H/C ratio & 0.8 & 0.2 & $ 0.1$& 0 \\
  \cline{2-5}
               & Mass & - &  0.2 &  0.02 & 0.006\\  \hline
  Detector  & Energy cuts & 0.8 & 0.2 & 0.1 & 0.1\\ \cline{2-5}
  Efficiency & Position cuts & 0.32 &   0.0 & 0.0 &0.0 \\ \cline{2-5}
             & Time cuts & 0.4 &  0.1 & 0.03 & 0.03\\ \cline{2-5}
             & H/Gd ratio & 1.0 & 0.1 & 0.1 & 0.0\\ \cline{2-5}
             & n multiplicity & 0.5 & 0.05 & 0.05  &0.05\\ \cline{2-5}
             & {Trigger}     & 0 & 0.01 & 0.01 & 0.01\\ \cline{2-5}
            & {Live time}   & 0 & $<$ 0.01  & $<$ 0.01& $<$ 0.01  \\
              \hline
 \multicolumn{2}{|l||}{Total detector-related uncertainty}
 & 1.7\% & 0.38\% & 0.18\% & 0.12\% \\ \hline
\end{tabular}
\caption{Comparison of detector-related systematic uncertainties
(all in percent, per detector module) of the Chooz experiment ({\it
absolute}) and projections for Daya Bay ({\it relative}). Baseline
values for Daya Bay are achievable through essentially proven
methods, whereas the goals should be attainable through additional
efforts described in the text. In addition, the additional
improvement from detector swapping is indicated in the last
column.} \label{tabsyserr}
\end{center}
\end{table}

\subsubsection{Target Mass and H/C Ratio}
\label{sssec:sys_mass}

The antineutrino targets are the free protons in the detector, so
the event rate in the detector is proportional to the total mass of
free protons. The systematic uncertainty in this quantity is controlled by
precise knowledge of the relative total mass of the central volumes
of the detector modules as well as filling the modules from a common
batch of scintillator liquid so that the H/C ratio is the same to
high precision.

The mass of the antineutrino target is accurately determined in
several ways. First the detector modules will be built to specified
tolerance so that the volume is known to $\sim 0.1\%$ (typically $<
$1~mm dimension out of a diameter of 3.2~meters). We will make a
survey of the detector geometry and dimensions after construction to
characterize the detector volumes to higher precision than 0.1\%.
Using optical measuring techniques and reflective survey targets
built into the detector modules and attached to the surfaces of the
acrylic vessels sub-mm precision is easily achievable with
conventional surveying techniques. A precision survey of each
detector module will be conducted after the assembly of the acrylic
vessels and the stainless steel tank in the surface assembly building
near the underground tunnel entrance.

\begin{figure}[htb!]
\begin{center}
\includegraphics[width=3.5in]{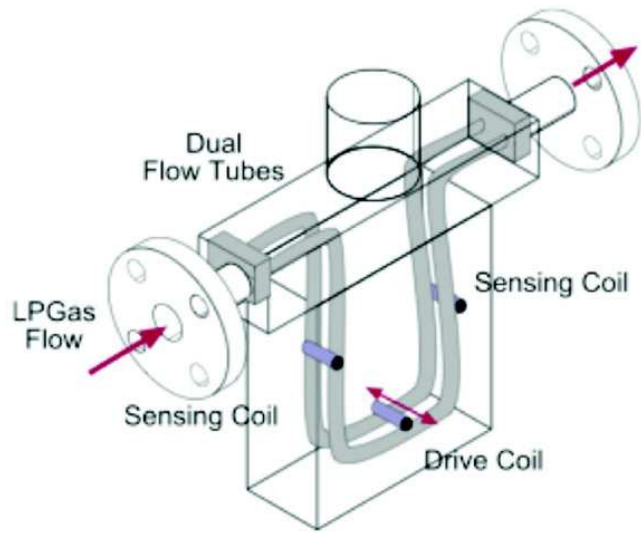}
\caption{Schematic drawing of a Coriolis mass flow meter. The driver
coil excites the tubes at 150~Hz, and a set of sensing coils
measures the tubes' amplitude and frequency while liquid is flowing.
\label{fig:coriolismeter}}
\end{center}
\end{figure}

Once the detectors are underground, we plan to fill each detector
from a common stainless tank the size of a detector volume using a
variety of instrumentation to directly measure the mass and volume
flow into the detector. A combination of Coriolis mass flow meters
(see Fig.~\ref{fig:coriolismeter}),
volume flow meters, and thermometers in the filling station and load
sensors in both the storage tank and possibly the antineutrino detector
will allow us to determine the mass of the liquid scintillator
reliably and with independent methods.

We will also measure the fluid flow using premium grade precision
volume flow meters with a repeatability specification of 0.02\%.
Several volume flowmeters will be run in series for redundancy.
Residual topping up of the detector module to a specified level
(only about $\sim$20~kg since the volume is known and measured) is
measured with these flowmeters as well.

During the filling process the temperature of the storage tank as well
as the ambient temperature of the filling station are controlled and
monitored. Liquid level sensors may be used during the filling process
to monitor and maintain the relative liquid levels in the antineutrino
detector volume and the $\gamma$-catcher region.

Coriolis mass flow meters are devices developed by the processing
industry to measure directly the mass flow of a liquid or gas
(compared to the volume flow of conventional flow meters). The
measured mass flow is independent of the liquid's density and
viscosity hence minimizing the need for environmental control of the
storage tank, filling station, and detector modules, and reducing
possible systematics due to ambient temperature fluctuations.
Coriolis meters use two U-shaped oscillating flow tubes. A sine
voltage is applied to an electromagnetic driver which produces an
oscillating motion of the tubes. The vibration of the tube causes a
slight angular rotation about its center. The fluid flow is then
deflected by the Coriolis force which changes the tube rotation, The
amplitude of this change is related to the mass flow and the
frequency is related to the product density. Coriolis meters measure
simultaneously the mass flow and density of the liquid. They are
commercially available with flow rates ranging from 1g/hr to
8000~kg/hr. The quoted absolute accuracy of the devices is 0.1--0.2\%,
and their repeatability is $\sim$0.1\%. Combined with a control
valve system Coriolis flow meters allow the precise and repeatable
filling of the detector modules with a chosen target mass.

The volume flowmeters have an {\it absolute} calibrated precision of
$0.2\%$, so we quote a baseline uncertainty of 0.2\% for the
detector mass. We use the 0.02\% repeatability performance of the
volume flowmeters to estimate the goal uncertainty of 0.02\%.

The absolute H/C ratio was determined by Chooz using scintillator
combustion and analysis to 0.8\% precision based on combining data
from several analysis laboratories. We will only require that the
{\it relative} measurement on different samples be known, so an
improved precision of 0.2\% or better is expected. We quote this as
the estimated baseline H/C systematic uncertainty.

We are presently engaged in a program of R\&D with the goal of
measuring the {\it relative} H/C ratio in different samples of
liquid scintillator to $\sim$0.1\% precision. We are exploring
three different methods to achieve this goal: precision NMR,
chemical analysis, and neutron capture. The neutron capture method
would need to be utilized before the introduction of Gd into the
scintillator, but could be used to precisely characterize the
organic liquids used in the liquid scintillator cocktail. In
principle, the other methods could be used on the final Gd-loaded
scintillator.

In addition, we will need to determine the {\it relative} H/C ratio
in the $\gamma$-catcher liquid scintillator to about 1\%. This is to
control the relative amount of ``spill-in'' events where a neutron
generated in the $\gamma$-catcher is captured in the Gd-loaded
scintillator after thermal diffusion. This should be much more
straightforward than the more demanding requirement on the Gd-loaded
scintillator but will be sufficient to achieve the goal H/C
systematic uncertainty in Table~\ref{tabsyserr}.

\subsubsection{Position Cuts}
\label{sssec:sys_position}

Due to the design of the detector modules, the event rate is
measured without resort to reconstruction of the event location.
Therefore the uncertainty in the event rate is related to the physical
parameters of the antineutrino volume. We do not anticipate employing
cuts on reconstructed position to select events, and there should be
no uncertainty related to this issue.

\subsubsection{Positron Energy Cut}
\label{sssec:sys_positron}

\begin{figure}[!htb]
 \centerline{\includegraphics[height=7cm]{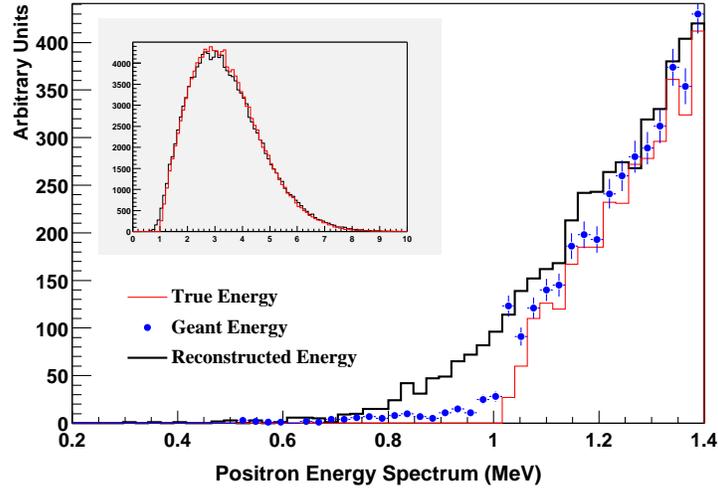}}
\caption{Energy spectra associated with the positron's true energy, simulated energy
(Geant Energy), and reconstructed energy at 1~MeV. The full
spectrum is shown in the  inset, where the red line corresponds to
the true energy and the black one corresponds to the reconstructed
energy.}
 \label{fig:posicut}
\end{figure}

Due to the high background rates at low energy, Chooz employed a
positron energy threshold of 1.3~MeV. This cut resulted in an
estimated uncertainty of 0.8\%. The improved shielding design of the
Daya Bay detectors makes it possible to lower this threshold to
below 1~MeV while keeping uncorrelated backgrounds as low as 0.1\%.
The threshold of visible energy of neutrino events is 1.022~MeV. Due
to the finite energy resolution of $\sim$12\% at 1~MeV, the
reconstructed energy will have a tail below 1~MeV. The systematic
uncertainty associated with this cut efficiency is studied by Monte Carlo
simulation. The tail of the simulated energy spectrum is shown in
Fig.~\ref{fig:posicut} with the full spectrum shown in the inset.
For this simulation, 200 PMTs are used to measure the energy
deposited in  a  20-ton module. The energy resolution is $\sim$15\%
at 1~MeV. The inefficiencies are 0.32\%, 0.37\%, and 0.43\% for cuts
at 0.98~MeV, 1.0~MeV, and 1.02~MeV, respectively. Assuming the
energy scale uncertainty is 2\% at 1~MeV, this inefficiency
variation will produce a 0.05\% uncertainty in the detected
antineutrino rate. The upper energy requirement for the positron signal will
be $E<8$~MeV and will also contribute a negligible uncertainty to the
positron detection efficiency.

\subsubsection{Neutron Detection Efficiency}
\label{sssec:sys_neutron}

The delayed neutron from the inverse beta decay reaction is produced
with $\sim$10~keV of kinetic energy. The neutron loses energy in
the first few interactions with H and C in the scintillator, and
reaches thermal energy in a few microseconds. The neutrons can
capture on either H or Gd during or after the thermalization
process. We will detect the neutrons that capture on Gd, yielding at
least 6~MeV of visible energy from the resulting capture $\gamma$ rays,
during the time period 0.3$< T <$200~$\mu$s.

The efficiency for detecting the neutron is given by
\begin{equation}
\epsilon_n = P_{Gd} \epsilon_E \epsilon_T
\end{equation}
in which $P_{Gd}$ is the probability to capture on Gd (as opposed to
H), $\epsilon_E$ is the efficiency of the $E>6$~MeV energy cut for
Gd capture, and $\epsilon_T$ is the efficiency of the delayed time
period cut. In order to measure the rates for two detectors (near
and far) with a precision to reach $\sin^2 2 \theta_{13} =0.01$ the
baseline requirement for the uncertainty on {\it relative} neutron detection
efficiencies is $0.25\%$. The $\epsilon_n$ for neutrons at the
center of a detector module can be determined directly by using a
tagged neutron source (either ${}^{252}$Cf, AmBe or both can be
used) and counting the number of neutrons using the time and energy
cuts after neutron producing event. (Corrections associated with
uniformly distributed neutrons are studied with spallation neutrons,
as discussed in Section~\ref{sec:cal}.) This will require
measurement of order 1 million neutron captures, and would likely
require several hours of measurement. This will be established
during the initial comprehensive calibration of each detector.

In addition, the individual components $P_{Gd}$, $\epsilon_E$, and
$\epsilon_T$ can be monitored separately as an additional check on
the measurement of $\epsilon_n$.

\smallskip
\noindent {\bf H/Gd ratio}

Neutrons are thermalized during their first $10~\mu$s of existence in
the detector central volume. Thus for times longer than $10~\mu$s the
delayed neutron capture events will exhibit an exponential time
constant, $\tau$, related to the average concentration of Gd in the
detector module. The rate of capture, $\Gamma \equiv 1/ \tau$, is
given by:
\begin{equation}
\Gamma = \Gamma_{Gd} + \Gamma_{H} = [n_{Gd} \sigma_{Gd} + n_{H}
\sigma_{H}] v \> 
\end{equation}
where $n_{H(Gd)}$ is the number density of hydrogen (Gd) in the liquid scintillator
and $\sigma_{H(Gd)}$ is the neutron capture cross section on hydrogen (Gd) and $v$ is
the thermal velocity.
The fraction of neutrons that capture on Gd rather than H is then
\begin{equation}
P_{Gd} = \frac{1}{1+\Gamma_H/\Gamma_{Gd}}
\end{equation}
and we would like to know this {\it relative} fraction between
different detector modules to $\sim$0.1\%. Thus we must measure the
time constants $\tau$ for different detector modules to a {\it
relative} precision of 0.2~$\mu$s. The value of $\tau$ is expected
to be about 30~$\mu$s, so we need to measure it to about 0.5\%
relative precision. Such a measurement requires measuring about
30,000 neutron captures, which can be done in a few minutes with a
neutron source. The Chooz experiment measured the ({\it absolute})
$\sim$30~$\mu$s capture time to $\pm$0.5~$\mu$s precision.

Measurement of $\tau$ to 0.5\% precision will provide a relative
value of $P_{Gd}$ to 0.1\% uncertainty, which is the baseline and
goal value in Table~\ref{tabsyserr}.

\smallskip

\noindent {\bf Energy cut efficiency}

\par
Another issue is the neutron detection efficiency associated with
the signal from capture of neutrons on Gd in the antineutrino detector
volume. An energy threshold of about 6~MeV will be employed to
select these delayed events, and the efficiency ($\sim$93\%) of
this criterion may vary between detector modules depending upon the
detailed response of the module. However, this can be calibrated
through the use of radioactive sources (see Section~\ref{sec:cal}) and
spallation neutron captures. The KamLAND detector gain is routinely
(every two weeks) monitored with sources, and a relative long-term
gain drift of $\sim 1 \%$ is readily monitored with a precision of
$0.05\%$. Monte Carlo simulations of the Daya Bay detector response
for the Gd capture $\gamma$s indicate that 1\% energy scale uncertainty
will lead to 0.2\% uncertainty in $\epsilon_E$, and we use this
value as the baseline systematic uncertainty.

\begin{figure}[!htb]
 \centerline{\includegraphics[width=0.85\textwidth]{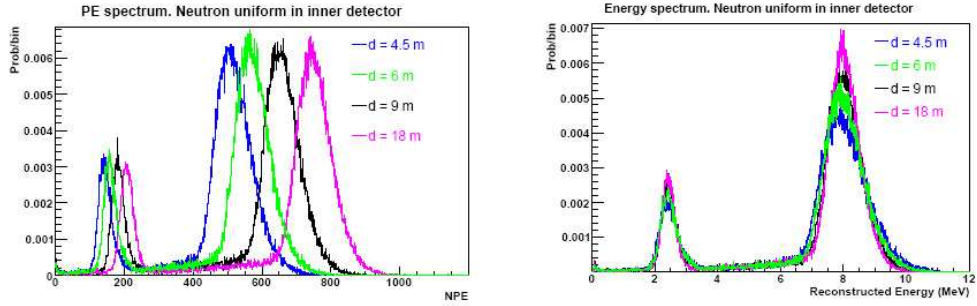}}
\caption{Spallation neutron response for detector modules with
scintillator optical attenuation lengths of $4.5 \le d \le 18$~m.
The left panel shows the raw photoelectron spectra, whereas the
right panel shows the spectra rescaled according to a non-linear
rescaling procedure we have developed. The rescaled 6~MeV effective
energy threshold produces a constant value of $\epsilon_E = 93\%$ to
within 0.4\% over this extreme range of attenuation length.}
 \label{fig:neutron_sim}
\end{figure}

We have also performed detailed Monte Carlo simulations of the
detector response to neutron sources and spallation neutrons. The
results of these studies indicate that we can indeed establish the
relative value of $\epsilon_E$ to $0.1\%$, even for reasonable
variations of detector properties (such as scintillator attenuation
length). As an example, Fig.~\ref{fig:neutron_sim} shows how the
source data can be used with uniform spallation neutrons to bootstrap
a non-linear energy scale that corrects the spectrum, independent of
attenuation length over the extreme range of 4.5--18~m.
Therefore, we estimate a value of 0.1\% for the goal systematic
uncertainty in $\epsilon_E$.

\smallskip

\noindent {\bf Time cuts}

\par
The time correlation of the prompt (positron) event and the delayed
(neutron) event is a critical aspect of the event signature.
Matching the time delays of the start and end times of this time
window between detector modules is crucial to reducing systematic
uncertainties associated with this aspect of the antineutrino signal. If
the starting time ($\sim$0.3~$\mu$s) and ending time ($\sim$200~$\mu$s)
of the delayed event window is determined to $\sim$10~ns
precision, the resulting uncertainty associated with missed events
is $\sim$0.03\%. We will insure that this timing is
equivalent for different detector modules by slaving all detector
electronics to one master clock. We estimate that with due care, the
relative neutron efficiency for different modules due to timing is
known to $\sim$0.03\%, and we use this value as the estimated goal
systematic uncertainty. We use a more conservative 0.1\% value for
the baseline value.

\subsubsection{Neutron Multiplicity}
\label{sssec:sys_mult}

Chooz required a cut on the neutron multiplicity to eliminate events
where it appeared that there were two neutron captures following the
positron signal, resulting in a 2.6\% inefficiency and associated
0.5\% systematic uncertainty. These multiple neutron events are due to
muon-induced spallation neutrons, and will be reduced to a much
lower level by the increased overburden available at the Daya Bay
site. For the near site at 500~m baseline, the muon rate relative to
the signal rate will be more than a factor nine lower than for the
Chooz site. Therefore, events with multiple neutron signals will be
reduced by this factor relative to Chooz, and should present a much
smaller problem for the Daya Bay site. We therefore estimate a
0.05\% value for this systematic uncertainty and use this for both
the baseline and goal values.

\subsubsection{Trigger}
\label{sssec:sys_trigger}

    The trigger efficiency can be measured to high precision (0.01\%)
using studies with pulsed light sources in the detector. (We note
that KamLAND has used this method to determine 99.8\% absolute
trigger efficiency~\cite{7kamland}.) In addition, we will employ
several redundant triggers so that they can be used to cross-check
each other to high precision. We estimate a systematic uncertainty of
0.01\% can be achieved, and use this for both the baseline and goal
values.

\subsubsection{Live Time}
\label{ssec:sys_LT}

The detector live time can be measured accurately by counting a
100~MHz clock using the detector electronics, and normalizing to
the number of clock ticks in a second (as defined by a GPS
receiver signal). The uncertainty associated with this procedure
should be extremely small, and certainly negligible relative to
the other systematic uncertainties. For example, SNO measured the
relative live times for their day/night analysis with a relative
fractional uncertainty of $5 \times 10^{-7}$.

\subsection{Cross-calibration and Swapping of Detectors}
\label{ssec:sys_swap}

\subsubsection{Detector Swapping}
\label{sssec:sys_swap}

The connection of the two near detector halls and the far hall by
horizontal tunnels provides the Daya Bay experiment with the
unique and important option of swapping the detectors between the
locations.  This could enable the further reduction of
detector-related systematic uncertainties in the measurement of
the ratio of neutrino fluxes at the near and far locations.
Although the estimated baseline and goal systematic uncertainties
in Table~\ref{tabsyserr} are sufficient to achieve a sensitivity
of 0.01 in $\sin^2 2 \theta_{13}$, implementation of detector
swapping could provide an important method to further reduce
systematic uncertainties and increase confidence in the
experimental results.

The swapping concept is easy to demonstrate for a simple scenario
with a single neutrino source and only two detectors deployed at two
locations, near and far. The desired measurement is the ratio of
event rates at the near and far locations: $N/F$. With detector \#1
(efficiency $\epsilon_1$) at the near location and detector \#2
(efficiency $\epsilon_2$)at the far location we would measure
\begin{equation}
\frac{N_1}{F_2} = \left( \frac{\epsilon_1}{\epsilon_2} \right)
\frac{N}{F} \> 
\end{equation}
By swapping the two detectors and making another measurement, we can
measure
\begin{equation}
\frac{N_2}{F_1} = \left( \frac{\epsilon_2}{\epsilon_1} \right)
\frac{N}{F} \> 
\end{equation}
where we have assumed that the detector properties (e.g.,
efficiencies) do not change when the detector is relocated. We can
now combine these two measurements to obtain a value of $N/F$ that
is, to first order, independent of the detector efficiencies:
\begin{equation}
\frac{1}{2} \left( \frac{N_1}{F_2} + \frac{N_2}{F_1} \right) =
\frac{N}{F} \left( 1 + \frac{\delta^2}{2}\right)
\end{equation}
where we have defined
\begin{equation}
\delta \equiv \frac{\epsilon_2}{\epsilon_1} -1 \>
\end{equation}
Note that even if the detector efficiencies are different by as much
as 1\%, we can determine $N/F$ to a fractional precision better than
$10^{-4}$.

The layout of the Daya Bay experiment involves two near sites with two
detectors each, and a far site with four detectors. The simplest plan
is to designate the eight detectors as four pairs: (1,2), (3,4), (5,6),
(7,8).  Using four running periods (designated I, II, III, IV,
separated by three detector swaps) we can arrange for each
detector to be located at the far site half the time and a near site
half the time by swapping two pairs between running periods, as shown
in Table~\ref{tab:swap}.
\begin{table}[!htb]
\begin{center}
\begin{tabular}[c]{|c||c|c|c|}
 \hline
 Run Period & Near(DB) & Near(LA) & Far \\ \hline  \hline
  I & 1,3 & 5,7 & 2,4,6,8 \\ \hline
 II & 2,3 & 6,7 & 1,4,5,8 \\ \hline
III & 2,4 & 6,8 & 1,3,5,7 \\ \hline
 IV & 1,4 & 5,8 & 2,3,6,7 \\ \hline
\end{tabular}
\caption{Swapping scheme with four running periods. The detectors
(labelled 1--8) are deployed at the Near(DB), Near(LA), and Far sites
during each period as indicated in this table. \label{tab:swap}}
\end{center}
\end{table}
Ratios of event rates can be combined in a fashion analogous to the
above discussion to provide cancellation of detector-related systematic
uncertainties and also reactor power systematic uncertainties. Careful calibration
of the detectors following each swap will be necessary to insure that
each detector's performance does not change significantly due to
relocation. In particular, all the parameters in Table~\ref{tabsyserr}
need to be checked and, if necessary, corrections applied to restore
the detection efficiency to the required precision through, e.g.,
changes in calibration constants.

Successful implementation of this swapping concept will lead to
substantial reduction in many of the detector-related systematic
uncertainties. The uncertainty associated with the H/C and H/Gd
ratios should be completely eliminated. By measuring the fluid
levels before and after swapping, we can insure that the detector
volume will be the same with negligible uncertainty. However, due to the
residual uncertainty in the monitored temperature of the detector
module ($0.1^\circ$~C), there will be a residual uncertainty in the
detector mass of 0.006\%, and this is the value quoted in
Table~\ref{tabsyserr}.

\subsubsection{Detector Cross-calibration}
\label{sssec:sys_cross}

Another important feature of the design of the Daya Bay experiment
is the presence of two detector modules at each near site. During a
single running period (I, II, III, or IV) each near detector module
will measure the neutrino rate with 0.23\% statistical precision. If
the systematic uncertainties are smaller than this, the two detectors at
the near site should measure the same rate, giving a detector
asymmetry of $0 \pm 0.34$\% (statistical uncertainty only).
Combining all the detector pairs in all 4 running periods will yield
an asymmetry of $0 \pm 0.04$\% (statistical uncertainty only). These
asymmetries are an important check to ensure that the detector-related
systematic uncertainties are under control. In addition, this analysis can
provide information on the the degree to which the detector-related
systematic uncertainties are correlated or uncorrelated so that we know how
to handle them in the full analysis including the far site.

Finally, the near detector data can provide important information on
the reactor power measurements. We will measure the ratio
\begin{equation}
R_{\rm near} = \frac{S_{DB}}{S_{LA}}
\end{equation}
where $S_{DB}$ ($S_{LA}$) is the detector signal (background
subtracted, normalized to the reactor power) for the Daya Bay (Ling
Ao) near site. If the reactor powers are correct (and the detector
systematic uncertainties are under control) then we expect $R_{\rm near} =
1.0 \pm 0.24\% \pm 0.51\%$, where the first uncertainty is statistical
(only 1 of the 4 running periods) and the second uncertainty is the
detector (baseline) systematic uncertainty. Note that these uncertainties are
small relative to the expected 2\% uncorrelated reactor power
uncertainty, so measurement of $R_{\rm near}$ will provide an
important check (and even perhaps additional information) on the
reactor powers. Furthermore, studies of the measured neutrino
spectra in the different near detectors during different parts of
the reactor fuel cycle can help provide constraints on the fuel
cycle effects on the spectrum.

\subsection{Backgrounds}
\label{ssec:sys_backgrounds}

\par
In the Daya Bay experiment, the signal events (inverse beta decay
reactions) have a distinct signature of two time-ordered signals: a
prompt positron signal followed by a delayed neutron-capture signal.
Backgrounds can be classified into two categories: correlated and
uncorrelated backgrounds. If a background event is triggered by two
signals that come from the same source, such as those induced by the
same cosmic muon, it is a correlated background event. On the other
hand, if the two signals come from different sources but satisfy the
trigger requirements by chance, the event is an uncorrelated
background.

\par
There are three important sources of backgrounds in the Daya Bay
experiment: fast neutrons, $^8$He/~$^9$Li, and natural
radioactivity. A fast neutron produced by a cosmic muon in the
surrounding rock or the detector can produce a signal mimicking the
inverse beta decay reaction in the detector: the recoil proton
generates the prompt signal and the capture of the thermalized
neutron provides the delayed signal. The $^8$He/~$^9$Li isotopes
produced by cosmic muons have substantial beta-neutron decay
branching fractions, 16\% for $^8$He and 49.5\% for $^9$Li. The beta
energy of the beta-neutron cascade
overlaps the positron signal of neutrino events, simulating the
prompt signal, and the neutron emission forms the delayed signal.
Fast neutrons and $^8$He/~$^9$Li isotopes create correlated
backgrounds since both the prompt and delayed signals are from the
same single parent muon. Some neutrons produced by cosmic muons are
captured in the detector without proton recoil energy. A single
neutron capture signal has some probability to fall accidentally
within the time window of  a preceding signal due to natural
radioactivity in the detector, producing  an accidental background.
In this case, the prompt and delayed signals are  from different
sources, forming an uncorrelated background.

\par
All three major backgrounds are related to cosmic muons. Locating
the detectors at sites with adequate overburden is the only way to
reduce the muon flux and the associated background to a tolerable
level. The overburden requirements for the near and far sites are
quite different because the signal rates differ by more than a
factor of 10. Supplemented with a good muon identifier outside the
detector, we can tag the muons going through or near the detector
modules and reject backgrounds efficiently.

\par
In this section, we describe our background studies and our
strategies for background management. We conclude that the
background-to-signal ratio will be around 0.3\% at the near sites
and around 0.2\% at the far site, and that the major sources of
background can be quantitatively studied {\it in-situ}.

\subsubsection{Cosmic Muons in the Underground Laboratories}
\label{sssec:sys_cosmic}

\par
The most effective and reliable approach to minimize the backgrounds
in the Daya Bay experiment is to have sufficient amount of
overburden over the detectors. The Daya Bay site is particularly
attractive because it is located next to a 700-m high mountain. The
overburden is a major factor in determining the optimal detector
sites. The location of detector sites has been optimized by using a
global $\chi^2$ analysis described in Section~\ref{sssec:sys_chi}.

\par
\begin{figure}[!htb]
\centerline{\includegraphics[width=0.7\textwidth]{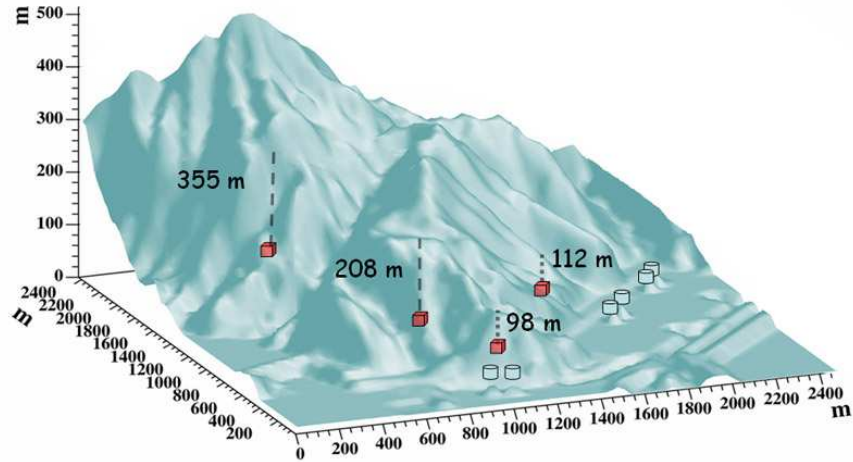}}
\caption{Three dimensional profile of Pai Ya Mountain,
where the Daya Bay experimental halls will be located, generated from
a 1:5000 topographic map of the Daya Bay area.} \label{fig:contour}
\end{figure}

Detailed simulation of the cosmogenic background requires accurate
information of the mountain profile and rock composition.
Figure~\ref{fig:contour} shows the mountain profile converted from a
digitized 1:5000 topographic map.
The horizontal tunnel and detector sites are designed to be about
$-$20~m PRD.\footnote{PRD is the height measured relative to the mouth
of the Zhu Jiang River (Pearl River), the major river in South China.}
Several rock samples at different locations of the Daya Bay site were
analyzed by two independent groups. The measured rock density ranges
from 2.58 to 2.68 g/cm$^3$. We assume an uniform rock density of 2.60
g/cm$^3$ in the present background simulation. A detailed description
of the topography and geology of the Daya Bay area is given in
Chapter~\ref{sec:civil}.

\par
\begin{figure}[!htb]
\begin{minipage}[t]{0.48\textwidth}
\includegraphics[width=0.95\textwidth]{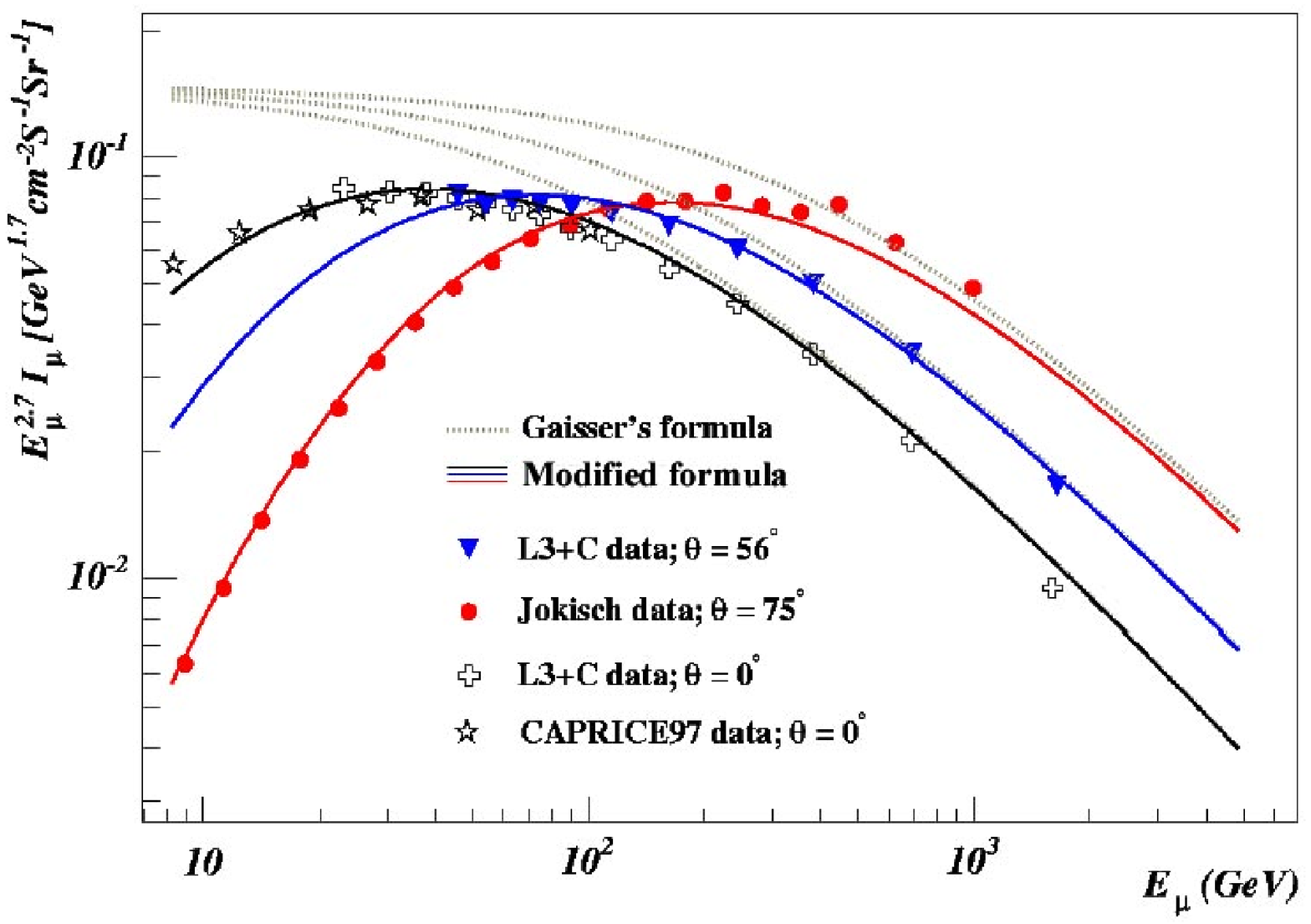}
\caption[Comparison of the modified formula (solid lines) with data.
Calculations with the standard Gaisser's formula are shown in dashed
lines.]{Comparison of the modified formula (solid lines) with data.
Calculations with the standard Gaisser's formula are shown in dashed
lines. The data are taken from Ref.~\protect\cite{6l3,6jokisch}.}
\label{fig:mfluxcom}
\end{minipage}
 \hfill
\begin{minipage}[t]{0.48\textwidth}
\includegraphics[width=\textwidth]{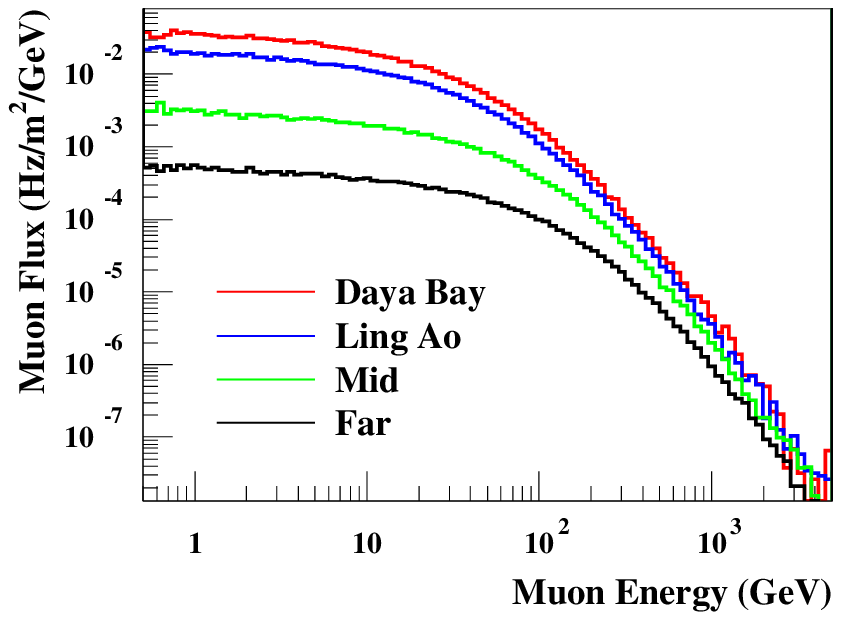}
\caption{Muon flux as a function of the energy of the surviving
muons. The four curves from upper to lower correspond to the Daya
Bay near site, the Ling Ao near site, the mid site and the far site,
respectively.} \label{fig:muonspec}
\end{minipage}
\end{figure}

The standard Gaisser formula~\cite{6gaisser} is known to poorly
describe the muon flux at large zenith angle and at low energies.
This is relevant for the Daya Bay experiment since the overburden at
the near sites is only $\sim$100~m. We modified the Gaisser formula as
\begin{equation}\label{modi}
 \frac{dI}{dE_{\mu}d\cos \theta}=
 0.14 \left( \frac{E_{\mu}}{\rm GeV}\left(1+\frac{3.64 {\rm \,GeV}}
 {E_{\mu}(\cos \theta^{\ast})^{1.29}}\right) \right) ^{-2.7}
 \left[\frac{1}{1+\frac{1.1E_{\mu}\cos \theta^{\ast}}{115{\rm \,GeV}}}
  + \frac{0.054}{1+\frac{1.1E_{\mu}\cos \theta^{\ast}}{850 {\rm \,GeV}}} \right]
\end{equation}
which is the same as the standard formula, except that the polar angle
$\theta$ is substituted with $\theta^\ast$,
\begin{equation}
\cos \theta^{\ast} = \sqrt{\frac{ (\cos
\theta)^{2}+P_{1}^{2}+P_{2}(\cos \theta)^{P_{3}}+P_{4}(\cos
\theta)^{P_{5}}}{1+P_{1}^{2}+P_{2}+P_{4}}}
\end{equation}
as defined in~\cite{Chirkin}. The parameters are determined to be
$P_{1}=0.102573,\;P_{2}=-0.068287,\; P_{3}=0.958633
,\;P_{4}=0.0407253$, and $P_{5}=0.817285$, by using CORSIKA to
simulate the muon production in the atmosphere. The comparison of
the modified formula with data is shown in Fig.~\ref{fig:mfluxcom},
where the calculations with the standard Gaisser formula are also
shown.
At muon energies of several tens of GeV, the standard Gaisser formula
has large discrepancies with data while the modified formula agrees
with data in the whole energy range.

\par
Using the mountain profile data, the cosmic muons are transported from
the atmosphere to the underground detector sites using the MUSIC
package~\cite{6music}. Simulation results are shown in
Table~\ref{tab.near-far} for the optimal detector sites.
\begin{table}[!htb]
\begin{center}
\begin{tabular} {|c||c|c|c|c|} \hline
 & DYB site & LA site & Mid site & Far site \\ \hline\hline
 Vertical overburden (m)   &  98 & 112 & 208 & 355 \\ 
 Muon Flux (Hz/m$^2$) & 1.16 & 0.73 & 0.17 & 0.041 \\ 
 Muon Mean Energy (GeV) & 55 & 60 & 97 & 138 \\ \hline
\end{tabular}
\caption{Vertical overburden of the detector sites and the
corresponding muon flux and mean energy. } \label{tab.near-far}
\end{center}
\end{table}
The muon energy spectra at the detector sites are shown in
Fig.~\ref{fig:muonspec}. The four curves from upper to lower
corresponds to the Daya Bay near site, the Ling Ao near site, the mid
site and the far site, respectively.

\subsubsection{Simulation of Neutron Backgrounds}
\label{sssec:sys_backgrounds}

\par
The neutron production rates will depend upon the cosmic muon flux
and average energy at the detector. However, the neutron
backgrounds in the detector also depend on the local detector
shielding. The neutrino detectors will be shielded by at least 2.5
meters of water. This water buffer will be used as a Cherenkov
detector to detect muons. Thus neutrons produced by muons in the
detector module or the water buffer will be identified by the muon
signal in the water Cherenkov detector. In addition, neutrons created
by muons in the surrounding rock will be effectively attenuated by
the 2.5~m water buffer. Together with another muon tracker outside
the water buffer, the combined muon tag efficiency is designed to be
99.5\%, with an uncertainty smaller than 0.25\%.

\par
From the detailed muon flux and mean energy at each detector site,
the neutron yield, energy spectrum, and angular distribution can be
estimated with an empirical formula~\cite{6wangyf} which has been
tested against experimental data whenever available. A full Monte
Carlo simulation has been carried out to propagate the primary
neutrons produced by muons in the surrounding rocks, the water
buffer, and the oil buffer layer of the neutrino detector, to the
detector. The primary neutrons are associated with their parent
muons in the simulation so that we know if they can be tagged by the
muon detector. Neutrons produced by  muons that pass through the
liquid scintillator neutrino detector will be tagged with 100\%
efficiency. Neutrons produced in the water buffer will be tagged
with an efficiency of 99.5\%, since their parent muons must pass
through the muon systems.  Neutrons produced in the rocks, however,
have to traverse at least 2.5 meters of water to reach a detector
module. About 70\% of the neutrons that enter the detector modules
from the surrounding rock arise from parent muons that leave a
signal in the muon system (i.e., ``tagged''). About 30\% of the
neutrons that enter the detector modules from the surrounding rocks
arise from muons that miss the muon system ($\equiv$~``untagged'').
The neutron background after muon rejection is the sum of the
untagged events and 0.5\% (due to veto inefficiency) of the tagged
events.

\par
\begin{figure}[!htb]
\begin{center}
\includegraphics[width=0.7\textwidth]{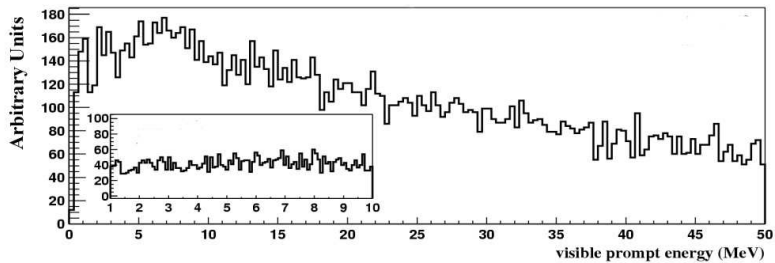}
\caption{The prompt energy spectrum of fast neutron background at
the Daya Bay far detector. The inset is an expanded view of the
spectrum from 1 to 10~MeV.} \label{fig:fastn}
\end{center}
\end{figure}

Some energetic neutrons will produce tertiary particles, including
neutrons. For those events that have energy deposited in the liquid
scintillator, many have a complex time structure due to multiple
neutron scattering and captures. These events are split into
sub-events in 50~ns time bins. We are interested in two kinds of
events. The first kind has two sub-events. The first sub-event has
deposited energy in the range of 1 to 8~MeV, followed by a sub-event
with deposited energy in the range of 6 to 12~MeV in a time window
of 1 to 200~$\mu$s. These events, called fast neutron events, can
mimic the antineutrino signal as correlated backgrounds. The energy
spectrum of the prompt signal of the fast neutron events, e.g.~at
the far site, is shown in Fig.~\ref{fig:fastn} up to 50~MeV.
The other kind of events has only one sub-event with deposited
energy in range of 6 to 12~MeV. These events, when combined with the
natural radioactivity events, can provide the delayed signal to form
the uncorrelated backgrounds. We call them single neutron events.
Most of the single neutron events are real thermalized neutrons
while others are recoil protons that fall into the 6--12~MeV energy
range accidentally. About 1.5\% thermalized neutrons will survive
the 200~$\mu$s cut, even though its parent muon is tagged. This
inefficiency will be also taken into account when calculating the
single neutron rate. The neutron simulation results are listed in
Table~\ref{tab_fastn}.
\begin{table}[!htb]
\begin{center}
\begin{tabular} {|c||c|c|c|c|} \hline
 \multicolumn{2}{|c|}{} & DYB site & LA site & far site \\ \hline\hline
 fast neutron    & tagged     & 19.6 & 13.1 & 2.0 \\ \cline{2-5}
 (/day/module)   & untagged   &  0.5 & 0.35 & 0.03 \\ \hline\hline
 single neutron  & tagged     & 476  & 320  & 45 \\ \cline{2-5}
 (/day/module)   & untagged   &  8.5 & 5.7  & 0.63 \\ \hline
\end{tabular}
\caption{Neutron rates in a 20-ton module at the Daya Bay sites.
The rows labelled "tagged" refer to the case where the parent muon
track traversed and was detected by the muon detectors, and thus it could be tagged.
Rows labelled "untagged" refer to the case where the muon track
was not identified by the muon detectors. \label{tab_fastn}}
\end{center}
\end{table}

\par
The rate and energy spectrum of the fast neutron backgrounds can be
studied with the tagged sample. 

\subsubsection{Cosmogenic Isotopes}
\label{sssec:sys_cosmogenic}

Cosmic muons, even if they are tagged by the muon identifier, can
produce radioactive isotopes in the detector scintillator which decay
by emitting both a beta and a neutron ($\beta$-neutron emission
isotopes). Some of these so-called cosmogenic radioactive isotopes
live long enough such that their decay cannot be reliably associated
with the last tagged muon. Among them, $^8$He and $^9$Li with
half-lives of 0.12~s and 0.18~s, respectively, constitute the most
serious correlated background sources. The production cross section of
these two isotopes has been measured with muons at an energy of
190~GeV at CERN~\cite{6hagner}. Their combined cross section is
$\sigma(^9{\rm Li}+^8{\rm He}) = (2.12\pm 0.35)~\mu{\rm barn}$.  Since
their lifetimes are so close, it is hard to extract individual cross
sections. About 16\% of $^8$He and 49.5\% of $^9$Li will decay by
$\beta$-neutron emission. Using the muon flux and mean energy at each
detector site (from Section~\ref{sssec:sys_cosmic}) and an energy
dependent cross section, $\sigma_{\rm tot}(E_\mu)
\propto E_\mu^\alpha$, with $\alpha = 0.73$, the estimated
$^8$He+$^9$Li backgrounds are listed in Table~\ref{tab:li9}.
\begin{table}[!htb]
\begin{center}
\begin{tabular}{|c||c|c|c|} \hline
 & DYB site & LA site & Far site \\  \hline\hline
($^8$He+$^9$Li)/day/module & 3.7 & 2.5 & 0.26 \\  \hline
\end{tabular}
\caption{ $^8$He+$^9$Li rates in a 20-ton module at the Daya Bay
sites.}\label{tab:li9}
\end{center}
\end{table}

\par
The recent Double Chooz paper~\cite{dc0606025} includes new
reactor-off data from Chooz~\cite{2chooz} that allow a better
separation of $^9$Li from fast neutron background. This basically
comes from including previously unreleased high energy data in the
fit. The extracted $^9$Li background level was 0.7$\pm$0.2
events/day. The mean muon energy in Chooz was $\sim$ 60 GeV, almost
the same as the Daya Bay near site (55~GeV) and the Ling Ao near
site (60~GeV). The fitting is based on the assumption that the fast
neutron background is flat in energy spectrum. Scaling from the
Chooz result, the Daya Bay experiment will have 8.0, 5.4, and 0.57
$^9$Li events per module per day at the Daya Bay near site, the Ling
Ao near site, and the far site, respectively. These estimates are
twice as large as the estimates from the CERN cross section.

\par
The KamLAND experiment measures this $^9$Li/~$^8$He background very
well by fitting the time interval since last muon. The muon rate is 0.3~Hz in
the active volume of KamLAND detector. The mean time interval of
successive muons is $\sim 3$ seconds, much longer than the lifetimes
of $^9$Li/~$^8$He. For the Daya Bay experiment, the target volume of
a 20~ton detector module has a cross section around 10~m$^2$, thus
the muon rate is around 10~Hz at the near sites, resulting in a mean
time interval of successive muons shorter than the lifetimes of
$^9$Li/$^8$He. With a modified fitting algorithm, we find that it is
still feasible to measure the isotope background {\it in-situ}.

\par
From the decay time and $\beta$-energy spectra fit, the contribution
of $^8$He relative to that of $^9$Li was determined by KamLAND to be
less than 15\% at 90\% confidence
level~\cite{6kamland0406}. Furthermore, the $^8$He contribution can be
identified by tagging the double cascade $^8$He $\rightarrow ^8$Li
$\rightarrow ^8$Be. So we assume that all isotope backgrounds are
$^9$Li. They can be determined with a maximum likelihood fitting even
at 10~Hz muon rate, by taking all contributions from the preceding
muons into account. The resolution of the background-to-signal ratio
can be determined to be~\cite{wenlj}
\begin{equation}
  \label{eqn:sigb}
\sigma_{b}=\frac{1}{\sqrt{N}}\cdot \sqrt{(1+\tau R_{\mu})^2-1} \,
\end{equation}
where $N$ is the total number of neutrino candidates, $\tau$ is the
lifetime of $^9$Li, and $R_{\mu}$ is the muon rate in the target
volume of detector. The resolution is insensitive to the $^9$Li
level since the statistical fluctuation of neutrino events dominates
the uncertainty. The background-to-signal ratio of $^9$Li background
can be measured to $\sim$ 0.3\% with two 20-ton modules at the near
sites of the Daya Bay experiment and $\sim$ 0.1\% at the far site
with four 20-ton modules, with the data sample of three years of
running. The fitting uses time information only. Inclusion of energy
and vertex information could further improve the precision.

\par
\begin{figure}[!htb]
\begin{minipage}[t]{0.48\textwidth}
 \includegraphics[width=0.85\textwidth]{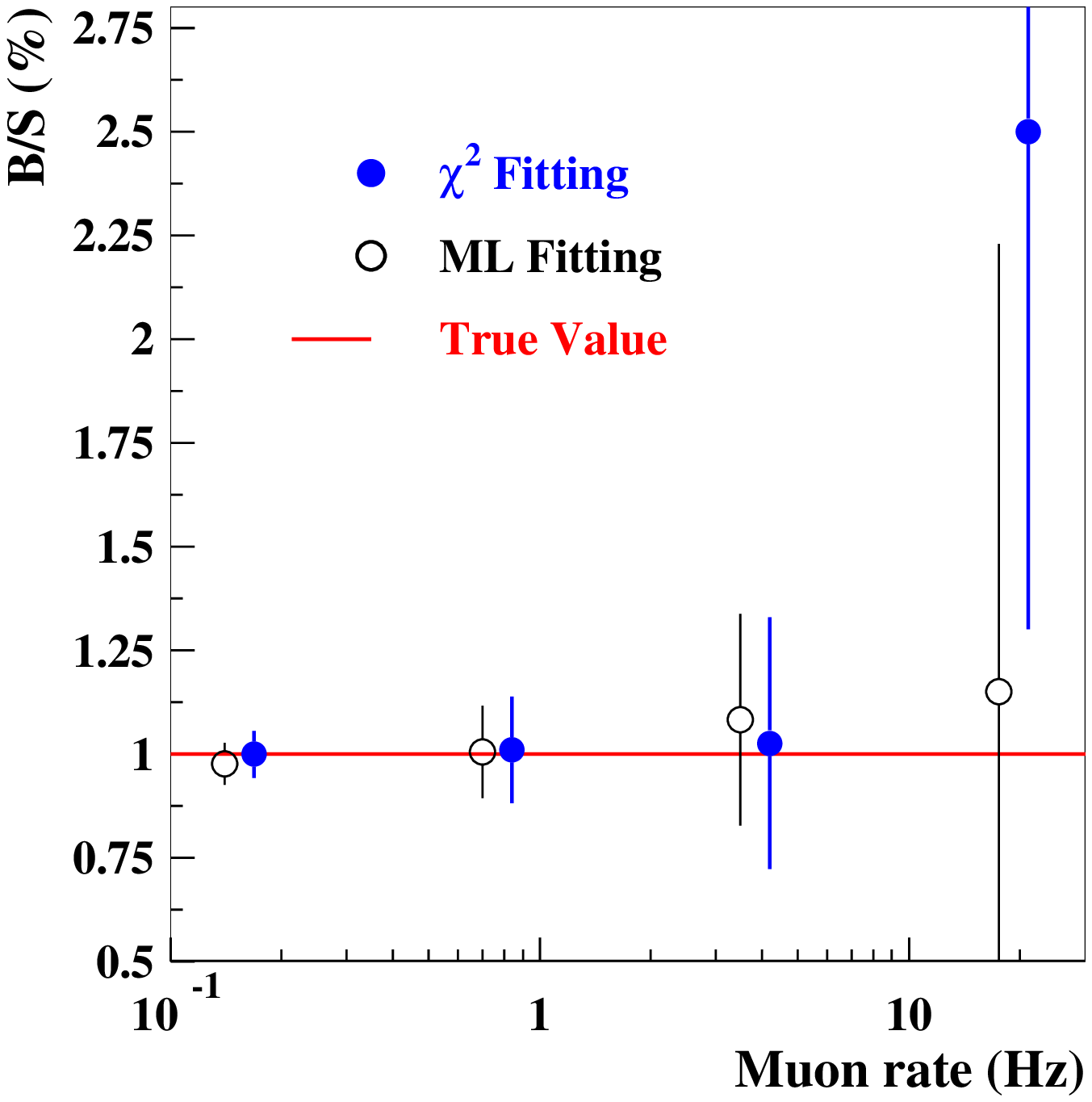}
\caption{Fitting results as a function of the muon rate. The uncertainty
bars show the precision of the fitting. The $\chi^2$ fitting uses
the same muon rate as the maximum likelihood fitting and is shown to the right of it.}
 \label{fig:licomp}
\end{minipage}\hfill
\begin{minipage}[t]{0.48\textwidth}
 \includegraphics[width=0.85\textwidth]{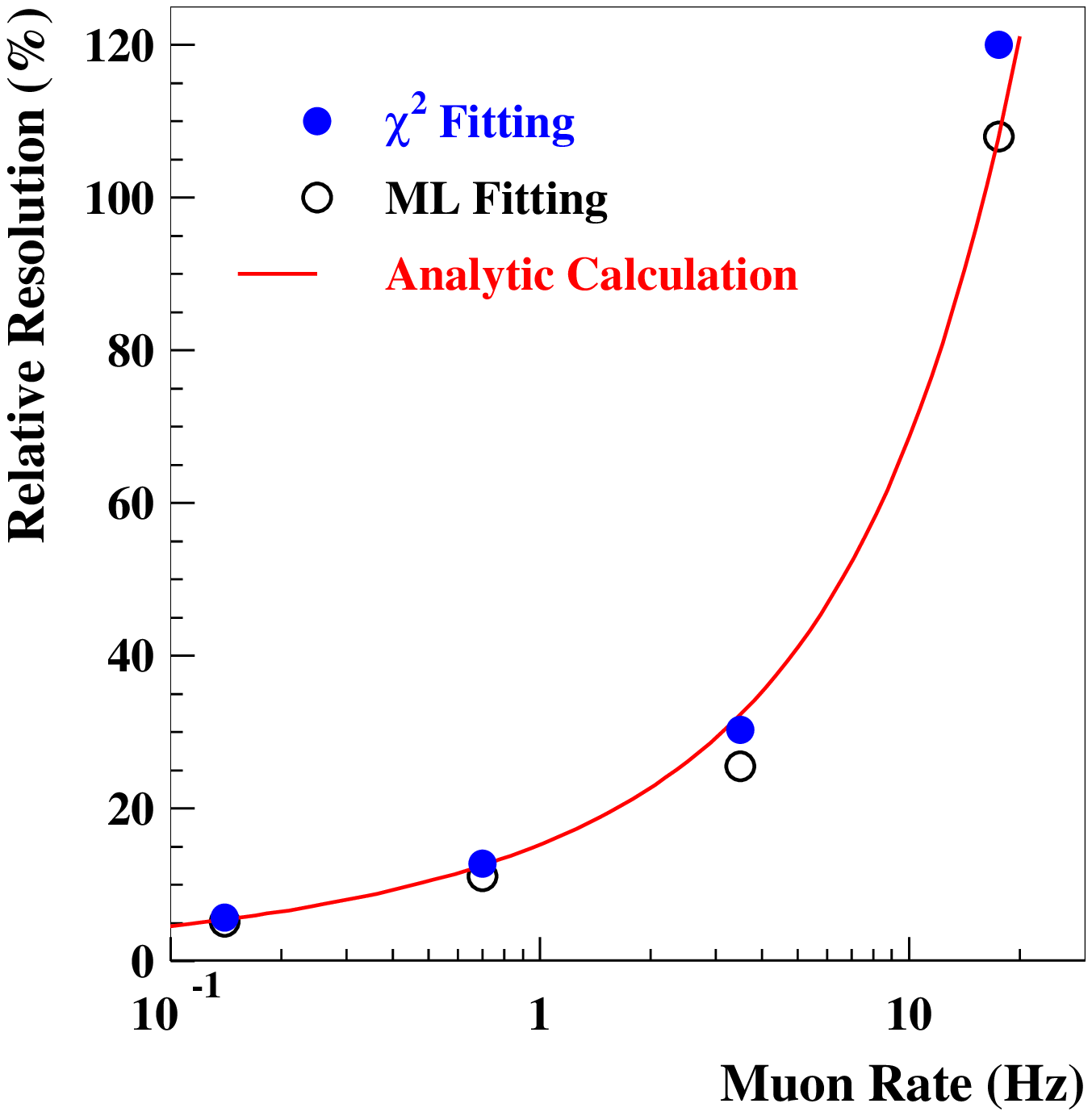}
\caption{The fitting precision as a function of the muon rate,
comparing with the analytic estimation of Eq.~\ref{eqn:sigb}. The
y-axis shows the relative resolution of the background-to-signal
ratio.}
 \label{fig:lisig}
\end{minipage}
\end{figure}

A Monte Carlo has been carried out to check the fitting algorithm.
The background-to-signal ratio is fixed at 1\%. The total number of
neutrino candidates is $2.5\times10^5$, corresponding to the far
site statistical uncertainty, 0.2\%. Figure~\ref{fig:licomp} shows the
fitting results as a function of muon rate.
The data sample generation and fitting were performed 400 times for
each point to get the fitting precision. In Fig.~\ref{fig:lisig} the
fitting precision is compared to the analytic formula
Eq.~\ref{eqn:sigb} with the same Monte Carlo samples. The Monte Carlo
results for minimizing $\chi^2$, the maximum likelihood fit, and the
simple analytical estimation are in excellent agreement.

\par
KamLAND also found that most $^8$He/~$^9$Li background are produced
by showering muons~\cite{6kamland0406}. A 2-second veto of the whole
detector is applied at KamLAND to reject these backgrounds. Roughly
3\% of cosmic muons shower in the detector. It is not feasible for
Daya Bay to apply a 2-second veto since the dead time of the near
detector would be more than 50\%. However, if the Daya Bay detector
is vetoed for 0.5~s after a showering muon, about 85\% of the
$^8$He/~$^9$Li backgrounds caused by shower muons can be rejected.
Approximately 30\% of the ${}^8$He/~${}^9$Li background will remain:
$\sim 15\%$ from non-showering muons and $\sim15\%$ from showering
muons. Although additional uncertainties may be introduced due to
the uncertainties in the relative contributions from showering and
non-showering muons and the uncertainties arising from the
additional cuts (e.g., increased dead time), this rejection method
can cross check the fitting method and firmly determine the
background-to-signal ratio to 0.3\% at the near sites and to 0.1\%
at the far site.

\par
Some other long-lived cosmogenic isotopes, such as $^{12}$B/$^{12}$N,
beta decay without an accompanying neutron. They can not form
backgrounds themselves but can fake the delayed `neutron' signal
of an accidental background if they have beta decay energy in the 6--10~MeV
range. The expected rates from these decays in the antineutrino
detector are listed in Table~\ref{tabiso}. The $^{12}$B/$^{12}$N cross
section is taken from KamLAND~\cite{6kamland0406} and the others are
taken from measurement at CERN~\cite{6hagner}. They are extrapolated
to Daya Bay mean muon energies using the power law $\sigma_{\rm
tot}(E_\mu) \propto E_\mu^{0.73}$. The total rates of all these
isotopes of visible energy in detector in the 6--10~MeV range, where they can
be misidentified as a neutron capture signal on Gadolinium, are 210,
141, and 14.6 events per module per day at the Daya Bay near site, the
Ling Ao near site, and the far site, respectively.  The dominant
contribution is from $^{12}$B/$^{12}$N. KamLAND found that $^{12}$N
yield is smaller than 1\% of $^{12}$B. Since the half-life of $^{12}$B
is short comparing to the mean muon interval, the rate can be well
determined {\it in situ} by fitting the time since last muon. Using
Eq.~\ref{eqn:sigb}, the yield can be determined to a precision of
0.34, 0.25, and 0.015 events per module per day at the Daya Bay near
site, the Ling Ao near site, and the far site, respectively, using
three years' data sample.  Therefore, we expect those isotopes will
introduce very little uncertainties in the background subtraction. On
the other side, these isotopes, uniformly produced inside the
detector, can be used to monitor detector response.
\begin{table}[!hbt]
\begin{center}
\begin{tabular}[c]{|c|c|c||c|c|c|} \hline
 isotopes    &  $E_{\rm max}$ & $T_{1/2}$(s) & DYB site & LA site & far site \\
 & (MeV) & (s) & (/day/module)& (/day/module)& (/day/module) \\\hline\hline
 $^{12}$B/$^{12}$N & 13.4 ($\beta^-$)& 0.02/0.01 & 396 & 267 & 27.5 \\
 $^9$C  &  16.0 ($\beta^+$) & 0.13 & 16.6 & 11.2 & 1.15 \\
 $^8$B  &  13.7 ($\beta^+$) & 0.77 & 24.5 & 16.5 & 1.71 \\
 $^8$Li &  16.0 ($\beta^-$) & 0.84 & 13.9 & 9.3 & 0.96 \\
 $^{11}$Be & 11.5 ($\beta^-$) & 13.8 & $<$8.0 & $<$5.4 & $<$0.56  \\ \hline
 \multicolumn{3}{|c|}{Total in 6-10 MeV} & 210 & 141 & 14.6  \\ \hline
\end{tabular}
\caption{Cosmogenic radioactive isotopes without neutron emission
but with beta decay energy greater than 6~MeV. Cross sections are
taken from KamLAND~\protect\cite{6kamland0406} ($^{12}$B/$^{12}$N)
and Hagner~\protect\cite{6hagner} (others).\label{tabiso}}
\end{center}
\end{table}

\subsubsection{Radioactivity}
\label{sssec:sys_radioactivity}

\par
Natural radioactivity and the single neutron events induced by
cosmic muons may occur within a given time window accidentally to
form an uncorrelated background. The coincidence rate is given by
$R_\gamma R_n \tau, $ where $R_\gamma$ is the rate of natural
radioactivity events, $R_n$ is the rate of spallation neutron, and
$\tau$ is the length of the time window. With the single neutron
event rate given in the previous section, the radioactivity should
be controlled to 50~Hz to limit the accidental backgrounds $<0.1$\%.
The accidental backgrounds can be well determined {\it in-situ} by
measurement of the individual singles rates from radioactivity and
the single neutrons. The energy spectrum can be also well
determined.

\par
Past experiments suppressed uncorrelated backgrounds with a
combination of carefully selected construction materials,
self-shielding, and absorbers with large neutron capture cross
section. However, additional care is necessary to lower the detector
energy threshold much below 1~MeV. A higher threshold will introduce a
systematic uncertainty in the efficiency of detecting the positron. In the
following, the singles rate is from radioactivity depositing $>$1~MeV
of visible energy in detector.

\par
Radioactive background can come from a variety of sources. For
simplicity, U, Th, K, Co, Rn, Kr in the following text always mean
their radioactive isotopes $^{238}$U, $^{232}$Th, $^{40}$K,
$^{60}$Co, $^{222}$Rn, $^{85}$Kr. The radioactive sources include
\begin{itemize}
\item U/Th/K in the rocks around the detector hall.
\item U/Th/K in the water buffer.
\item Co in the detector vessel and other supporting structures.
\item U/Th/K in weld rods.
\item U/Th/K in the PMT glass.
\item U/Th/K in the scintillator.
\item U/Th/K in materials used in the detector.
\item Dust and other impurities
\item Rn and Kr in air.
\item Cosmogenic isotopes.
\end{itemize}

\par
The radioactivity of rock samples from the Daya Bay site has been
measured by several independent groups, including the Institute for
Geology and Geophysics (IGG). The concentrations are: $\sim$10~ppm for
U, $\sim$30~ppm for Th, and $\sim$5~ppm for K.  The effect of the rock
radioactivity on the antineutrino detectors has been studied with
Monte Carlo. With the shielding of 2.5-meter water buffer and 45~cm
oil buffer, there are 0.65~Hz, 2.6~Hz, and 0.26~Hz singles rates with
visible energy greater than 1~MeV in each antineutrino detector module
for U/Th/K, respectively. The total rate is $\sim$3.5~Hz.

\par
\begin{figure}[!htb]
  \begin{center}
 \includegraphics[height=7cm]{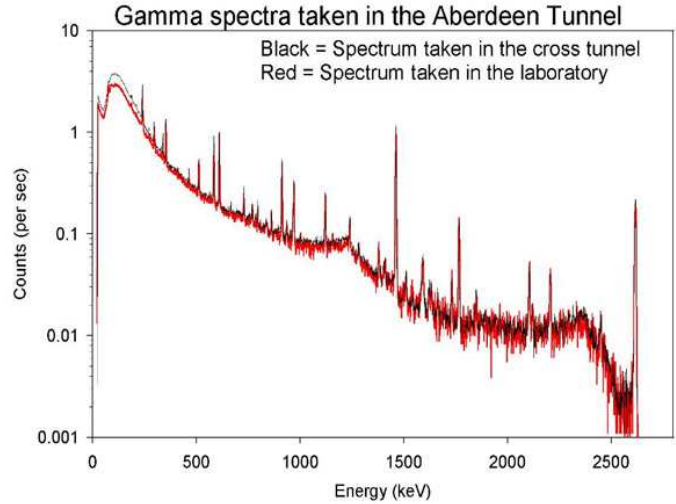}
  \caption{Spectrum of natural radioactivity measured with a Ge crystal
  in the Hong Kong Aberdeen Tunnel. Prominent peaks for $^{40}$K 
(1.461~MeV) and $^{208}$Tl  (2.615~MeV) are clearly evident along with many 
other lines associated with the U/Th series.}\label{fig:natspec}
  \end{center}
\end{figure}

The geological environment and rock composition are very similar in
Hong Kong and Daya Bay. The spectrum of natural radioactivity that we
have measured of the rock in the Aberdeen Tunnel in Hong Kong is shown
in Fig.~\ref{fig:natspec}.

\par
The water buffer will be circulated and purified to achieve a long
attenuation length for Cherenkov light as well as low
radioactivity. Normally tap water has 1~ppb U, 1~ppb Th, and also
1~ppb K. If filling with tap water, the water buffer will contribute
1.8~Hz, 0.4~Hz, and 6.3~Hz singles rates from U/Th/K, respectively.
Purified water in the water pool will have much lower
radioactivity. Thus the radioactivity from water buffer can be ignored.

\par
The Co in stainless steel varies from batch to batch and should be
measured before use as detector material, such as the outer
vessel. U/Th/K concentration in normal weld rods are very high.
There are non-radioactivity weld rods commercially available. Weld
rods TIG308 used in KamLAND were measured to have $<$1~ppb Th,
$0.2\pm0.08$~ppb U, $0.1\pm0.03$~ppb K, and $2.5\pm0.04$~mBq/kg
Co, five orders of magnitude lower than normal weld rods. The
welded stainless steel in KamLAND has an average radioactivity of
3~ppb Th, 2~ppb U, 0.2~ppb K, and 15~mBq/kg Co. Assuming the same
radioactivity for the vessel of the Daya Bay neutrino detector
module, the corresponding rate from a 20-ton welded stainless
steel vessel are 7~Hz, 4.6~Hz, 1.5~Hz, 4.5~Hz for U/Th/K/Co,
respectively for a total of 17.6~Hz.

\par
A potential PMT candidate is the Hamamatsu R5912\footnote{The R5912 is a
newer version of the R1408 used by SNO~\cite{NIMa449}.} with low
radioactivity glass. The concentrations of U and Th are both less
than 40~ppb in the glass, and that of K is 25~ppb. The Monte Carlo
study shows that the singles rate is 2.2~Hz, 1~Hz, 4.5~Hz for
U/Th/K, respectively, with a 20~cm oil buffer from the PMT surface
to the liquid scintillator. The total rate from the PMT glass is
7.7~Hz.

\par
Following the design experience of Borexino and Chooz, backgrounds
from impurities in the liquid scintillator can be reduced to the
required levels. A major source is the U/Th contamination in the
Gadolinium, which can be purified before doped into liquid
scintillator. The U/Th/K concentration of $10^{-12}g/g$ in liquid
scintillator will contribute only 0.8~Hz of background in a 20-ton
detector module.

\par
Radon is one of the radioactive daughters of $^{238}$U, which can
increase the background rate of the experiment. The Radon
concentration in the experimental halls can be kept to an
acceptable level by ventilation with fresh air from outside. Since
the neutrino detector modules are immersed in a 2.5-meter thick
water buffer, it is expected that the radon contribution, as well
as the krypton, can be safely ignored for the water pool design.

\par
The $\beta$ decay of long lived radioactive isotopes produced by
cosmic muons in the scintillator will contribute a couple of Hz at
the near detector, and less than 0.1~Hz at the far detector. The
rate of muon decay or muon capture are 2--6\% of the muon rate. So
they can be ignored when viewed as a source of singles.

\subsubsection{Background Subtraction Uncertainty}
\label{sssec:sys_subtraction}

\par
There are other sources of backgrounds, such as cosmogenic nuclei,
stopped-muon decay, and muon capture. While they are important for a
shallow site, our study shows that they can be safely ignored at
Daya Bay.

\par
Assuming a muon efficiency of 99.5\%, the three major backgrounds are
summarized in Table~\ref{tabbkg} while the other sources are
negligible (the signal and singles rates are also included).
\begin{table}[!hbt]
\begin{center}
\begin{tabular}[c]{|c||c|c|c|} \hline
    & DYB site & LA site & far site \\  \hline\hline
 Antineutrino rate (/day/module)   & 930 & 760 & 90   \\
 Natural radiation (Hz) & $<$50 & $<$50 & $<$50 \\
 Single neutron (/day/module)  & 18   & 12 & 1.5   \\
 $\beta$-emission isotopes  & 210 & 141 & 14.6 \\
 Accidental/Signal   & $<$0.2\% & $<$0.2\% & $<$0.1\% \\
 Fast neutron/Signal & 0.1\% & 0.1\% & 0.1\% \\
 $^8$He$^9$Li/Signal & 0.3\% & 0.2\% & 0.2\% \\ \hline
\end{tabular}
\caption{Summary of signal and background rates in the antineutrino 
detectors at Daya Bay. A neutron detection efficiency of
78\% has been applied to the antineutrino and single-neutron
rates.\label{tabbkg}}
\end{center}
\end{table}
In our sensitivity study, the uncertainties were taken to be 100\% for the
accidental and fast neutron backgrounds. The ${}^8$He/~${}^9$Li
background can be measured to an uncertainty of 0.3\% and 0.1\% at the
near and far sites, respectively.

\begin{figure}[!htb]
  \begin{center}
 \includegraphics[height=7cm]{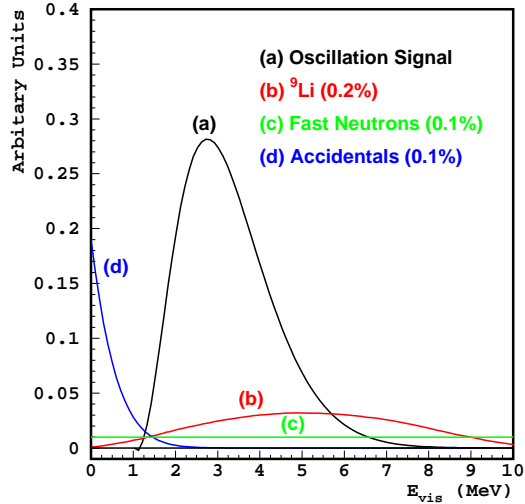}
\caption{Spectra of three major backgrounds for the Daya Bay
experiment and their size relative to the oscillation signal,
which is the difference of the expected neutrino signal without
oscillation and the `observed' signal with oscillation for
$\sin^22\theta_{13}=0.01$.}\label{fig:bkgspec}
  \end{center}
\end{figure}

The rates and energy spectra of all three major backgrounds can be
measured {\it in-situ}. Thus the backgrounds at the Daya Bay
experiment are well controlled. The simulated energy spectra of
backgrounds are shown in Fig.~\ref{fig:bkgspec}. The
background-to-signal ratios are taken at the far site.

\subsection{Sensitivity}
\label{ssec:sys_sensitivity}

\par
If $\theta_{13}$ is non-zero, a relative rate deficit will be present at the
far detector due to oscillation. At the same time, the
energy spectra of neutrino events at the near and far detectors will
be different because neutrinos of different energies oscillate at
different frequencies. Both the rate deficit and spectral distortion of
neutrino signal will be exploited in the final analysis to obtain
maximum sensitivity. When the neutrino event statistics are low
($<$400~ton$\cdot$GW$\cdot$y), the sensitivity is dominated by the
rate deficit. For luminosity higher than 8000~ton$\cdot$GW$\cdot$y,
the sensitivity is dominated by the spectral
distortion~\cite{7huber}. The Daya Bay experiment will have
$\sim$4000~ton$\cdot$GW$\cdot$y exposure in three years, so both
rate deficit and shape distortion effects will be important to the
analysis.

\subsubsection{Global $\chi^2$ Analysis}
\label{sssec:sys_chi}

\par
Many systematic uncertainties will contribute to the final
sensitivity of the Daya Bay experiment, and many of them are
correlated. The correlation of the uncertainties must be taken into
account correctly. A rigorous analysis of systematic uncertainties
can be done by constructing a $\chi^2$ function with pull terms,
where the uncertainty correlations can be introduced
naturally~\cite{7huber,7stump,7superk,7sugiyama}:
\begin{eqnarray}  \label{eqn:chispec}
 \chi^2 &=&
 \min_{\gamma}\sum_{A=1}^{8}\sum_{i=1}^{N_{bins}}
 \frac{\left[M_i^A-T_i^A\left(1+\alpha_c +
 \sum_r\omega_r^A\alpha_r + \beta_i + \varepsilon_D
 + \varepsilon_d^A\right) - \eta_f^A F^A_i -
  \eta_n^A N^A_i - \eta_s^A S^A_i \right]^2}
 {T_i^A + \sigma_{b2b}^2}  \nonumber \\
 &+&
   \frac{\alpha_c^2}{\sigma_c^2}
 + \sum_r\frac{\alpha_r^2}{\sigma_r^2}
 + \sum_{i=1}^{N_{bins}}\frac{\beta_i^2}{\sigma_{\rm shp}^2}
 + \frac{\varepsilon_D^2}{\sigma_D^2}
 + \sum_{A=1}^{8}\left[
 \left( \frac{\varepsilon_d^A}{\sigma_d} \right)^2
 + \left( \frac{\eta_f^A}{\sigma_f^A} \right)^2
 + \left( \frac{\eta_n^A}{\sigma_n^A} \right)^2
 + \left( \frac{\eta_s^A}{\sigma_s^A} \right)^2
 \right] \ 
\end{eqnarray}
where $A$ sums over detector modules, $i$ sums over energy bins, and
$\gamma$ denotes the set of minimization parameters,
$\gamma=\{\alpha_c, \alpha_r, \beta_i, \varepsilon_D,
\varepsilon_d^A, \eta_f^A, \eta_n^A, \eta_s^A \}$. The $\gamma$'s
are used to introduce different sources of systematic uncertainties.
The standard deviations of the corresponding parameters are $\{
\sigma_c, \sigma_r, \sigma_{\rm shp}, \sigma_D, \sigma_d,
\sigma_f^A, \sigma_n^A, \sigma_s^A \}$. They will be described in
the following text. $T_i^A$ is the expected events in the $i$-th
energy bin in detector $A$, and $M_i^A$ is the corresponding
measured events. $F^A_i, N^A_i, S^A_i$ are number of fast neutron,
accidental, and ${}^8$He/~${}^9$Li backgrounds, respectively. For
each energy bin, there is a statistical uncertainty $T_i^A$ and a
bin-to-bin systematic uncertainty $\sigma_{b2b}$. For each point in
the oscillation space, the $\chi^2$ function is minimized with
respect to the parameters $\gamma$.

\par
Assuming each uncertainty can be approximated by a Gaussian, this form of
$\chi^2$ can be proven to be strictly equivalent to the more
familiar covariance matrix form $\chi^2 = (M-T)^T V^{-1} (M-T)$,
where $V$ is the covariance matrix of $(M-T)$ with systematic uncertainties
included properly~\cite{7stump}.

\par
To explore the sensitivity to $\theta_{13}$, we use the single
parameter raster scan method. We make an assumption of no
oscillations so that $T_i^A$ are the event numbers without
oscillation. For each given $\Delta m^2_{31}$, the "measured" event
numbers $M_i^A$ are calculated with different $\sin^22\theta_{13}$.
The $\sin^22\theta_{13}$ value corresponding to $\chi^2=2.71$ is the
limit of the experiment to exclude the "no oscillation" assumption
at 90\% confidence level.

 The systematic uncertainties are described in detail:

\begin{itemize}
\item
 The reactor-related correlated uncertainty is $\sigma_c\approx2\%$.
This fully correlated uncertainty will be cancelled by the near-far
relative measurement and has little impact on the sensitivity.

\item The reactor-related uncorrelated uncertainty for core $r$ is
$\sigma_r\approx2\%$. These enter the normalization of the predicted
event rate for each detector $A$ according to the weight fractions
$\omega_r^A$. After minimization, the $\sigma_r$ contribute a total
of  $\sim$0.1\% to the relative normalization of neutrino rate. This
is essentially equivalent to the analysis described in
Section~\ref{ssec:sys_reactor}, and takes into account the
correlations of this uncertainty with the others (like the detector
efficiencies $\varepsilon_d^A$).

\item The spectrum shape uncertainty is $\sigma_{\rm shp}\approx2\%$:
The shape uncertainty is the uncertainty in the neutrino energy
spectra calculated from the reactor information. This uncertainty is
uncorrelated between different energy bins but correlated between
different detectors. Since we have enough statistics at near
detector to measure the neutrino energy spectrum to much better than
2\%, it has little effect on the Daya Bay sensitivity.

\item The detector-related correlated uncertainty is $\sigma_D \approx 2\%$.
Some detection uncertainties are common to all detectors, such as
H/Gd ratio, H/C ratio, neutron capture time on Gd, and the edge
effects, assuming we use the same batch of liquid scintillator and
identical detectors. Based on the Chooz experience, $\sigma_D$ is
(1--2)\%. Like other fully correlated uncertainties, it has little
impact on sensitivity.

\item The detector-related uncorrelated uncertainty is $\sigma_d =
0.38\%$. We take the baseline systematic uncertainty as described in
Section~\ref{ssec:sys_detector}. The goal systematic uncertainty with
swapping is estimated to be 0.12\%.

\item The background rate uncertainties $\sigma_f^A, \sigma_n^A$, 
and $\sigma_s^A$, corresponding to the rate uncertainty of fast
neutron, accidental backgrounds, and ${}^8$He/~${}^9$Li isotopes. They
are listed in Table~\ref{tabbkg}.

\item Bin-to-bin uncertainty $\sigma_{b2b}$:
The bin-to-bin uncertainty is the systematic uncertainty that is
uncorrelated between energy bins and uncorrelated between different
detector modules.  The bin-to-bin uncertainties normally arise from
the different energy scale at different energies and uncertainties
of background energy spectra during background subtraction. The only
previous reactor neutrino experiment that performed spectral
analysis with large statistics is Bugey, which used a bin-to-bin
uncertainty of order of 0.5\%~\cite{7bugey,7yasuda}. With better
designed detectors and much less background, we should have much
smaller bin-to-bin uncertainties than Bugey. The bin-to-bin
uncertainty can be studied by comparing the spectra of two detector
modules at the same site. We will use 0.3\%, the same level as the
background-to-signal ratio, in the sensitivity analysis. The
sensitivity is not sensitive to $\sigma_{b2b}$ at this level. For
example, varying $\sigma_{b2b}$ from 0 to 0.5\% will change the
$\sin^22\theta_{13}$ sensitivity from 0.0082 to 0.0087 at the best
fit $\Delta m^2_{31}$.

\end{itemize}

\par
There are other uncertainties not included in the $\chi^2$ function.
1) Due to the energy resolution, the spectra are distorted. However,
the energy bins used for sensitivity analysis ($\sim 30$ bins) is 
2$\sim$6 times larger than the energy resolution, and the distortion
happens at all detectors in the same way. It has little impact on
the final sensitivity. 2) Detector energy scale uncertainty has
significant impact on detection uncertainties (neutron efficiency
and positron efficiency) which has been taken into account in
$\sigma_d$. An energy scale uncertainty will shift the whole
spectrum, thus directly impacting the analysis, especially on the
best fit values. However, this shift  has very little impact on our
sensitivity computations. 3) Current knowledge on $\theta_{12}$ and
$\Delta m_{21}$ has around 10\% uncertainties. Although the primary
oscillation effect at the Daya Bay baseline is related to
$\theta_{13}$ only, the subtraction of $\theta_{12}$ oscillation
effects introduce very small uncertainties (see
Section~\ref{sssec:physics_reactor_mixing}). We have studied the
above three sources of uncertainty and found that none of them have
a significant impact on the sensitivity of the Daya Bay experiment.
For simplicity, they are ignored in our $\chi^2$ analysis of
sensitivity.

\subsubsection{$\theta_{13}$ Sensitivity}
\label{sssec:sys_sensitivity}

\begin{figure}[!hbt]
\begin{minipage}[t]{0.49\textwidth}
 \includegraphics[width=\textwidth]{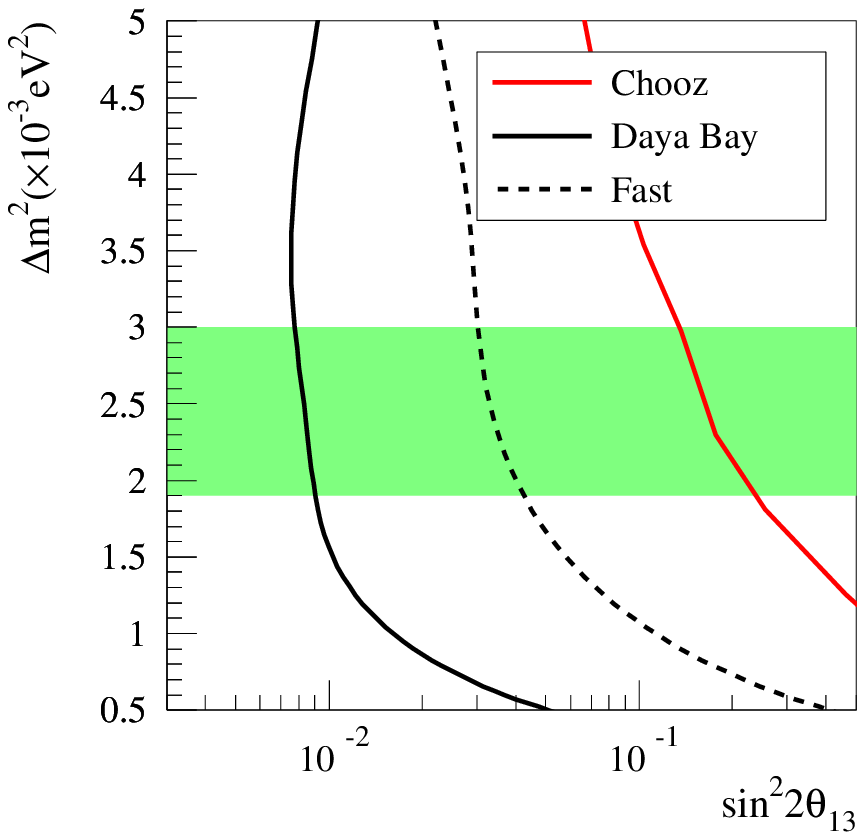}
 \caption{Expected $\sin^22\theta_{13}$ sensitivity at 90\% C.L.
with 3 years of data, as shown in solid black line. The dashed line
shows the sensitivity of a fast measurement with the DYB near site
and mid site only. The red line shows the current upper limit
measured by Chooz.}
 \label{fig:sens3year}
 \end{minipage}
\hfill
 \begin{minipage}[t]{0.49\textwidth}
 \includegraphics[width=\textwidth]{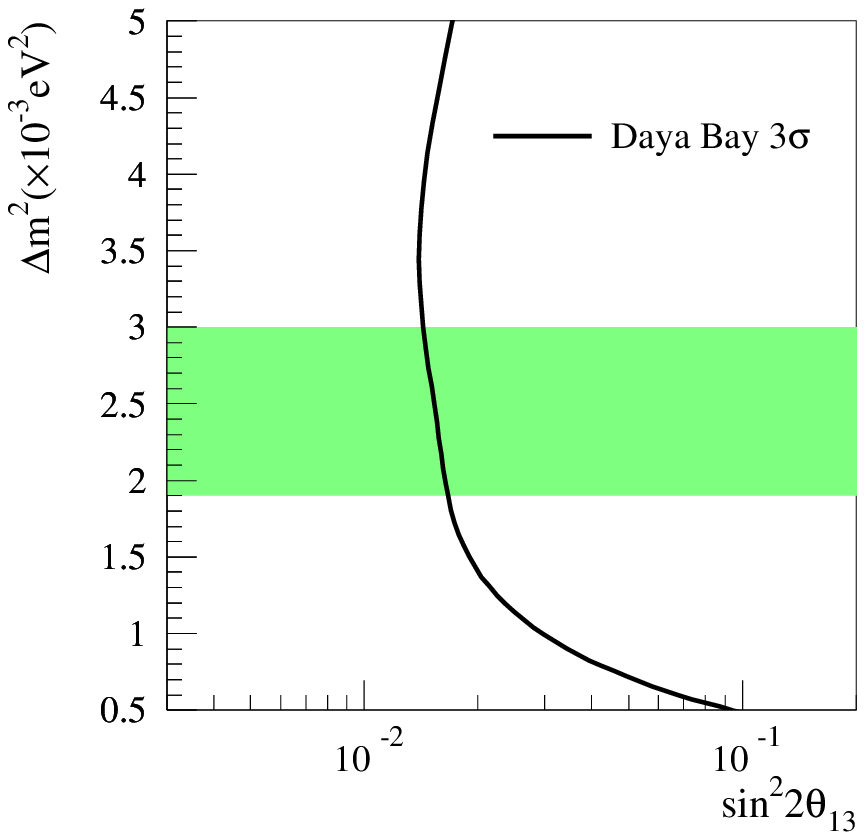}
  \caption{Expected 3$\sigma$ discovery limit for
  $\sin^22\theta_{13}$ at Daya Bay with 3 years of data.}
  \label{fig:3sig}
 \end{minipage}
\end{figure}

Figure~\ref{fig:sens3year} shows the sensitivity contours in the
$\sin^22\theta_{13}$ versus $\Delta m^2_{31}$ plane for three years
of data, using the global $\chi^2$ analysis and the baseline values
for detector-related systematic uncertainties.
The green shaded area shows the 90\% confidence region of $\Delta
m^2_{31}$ determined by atmospheric neutrino experiments. Assuming
four 20-ton modules at the far site and two 20-ton modules at each
near site, the statistical uncertainty is around 0.2\%. The
sensitivity of the Daya Bay experiment with this design can achieve
the challenging goal of 0.01 with 90\% confidence level over the entire
allowed (90\% CL) range of $\Delta m^2_{31}$. At the best
fit $\Delta m^2_{31}=2.5\times10^{-3}$ eV$^2$, the sensitivity is
around 0.008 with 3 years of data. The corresponding values for
other assumptions of systematic uncertainties are shown in
Table~\ref{tab:sens}.

\begin{table}[!hbt]
\begin{center}
\begin{tabular}[c]{|c||c|c|c|} \hline
   Systematic Uncertainty Assumptions: & Baseline & Goal & Goal \\
  & & & {with swapping} \\ \hline\hline
 90\% CL Limit:  & 0.008 & 0.007 & 0.006  \\ \hline
\end{tabular}
\caption{90\% CL sensitivity limit for $\sin^2 2 \theta_{13}$ at
$\Delta m^2_{31}=2.5\times10^{-3}$ eV$^2$ for different assumptions
of detector related systematic uncertainties as considered in
Section~\ref{ssec:sys_detector}. We assume 3 years running for each
scenario. \label{tab:sens}}
\end{center}
\end{table}

\par
\begin{figure}[!hbt]
\begin{center}
 \includegraphics[width=0.48\textwidth]{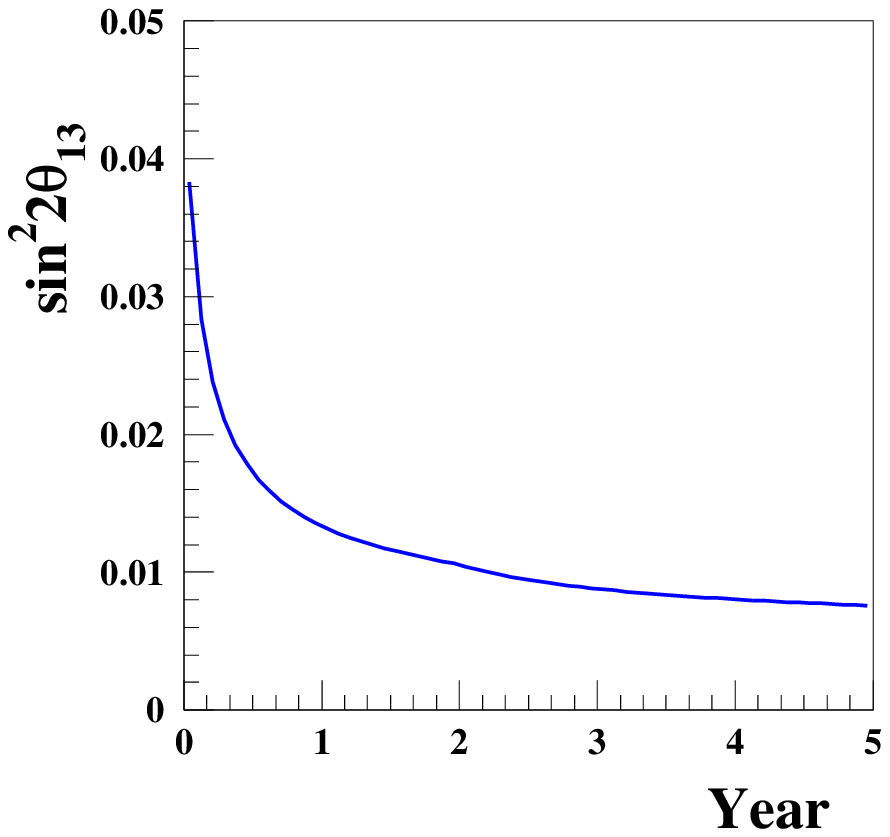}
  \caption{Expected $\sin^22\theta_{13}$ sensitivity at 90\% C.L.
versus year of data taking of the full measurement, with two near
sites and one far site. The value of $\Delta m^2_{31}$ is taken to be
$2.5\times10^{-3}$ eV$^2$.}
  \label{fig:year}
\end{center}
\end{figure}

Figure~\ref{fig:3sig} shows the 3$\sigma$ discovery limit for
$\sin^22\theta_{13}$ at Daya Bay with 3 years of data. At $\Delta
m^2_{31}=2.5\times10^{-3}$ eV$^2$, the corresponding
$\sin^22\theta_{13}$ discovery limit is 0.015. Figure~\ref{fig:year}
shows the sensitivity versus time of data taking. After one year
of data taking, $\sin^22\theta_{13}$ sensitivity will reach 0.014
(1.4\%) at 90\% confidence level.

\par
The tunnel of the Daya Bay experiment will have a total length around
3 km. The tunnelling will take $\sim 2$ years. To accelerate the
experiment, the first completed experimental hall, the Daya Bay near
hall, can be used for detector commissioning. Furthermore, it is
possible to conduct a fast experiment with only two detector sites,
the Daya Bay near site and the mid site. For this fast experiment, the
`far detector', which is located at the mid hall, is not at the
optimal baseline. At the same time, the reactor-related uncertainty
would be 0.7\%, very large compared with that of the full
experiment. However, the sensitivity is still much better than the
current best limit of $\sin^22\theta_{13}$. It is noteworthy that the
improvement comes from better background shielding and improved
experiment design. The sensitivity of the fast option for one year of
data taking is shown in the dashed line in Fig.~\ref{fig:sens3year}.
With one year of data, the sensitivity is $\sim$0.035 for $\Delta
m^2=2.5\times10^{-3}$ eV$^2$, compared with the current limit of 0.17
from the Chooz experiment. This fast option will allow us to gain
valuable experience and a preliminary physics result while
construction is being completed. The higher precision of the completed
experiment will be necessary to fully complement the future long
baseline accelerator experiments as discussed in
Section~\ref{sec:synergy}.

\newpage
\renewcommand{\thesection}{\arabic{section}}
\setcounter{figure}{0}
\setcounter{table}{0}
\setcounter{footnote}{0}

\section{Experimental Site and Laboratories}
\label{sec:civil}

The Daya Bay site is an ideal place to search for oscillations
of antineutrinos from reactors.
The nearby mountain range provides
excellent overburden to suppress cosmogenic background at the
underground experimental halls. 
Since the Daya Bay nuclear power complex consists of multiple
reactor cores, there will be two near detector sites to monitor the yield
of antineutrinos from these cores and one far detector site to look for
disappearance of antineutrinos. It is possible to instrument  another
detector site about half way between the near and far detectors to provide
independent consistency checks.

The proposed experimental site is located at the east side of the
Dapeng peninsula, on the west coast of Daya Bay, where the coastline
goes from southwest to northeast (see Fig.~\ref{fig:DayaVicinity}). It
is in the Dapeng township of the Longgang Administrative District,
Shenzhen Municipality, Guangdong Province. Two mega cities, Hong Kong
and Shenzhen are nearby.  Shenzhen City\footnote{Shenzhen is the first
Special Economic Zone in China. With a total population of about 7
million, many international corporations have their Asian headquarters
there. It is both a key commercial and tourist site in South China.}
is 45~km to the west and Hong Kong is 55~km to the southwest.  The
geographic location is east longitude 114$^\circ$33'00" and north
latitude 22$^\circ$36'00". Daya Bay is semi-tropical and the climate
is dominated by the south Asia tropical monsoon. It is warm and rainy
with frequent rainstorms during the typhoon season in one half of the
year, while relatively dry in the other half. Frost is rare.

The Daya Bay Nuclear Power Plant (NPP) is situated to the southwest
and the Ling Ao NPP to the northeast along the coastline. Each NPP
has two cores that are separated by 88~m. The distance between the
centers of the two NPPs is about 1100~m.  The thermal power,
$W_{\rm th}$, of each core is 2.9~GW.  Hence the total thermal
power available is $W_{\rm th}=11.6$~GW. A third NPP, Ling
Ao II, is under construction and scheduled to come online by
2010--2011. This new NPP is built roughly along the line extended
from Daya Bay to Ling Ao, about 400~m northeast of Ling Ao. The
core type is the same as that of the Ling Ao NPP but with slightly
higher thermal power. When the Daya Bay---Ling Ao---Ling Ao II NPP are
all in operation, the complex can provide a total thermal power of
17.4~GW.

The site is surrounded to the north by a group of hills which slope
upward from southwest to northeast. The slopes of the hills vary from
10$^\circ$ to 45$^\circ$. The ridges roll up and down with smooth
round hill tops. Within 2~km of the site the elevation of the hills
generally vary from 185~m to 400~m.  The summit, called Pai Ya Shan,
is 707~m PRD. 
Due to the construction of the Daya Bay and Ling Ao NPPs,
the foothills along the coast from the southwest to the northeast have
been levelled to a height of 6.6~m to 20~m PRD. Daya Bay experiment
laboratories are located inside the mountain north of the Daya Bay and
Ling Ao NPPs.

There is no railway within a radius of 15~km of the site. The
highway from Daya Bay NPP to Dapeng Township (Wang Mu) is
of second-class grade and 12~m wide. Dapeng Town is connected to
Shenzhen, Hong Kong, and the provincial capital Guangzhou by
highways which are either of first-class grade or expressways.

There are two maritime shipping lines near the site in Daya Bay,
one on the east side and the other on the west side.
Oil tankers to and from Nanhai Petrochemical use the
east side. Huizhou Harbor, which is located in Daya Bay is
13~km to the north. Two general-purpose 10,000-ton docks were
constructed in 1989. Their functions include transporting
passengers, dry goods, construction materials, and petroleum
products. The ships using these two docks take the west line. The
minimum distance from the west line to the power plant site is
about 6~km. Two restricted docks of 3000-ton and 5000-ton
capacity, respectively, have been constructed on the power plant
site during the construction of the Daya Bay
NPP~\cite{BINEReport}.

\subsection{General Laboratory Facilities}
\label{ssec:civil_facilities}

\begin{figure}[!htb]
\centerline{\includegraphics[width=0.9\textwidth]{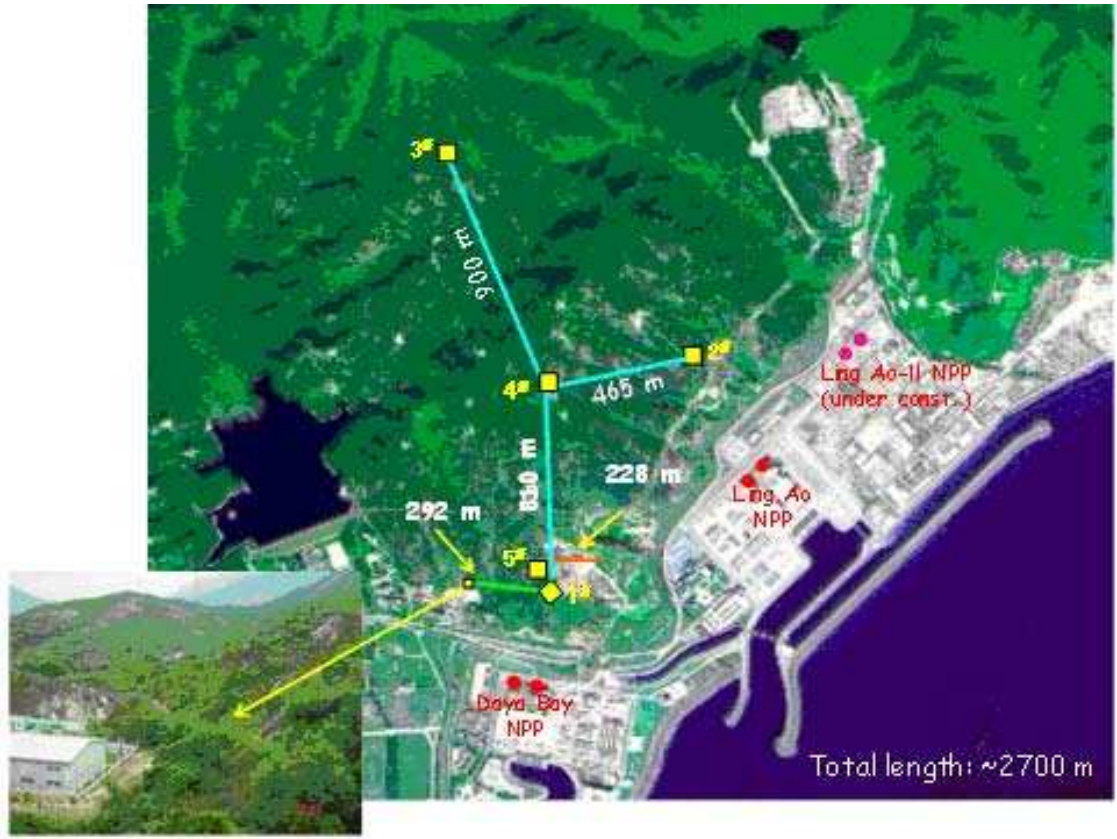}}
\caption{Layout of the Daya Bay and Ling Ao cores, the future Ling Ao
II cores and possible experiment halls.
The entrance portal is shown at the bottom-left. Five experimental
halls marked as \#1 (Daya Bay near hall), \#2 (Ling Ao near hall),
\#3 (far hall), \#4 (mid hall), \#5 (LS filling hall) are shown. The green line
represents the access tunnel, the blue lines represent the main tunnels and the pink
line represents the construction tunnel. The total tunnel length is
about 2700~m}. \label{fig:design}
\end{figure}

The laboratory facilities include access tunnels connected to the
entrance portal, a construction tunnel for waste rock transfer, a
main tunnel connecting all the four underground detector
halls, a LS filling hall, counting rooms, water and electricity
supplies, air ventilation, and communication. There is an assembly
hall and control room near the entrance portal on surface. The
approximate location of the experiment halls and the layout of the
tunnels are shown in Fig.~\ref{fig:design}. All experimental halls
are located at similar elevations, approximately $-$20~m PRD.

\subsubsection{Tunnels}
\label{sssec:civil_tunnels}

\begin{figure}[!h]
\centerline{\includegraphics[width=0.6\textwidth]{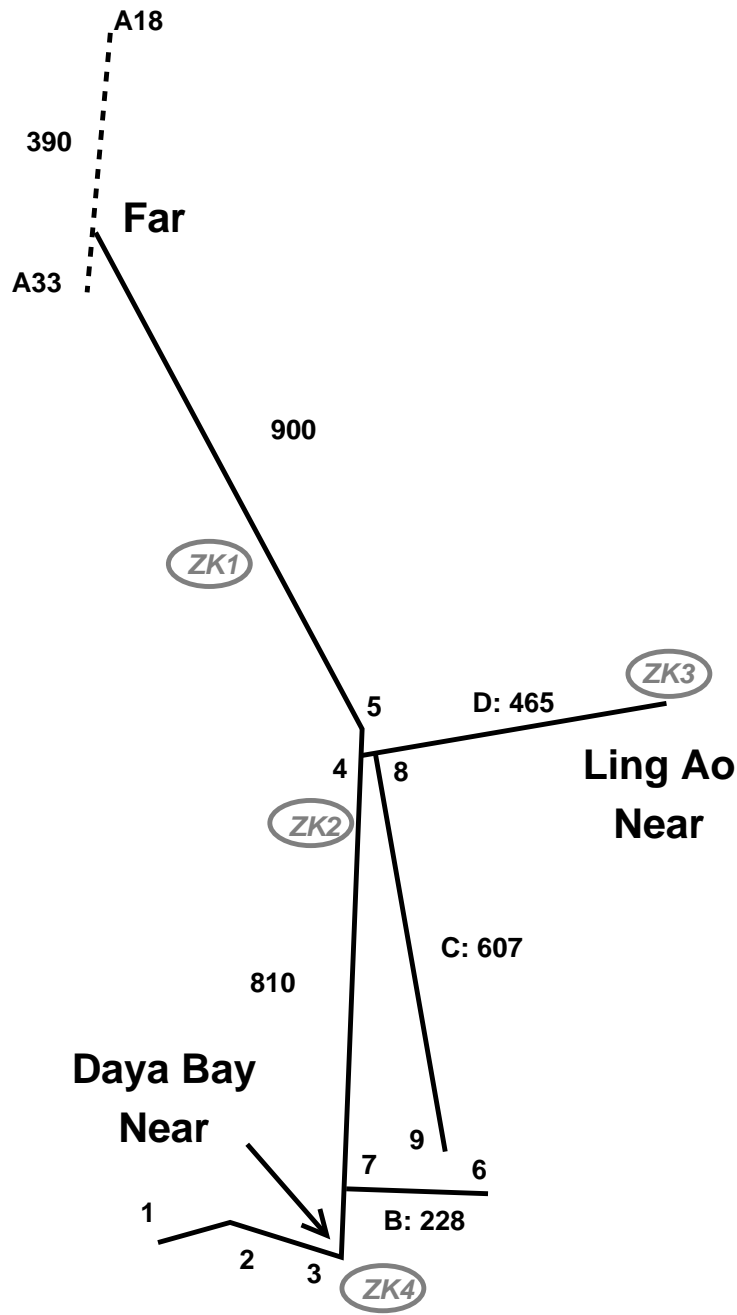}}
\caption{Plan view of the experimental halls and tunnels from the site survey
(not a detailed tunnel design). All distances are in meters.
Line A\{1-2-3-7-4-5-far site\} has a total length of 2002~m; Line
B\{7-6\} has a total length of 228~m; Line C\{8-9\} has a total length of 607~m;
Line D\{4-8-Ling Ao near\} has a total length of 465~m. Line E is the
dashed line on the top across far site. The four bore hole sites are marked
as ZK1, ZK2, ZK3, ZK4 from north to south.} \label{fig:tunnel}
\end{figure}

A sketch of the layout of the tunnels is shown in
Fig.~\ref{fig:tunnel}. There are three branches, which are
represented by line\{3-7-4-5\}, line\{4-8-Ling Ao near\} and
line\{5-far site\}, form the horizontal main tunnel extending from
a junction near the mid hall to the near and far underground
detector halls. The lines marked as A, B, C, D and E are for the
geophysical survey. Line E, which is a dashed line on the top of
figure across the far site, is the geophysical survey line
investigated if the far site needs to be pushed further from the cores
as a result of future optimizations.
Line\{1-2-3\} is the access tunnel with a length of 292~m. Lines B
and C are from the survey for the design of the construction
tunnel (which may have different options for cost optimization).

Figure~\ref{fig:design} shows the entrance portal of the
access tunnel behind the on-site hospital and to the west of
the Daya Bay near site. From the portal to the Daya Bay near site is a
downward slope with a grade of
less than 10\%.  A sloped access tunnel will allow the underground
facilities to be located deeper with more overburden.

The access and main tunnels will be able to accommodate vehicles
transporting equipment of different size and weight. The grade of the
main tunnel will be 0.3\% upward from the Daya Bay near hall to the
mid hall, and from the mid hall to both the Ling Ao hall and the far
hall. The slightly sloped tunnel has two important functions: to
ensure a nearly level surface for the movement of the heavy detectors
filled with liquid scintillator inside the main tunnel and to channel
any water seeping into the tunnel to a collection pit which is
located at the lowest point near the Daya Bay near site. The collected
water will be pumped to the surface.

The entrance portal of the construction tunnel is near the lower
level of the Daya Bay Quarry. The length of this tunnel is 228~m
from the entrance to the junction point with the main tunnel if
the shortest construction tunnel option is chosen (see
Fig.~\ref{fig:design}). During most of the tunnel construction,
all the waste rock and dirt is transferred through this tunnel
to the outside in order to minimize the interference with the
operation of the hospital and speed up the tunnel construction.
We expect the access tunnel and the Daya Bay near hall to be finished
earlier than the far and Ling Ao halls since it requires much less
tunnelling. After the work on this section of tunnel is finished, the Daya Bay
near hall will be available for detector installation. Since the
construction tunnel is far from the access tunnel and the
Daya Bay near hall, we can therefore avoid interference with the rest
of the excavation activities and the assembly of detectors in the Daya Bay
near site can proceed in parallel. The cross section of the construction tunnel can be
smaller than the other tunnels; it is only required to be large enough for rock and
dirt transportation. The grade and the length of this construction
tunnel will be determined later to optimize the construction cost and 
schedule.

Excavation will begin from the construction portal. Once it
reaches the intersection of the main tunnel, the excavation will
proceed in parallel in the directions of the Daya Bay near hall and the mid hall.
Once the tunnelling reaches the the mid hall, it will proceed parallel in the
direction of the far hall and the Ling Ao hall.

The total length of the tunnel is about 2700~m. The amount of
waste to be removed will be about 200,000~m$^3$. About half of the
waste will be dumped in the Daya Bay Quarry to provide
additional overburden to the Daya near site which is not far away
from the Quarry. This requires additional protection slopes and
retaining walls. The rest of the waste could be disposed of along
with the waste from the construction of the Ling Ao II NPP. Our tunnel waste is
about one tenth of the Ling Ao II NPP waste.

\subsection{Site Survey}
\label{ssec:civil_survey}

The geological integrity of the Daya Bay site was studied in order to
determine its suitability for the construction of the underground
experimental halls and the tunnels connecting them. The survey
consisted of a set of detailed geological surveys and studies: (1)
topographic survey, (2) engineering geological mapping, (3)
geophysical exploration, (4) engineering drilling,
(5) On-site tests at boreholes and (6) laboratory tests. The site survey
has been conducted by the Institute of Geology and Geophysics (IGG)
of Chinese Academy of Sciences (CAS). The work started in May 2005
and was completed in June 2006.

\subsubsection{Topographic Survey}
\label{sssec:civil_topographic}

The topographic survey is essential for determining the position
of the tunnels and experimental halls. From the topographic survey the
location of the cores relative to the experimental halls is determined, as is the overburden
above each of the experimental halls. This measurement of the overburden
was input to the optimization of the experimental sensitivity.
It is also needed for the
portal design and construction. Appropriate maps are constructed
out of this measurement. The area surveyed
lies to the north of the Daya Bay complex
The area of the survey extends 2.5~km in the north-south direction and varies from
450~m to 1.3~km in the east-west direction as determined by the
location of the experimental halls and tunnels. The total area
measured is 1.839~km$^2$. The results of the survey are plotted at a scale of 1:2000.

\begin{figure}[!htb]
\begin{center}
 \includegraphics[width=0.85\textwidth]{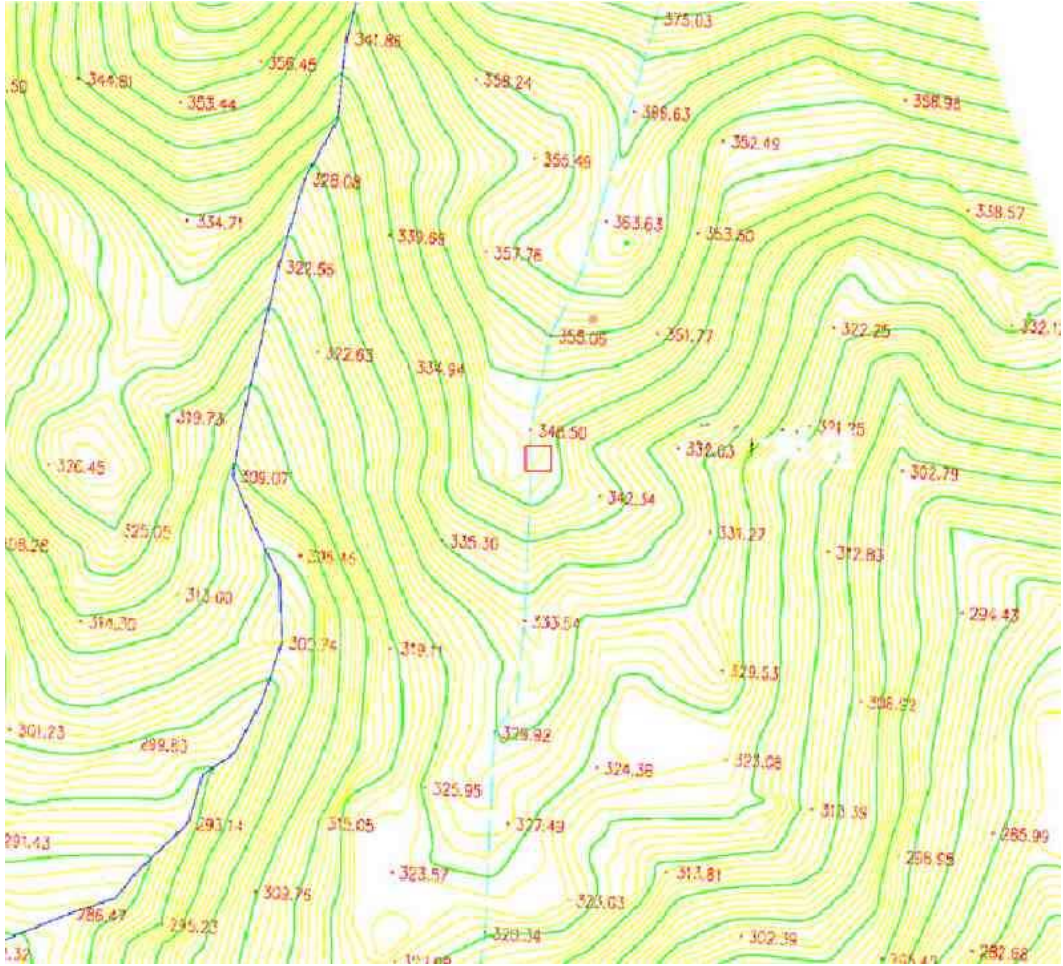}
 \caption{Topographical map in the vicinity of the far site. The location of the
far detector hall is marked by a red square in the middle of the
map.}
 \label{fig:topog}
\end{center}
\end{figure}

The instrument used for the topographic measurement is a LEICA
TCA2003 Total Station, with a precision of $\pm$0.5" in
angle and $\pm$1~mm in distance.
Based on four very high standard control points that exist in the
area, twenty-six high grade control points and forty-five map baseline points are
selected. In total, 7000 points are used to obtain the topographic
map. As an example, Fig.~\ref{fig:topog} shows the topographic map
around the far site.
The altitude difference between adjacent contour lines is one
meter. The area around the entrance portal, which is behind the
local hospital, and the two possible construction portals are
measured at the higher resolution of 1:500. The cross sections along
the tunnel line for the access and construction portals are measured
at an even higher resolution of 1:200. The positions of the
experimental halls, the entrance portal, and the construction
portal are marked on the topographic map.

\subsubsection{Engineering Geological Mapping}
\label{sssec:civil_mapping}

Geological mapping has been conducted in an area extending about 2.5~km
in the north-south direction and about 3~km in the east-west
direction. From an on-the-spot survey to fill in the
geological map of the area, a listing of the geological faults,
underground water distribution and contact interface between
different rocks and weathering zones could be deduced. The
statistical information on the orientation of joints is
used to deduce the general property of the underground rock, and
the determination of the optimal tunnel axes. The
survey includes all the areas through which the tunnels will pass and
those occupied by the experimental halls. Reconnaissance has been
performed along 28 geological routes, of 18.5~km total length.
Statistics of 2000 joints and rock mine skeletons are made at 78
spots. Rock mine appraisals are done with 36 sliced samples.

Surface exploration and trenching exposure show that the landforms
and terrain are in good condition.  There are no karsts,
landslides, collapses, mud slides, empty pockets, ground sinking
asymmetry, or hot springs that would affect the stability of the
site. There are only a few pieces of weathered granite scattered around
the region.

The mountain slopes in the experimental area, which vary from $10^\circ$ to
$30^\circ$, are stable and the surface consists mostly of lightly
effloresced granite. The rock body is comparatively
integrated.  Although there is copious rainfall which can cause
erosion in this coastal area, there is no evidence of large-scale
landslide or collapse in the area under survey. However, there are
small-scale isolated collapses due to efflorescence of the
granite, rolling and displacement of effloresced spheroid rocks.

\begin{figure}[!htb]
\begin{center}
 \includegraphics[width=0.85\textwidth]{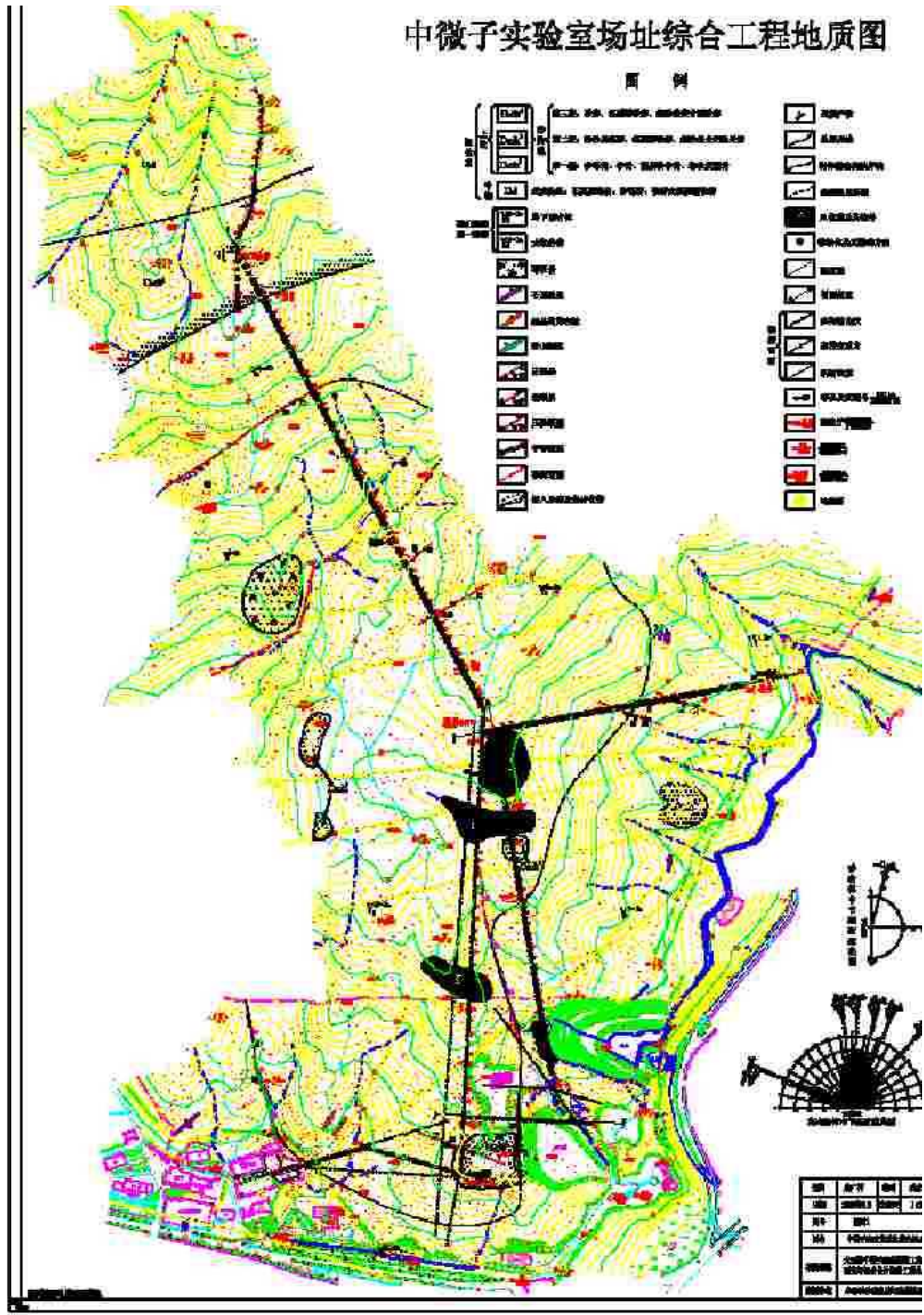}
 \caption{Geological map of the experimental site.}
 \label{fig:faults-etc}
\end{center}
\end{figure}

The engineering geological survey found mainly four types of rocks
in this area: (1) hard nubby and eroded but hard nubby mid-fine
grained biotite granite, (2) gray white thick bedding conglomerate
and gravel-bearing sandstone, (3) siltstone, (4) sandy
conglomerate sandstone. Most of the areas are of hard nubby
granite, extended close to the far detector site in the north and
reaching to the south, east, and west boundaries of the
investigated area. There exists a sub-area, measured about 150~m
(north-south) by 100~m (east-west), which contains eroded but
still hard nubby granites north of a conspicuous valley existing
in this region.\footnote{The valley extends in the north-east
direction from the north-east edge of the reservoir. The valley
can be seen in Fig.~\ref{fig:design}, as a dark strip crossing
midway along the planned tunnel connecting the mid hall and the far
hall.} Mildly weathered and weathered granites lie on top
of the granite layer. Devonian sandstones are located in the
north close to the far detector site. There are also scattered
sandstones distributed on the top of the granites. The granites
are generally very stable, and there exist only three small areas
of landslide found around the middle of the above mentioned
valley. The total area of the slide is about 20~m$^2$ and the
thickness is about 1~m. Four faults (F2, F6, F7, F8) and two weathering
bags have been identified, as shown in Fig.~\ref{fig:faults-etc}

The accumulation and distribution of underground water depends
generally on the local climate, hydrology, landform, lithology of
stratum, and detailed geological structure. In the investigated
area of the Daya Bay site, the amount of underground water flux
depends, in a complicated way, on the atmospheric precipitation
and the underground water seeping that occurs. The sandstone area
is rich in underground water seeping in, mainly through joints
caused by weathering of crannies that formed in the structure. No
circulation is found between the underground water and outside
boundary water in this area. Underground water mainly comes from
the atmospheric precipitation, and emerges in the low land and
is fed into the ocean.

Table~\ref{tab:weatherele} gives the values of various aspects of the
meteorology of the Daya Bay area. 
\begin{table}[!htb]
\begin{center}
\begin{tabular}{|l||c|c|} \hline
 Meteorological Data & Units & Magnitude \\ \hline\hline
 Average air speed & m/s & 3.29 \\ \hline
 Yearly dominant wind direction & ~ & E \\ \hline
 Average temperature & $^\circ$C & 22.3 \\ \hline
 Highest temperature & $^\circ$C & 36.9 \\ \hline
 Lowest temperature & $^\circ$C & 3.7 \\ \hline
 Average relative humidity & \% & 79 \\ \hline
 Average pressure & hPa & 1012.0 \\ \hline
 Average rainfall & mm & 1990.8 \\ \hline
\end{tabular}
\caption{Average values of meteorological data from the Da Ken station in 1985.}
\label{tab:weatherele}
\end{center}
\end{table}
A direct comparison shows that the
weather elements in Daya Bay are similar to those in the Hong
Kong---Shenzhen area.

According to the historical record up to December 31, 1994, there
have been 63 earthquakes above magnitude 4.7 on the Richter scale (RS),
including aftershocks, within a radius of 320~km of the
site.\footnote{The seismic activity quoted here is taken from a
Ling Ao NPP report~\cite{Ling AoSaftyReport}.} Among the stronger
ones, there was one 7.3~RS, one 7.0~RS, and ten 6.0--6.75~RS.
There were 51 medium quakes between 4.7 and 5.9~RS. The strongest,
7.3~RS, took place in Nan Ao, 270~km northeast of Daya Bay, in 1918. The most recent one
in 1969 in Yang Jiang at 6.4~RS.  In addition, there have been
earthquakes in the southeast of China and one 7.3~RS quake
occurred in the Taiwan Strait on Sept. 16, 1994. The epicenters of
the quakes were at a depth of roughly 5 to 25~km. These statistics
show that the seismic activities in this region originate
from shallow sources which lie in the earth crust. The strength of
the quakes generally decreases from the ocean shelf to inland.

Within a radius of 25~km of the experimental site, there is no
record of earth quakes of ${\rm M_s}\geq 3.0~ ({\rm M_L}\geq
3.5)$\footnote{$M_s$ is the magnitude of the seismic surface wave
and $M_L$ the seismic local magnitude. $M_s$ provides the
information of the normal characteristics of an earthquake. There
is a complicated location-dependent relationship between $M_s$ and
$M_L$. In Daya Bay ${\rm M_s}\geq 3.0$ is equivalent to ${\rm
M_L}\geq 3.5$.}, and there is no record of even weak quakes within
5~km of the site. The distribution of the weak quakes is isolated
in time and separated in space from one another, and without any
obvious pattern of regularity.

According to the Ling Ao NPP site selection report~\cite{Ling AoNPP},
activity in the seismic belt of the southeast sea has shown a
decreasing trend.  In the next one hundred years, this region will
be in a residual energy-releasing period to be followed by a calm
period. It is expected that no earthquake greater than 7~RS will
likely occur within a radius of 300~km around the site; the
strongest seismic activity will be no more than 6~RS. In
conclusion, the experimental site is in a good region above the
lithosphere, as was argued when the NPP site
was selected.

\subsubsection{Geophysical Exploration }
\label{sssec:civil_exploration}

Three methods are commonly used in geophysical prospecting:
high density electrical resistivity method, high
resolution gravity method, and seismic refraction image method
using a mechanical hammer. The first two methods together with the
third as supplement have been used for the Daya Bay
geophysical study\footnote{In order not to affect the construction work
of Ling Ao II, a heavy blaster cannot be used as a source of the seismic refraction
measurement, as required for deep underground measurement.
Therefore seismic refraction cannot be used as a major
tool for the Daya Bay prospecting.}. The combination of these
three methods reveal the underground structure, including: faults, type
of granite, rock mine contact interface, weathering zone
interface and underground water distribution.

\begin{figure}[!htb]
\begin{center}
 \includegraphics[width=1.0\textwidth]{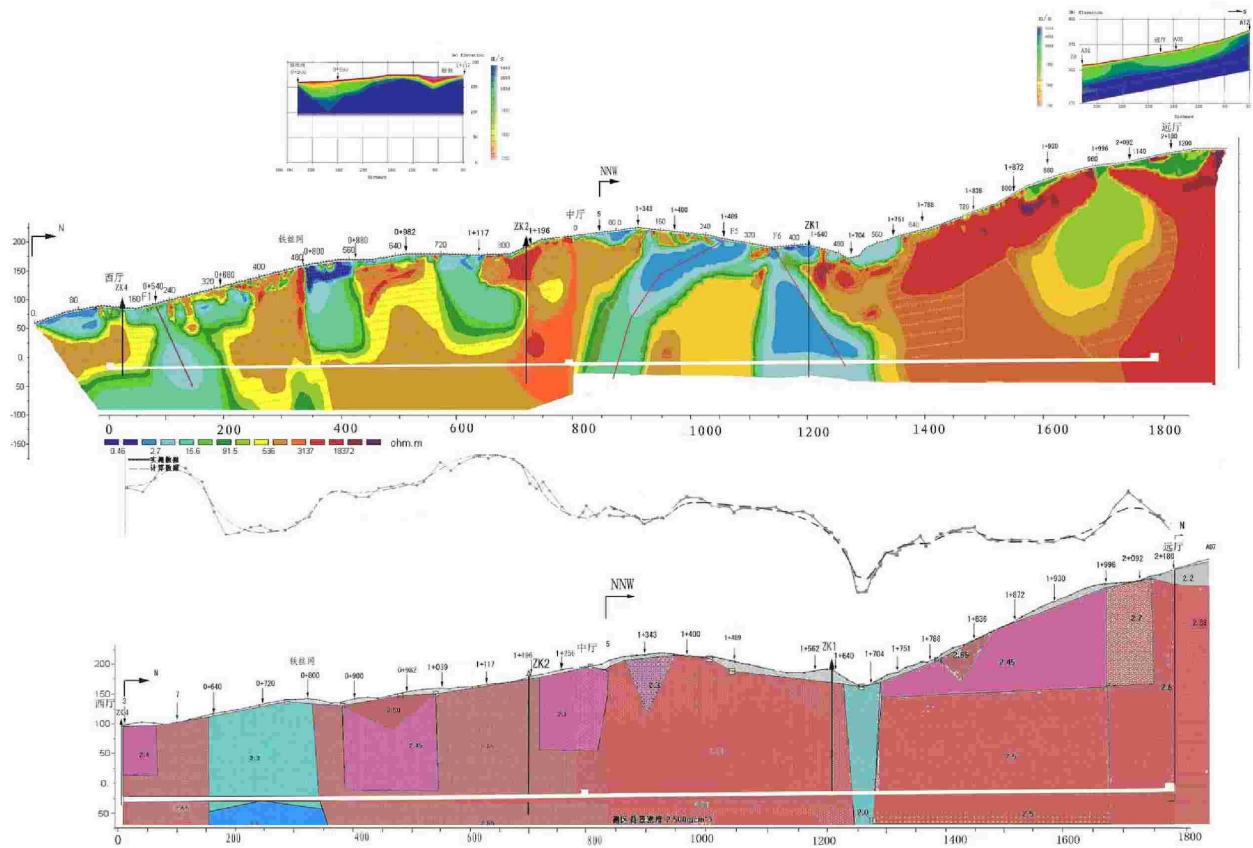}
 \caption{Seismic refraction, electrical resistivity and high
 resolution density maps along the tunnel cross section from
 the Daya Bay experimental hall (left end) to the
 far hall (right end).}
 \label{fig:resistance}
\end{center}
\end{figure}

Geophysical exploration revealed another four faults (F1, F3, F4, F5
shown in Fig.~\ref{fig:faults-etc}) along the tunnel lines.
Figure~\ref{fig:resistance} shows the regions of the
geophysical survey, including the experimental halls and tunnel
sections from the Daya Bay near hall to the mid hall and the far hall.
The experimental halls, tunnel sections, faults and weathering
bags are marked explicitly in the figure. 
The electrical
resistivity measurements are shown in the middle of the figure,
the high resolution density measurements on the bottom, and two
sections of seismic refraction measurement in the corresponding
part on the top. Because of the complexity and variety of
underground structures, the electrical resistivity was measured in
boreholes ZK1 and ZK2. The resistivity and density of the rock
samples from the boreholes were used for calibration of the resistivity 
map. Depending on the characteristics of
the granite and its geological structure, the electrical
resistivity of this area can vary from tens of ohm-m to more than
10k ohm-m. The non-weathered granite has the highest electrical
resistivity, whereas the sandstone has medium resistivity due to
trapped moisture. The weathered zone, consisting of weathered
bursa and faults, has relatively low resistivity.

\subsubsection{Engineering Drilling}
\label{sssec:civil_drilling}

\begin{figure}[!htb]
\begin{center}
 \includegraphics[width=0.85\textwidth]{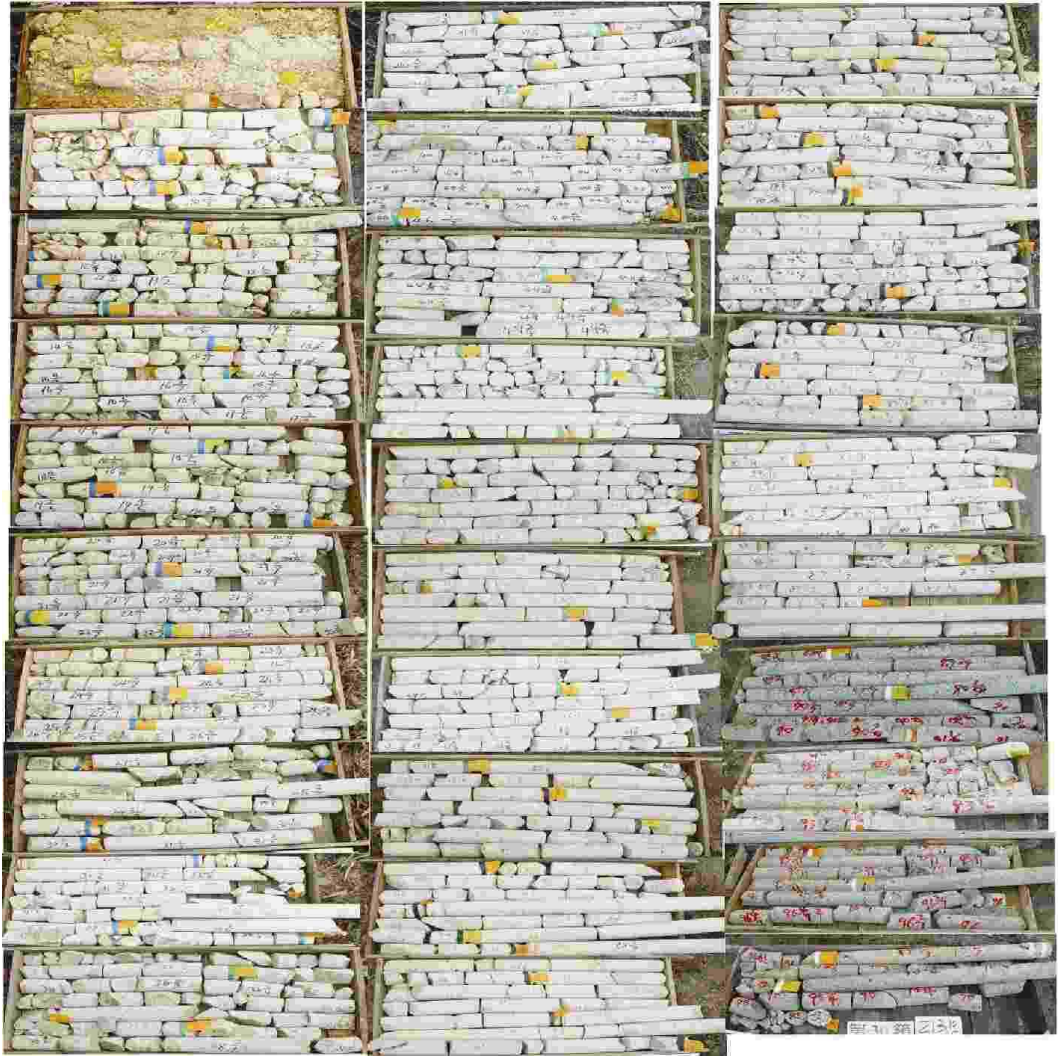}
 \caption{Rock samples from borehole ZK1.}
 \label{fig:ZKX}
\end{center}
\end{figure}

Based on the information about faults, zones with relative high
density of joints, weathering bags, low resistivity areas revealed
from previous geological survey, four borehole positions were
determined. The purpose of the boreholes was essentially to prove or
exclude the inferences from the previous survey approaches above
ground. These four boreholes are labelled as ZK1, ZK2, ZK3, ZK4 from
north to south in Fig.~\ref{fig:tunnel}. The depth of the four
boreholes are 213.1~m, 210.6~m, 130.3~m, 133.0~m respectively (all to at least the tunnel depth).
Figure~\ref{fig:ZKX} shows sections of rock samples obtained from
borehole ZK1. 
Similar samples are obtained in the other three boreholes. The samples
are used for various laboratory tests.

\subsubsection{On-site Test at Boreholes}
\label{sssec:civil_boreholes}

\begin{figure}[!htb]
\begin{center}
 \includegraphics[width=0.8\textwidth]{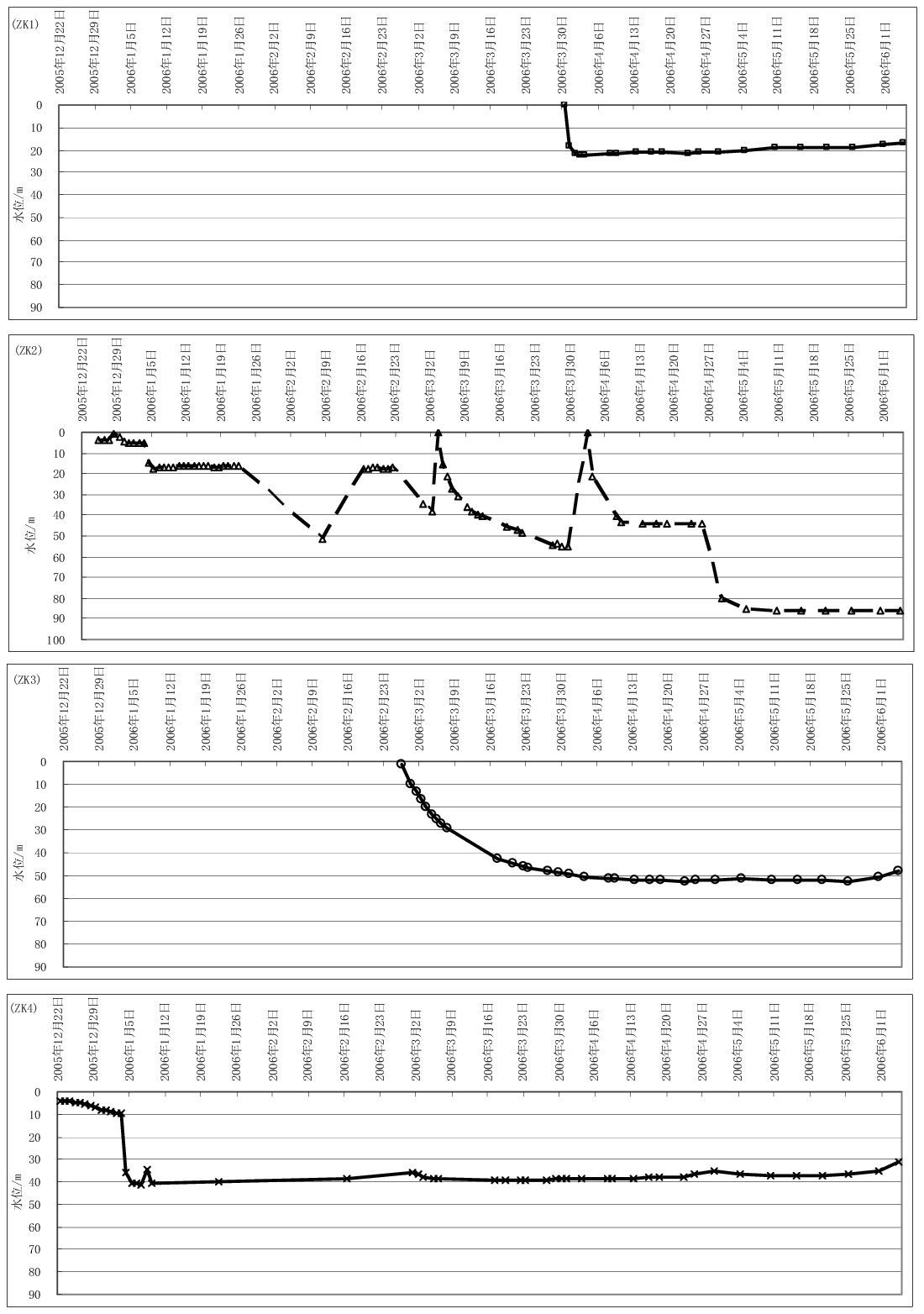}
 \caption{Water level variation vs time in the four boreholes.
 There is no measurement during holidays in January 2006 in ZK2. The
 cause of the sudden drop of the water level in April 2006 is unknown. The
 unit of the $x$ coordinate is 7 days, the date reads as year/month/day.
 The unit of the $y$ coordinate is the water level in meters down the
 borehole.}
 \label{fig:Kvalue}
\end{center}
\end{figure}

There are many on-site tests performed at the boreholes: (1) High
density electrical resistivity measurement in boreholes ZK1 and
ZK2. (2) Permeability tests at different time and depth are made
in the boreholes during borehole drilling and at completion. The
test shows that all measured values of the permeability parameter
K are less than $0.0009~{\rm m/d}$. The K values in ZK2, ZK3 are
smaller than that in ZK1 and ZK4. Figure~\ref{fig:Kvalue} shows
the water level variation vs time from pouring tests in
the four boreholes. 
(3) Acoustic logging, which is tested at
different segments separated by 0.5~m. There are 66, 26, 34, 23
segments tested in ZK1, ZK2, ZK3, ZK4 respectively. The combined
results give the velocity of longitudinal wavelength $V_{\rm
p}=5500~{\rm m/s}$ in the fresh granite. (4) Geo-stress test.  (5)
Digital video.
(6) The radon emanation rate inside the borehole ZK4
was measured up to a depth of 27~m  with an electronic
radon dosimeter inserted into the borehole. An average rate of
$0.58\times 10^{-3}$~Bq~m$^{-2}$~s$^{-1}$
was determined at depths of 14--27~m 
after correction for back diffusion.
These values generally agree with the rates
(0.13--2.56))~$\times 10^{-3}$~Bq~m$^{-2}$~s$^{-1}$
measured directly from the rock samples extracted
from the borehole.
(7) Measurements of the rock chemical composition. The
chemical elements of the rock were measured, among
these elements, the amount of radioactive U was measured to be 10.7, 16.6,
14.5 and 14.2~ppm from the samples in each of the four boreholes,
respectively. The Th concentrations were measured to be 25.2, 49.6,
29.4 and 41.9~ppm in each of the borehole respectively. (8) Water
chemical analysis.  Water samples from the four boreholes and a surface
stream have a pH slightly smaller than 7.5, considered neutral.
The water hardness is smaller than 42~mg/l which is considered to be very soft.
The underground water is thus very weakly corrosive to the
structure of steel, but is not corrosive to reinforced
concrete.

\subsubsection{Laboratory Tests}
\label{sssec:civil_laboratory}

Laboratory tests performed includes: rock chemical properties,
mineral elements, physical and mechanical property tests. The
following data are some of the physical properties of slightly
weathered or fresh rock which are the most comment type of rocks
in the tunnel construction:
\begin{itemize}{\setlength\itemsep{-0.4ex}
\item Density of milled rock: $2.609 \sim 2.620~g/cm^3$ 
\item Density of bulk rock: $2.59 \sim 2.60~g/cm^3$ 
\item Percentage of interstice: $0.765\% \sim 1.495\%$ 
\item Speed of longitudinal wave $(V_p): 4800 \sim 5500~m/s$ 
\item Pressure resistance strength of a saturated single stalk: $85.92 \sim 131.48~MPa$
\item Pressure resistance strength of a dry single stalk: $87.88 \sim 125.79~MPa$ 
\item Softening coefficient: $0.924 \sim 1.000$
\item Elastic modulus: $32.78 \sim 48.97~GPa$ 
\item Poisson ratio: $0.163 \sim 0.233$}
\end{itemize}

\subsubsection{Survey Summary}
\label{sssec:civil_summary}

Based on the combined analyses of the survey and tests described
above, IGG concludes that the geological structure of the proposed
experimental site is rather simple, consisting mainly of massive,
slightly weathered or fresh blocky granite. There are only a few
small faults with widths varying from 0.5~m to 2~m, and the
affected zone width varies from 10~m to 80~m. There are a total of four
weathering bags along the tunnel from the Daya Bay near site to
the mid site and on the longer construction tunnel option from the
Daya Bay quarry to the mid site. The weathering depth and width
are 50--100~m. 
Just below the surface, the granite is mild to mid weathered.
These weathered zones are well above the tunnel,
more than three times the tunnel diameter away, so the tunnel is not expected
to be affected by these weathering bags.
Nevertheless, there are joints around this region and some sections
of the tunnel will need extra support.

The far hall at a depth of 350~m is thought to consist of lightly effloresced or
fresh granites; the far hall is most likely surrounded by hard
granite. The distance to the interface with Devonian sandstone is
about 100~m (to the North) from the present analysis estimate.

\begin{figure}[!htb]
\begin{center}
 \includegraphics[width=0.85\textwidth, height=0.85\textheight,angle=180]{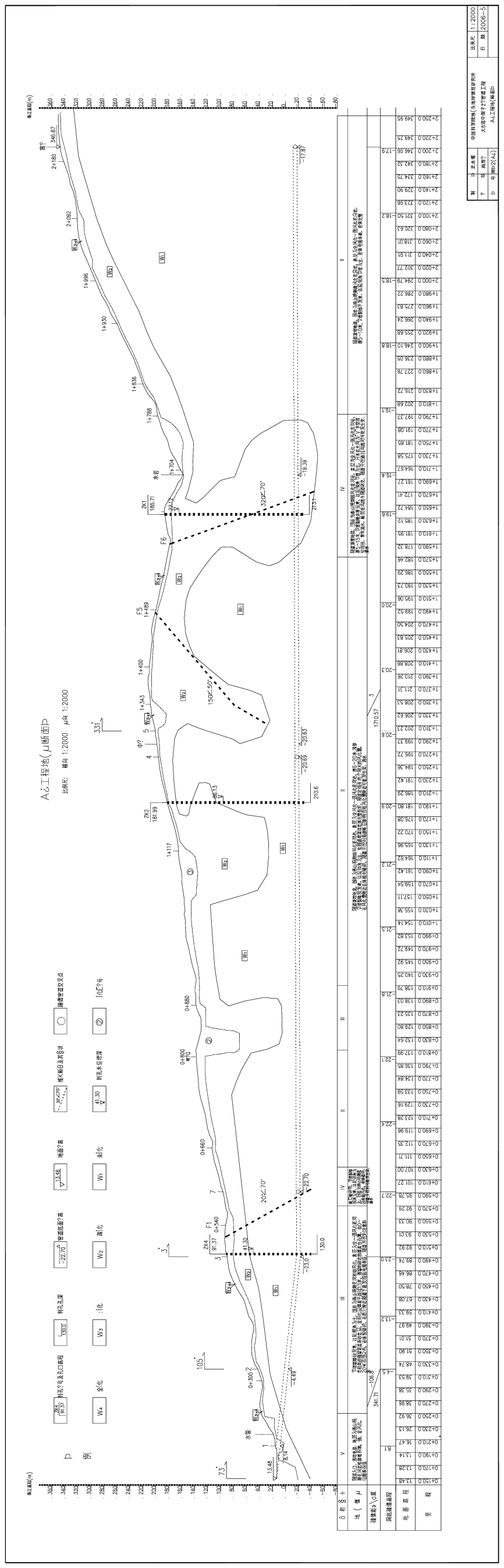}
 \caption{Engineering geological section in line A: the faults,
 weathering bags and tunnel are shown on the figure. The first curve down from the surface shows the boundary of the weathered granite and the second curve down shows the boundary of the slightly weathered granite. The tunnel passes through one region of slightly weathered granite.}
 \label{fig:lineA}
\end{center}
\end{figure}

The rock along the tunnel is lightly effloresced or fresh granite,
and mechanical tests found that it is actually hard rock. No
circulation is found between the underground water and the outside
boundary water in this area, underground water mainly comes from
the atmospheric precipitation. Water borehole permeability tests
show that underground water circulation is poor and there is no uniform
underground water level at the tunnel depth. At the tunnel depth
the stress is 10~MPa, which lies in the normal stress regime.  The
quality of most of the rock mass varies from grade II to grade III
(RQD around 70\% which indicates good and excellent rock quality).
From the ZK1 and ZK2 stress measurements and structure analysis,
the orientation of the main compressive stress is NWW. For the
east-west oriented excavation tunnel, this is a favorable
condition for tunnel stability. For the 810~m segment of
the main tunnel from the Daya Bay near hall
(\#1) to the mid hall (\#4) the tunnel orientation will run
sub-perpendicular to the orientation of the maximum principal stress
and it will thus be subject to higher stress levels at the excavation
perimeter. These higher stress levels are not expected to cause
significant stability problems due to the strength of the granite rock
mass.  There are some tunnel sections, including the access tunnel,
where the rock mass quality belongs to grade IV, and some belongs to
grade V. Figure~\ref{fig:lineA} shows the details of the engineering
geological section along Line A.
Detailed results from the site survey by IGG can be found
in references~\cite{IGGRep0,IGGRep2,IGGRep3,IGGRep4,IGGRep5,IGGRep6,IGGRep7}.

\subsection{Conceptual Design}
\label{ssec:civil_conceptual}

In June 2006 we organized a bid for a conceptual design of the civil
construction. The purpose of this effort was to further refine our
understanding of the cost of various options and to make sure that
we do not leave any important points out of the final design specifications.
The major items of the conceptual design included: (1)
the underground experimental halls, the connecting tunnel, access
tunnel, and construction tunnel; (2) the infrastructure buildings
above ground; (3) the electric power, communication, monitor,
ventilation system, water supply, and drainage, safety, blast control,
and environmental protection. Two design firms were selected: the
Fourth Survey and Design Institute of China Railways (TSY) and the
Yellow River Engineering Consulting Co. Ltd. (YREC).  
TSY has expertise in the design of railway tunnels, and YREC
has a great deal of experience in underground hydroelectric
engineering projects. They completed their designs in the end
of July and beginning of August 2006. These two reports will
aide the writing of the specifications of the bid for the detailed tunnel design.

\subsubsection{Transportation Vehicle for the Antineutrino Detectors}
\label{sssec:civil_trucks}

The biggest item to transport in the tunnel is the antineutrino
detector module. Each module is a cylinder of 100~T with
an outer diameter 5~m and a height of 5~m, with ports extending above. The
transportation of the antineutrino detector determines the cross
section of the tunnel and directly affects the total tunnelling
construction plan.

The space in the tunnel is limited, so the transportation vehicle for
the heavy antineutrino detector should be easy to operate and very stable and smooth
during movement. TSY has investigated two kinds of transportation vehicles:
(1) heavy-truck with a lowboy trailer, and (2) truck with a platform
on top.  The bed of the lowboy trailer is 40~cm off the ground and the
loading height is 80~cm. The total length of the truck plus the
trailer is more than 20~m long, the turnaround radius is 50~m.  This
turnaround radius makes it impossible to turn the vehicle around
without significantly increase the total length of the tunnel.  So TSY
recommends the use of a truck with a platform on top and the
specifications of this platform vehicle available in two manufacturing
companies in China are listed in Table~\ref{tab:platform}.
\begin{table}[!htb]
\begin{center}
\begin{tabular}{|l||c|c|} \hline
 Manufacturer & QinHuangDao Heavy &  WuHan TianJie special\\
  ~  & Engineering Union Co. Ltd. & transportation Co. Ltd.\\ \hline
 Model & TLC100A & TJ100 \\ \hline\hline
 Full loading(t) & 100 & 100 \\ \hline
 Out dimension L x W (m) & 11.0 x 5.0 & 11.0 x 5.5 \\ \hline
 Height of loading (mm) & 1700($\pm$300) & 1750($\pm$300) \\ \hline
 Self weight(t) & 28 & 28 \\ \hline
 Axles and Wheels/axle & 4/8 & 4/8 \\ \hline
 Speed & full loading (on flat): 6~km/h & full loading (on flat): 5~km/h \\ \hline
 Slope & Vertical 6\% Horizontal 4\% & Vertical 8\% Horizontal 2\% \\ \hline
 Power & 168~kW & 235~kW \\ \hline
\end{tabular}
\caption{Technical parameters of platform trucks.}
\label{tab:platform}
\end{center}
\end{table}

\begin{figure}[!htb]
\begin{center}
 \includegraphics[width=0.7\textwidth]{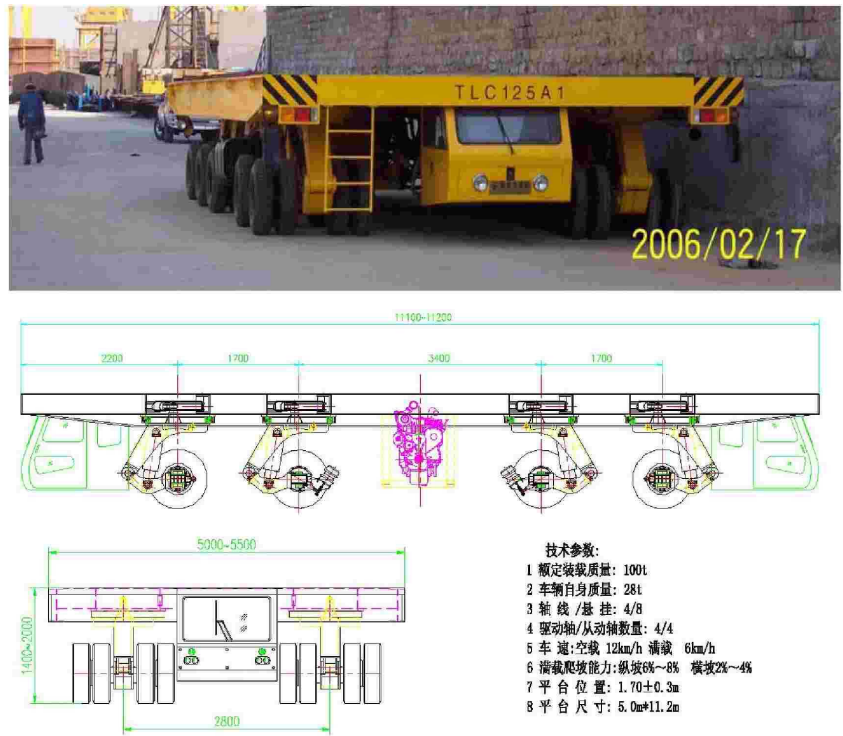}
 \caption{Photo of a platform truck with schematic diagrams
 of wheel rotations. The specifications written in Chinese
 on the right-bottom are the same as in Table~\ref{tab:platform}}
 \label{fig:platformT}
\end{center}
\end{figure}

An example of the platform truck is shown in
Fig.~\ref{fig:platformT}. 
It has an easy rotating system with the wheels rotatable in any
directions. It has two driving cabs, one in the front and one in the
back which makes turning around in the tunnel unnecessary. Its
movement is more steady than the lowboy trailer which is very
important for transporting the antineutrino detector modules.

\begin{figure}[!htb]
\begin{center}
 \includegraphics[width=0.7\textwidth]{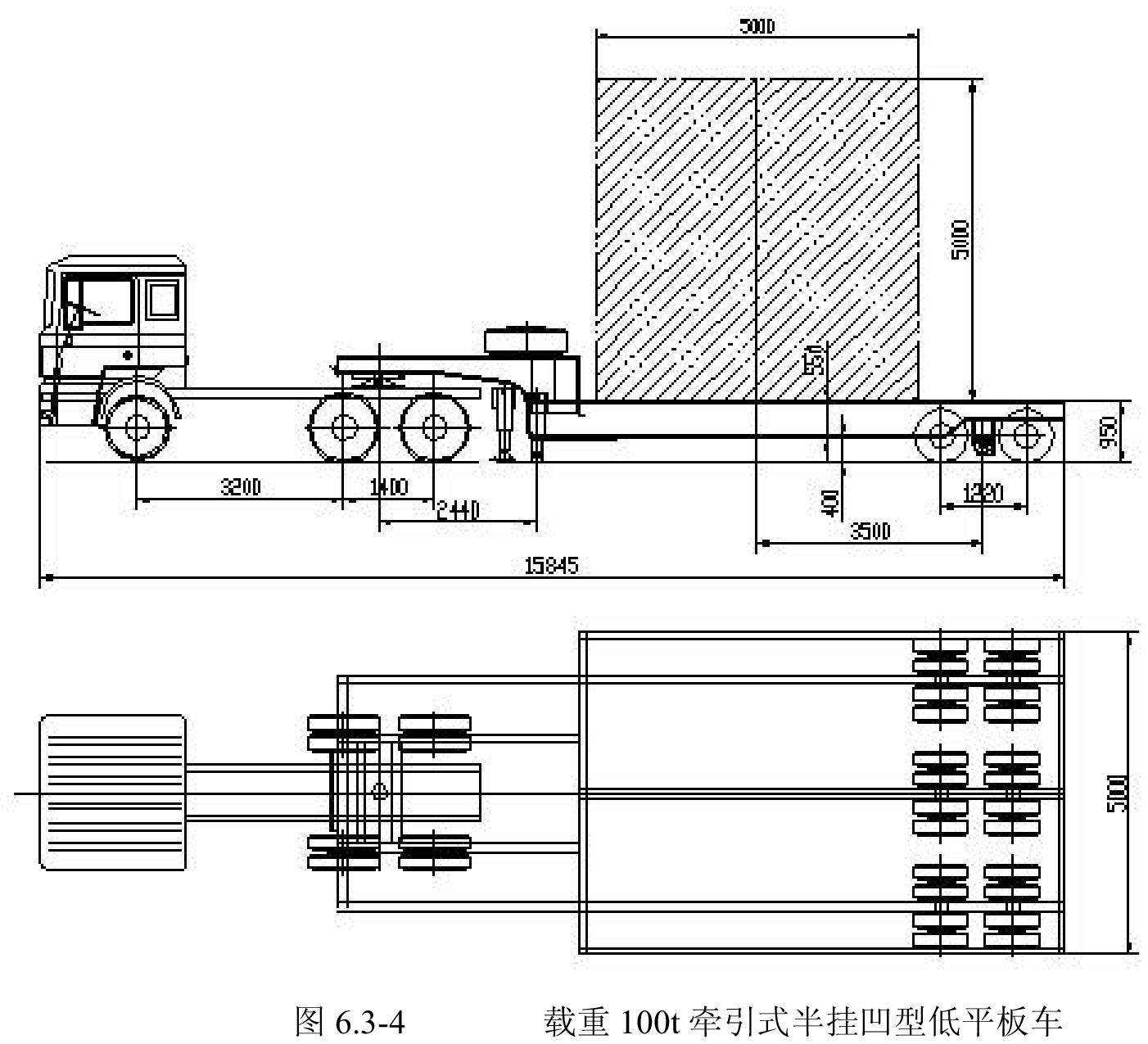}
 \caption{Schematic diagram of a semi-trailer. The dimensions, length,
 width, and height are in mm.}
 \label{fig:semiT}
\end{center}
\end{figure}

YREC also investigated the above mentioned transporting vehicles
with similar specifications. In addition, they have investigated
an electric railway transportation system which consists of a transport
framework, support frame, cable winding, and control desk.
However, the loading height is 1~m, and laying the rail is
expensive and time consuming. Finally, YREC recommend the use of a
semi-trailer with a platform loading, as shown in
Fig.~\ref{fig:semiT}. 
The total length is 15.8~m and the loading
height is 95~cm. Since it is not very long, this semi-trailer will
drive forward and backward in the tunnel without turning around.
The ventilation speed in the tunnel has to be increased during the
transportation of the detector modules to vent the exhaust
discharged in the tunnel.

Further investigation about the transportation vehicle with lower
height of the loading platform is needed in order to lower the
required height of the tunnel. It is also necessary to find a
suitable electric powered vehicle instead of one powered by
petroleum. Current studies are focussing on custom low-boy trailers
with electric tugs and on sophisticated, low-profile computer
controlled transporters.

\subsubsection{Lifting System for the Antineutrino Detectors}
\label{sssec:civil_cranes}

\begin{figure}[!htb]
\begin{center}
 \includegraphics[width=0.5\textwidth, angle=-90]{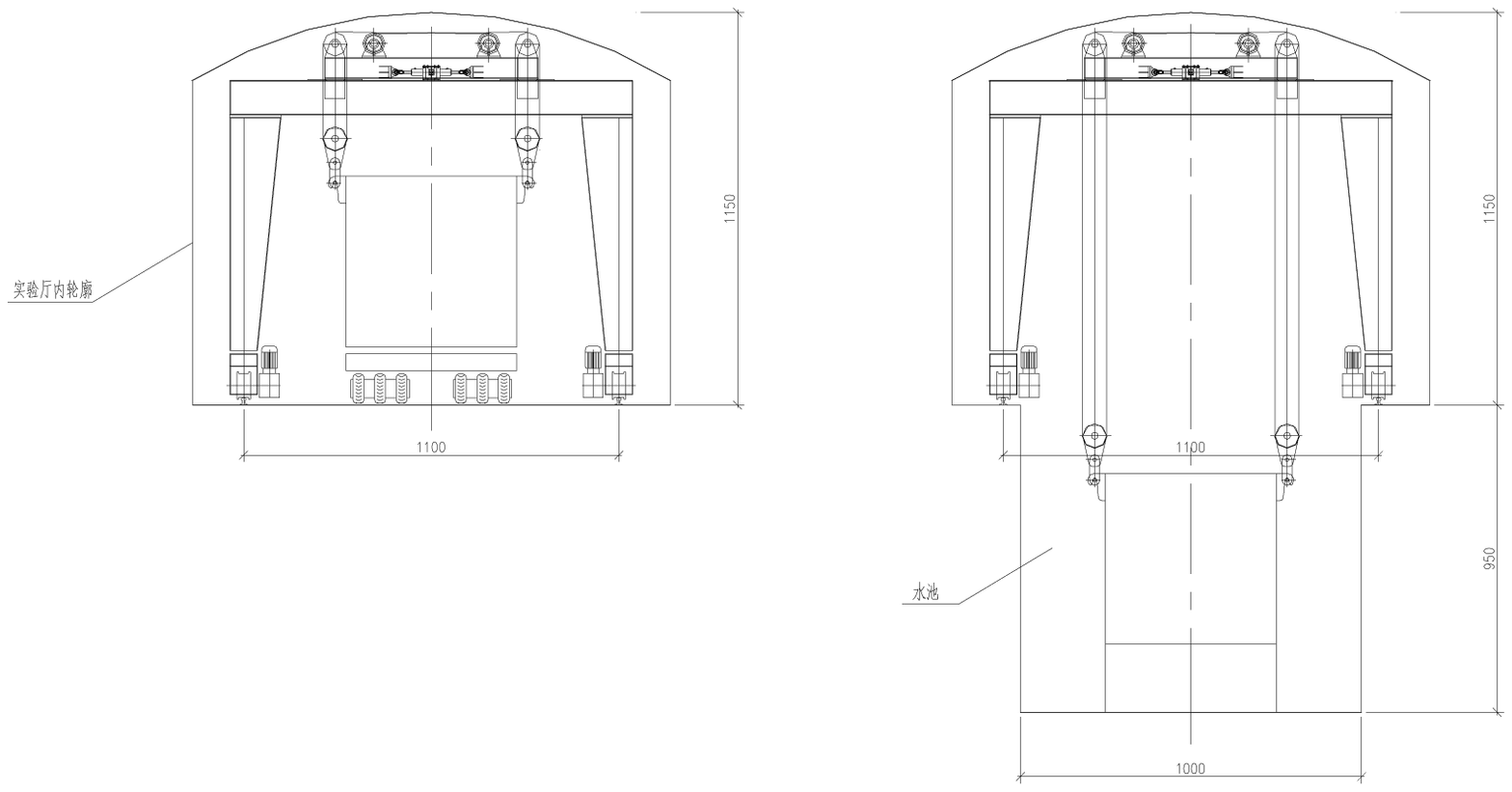}
 \caption{Schematic diagrams of a gantry crane in the experimental
 hall to lift the antineutrino detector (left panel) and lower it into
 the water pool (right panel).}
 \label{fig:gantryTSY}
\end{center}
\end{figure}

\begin{figure}[!htb]
\begin{center}
 \includegraphics[width=0.7\textwidth]{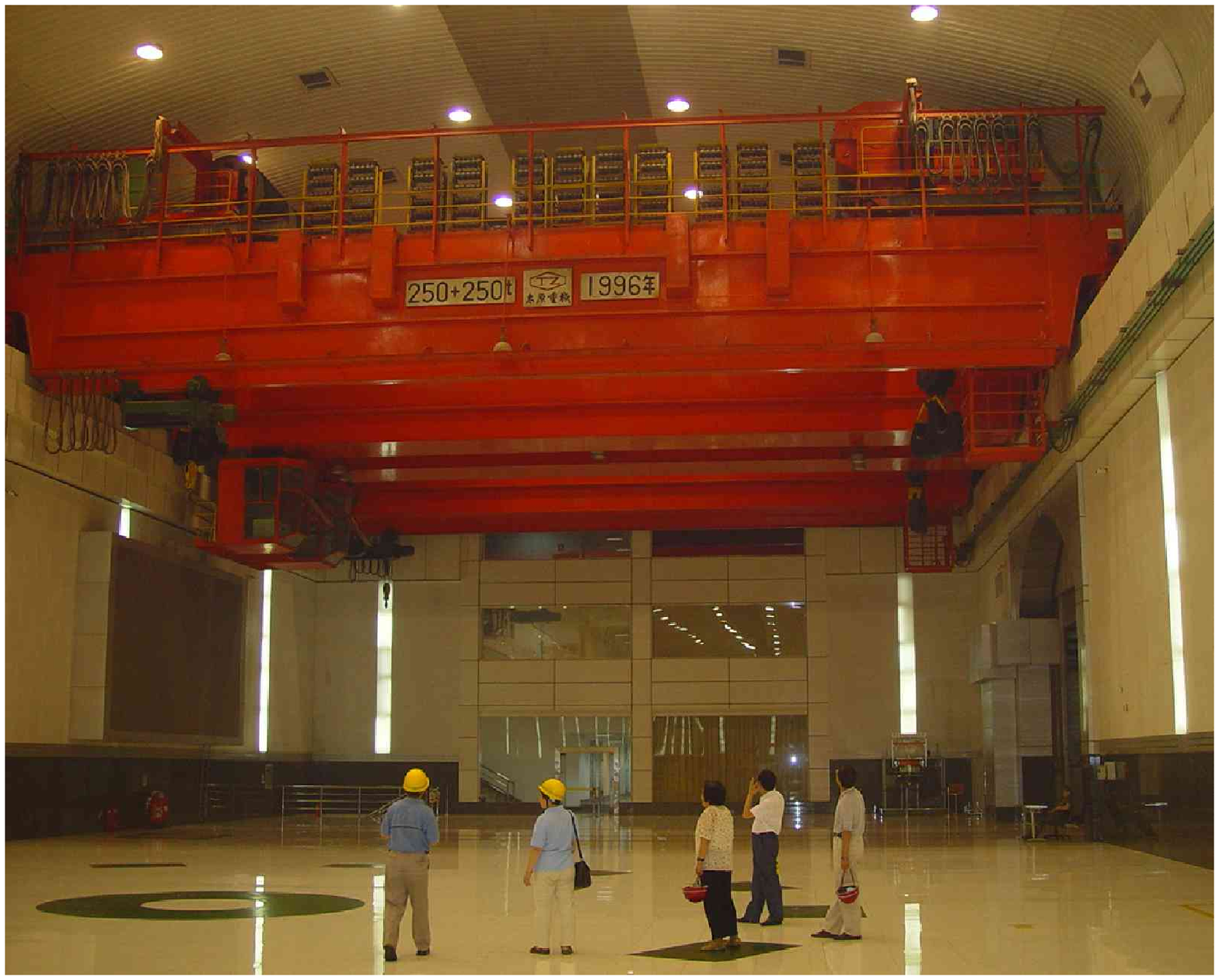}
 \caption{A photo of a bridge style crane, the crane rail
 is fixed to the wall of the experimental hall.}
 \label{fig:bridgeYREC}
\end{center}
\end{figure}

Lifting systems, mainly for handling the antineutrino detectors, have
been investigated. The lifting system should be low in order to
minimize the height of the experimental hall and to gain
overburden. Both gantry cranes (suggested by TSY) and bridge style
cranes (suggested by YREC) satisfy our requirement. 
The heights of the experimental halls required to install and lift the
antineutrino detector with these two types of cranes are similar:
about 12--13~m.  Figures~\ref{fig:gantryTSY} and \ref{fig:bridgeYREC}
show these two kind of cranes.
Both cranes have two hooks working during the lifting which will
greatly decrease the height of the hall and can be operated more
steadily. The final choice of one vs. two hooks needs further
study. The rails of the bridge crane are supported on the two side
walls of the experiment hall. The final choice of a crane system
needs further study.

\subsubsection{Experimental Hall Layout}
\label{sssec:civil_hall}

\begin{figure}[!htb]
\begin{center}
 \includegraphics[width=0.7\textwidth]{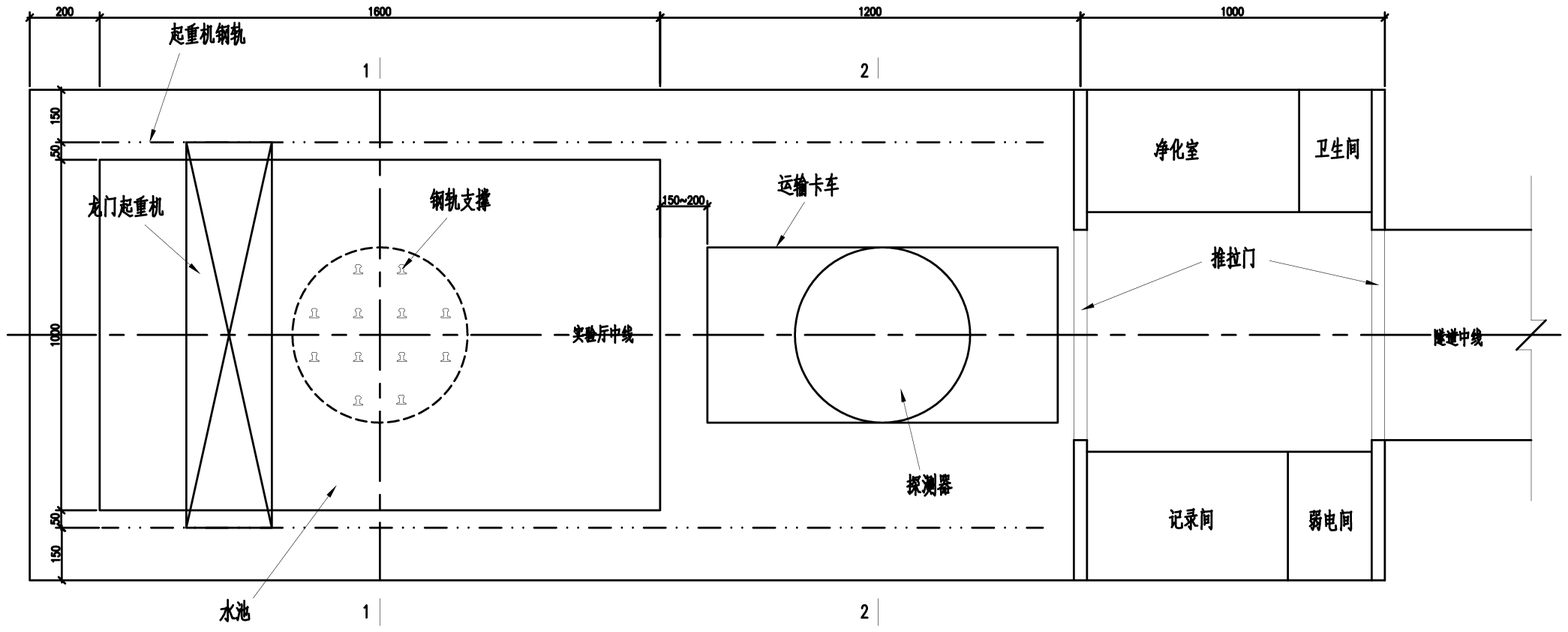}
 \caption{Layout of the experimental hall where the counting room,
 etc., are laid out in series along the hall (as proposed by TSY).}
 \label{fig:HallTSY}
\end{center}
\end{figure}

\begin{figure}[!htb]
\begin{center}
 \includegraphics[width=0.7\textwidth]{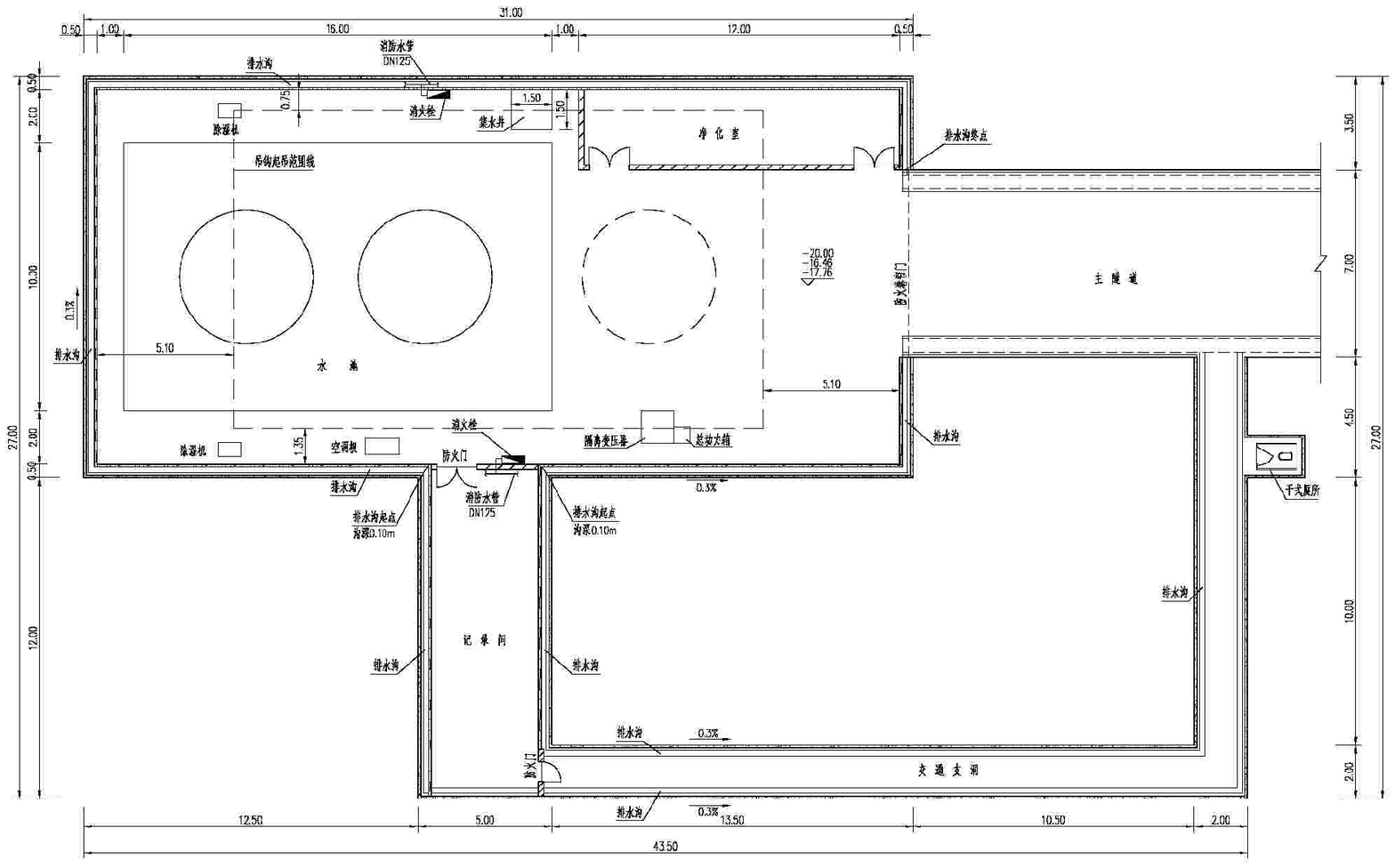}
 \caption{Layout of the experimental hall where the counting room,
 etc., are along one side of the hall (as proposed by YREC).}
 \label{fig:HallYREC}
\end{center}
\end{figure}

The experimental hall layout can not be fixed before we know how to
install the antineutrino detectors, how to lay the muon detectors on
top, and what auxiliary facilities are needed. The two designers
presented two sketches which include antineutrino detector
transportation, lifting space, and rooms for auxiliary facilities, see
Fig.~\ref{fig:HallTSY} (designed by TSY)
 and Fig.~\ref{fig:HallYREC} (designed by YREC). 
The auxiliary facilities rooms are at the side of the hall in
Fig.~\ref{fig:HallYREC} which may reduce the length of the electronics
cables from the detector to the counting room, and other auxiliary
facility rooms, which could be arranged parallel to the counting room,
are more flexibly arranged. A side
tunnel links the main tunnel with the control room and the other possible
rooms.

The longitudinal direction of the Daya Bay (\#1) and mid (\#4) experimental halls 
is preferred to be along the tunnel direction for
construction convenience. The Ling Ao near hall (\#2) is the only
one with its longitudinal direction to be about 90$^\circ$ with
the accessing tunnel in order to keep all the halls in the same
orientation.

The LS filling hall (\#5) will be decided upon once we settle on
the LS mixing and filling procedures. We expect no special
questions about the design and construction of this hall.  At this
stage, we put it near the Daya Bay hall (\#1).

\subsubsection{Design of Tunnel}
\label{sssec:civil_tunneldesign}

According to the size of the selected transportation vehicles, the
cross section of the main tunnel will be relatively easy to
define:
\begin{itemize}{\setlength\itemsep{-0.4ex}
\item Width of the roadway: 5.0~m.
 \item Width of safety distance to side wall: 1.0~m x 2.
 \item Width of drainage channel: 0.25~m x 2.
 \item The total width of the tunnel is: 7.5~m (YREC has 7.0~m because
they have a narrower space for safety).
 \item Height of the transporting vehicle plus height of antineutrino detector: 6.4~m.
 \item Duct diameter: 1.5~m.
 \item Safety distance between detector module to the duct: 0.5~m.
 \item Total height of the tunnel is: 8.4~m (YREC has 8.5~m)
}\end{itemize}

\begin{figure}[!htb]
\begin{center}
 \includegraphics[width=0.5\textwidth]{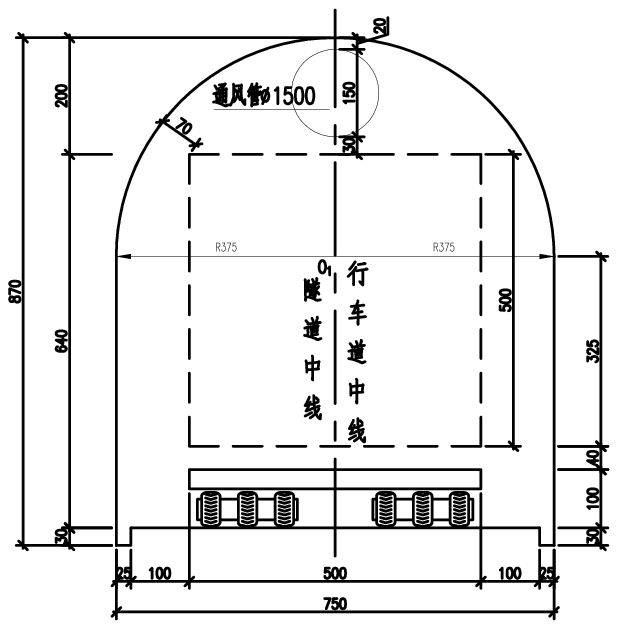}
 \caption{An engineering schematic diagram of the tunnel layout
 proposed by TSY. The dimensions are in cm.}
 \label{fig:TunnelTSY}
\end{center}
\end{figure}

\begin{figure}[!htb]
\begin{center}
 \includegraphics[width=0.5\textwidth]{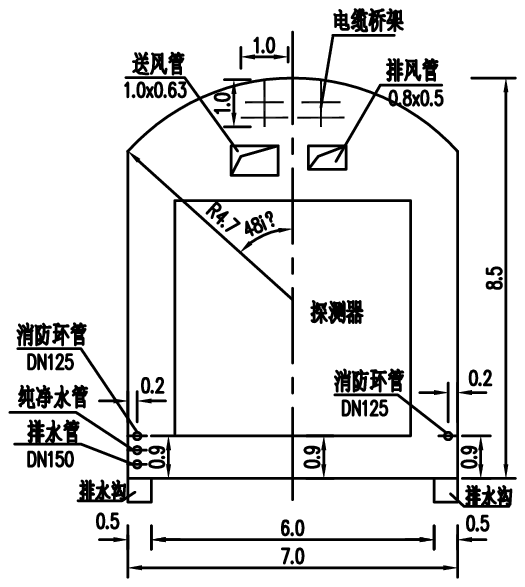}
 \caption{An engineering schematic diagram of the tunnel layout
Proposed by YREC. The dimensions are in meters.}
 \label{fig:TunnelYREC}
\end{center}
\end{figure}

Figures~\ref{fig:TunnelTSY} and \ref{fig:TunnelYREC}
describe the cross sections of the main tunnel. The lining of the
tunnel depends on the rock quality. The
rock quality varies from grades I to V, grade I being excellent and
grade V poor. According to the site survey, more than 90\% of the
rock belongs to grade I, II or III which are stable rocks. Some
very short section of the tunnel have grade IV rock and the only 
grade V rock is in the first tens of meters at the main portal. The
lining for different quality of rocks are giving by two designers
in their report~\cite{TSYReport}~\cite{YRECReport}.

The access tunnel has the same cross section as the main tunnel
to enable transportation of the antineutrino detector. This tunnel section has a
slope of up to 10\%. The antineutrino detector is not yet filled
with LS when it is transported down the access tunnel.
The length of the tunnel is less than 300~m and modern
mining/industrial equipment will have no difficulties in moving on the
9.6\% slope of the access tunnel (in the YREC design).

There are two possible design strategies for the construction tunnel.
One is to transport the dirt by heavy truck, another one is by
tram. In the truck option, the allowed slope is up to 13\%
(TSY), the width of this tunnel is 5.0~m and height 5.8~m. There
will be a passing section in every 80~m along the tunnel for
two trucks to cross into the opposite directions. The total length
of such a tunnel is 528~m. If a tram is used for dirt
transportation, the tunnel can tolerate a much steeper slope,
up to 42\% ( $< 23^\circ$ ). The tunnel length can be as short as
200~m and the cross section is 4.6~m wide by 4.08~m high.
Construction with a tram will allow for a shorter tunnel,
therefore saving both time and money. The dirt removal with
a tram is more complicated than using heavy trucks, which
will take more time and money.  Let us note that in
the case of a tram, since special tools are needed,
the number of construction companies 
bidding on the tunnel construction contract may be more limited.

\begin{figure}[!htb]
\begin{center}
 \includegraphics[width=0.95\textwidth]{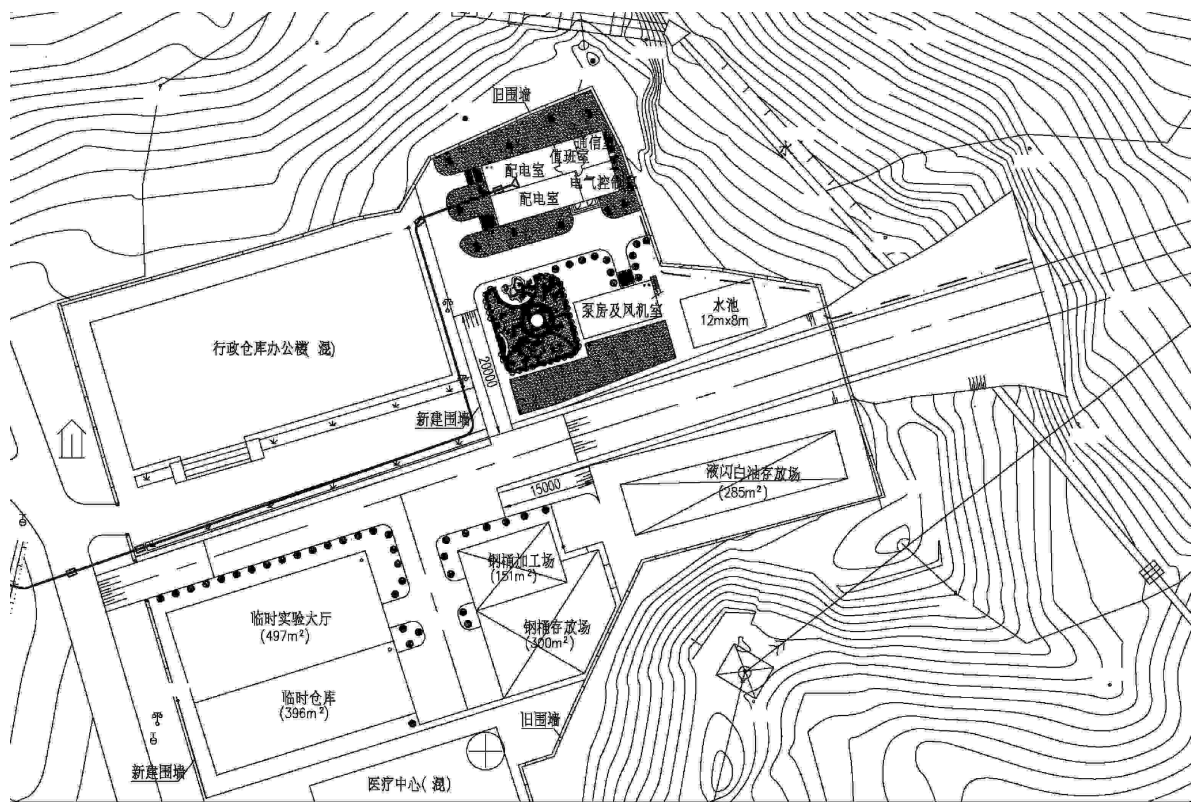}
 \caption{A schematic diagram of the the main portal and the layout
 of auxiliary buildings.}
 \label{fig:mainportal}
\end{center}
\end{figure}

A possible layout of the main portal behind the local hospital is shown
in Fig.~\ref{fig:mainportal} (YREC's design).

\subsubsection{Other Facilities}
\label{sssec:civil_facilities}

Other facilities are also included in the conceptual
design reports submitted by TSY~\cite{TSYReport} and
YREC~\cite{YRECReport}. They include: (1) electricity, (2)
ventilation system (3) water supply and drainage, (4)
communication, (5) monitoring systems, (6) blast control, and (7)
environmental effect evaluations.

\subsection{Civil Construction Overview}
\label{ssec:civil_civil}

Based in part on the two conceptual design reports, we are optimizing
the construction tunnel layout, the crane system, the transport system
and the tunnel cross section. Once this process is completed a specification
for the final civil construction design package will be drafted.
The final tunnel design and civil construction contractors will be
selected via a bidding process. Most likely the detailed design and
civil construction team will be separated. An oversight agency
is needed for the construction. The time needed to complete the
final design will be 4--5 months once all of the specifications are laid
out. The civil construction will last 1.5--2 years as estimated
by the conceptual designers.

The main civil construction work items are listed in
Table~\ref{tab:civiltotal}.
\begin{table}[!htb]
\begin{center}
\begin{tabular}{|l||r|} \hline
 Construction item         & Volume ($m^3$) \\ \hline\hline
 Excavation dirt in open   & 17,068 \\ \hline
 Excavation dirt in tunnel & 202,745 \\ \hline
 Concrete                  & 8,740 \\ \hline
 Eject concrete            & 7,774 \\ \hline
\end{tabular}
\caption{Table of the main civil construction work items.}
\label{tab:civiltotal}
\end{center}
\end{table}

\newpage
\setcounter{figure}{0}
\setcounter{table}{0}
\setcounter{footnote}{0}

\section{Antineutrino Detectors}
\label{sec:det}

\subsection{Overview}
\label{ssec:det_overview}

The measurement of $\sin^22\theta_{13}$ to 0.01 or better is an
experimental challenge.  A value of 0.01 for $\sin^22\theta_{13}$
yields a tiny oscillation effect. This corresponds to a small
difference in the number of antineutrino events observed at the far
site from the expectation based on the number of events detected at
the near site after correcting for the distance under the
assumption of no oscillation. To observe such a small change, the
detector must be carefully designed following the guidelines discussed
in Chapter~\ref{sec:over}, and possible systematic uncertainties
discussed in Chapter~\ref{sec:sys}.  The following requirements should
be satisfied in the design of the antineutrino detector modules and
related components:
\begin{enumerate}
   \item The detector modules should be homogeneous to minimize edge effects.
   \item The energy threshold should be less than 1.0~MeV to be fully
   efficient for positrons of all energies.
   \item The number of protons in the target liquid scintillator
should be well known, implying that the scintillator mass and the
proton to carbon ratio should be precisely determined. The target
scintillator should come from the same batch for each pair of near-far
detector modules, and the mixing procedure should be well controlled to
ensure that the composition of each antineutrino target is the same.
\item The detector module should not be too large; otherwise, it would be
   difficult to move from one detector site to another for a cross check
   to reduce systematic effects. In addition, beyond a certain size,
   the rate of cosmic-ray muons passing through the detector module is too
   high to be able to measure the ${}^9$Li background.  
\item The event time should be determined to be better than 25~ns for
   studying backgrounds.
\item The energy resolution should be better
   than 15\% at 1~MeV. Good energy resolution is desirable for
   reducing systematic uncertainty (see Chapter~\ref{sec:sys}).  It is also important for the
   study of spectral distortion as a signal of neutrino oscillation.
\end{enumerate}

\subsubsection{Module Geometry}
\label{sssec:det_energy}

Several previous neutrino experiments have designed spherical or
ellipsoidal detectors to insure uniform energy response in the entire
volume. This type of detector vessel is expensive and requires many
PMTs for 4$\pi$ coverage. Two types of alternative
detector geometries have been investigated: cubic and
cylindrical. Both are attractive from the viewpoint of construction.
Monte Carlo simulation shows that a cylindrical shape, as shown in
Fig.~\ref{fig:detector}, can deliver a better energy and position
resolution while maintaining good uniformity of light response over
the volume, similar to that of a sphere or ellipsoid. This design is
verified by our prototype tests as discussed in
section~\ref{ssec:det_prototype}.  An optical reflector can be put at
the top and bottom of the cylinder, so that PMTs are only positioned
on the circumference of the cylinder, to reduce the number of PMTs by
half.

This design, which allows a tremendous reduction of the detector cost including
savings on the PMT readout, steel and acrylic vessel construction, 
is practical due to the following considerations:
\begin{enumerate}
 \item The event vertex is determined by the
center of gravity of the charge, without reliance on time-of-flight,
so that the light reflected from the top and bottom of the cylinder
will not worsen the performance of the detector module. The individual PMT hit times
are measured to a resolution of 0.5~ns for background studies.
 \item The fiducial volume is well defined with a
three-zone-structure as discussed below wherein no accurate vertex
information is needed.
\end{enumerate}

\subsubsection{Target Mass}
\label{sssec:mass}

\begin{figure}[!htb]
\begin{center}
\includegraphics[height=7cm]{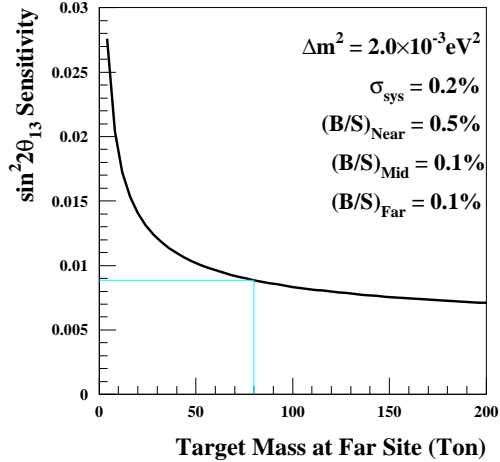}
\caption{Sensitivity of $\sin^22\theta_{13}$ at the 90\% C.L. as a
function of the target mass at the far site.
\label{fig:fig7_mass}}
\end{center}
\end{figure}

The total target mass at the far site is determined by the sensitivity
goal as is shown in Fig.~\ref{fig:fig7_mass} as a function of the far site
detector mass.
To measure $\sin^22\theta_{13}$ to better than 0.01, a total target
mass of 80--100~tons is needed, which corresponds to a statistical
uncertainty of $\sim$0.2\% after three years data taking. A larger target
mass is not attractive since the sensitivity improves rather slowly
when the target mass goes beyond 100~tons. By adopting a
multiple-module-scheme as discussed in Chapter~\ref{sec:over}, two
modules are chosen for each near site to allow a cross check of the
module behavior (within the limit of statistic uncertainties  at the near
site). For the far detector site, at least four modules are needed for
sufficient statistics to reach the designed sensitivity while
maintaining the number of modules at a manageable level. A detector
scheme of eight identical modules, each with a target mass of 20~tons,
is chosen. About 600 to 1200 events per day per module will be
detected at the Daya Bay near site (300--500~m) with about 90 events
per day per module at the far site ($>$1800~m).

\subsubsection{Three-zone Antineutrino Detector}
\label{sssec:det_3zone}

\begin{figure}[!htb]
\begin{center}
\includegraphics[height=7cm]{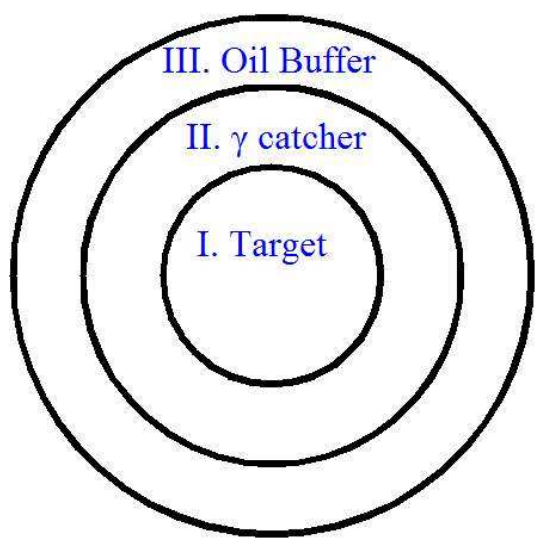}
\caption{Cross section of a simple detector module showing the
three-zone antineutrino detector.
\label{fig:fig7_3zone}}
\end{center}
\end{figure}

 A Chooz-type detector with suitable upgrades can in principle fulfill
the requirements although completely new concepts are not
excluded. The energy threshold of a Chooz-type scintillator detector
can be reduced by a three-zone structure as shown in
Fig.~\ref{fig:fig7_3zone}.
The inner-most zone (region I) is the Gd-loaded liquid scintillator
antineutrino target. The second zone (region II) is filled with
normal liquid scintillator and serves as a $\gamma$-catcher to contain the energy of
$\gamma$s from neutron capture or positron annihilation. This zone does not
serve as an antineutrino target as neutron-capture on hydrogen
does not release sufficient energy to satisfy the 6~MeV neutron detection threshold.
The outer-most zone (region III) contains mineral oil
that shields radiation from the PMT glass from entering the fiducial
volume. This buffer substantially reduces the singles rates and allows
the threshold to be lowered below 1.0~MeV. The three regions are
partitioned with transparent acrylic tanks so that the target mass
contained in region I can be well determined without the need for
event vertex reconstruction and a position cut.

\subsubsection{$\gamma$-Catcher}
\label{sssec:det_catcher}

\begin{figure}[!htb]
\begin{center}
\includegraphics[height=7cm]{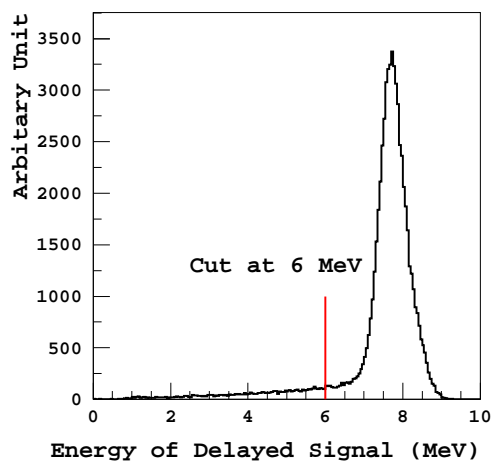}
\caption{The neutron capture energy spectrum in gadolinium as
obtained from the GEANT3 simulation. The long tail at low energies
corresponds to the escaped events.
\label{fig:fig7_neff}}
\end{center}
\end{figure}

The $\gamma$ rays produced in the target region by positron
annihilation or neutron capture will undergo many collisions with the
LS molecules to transfer most of their energy to the liquid
scintillator before converting to visible scintillation light.
However, the $\gamma$ rays can also escape from this target region and
deposit energy outside of this region. To capture the escaping $\gamma$
rays a layer of undoped liquid scintillator surrounding the target
zone is added, significantly reducing this energy loss mechanism.
The energy
spectrum of the delayed neutron capture signal is shown in
Fig.~\ref{fig:fig7_neff}.
The tail to low energies is from events with an escaping $\gamma$. The
Gd capture peak at 8~MeV is from the two most abundant isotopes of
gadolinium, ${^{155}}$Gd and ${^{157}}$Gd, with total $\gamma$
energies of 7.93 and 8.53~MeV, respectively.

\begin{figure}[!htb]
\begin{center}
\includegraphics[height=7cm]{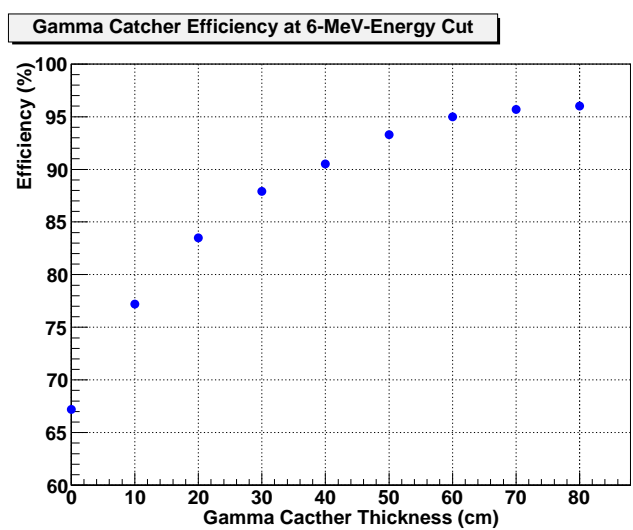}
\caption{The neutron detection efficiency as a function of the $\gamma$-catcher (GCAT).
The neutron energy cut is set at 6~MeV. The thickness
of the middle zone of the Daya Bay experiment will be 45~cm.
\label{fig:fig7_gcat}}
\end{center}
\end{figure}

A threshold of 6~MeV cleanly separates the 8~MeV neutron capture signal
from the background due to natural radioactivity. However, this
threshold will cause a loss of some neutron capture events and a corresponding loss
of detection efficiency. A simulation of the detector module giving the
correlation between the thickness of the $\gamma$-catcher region and the
neutron detection efficiency is shown in
Fig.~\ref{fig:fig7_gcat}.
The figure shows that with a $\gamma$-catcher thickness of 45~cm the
neutron detection efficiency is 92\%.  Chooz had a smaller detector
and a $\gamma$-catcher thickness of 70~cm, and neutron source test
showed a (94.6$\pm$0.4)\% detection efficiency~\cite{apollonio}. The
uncertainty includes a vertex selection uncertainty that Daya Bay will not
have. Chooz, Palo Verde and KamLAND all claimed an uncertainty on the
energy scale at 6~MeV of better than 1\%. Our detector simulation
shows that a 1\% uncertainty in energy calibration will cause a 0.2\% uncertainty in
the relative neutron detection efficiencies of different detector modules for a 6~MeV
threshold. After subtracting the vertex selection uncertainty, the
results of the efficiency test are consistent with simulation. After a
comprehensive study of detector size, detection efficiency, and
experimental uncertainties, we choose 45~cm as the thickness of the
$\gamma$-catcher.

\subsubsection{Oil Buffer}
\label{sssec:det_buffer}

\begin{figure}[!htb]
\begin{center}
\includegraphics[height=9cm]{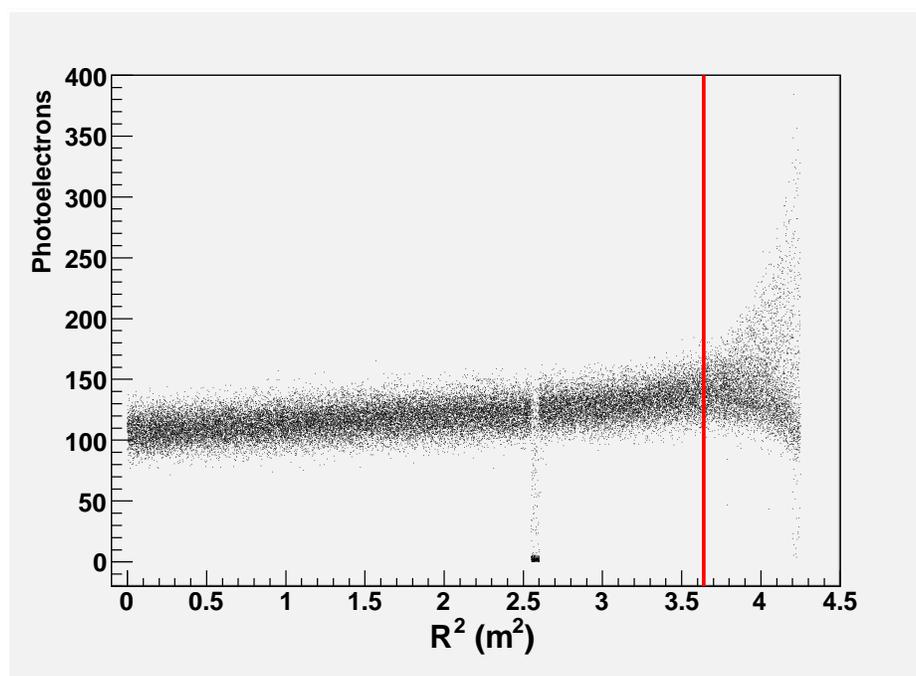}
\caption{Antineutrino detector response (in number of photoelectrons) as a function of 
radial location of a 1~MeV electron energy deposit. The mineral oil
volume has been removed and the PMTs are positioned directly outside
the $\gamma$-catcher volume.  The vertical red line is 15~cm from the
PMT surface and indicates the need for 15~cm of buffer between the PMT
surface and the region of active energy deposit in order to maintain
uniform detector response.
\label{fig:fig7_echgtot}}
\end{center}
\end{figure}

The outermost zone of the detector module is composed of mineral oil. The
PMTs will be mounted in the mineral oil next to the stainless steel
vessel wall, facing radially inward.  This mineral oil layer is
optically transparent and emits very little scintillation light. There
are two primary purposes for this layer: 1) to attenuate radiation
from the PMT glass, steel tank and other sources outside of the
module; and 2) to ensure that the PMTs are sufficiently far from the
liquid scintillator so that the light yield is quite uniform.
Simulations indicate that the location of light emission should be at
least 15~cm away from the PMT surface, as indicated in
Fig.~\ref{fig:fig7_echgtot}.
Simulation shows that with 20~cm of oil
buffer between the PMT glass and the liquid scintillator (which
corresponds to a 45~cm total oil buffer thickness), the radiation from
the PMT glass detected in the liquid scintillator is 7.7~Hz, as
summarized in Table~\ref{tab:thickness}.
\begin{table}[!htb]
\begin{center}
\begin{tabular}[c]{|l||r|r|r|r|r|} \hline
             &  & \multicolumn{4}{|c|}{Buffer Oil Thickness}  \\
  Isotope    & Concentration & 20~cm & 25~cm & 30~cm & 40~cm  \\
             &               &  (Hz) &  (Hz) &  (Hz) & (Hz)  \\ \hline\hline
${^{238}}$U  &     40~ppb    &  2.2  &  1.6  &  1.1  &  0.6  \\ \hline
${^{232}}$Th &     40~ppb    &  1.0  &  0.7  &  0.6  &  0.3  \\ \hline
${^{40}}$K   &     25~ppb    &  4.5  &  3.2  &  2.2  &  1.3  \\ \hline\hline
Total        &               &  7.7  &  5.5  &  3.9  &  2.2  \\ \hline
\end{tabular}
\caption{Radiation from the PMT glass detected in the
Gd-scintillator (in Hz) as a function of the oil-buffer thickness
(in cm). A 45~cm thick oil buffer will provide 20~cm of shielding
against radiation from teh PMT glass.
\label{tab:thickness}}
\end{center}
\end{table}

The welded stainless steel in KamLAND has an average radioactivity of
3~ppb Th, 2~ppb U, 0.2~ppb K, and 15~mBq/kg Co. Assuming the same
radioactivity levels for the vessel of the Daya Bay antineutrino
detector module, the corresponding rate from a 10-ton welded stainless
steel vessel shielding by 45~cm of oil buffer are 3.5~Hz, 2.3~Hz,
0.8~Hz and 2.2~Hz for U/Th/K/Co, respectively at a threshold of
1~MeV. The total is 8.8~Hz. The natural radioactivity of rock, buffer
water, mineral oil, dust, radon and krypton in air play a minor role,
as described in section \ref{sssec:sys_radioactivity}. The total
$\gamma$ rate is $<$50~Hz.

Since the PMTs are placed in the mineral oil, and the length of PMT
plus its base is about 25--30~cm, a 45-cm thick oil buffer will be
sufficient to suppress the $\gamma$ rate and the subsequent
uncorrelated backgrounds to an acceptable level.  The dimensions of
the antineutrino detector modules are shown in Table~\ref{tab:dimensions}.
\begin{table}[!htb]
\begin{center}
\begin{tabular}[c]{|l||r|r|r|r|r|l|} \hline
 Region & IR(m) & OR(m) & inner height(m) & 
 outer height(m)  & thickness(mm) & material \\ \hline\hline
 target           & 0.00 & 1.60 & 0.00 & 3.20 & 10.0 & Gd-LS \\ \hline
 $\gamma$-catcher & 1.60 & 2.05 & 3.20 & 4.10 & 15.0 & LS \\ \hline
 buffer           & 2.05 & 2.50 & 4.10 & 5.00 & 8.0--10.0 & Mineral oil \\ \hline
\end{tabular}
\caption{Dimensions of the mechanical structure and materials of
the antineutrino detector modules. IR (OR) refers to the inner (outer)
radius of each volume and thickness refers to the wall thickness. \label{tab:dimensions}}
\end{center}
\end{table}

The neutrino target is a cylinder of 3.2~m height and 1.6~m radius.
The $\gamma$-catcher and oil buffer are both 0.45~m thick. The
diameter of the stainless steel vessel is 5.0~m, with a height of
5.0~m and a total mass of 100~tons.

\subsubsection{2-zone vs. 3-zone Detector}
\label{sssec:det_2zone}

The possibility of adopting a detector module design  with a 2-zone structure, by
removing the $\gamma$-catcher from the current 3-zone design, has been
carefully studied. A 2-zone detector module with the same outer dimension as
the 3-zone structure has a target mass of 40~ton (keeping the same oil
buffer and $\gamma$-catcher thicknesses). The efficiency of the neutron
energy cut at 6~MeV will be $\sim$70\%, compared to $\sim$90\% with
the $\gamma$-catcher and the 2-zone 40~ton detector module will have only
$\sim$60\% more detected events than the 3-zone 20~ton detector module. The reduction
of efficiency in the neutron energy cut will introduce a larger uncertainty due to
the energy scale uncertainty. This uncertainty is irreducible, not removable
by the near/far relative measurement, in the different detector
modules due to differences in the energy scales.

The energy scale is possibly site-dependent due to variation of
calibration conditions in the different sites. According to the
experience gained from KamLAND, a 1\% energy scale stability at 8~MeV
and 2\% at 1~MeV can be readily achieved. The uncertainties in neutron
detection efficiency for a 1\% relative energy scale uncertainty have
been studied by Monte Carlo for the 2-zone 40-ton detector module and the
3-zone 20-ton detector module. The uncertainty in the relative neutron
detection efficiency for the 2-zone detector module is 0.4\% at 6~MeV as
compared with 0.22\% for the 3-zone detector module. Similar uncertainties at
4~MeV have also been studied, see Table~\ref{tab:3zone}.
\begin{table}[!htb]
\begin{center}
\begin{tabular}[c]{|l||r|r|} \hline
 Configuration & 6~MeV & 4~MeV \\ \hline\hline
 2-zone & 0.40\% & 0.26\% \\ \hline
 3-zone & 0.22\% & 0.07\% \\ \hline
\end{tabular}
\caption{Uncertainty of the neutron energy threshold efficiency caused
by uncertainty in the energy scale for 2-zone and 3-zone detector modules. The
energy scale uncertainty is taken to be 1\% and 1.2\% at 6~MeV and
4~MeV, respectively. \label{tab:3zone}}
\end{center}
\end{table}
This uncertainty will be the dominant residual detector uncertainty (see
Table~\ref{tabsyserr}), while other uncertainties are cancelled by
detector module swapping this one is not (e.g., a doubling of this uncertainty
will significantly degrade the $\sin^22\theta_{13}$ sensitivity that
can be achieved).

\begin{figure}[!htb]
\begin{center}
\includegraphics[width=10cm]{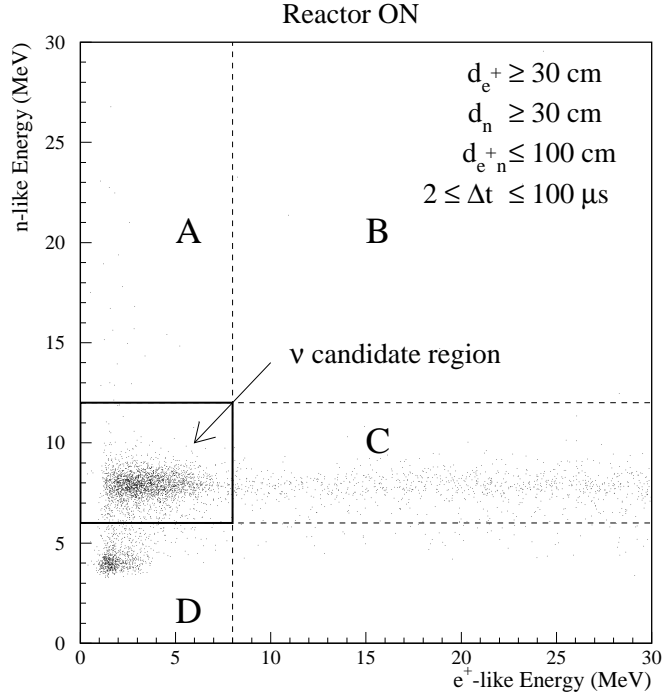}
\caption{The energy distribution observed by Chooz, horizontal
axis is the prompt signal energy; the vertical axis is the delayed
signal energy. In the region labelled D there are many background
events with delayed signal falling into the 3--5~MeV energy range.
\label{fig:fig7_chooz}}
\end{center}
\end{figure}

As shown in Table~\ref{tab:3zone}, lowering the energy cut to 4~MeV
can reduce the neutron energy threshold efficiency uncertainty.  However,
the intrinsic radioactivity from the Gd-doped liquid scintillator and
the acrylic vessel will cause a significant increase of the accidental
background rate. For external sources (such as radioactivity from the
PMTs and the rock) only $\gamma$s, with an upper limit of
$\sim$3.5~MeV, can enter the detector module. For internal sources, however,
$\gamma$s, $\beta$s, and $\alpha$s contribute --- these can produce
significant rates of signals above 3.5~MeV (e.g. ${^{208}}$Tl has
an endpoint of 5~MeV) as observed by KamLAND. Chooz has also observed
a significant number of events of delayed energy of 4--6~MeV (see
Fig.~\ref{fig:fig7_chooz}).
In addition, gadolinium has contamination from ${^{232}}$Th which
increases the rate of ${^{208}}$Tl decay in the scintillator. All of
these factors make a reduction of the neutron threshold from 6~MeV to
4~MeV undesirable. The accidental background rate would be a couple of
orders of magnitude larger with the lower threshold at 4~MeV.

\subsubsection{Expected Performance}
\label{sssec:performance}

\begin{figure}[!htb]
\begin{minipage}[t]{0.48\textwidth}
\includegraphics[width=\textwidth]{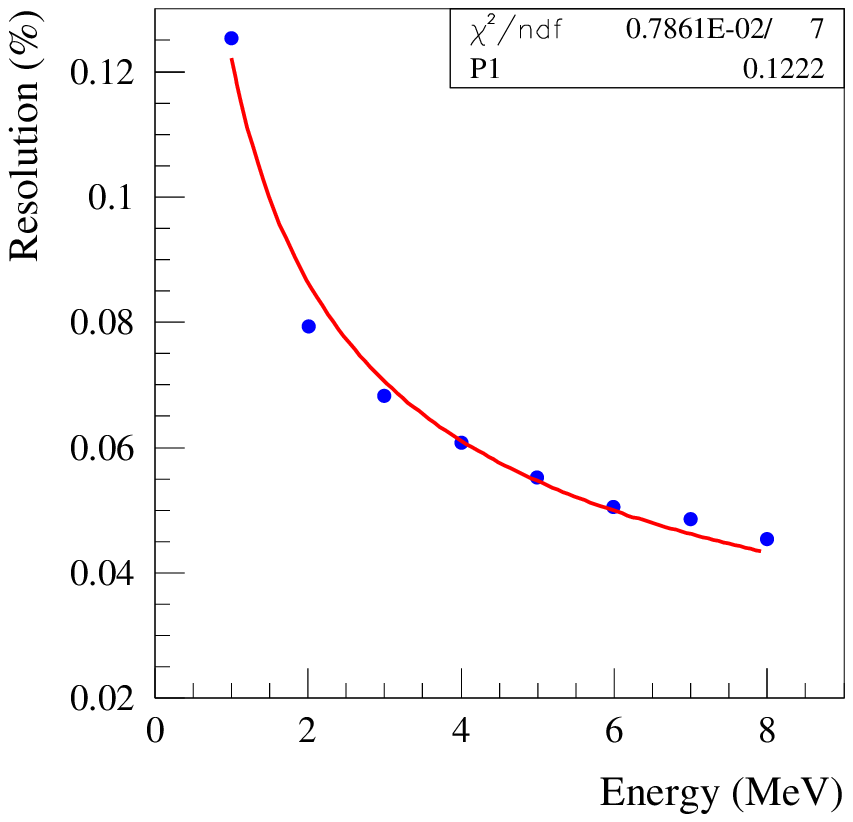}
\end{minipage}
 \hfill
\begin{minipage}[t]{0.48\textwidth}
\includegraphics[width=\textwidth]{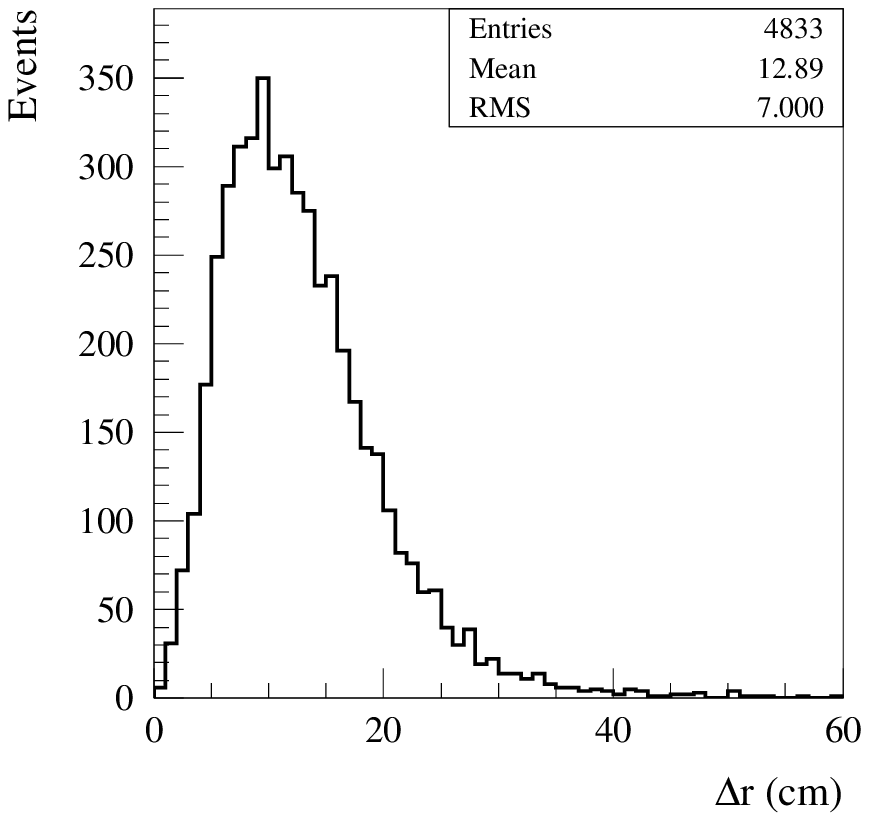}
\end{minipage}
\caption{Left: The energy reconstruction resolution for electron
events uniformly generated in the target region follows
$12.2\%/\sqrt{E(MeV)}$. Right: The vertex reconstruction
resolution for 8~MeV electron events uniformly generated in the
target region using maximum likelihood fitting. The x-axis is the
distance of the reconstructed vertex to the true vertex and the
y-axis is the number of events. \label{fig:fig7_posit}}
\end{figure}

With reflectors at the top and bottom the effective photocathode
coverage is 12\% with 224 PMTs, the light yield is $\sim$100~p.e./MeV
and the energy resolution is around
5.4\% at 8~MeV when the total-charge method is used, or 4.5\% with
a maximum likelihood fit approach. The vertex can also be
reconstructed with a resolution similar to a design with 12\% PMT
coverage on all surfaces. The vertex reconstruction resolution is
$\sim$13~cm for a 8~MeV electron event using the maximum
likelihood fit, as shown in Fig.~\ref{fig:fig7_posit}.
The horizontal axis is the distance of the reconstructed vertex to
the true vertex and the vertical axis is the number of events. Such a
vertex resolution is acceptable since the neutron capture vertex
has $\sim$20~cm intrinsic smearing, as found by Chooz~\cite{apollonio}
and by our Monte Carlo simulation as well. The intrinsic smearing of
the neutron capture vertex is caused by the energy deposition of the
$\gamma$s released from neutron-capture on Gd.

\subsection{Containers and Calibration Ports}
\label{ssec:det_containers}

The stainless steel vessel is the outer tank of the antineutrino
detector module, and surrounds the buffer oil region. It will be built
with low radioactivity 304L stainless steel and will satisfy the
following requirements:
\begin{itemize}{\setlength\itemsep{-.4ex}}
 \item[a.] leak-tight against mineral oil and water over a long period of time (10~years);
 \item [b.] chemically compatible with the mineral oil buffer;
 \item[c.] mechanical strength to support the hydrostatic pressure 
of the liquids, to support the PMT structure and to handle the stresses 
induced by transporting, lifting and handling;
 \item[d.]  minimal material so as to reduce backgrounds from radioactivity in the steel and welds.
\end{itemize}

\begin{figure}[!htb]
\begin{center}
\includegraphics[height=6cm]{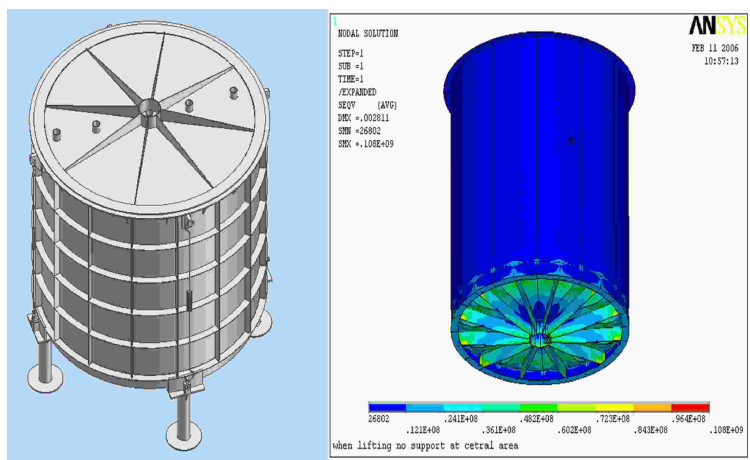}
\end{center}
\caption{3D view of the stainless steel Buffer vessel.
\label{fig:fig7_ss}}
\end{figure}

The stainless steel vessel is a cylinder of 5000~mm height and 5000~mm
diameter (external dimensions) with a 10~mm wall thickness (304L
stainless) as depicted in Fig.~\ref{fig:fig7_ss}.
It weighs about 20~tons (including the support structures) and has a
volume of $\sim$95~m$^3$ (without the chimney).

\subsubsection{Acrylic Vessel} \label{ssec:det_acrylic}

\begin{figure}[!htb]
\begin{center}
\includegraphics[height=6cm]{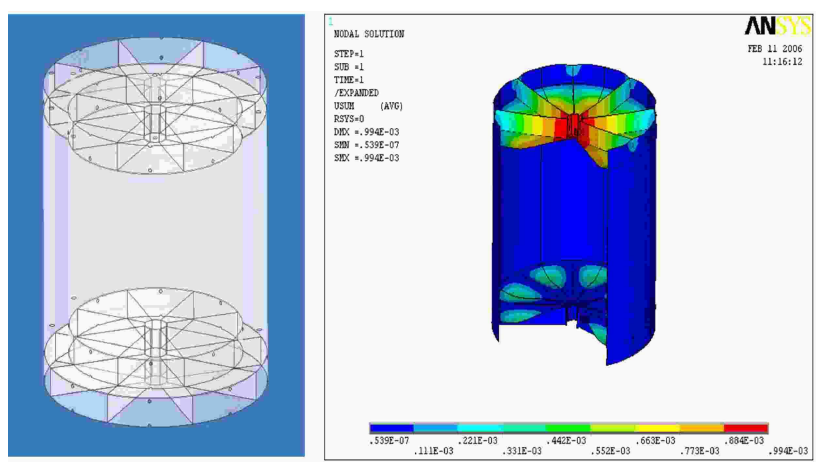}
\caption{The design of the double vessel.
\label{fig:fig7_acrylicvessel1}}
\end{center}
\end{figure}

The target vessel is a cylinder of 3200~mm height and 3200~mm diameter
(external dimensions) with 10~mm wall thickness (acrylic). It weighs
$\sim$580~kg, and contains a volume of $\sim$25~m$^3$ (without
chimneys). The $\gamma$-catcher vessel surrounding the Target is a
cylinder of 4100~mm height and 4100~mm diameter (external dimensions)
with a 15~mm wall thickness (acrylic). It weighs 1420~kg, and contains
a volume of 28~m$^3$ ($53~m^3 - 25~m^3$) (without the chimneys). At the top of
the target vessel, there are two or three chimneys for injecting the
LS and for passage of radioactive calibration sources. There will be
one or two chimneys for the $\gamma$-catcher as well. The chimneys
diameter will be $\sim$50--100~mm. Drawings of the target and the
$\gamma$-catcher vessels are shown in Fig.~\ref{fig:fig7_acrylicvessel1}.

The target and $\gamma$-catcher vessels will be built of acrylic which
is transparent to photons with wavelength above $\sim$300~nm (50\% at
300~nm~\cite{acrylic}). Both vessels are designed to contain aromatic
liquids with a long term leak-tightness (free from leakage for ten
years) and stability. The critical constraint is the chemical
compatibility between the vessel and the scintillating liquids, for at
least five years. There must be no degradation of the liquid
properties (scintillation efficiency, absorption length) nor any
significant degradation of the acrylic material (yellowing or crazing
of more than a few percent of the acrylic surface area). The
$\gamma$-catcher vessel will also be chemically compatible with the
mineral oil in the buffer region.

Acrylic is normally PMMA plus additional ingredients to prevent
aging and UV light absorption. Different manufacturing companies
have different formulas and trade secrets for the additional
ingredients, resulting in  different appearance, chemical
compatibility, and aging effects. For the material choice, we have
surveyed many kinds of organic plastic. We have identified two
possible sources for fabrication of the acrylic vessels: the Jiang
Chuan Organic Plastic Ltd. Corp, located in city of Lang Fang, Hebei
Province, China, and the Gold Aqua System Technical Co. in
Kaoshiung, Taiwan (subsidiary of the Nakano company).

The Jiang Chuan Corp. uses a centrifugal casting method for their
construction of the vessel.
The approach of Nakano's subsidiary company uses bent plate sheets
to be glued together by the polymerization method. It appears at
this time that this method will be preferable as it should provide a
higher quality vessel.

\begin{figure}[!htb]
\begin{center}
\includegraphics[height=7cm]{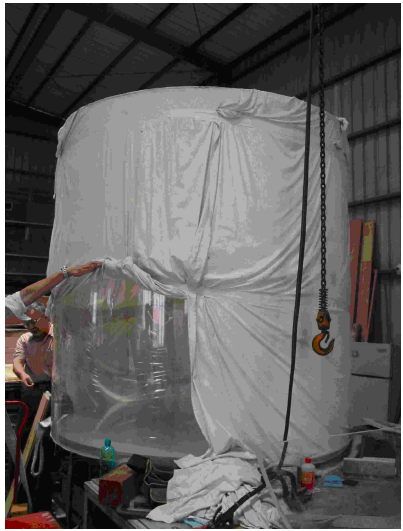}
\caption{A sample acrylic vessel produced at the Gold Aqua Technical Co. in Taiwan.
The diameter and
the height are both 2~m, with high precision.
\label{fig:fig7_acrylicvessel}}
\end{center}
\end{figure}

In the polymerization gluing method, they add the same raw materials
as the acrylic (PMMA + ingredients) into the gap between the plates.
Thus the joints consist of exactly the same acrylic material as the
joined plates, and there is no difference in their mechanical,
chemical and optical properties. During polymerization, UV light is
used instead of heating, in order to prevent the bent sheets from
rebounding. The speed of polymerization is controlled to minimize the
remaining stress. Once the tank is fabricated in shape, it will be put
in a thermally insulated enclosure for up to a month ($\sim$1~week in
our case) to be heated for releasing the stresses.  The temperature
will be controlled within $\pm1^\circ$~C. Different acrylic types,
shapes, thicknesses, etc., need different temperature curves for
bending and curing. Hence experience is very important.  The geometric
precision can be controlled to $\pm$2~mm for a 2~m-diameter tank. The
tank can have reinforcement structures at both the top and bottom;
therefore the mechanical strength is not a problem for a very thin
tank ($\sim$1~cm). However, a thin sheet tends to have more residual
stress which may be problematic for chemical compatibility. The
minimum thickness of our tank is to be discussed after the
compatibility tests of acrylic sheets with liquid scintillator and
mineral oil are completed. Figure~\ref{fig:fig7_acrylicvessel} shows a
example of the acrylic plastic vessel.
Mechanically, the double vessels must be strong and stable enough to
ensure identical shapes between near and far target vessels. 

The manufacture and transportation of the detector vessels can cause
complications to the experiment and they need to be studied in great
detail. Simulation has shown that the transportation phase is
hazardous for a double acrylic vessel which has been completely
assembled.
The vessel design, and the design of a transportation system to
isolate large shock and vibration loads, requires further analysis.
This problem could also be solved, without changing the baseline
design, by transporting the target and $\gamma$-catcher vessels
separately, and integrate and glue the $\gamma$-catcher top lid and
the chimney in the filling hall.

All three regions within the antineutrino detector module have to be filled
simultaneously. The filling phase generates constraints related to the
differences in height of the liquid. According to mechanical
simulations, if we neglect density variations, the acceptable
difference in relative fluid heights is 30~cm.

\subsubsection{Calibration Ports}
\label{sssec:det_ports}

In addition to the central chimney port, the buffer vessel lid will
have several (3--6) ports, each 100--200~mm diameter, to facilitate
the deployment of radioactive calibration sources and light sources.
These ports will have gate valves to isolate the calibration devices
when they are not in use and facilitate their removal. Around the
side wall of the stainless steel vessel there will be 32 ports of 5--10~cm
diameter for high voltage, signal, and instrumentation cables.

The cables will be routed down to the bottom of the water pool and
up the side so as to minimize interference with the water Cherenkov
system. The cables may either be contained in pipes, or we will
design a fail-safe isolation connector to allow the cables to be in
the water.

\subsection{Liquid Scintillator}
\label{ssec:det_LS}

The gadolinium-loaded organic liquid scintillator, Gd-LS, is a crucial
component of the antineutrino detector. The H atoms ("free protons")
in the LS serve as the target for the inverse beta-decay (IBD)
reaction, and the Gd atoms produce the delayed coincidence, so
important for background reduction, between the prompt positron and
the delayed neutron from the IBD. The LS contains $\sim$10\%
hydrogen. Gd has a very large neutron-capture cross section; the
$\sigma$ of natural abundance Gd is 49,000 barns so that isotopic
enrichment of the Gd is not required. Two stable isotopes of Gd
contribute most of this cross section: $\sigma({^{155}}{\rm
Gd})=61,400$ barns and $\sigma({^{157}}{\rm Gd})=255,000$ barns.
Furthermore, neutron-capture on Gd leads to emission of $\gamma$ rays
with a total energy of $\sim$8~MeV, that is much higher than the
energies of the $\gamma$ rays from natural radioactivity which are
normally below 3.5~MeV. Hence, organic LS doped with a small amount of
Gd is an ideal antineutrino target and detector. Both
Chooz~\cite{apollonio} and Palo Verde~\cite{boehm} used 0.1\%
Gd-doping (1~g Gd per kg LS) that yielded a capture time of
$\tau\sim$28~$\mu$s, about a factor of seven shorter than that on
protons in undoped liquid scintillator, ($\tau\sim$180~$\mu$s). This
shorter capture time reduces the backgrounds from random coincidences
by a factor $\sim$7.

To detect reactor antineutrinos with high precision, the Gd-LS must
have the following key properties: a) high optical transparency = long
optical attenuation length, (b) high photon production (high light
yield) by the scintillator, (c) ultra-low impurity content, mainly of
the natural radioactive contaminants, such as U, Th, Ra, K, and Rn,
and (d) long-term chemical stability, over a period of several
years. It is necessary to avoid any chemical decomposition,
hydrolysis, formation of colloids, or polymerization, which can lead
over time in the LS to development of color, cloudy suspensions, or
formation of gels or precipitates, all of which can degrade the
optical properties of the LS. To achieve these criteria, R\&D is
required on a variety of topics, such as: (1) selection of the proper
organic LS, (2) development of chemical procedures to synthesize an
organo-Gd complex that is soluble and chemically stable in the LS, (3)
purification of the components of the Gd-LS, and (4) development of
analytical methods to measure these key properties of the Gd-LS over
time. These topics will be discussed in the subsections below.

Major R\&D efforts on LS and Gd-LS are being carried out at BNL in the
U.S., IHEP in the Peoples Republic of China and JINR in Russia:
\begin{itemize}
 \item[A.] The Solar-Neutrino/Nuclear-Chemistry Group in the BNL
Chemistry Department has been involved since 2000 in R\&D of chemical
techniques for synthesizing and characterizing organic liquid
scintillators loaded with metallic elements, M-LS. They helped to
develop a proposed new low-energy solar-neutrino detector,
LENS/Sol~\cite{LENS/Sol}. Concentrations of M in the LS $\sim$5-10\%
by weight were achieved to serve as targets for solar neutrino
capture, with M being ytterbium (Yb$^{3+}$) and indium (In$^{3+}$). It
was obvious that these chemical results could readily be extended to
the new reactor antineutrino experiments, to prepare Gd-LS (with
Gd$^{3+}$) at the much lower concentrations required for neutron
detection, $\sim$0.1\%. BNL began R\&D in 2004 on solvent extraction
methods to synthesize Gd-LS.  \item[B.] Nuclear chemists at IHEP also
began their R\&D on Gd-LS in 2004. They have tended to focus on
preparing solid organo-Gd complexes, the idea being that the solid
should be readily dissolvable in the LS, to allow preparation of the
Gd-LS at the Daya Bay reactor site.  \item[C.] The JINR chemists, who
have long experience in the development of plastic scintillators, are
currently studying the characteristics of different LS solvents,
especially Linear Alkyl Benzene. They have also began some
collaborative work on Gd-LS with chemists at the Institute of Physical
Chemistry of the Russian Academy of Sciences, who also did R\&D on
In-LS for LENS/Sol starting in about 2001.
\end{itemize}

It should be noted that the general approach of these different groups
is pretty much the same, to prepare organo-Gd complexes that are
soluble and stable in the LS organic solvent. However, the chemical
details of their R\&D programs do differ in significant respects at
present, such as in the purification procedures, the control of pH,
and reliance on either solvent-extraction methods or formation of
Gd-precipitates to isolate the Gd organo-complex. We discuss the
current status of the major similarities and differences of these
approaches. We note that the differences cited do not seriously affect
the general goals of the experiment and are acceptable at this stage
of development of the experiment.  As closer cooperative R\&D ties
between these groups develop in the coming months, it is expected that
these issues will be resolved.

\subsubsection{Selection of Solvents}
\label{ssssec:det_LS_solvents}

Several aromatic (organic compounds based on benzene) scintillation
liquids were studied at BNL to test their applicability as solvents
for Gd-LS. (1) Pseudocumene (PC), which is the 1,2,4-isomer of
trimethylbenzene (and mesitylene, the 1,3,5-isomer), has been the most
commonly used solvent for Gd-LS in previous neutrino experiments. But
it has a low flash point (48$^\circ$~C) and aggressively attacks
acrylic plastic. (2) Phenylcyclohexane (PCH) has a lower reactivity
than PC, but only half of the light yield. (3) Both
di-isopropylnaphthalene (DIN) and 1-phenyl-1-xylyl ethane (PXE) have
optical absorption bands in the UV region below 450-nm that cannot be
removed by our purification procedure (although we note that Double
Chooz has chosen PXE as a satisfactory solution for their
requirements). (4) Recently, a new LS solvent, Linear Alkyl Benzene
(LAB)~\cite{SNO+}, has been identified as a potentially excellent
solvent for Gd-LS. LAB is composed of a linear alkyl chain of 10--13
carbon atoms attached to a benzene ring, and has a light yield
comparable to PC. LAB also has a high flash point, which significantly
reduces the safety concerns. It is claimed by the manufacturers to be
biodegradable, and is relatively inexpensive, since it is used in the
industrial manufacturing of detergents. (5) Mineral oil (MO) and
dodecane (DD) both have very good light transmission in the UV-visible
region so that no further purification is required. They produce
little or no scintillating light. It has been reported that mixtures
of PC + mineral oil will not attack acrylic.

PC and LAB, as well as mixtures of PC with DD and of LAB with PC, have
been selected as the candidate scintillation liquids for loading Gd in
the Daya Bay neutrino detector. In China, an unpurified LAB sample
obtained from Fushun Petroleum Chemical, Inc. has an attenuation
length longer than 30 m; if its quality is uniform from production
batch to batch, it can be used directly as the required solvent
without further purification. In the U.S., pure LAB has been obtained
from the Petresa Company in Canada. Even though this LAB is quite
pure, BNL routinely uses purification procedures to ensure that all of
its LAB samples have uniform properties.  The chemical properties and
physical performance of these scintillation solvents, plus mineral oil
and dodecane, are summarized in Table~\ref{tab:BNL_LS_Table}.
\begin{table}[!htb]
\begin{center}
\begin{tabular}[c]{|l|c|c|c|c|c||r|} \hline
LS & Gd Loading & Density  & abs$_{430}$ &
Purification & Relative & Flash Point\\
  & in LS  &  (g/cm$^3$)&  & Method & Light Yield  &  \\ \hline\hline
 PC  & Yes  & 0.889 &  0.002 & Distillation & 1 & 48$^o$C   \\ \hline
 PCH  & Yes  & 0.95 &  0.001 & Column & 0.46 & 99$^o$C   \\ \hline
 DIN  & Yes  & 0.96 &  0.023 & Column & 0.87 & $\ge$140$^o$C   \\ \hline
 PXE  & Yes, but is not stable  & 0.985 &  0.022 & Column &
 0.87 & 167$^o$C \\ \hline
 LAB  & Yes  & 0.86 &  0.000 & Column & 0.98 & 130$^o$C     \\ \hline
 MO  & No  & 0.85 &  0.001 & Not needed & N\/A & 215$^o$C     \\ \hline
 DD  & No  & 0.75 &  0.000 & Not needed & N\/A & 71$^o$C     \\ \hline
\end{tabular}
\caption{ Properties of Selected Liquid Scintillators, as compiled at BNL
\label{tab:BNL_LS_Table}}
\end{center}
\end{table}

\subsubsection{Preparation of Gadolinium Complexes}
\label{ssssec:det_LS_prep}

One of the major research challenges is how to dissolve the Gd into
the liquid scintillator. Since the LS detector is made of an
aromatic organic solvent, it is
difficult to add inorganic salts of Gd into the organic LS. The only
solution to this problem is to form organometallic complexes of Gd
with organic ligands that are soluble in the organic LS.

The recent Chooz and Palo Verde antineutrino experiments used
different methods to produce their Gd-doped liquid scintillator. In
the Chooz experiment, Gd(NO$_3$)$_3$ was directly dissolved in the LS,
resulting in a scintillator whose attenuation length decreased at a
relatively rapid rate, 0.4\% per day. As a result, Chooz had to be
shut down prematurely.  On the other hand, the Palo Verde experiment
used the organic complex, Gd-ethylhexanoate, yielding a
scintillator which aged at a much slower rate, 0.03\% per day.

In the Periodic Table, Gd belongs to the lanthanide (Ln) or rare-earth
series of elements. Lanthanides such as Gd can form stable
organometallic complexes with ligands that contain oxygen, nitrogen,
and phosphorus, such as carboxylic acids, organophosphorus compounds,
and beta-diketones.  Several recipes for Ln-LS have been developed
based on these organic ligands. For example, Gd-ethylhexanoate
(Palo Verde, Univ. Sheffield, Bicron), In-, Yb- and Gd-carboxylates (BNL for
LENS and Daya Bay), Gd-triethylphosphate (Univ. Sheffield),
Yb-dibutyl-butylphosphonate (LENS), and Gd-acetylacetonate (Double-Chooz).

Complexants that have been studied at BNL are (i) carboxylic acids
(R-COOH) that can be neutralized with inorganic bases such as NH$_4$OH
to form carboxylate anions that can then complex the Ln$^{3+}$ ion,
and (ii) organic phosphorus-oxygen compounds, "R-P-O", such as
tributyl phosphate (TBP), or trioctyl phosphine oxide (TOPO), that can
form complexes with neutral inorganic species such as
LnCl$_3$~\cite{MLS}.  Initially, work was done with the R-P-O
compounds. The extraction of Ln is effective, but the attenuation
length is only a few meters and the final Ln-LS was not stable for
more than a few months. On the other hand, the carboxylic acids,
"RCOOH", form organic-metal carboxylate complexes that can be loaded
into the LS with more than 95\% efficiency using solvent-solvent
extraction.  Moreover the carboxylic acids are preferable because they
are less expensive and easier to dispose of as chemical waste,
compared to the phosphorus-containing compounds. In principle, the
chemical reactions are (a) neutralization, RCOOH + NH$_4$OH
$\rightarrow$ RCOO$^{-}$ + NH$_{4}^{+}$+ H$_2$O in the aqueous phase,
followed by (b) Ln-complex formation, Ln$^{3+}$ + 3RCOO$^-
\rightarrow$ Ln(RCOO)$_3$, which is soluble in the organic LS. These
reactions are very sensitive to pH: the neutralization step to form
the RCOO$^-$ depends on the acidity of the aqueous solution, and
hydrolysis of the Ln$^{3+}$ can compete with formation of the
Ln(RCOO)$_3$ complex.

A range of liquid carboxylic acids with alkyl chains containing from 2
to 8 carbons was studied.  It was found that acetic acid (C2) and
propionic acid (C3) have very low efficiencies for extraction of Ln
into the organic phase. Isobutyl acid (C4) and isovaleric acid (C5)
both have strong unpleasant odors and require R-P-O ligands to achieve
high extraction efficiencies for Ln. Carboxylic acids containing more
than 7 carbons are difficult to handle because of their high
viscosity; also as the number of carbon atoms increases in the
carboxylate complex, the relative concentration by weight of Ln
decreases. The best complexant found to date is the C6 compound,
2-methylvaleric acid, C$_5$H$_{11}$COOH or "HMVA".

Several instrumental and chemical analytical techniques have been used
at BNL as guides for optimization of the synthesis procedures for
Gd-LS.  Besides the measurements of light yield and optical
attenuation length to be described below, are measurements in the LS
of the concentrations of: (1) Gd$^{3+}$ by spectrophotometry, (2) the
total carboxylic acid, R-COOH, by acid-base titrations, (3) the
uncomplexed R-COOH by IR spectroscopy, (4) the different organo-Gd
complexes in the organic liquid by IR spectroscopy, (5) the H$_2$O by
Karl-Fischer titration, and (6) the NH$_4^+$ and Cl$^-$ by
electrochemistry with specific ion-sensitive electrodes. These
measurements produced very interesting results that indicated that the
chemistry of the Gd-LS is more varied and complicated than what is
expected from the simple chemical reactions (a) and (b) listed
above. The Gd molecular
complex in the LS is not simply Gd(MVA)$_3$, but contains some OH as
well, and the form of this complex changes with changing pH. So, even
though the long-term studies consistently show that the Gd-LS is
chemically stable for periods $\ge$1 year, there is the lingering
concern that hydrolysis reactions might occur over long times in the
LS. Careful attention to chemical details, especially pH control, is
crucial here, as is long-term monitoring of the Gd-LS. 

To date at
BNL, many hundreds of Ln-LS samples have been synthesized, including
scores of Gd-LS.~\cite{GdLS} There are two approaches for preparing batches of the
Gd-LS: (i) synthesizing each batch at the desired final Gd
concentration, 0.1--0.2\%, or (ii) synthesizing more concentrated
batches, $\ge$1--2\% Gd, and then diluting with the organic LS by a
factor $\ge$10 to the desired concentration. The two approaches are
not identical, with regard to possible long-term effects such as
hydrolysis and polymerization. Approach (ii) is currently favored because it
simplifies the logistics of preparing and transporting very large
volumes of Gd-LS. At IHEP, thirteen organic ligands including four
organophosphorus compounds, five carboxylic acids, and four
$\beta$-diketones have been tested. The carboxylic acids seem most
suitable; three of them have been used for further study. The Gd
carboxylate can be synthesized by the following methods:
[a.] Carboxylic acids are neutralized by ammonium hydroxide and
 reacted with GdCl$_3$ to form a precipitate. The solid is collected
 by filtration, washed with distilled water, and dried at room temperature.  
[b.] Carboxylic acids are dissolved in an organic solvent that is also the
 LS and mixed with a GdCl$_3$ water solution. Then the pH of the
 solution is adjusted with ammonium hydroxide. The Gd-carboxylate
 is simultaneously formed and extracted into the LS solvent.
Method [a], the preparation of the solid Gd complex, is currently
being emphasized at IHEP.

After the Gd-complex is synthesized and dissolved in the LS, a primary
fluorescent additive and a secondary spectrum shifter (both called
"fluors") are added. At IHEP, the final concentration of the solutes
includes 1~g/L Gd, 5~g/L PPO (primary), and 10~mg/L bis-MSB
(secondary). The resulting liquid is then pumped through a 0.22-$\mu$m
filter and bubbled with nitrogen for the removal of air. At BNL, the
fluors, butyl-PBD (3~g/L) and bis-MSB (15~mg/L), are used. No
filtration is applied.

\subsubsection{Purification of Individual Components for Gd-LS}
\label{ssssec:det_LS_purification}

Most purification steps developed at BNL are applied before and during
the synthesis of the Gd-LS~\cite{th-rev}. Chemical separation schemes
that would be used after the Gd-LS has been synthesized are usually
unsuitable because they would likely remove some of the Gd as well as
other inorganic impurities.

\begin{figure}[!htb]
\begin{center}
\includegraphics[height=7cm]{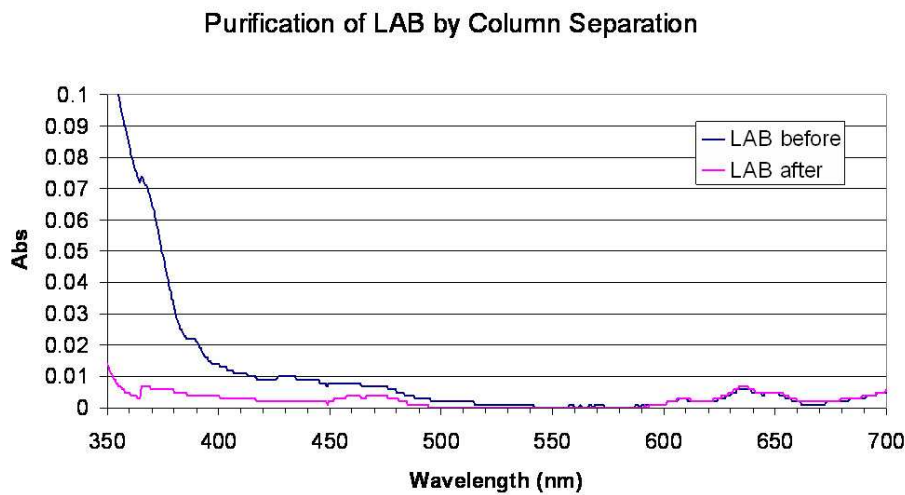}
\caption{UV-visible spectra of LAB before and after purification
\label{fig:fig7_BNL_LS1}}
\end{center}
\end{figure}

The removal of non-radioactive chemical impurities can increase the
transmission of the light in the LS and enhance the long-term
stability of the Gd-LS, since some impurities can induce slow chemical
reactions that gradually reduce the transparency of the
Gd-LS. Chemical purification steps have been developed for use prior
to or during the chemical synthesis: (1) The purification of chemical
ingredients in the aqueous phase, such as ammonium hydroxide and
ammonium carboxylate, is done by solvent extraction with toluene mixed
with tributyl phosphine oxide (TBPO). (2) LAB, which has low volatility, is purified by
absorption on a column of activated Al$_2$O$_3$. (3) High-volatility
liquids, such as the carboxylic acids and PC, are purified by
temperature-dependent vacuum distillation at $\le$0.04~bar.  
Vacuum distillation should remove any radioisotopic impurities, including radon.
Figure~\ref{fig:fig7_BNL_LS1}
compares the optical spectra for LAB before and after purification with
the activated Al$_2$O$_3$ column.

Two methods, cation exchange and solvent extraction, are being
considered at BNL for the purification of radioactive impurities
associated with Gd, mainly the U and Th decay chains.  The contents of
the radioactive impurities in the commercially obtained 99\% and
99.999\% GdCl$_3$.6H$_2$O solids that are used as starting materials
were measured by low-level counting at BNL and at the New York State
Department of Health and found to be less than the detectable limits
(10$^{-8}$~g/g). More sophisticated radioactivity measurement steps
will have to be developed to quantify these radioactive species at
concentrations of 10$^{-9}$~g/g in the Gd (implying impurity levels of
10$^{-12}$~g/g in the final 0.1\% Gd-LS) to fulfill our criterion 
of a random singles rate below 50~Hz (with 0.8~Hz from radioactive 
contamination of ${^{238}}$U, ${^{232}}$Th, and
${^{40}}$K in the Gd-LS). Although this goal is achievable
routinely for unloaded LS (i.e., without added Gd),~\cite{Kam},
special care is required for Gd-loaded LS since the Gd (and other
lanthanides) obtained in China usually contain ${^{232}}$Th at a
level of $\sim$0.1~ppm. For Gd loading of 0.1\% by weight in the
antineutrino detector, the Gd starting material has to be purified to a level 
$\le$10$^{-10}$g/g. In order to eliminate the Th, Gd$_2$O$_3$ powder at
IHEP is dissolved in hydrochloric acid and passed through a
cation-exchange resin column. Preliminary assays at IHEP
showed that this Gd purification procedure reduced the Th content at
the ppb level by a factor of four.

\subsubsection{Characterization of Gd-LS}
\label{ssssec:det_GdLS_char}

\begin{figure}[!htb]
\begin{center}
\includegraphics[height=7cm]{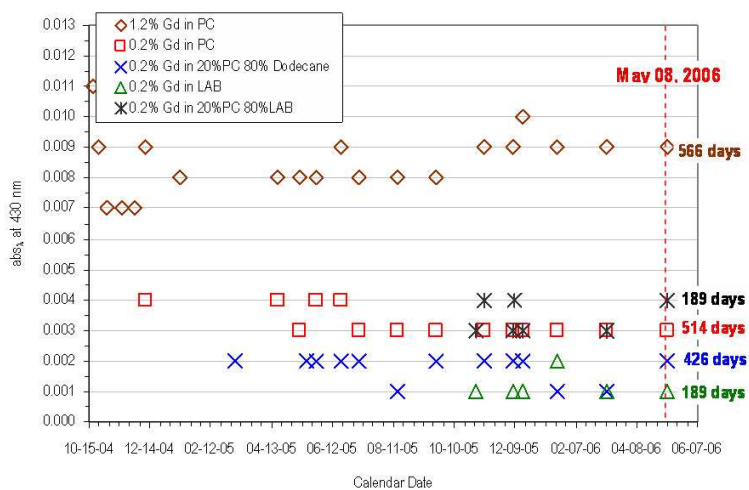}
\caption{The UV absorption values of BNL Gd-LS samples at 430~nm as a function of time
\label{fig:fig7_BNL_LS3}}
\end{center}
\end{figure}

The long-term stability of the Gd-LS preparations is periodically
monitored in a "QC", quality control, program, by measuring their
light absorbance and light yield. Samples from the same synthesis
batch are sealed respectively in 10-cm optical glass cells for UV
absorption measurements, and in scintillation vials for light yield
measurements. Monitoring the UV absorption spectrum as a function of
time gives a more direct indication of chemical stability than does
the light yield. In Fig.~\ref{fig:fig7_BNL_LS3},
the UV absorption values for a wavelength of 430~nm (in the UV
spectrometer) are plotted for BNL Gd-LS samples as a function of
calendar date, until May 2006, for different concentrations of Gd from
0.2\% to 1.2\% by weight in a variety of solvent systems --- pure PC,
pure LAB, and mixtures of PC+dodecane and of PC+LAB.
The figure shows that, since synthesis, samples of: (a) the 1.2\%
and 0.2\% of Gd in pure PC have so far been stable for more than {1.5
and 1}~years, respectively; (b) the 0.2\% of Gd in the mixture of 20\%
PC and 80\% dodecane has so far been stable for more than {a year};
and (c) the recently developed 0.2\% of Gd in pure LAB and in 20\% PC + 80\% LAB have been stable
so far for approximately 6~months.

The value of the optical attenuation length, $L$, is extrapolated from
the UV absorption data. It is defined as the distance at which the
light transmitted through the sample has its intensity reduced to
$1/e$ of the initial value: {$L$} = 0.434 {d/a}, where {$a$} is the
absorbance of light (at a reference wavelength, usually 430~nm)
measured in an optical cell of length {d}. Note that for {d} = 10~cm,
a value of {$a$} = 0.004 translates into an attenuation length
{$L$}$\sim$11~m.  However, it is difficult to extract accurate optical
attenuation lengths from these short pathlength measurements because
the {$a$} values are close to zero. Measurements over much longer
pathlengths are needed. BNL has constructed a system with a
1-meter-pathlength, horizontally aligned quartz tube.  The light
source is a He-Cd, blue laser with $\lambda$ = 442~nm.  The light beam
is split into two beams with 80\% of the light passing through the 1-m
tube containing the Gd-LS before arriving at a photodiode
detector. The remaining 20\% of the light passes through an air-filled
10-cm cell and reaches another photodiode detector to measure the
fluctuations of the laser intensity, without any interactions in the
liquid.  Use of this dual-beam laser system with 1-m pathlength
confirmed the values of the attenuation length extrapolated from the
measurements with the 10-cm cell in the UV Spectrometer. For 0.2\% Gd
in a 20\% PC + 80\% dodecane mixture without fluors, the 1-m
measurement gave 99.54\% transmission, corresponding to attenuation
length $\sim$22~m.  This agreed with the value $\sim$21.7~m that was
extrapolated from the measured {$a$}= 0.002 in the 10-cm cell.

\begin{figure}[!htb]
\begin{center}
\includegraphics[height=7cm]{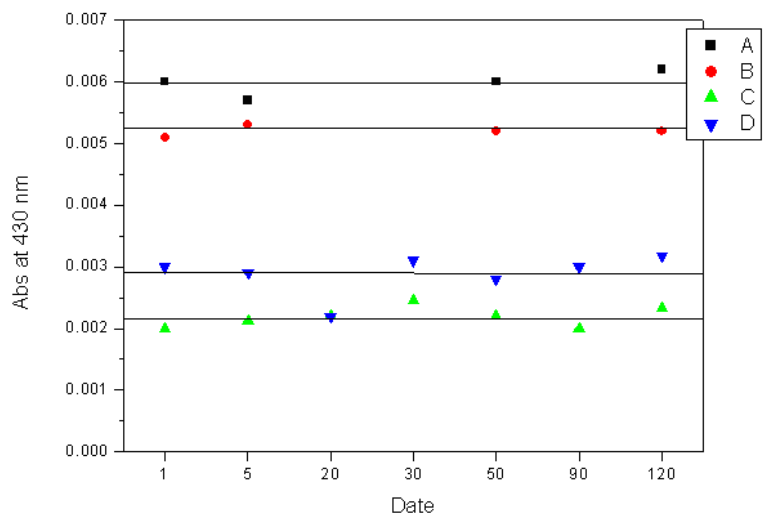}
\caption{Long-term Stability Test: 2~g/L IHEP Gd-LS as a Function of Time.
\label{fig:fig7_stability}}
\end{center}
\end{figure}

The long-term stability of the Gd-LS developed at IHEP has also been
investigated with a UV-Vis spectrophotometer using a 10-cm optical
cell. IHEP also has developed an optical system with variable vertical
pathlengths up to $\sim$2 m to measure the optical attenuation more
accurately.  Figure~\ref{fig:fig7_stability} shows the long term
stability over time of four IHEP Gd-LS samples as measured by optical
absorption at 430~nm.
In all of these IHEP samples, fluors were added, 5~g/L PPO and 10~mg/L
bis-MSB. The concentrations of Gd, complexing ligand, and solvent in
the four samples are:
[A.] 2~g/L Gd, isonanoate as ligand, 4: 6 Mesitylene/dodecane;
[B.] 2~g/L Gd, ethylhexanoate as ligand, 2: 8 Mesitylene/dodecane;
[C.] 2~g/L Gd, isonanoate as ligand, LAB;
[D.] 2~g/L Gd, 2:8 ethylhexanoate as ligand, 2: 8 Mesitylene/LAB;
The IHEP results show that the variations of the absorption values are
very small during the four-month period for all four samples. The
attenuation lengths of samples C and D are longer than 10~m.  The IHEP
QC program of long-term stability testing will continue for
$>$1~year. At JINR, optical properties and light yields of samples of
LAB are being measured.

The light yield of the Gd-LS is also measured at BNL and at IHEP. At
BNL, a scintillation vial containing ten mL of Gd-LS plus the
wavelength-shifting fluors, butyl-PBD (3 g/L) and bis-MSB (15~mg/L),
is used for measurement of the photon production.  The value of the
Gd-LS light yield, which is determined from the Compton-scattering
spectrum produced by an external $^{137}$Cs $\gamma$-ray source that
irradiates the sample, is quoted in terms of S\%, relative to a value
of 100\% for pure PC with no Gd loading. Measured S\% values are
respectively 95\% for 95\% LAB + 5\% PC, and 55\% for 0.1\% Gd in 20\%
PC + 80\% dodecane.

Table~\ref{tab:lightyield} lists the light yield for several IHEP
Gd-LS samples, relative to a value of 100\% for an anthracene crystal.
\begin{table}[!htb]
\begin{center}
\begin{tabular}[c]{|l|c|c|c||r|} \hline
 Gd(g/L) & Scintillator & Complex               & Solvent
 & Light Yield  \\ \hline\hline
      ---  & PPO bis-MSB  &    ---                  & PC:dodecane
      & 0.459        \\ \hline
      ---  & PPO bis-MSB  &    ---                  & LAB
      & 0.542        \\ \hline
   1.5   & PPO bis-MSB  & Gd-ethylhexanoate & 2:8 PC:LAB
   & 0.538        \\ \hline
   2.0   & PPO bis-MSB  & Gd-ethylhexanoate & 2:8 PC:LAB
   & 0.528        \\ \hline
   1.5   & PPO bis-MSB  & Gd-isononanoic acid   & LAB
   & 0.492        \\ \hline
   2.0   & PPO bis-MSB  & Gd-isononanoic acid   & LAB
   & 0.478        \\ \hline
\end{tabular}
\caption{Light yield for several Gd-LS samples prepared at 
IHEP, measured relative to an anthracene crystal.
\label{tab:lightyield}}
\end{center}
\end{table}
It is seen that the concentration of Gd loading has very little effect
on the light yield.

\subsubsection{Comparisons with Commercial Gd-LS}
\label{ssssec:det_LS_BNL_comparisons}

At BNL, a sample of commercially available Gd-LS, purchased from
Bicron, BC-521, containing 1\% Gd in pure PC, has been compared with a
BNL Gd-LS sample containing 1.2\% Gd in PC. BC-521 is the concentrated
Gd-doped scintillator with organic complexing agent in PC that was
used in the Palo Verde reactor experiment after it was diluted to 40\%
PC + 60\% mineral oil. The light yields of the respective BNL and
Bicron samples were found to be comparable, 82\% vs. 85\%, when
measured at BNL relative to 100\% PC, and, as quoted by Bicron, 57\%
relative to anthracene. However, the attenuation length for the
BNL-prepared Gd-LS was $\sim$2.5 times longer than the value for the
Bicron BC-521 sample, 6.2~m vs. 2.6~m as measured at BNL; Bicron
quoted a value $>$4.0 m for its sample.  This significant difference
in attenuation may reflect the care put into the BNL pre-synthesis
purification steps.

The chemical stability of these BNL and Bicron BC-521 samples are
being followed in our QC program. No perceptible worsening of the
optical properties of these samples has been observed over periods of
1.5 and $\sim$1~years, respectively. Note that Bicron simply
characterizes the stability of its BC-521 as being "long term". 

\subsubsection{Large Scale Production of Gd-LS}
\label{ssssec:det_LS_BNL_RD}

Tasks that have begun or will be undertaken in the next several months
are as follows: (1) to continue the QC program of long-term stability
of different Gd-LS preparations; (2) to determine the quality,
quantity, and types of fluors required to add to the Gd-LS to optimize
photon production and light attenuation, in order to decide upon a
final recipe for the Gd-LS synthesis; (3) to build a closed synthesis
system that eliminates exposure of the Gd-LS to air; (4) to scale the
chemical procedures for Gd-LS synthesis up from the current lab-bench
scale to volumes of several hundred liters, for prototype detector
module studies, and as a prelude to industrial-scale production on the
level of 160~tons; (5) to automate many of these chemical procedures,
which are currently done by hand; (6) to use standardized ASTM-type
tests to study the chemical compatibility of the LS with the materials
that will be used to construct the detector vessel, e.g., acrylic; (7)
to find methods to measure accurately, with high precision, the
concentration ratios, C/H and Gd/H, and the H and C concentrations.

\subsubsection{Storing and Handling of LS at Daya Bay}
\label{sssec:det_LS_stor-mix}

   The basic assumptions that underlie the following discussion are
that we will select LAB as our LS choice. If another scintillator is
chosen, the procedures will be modified appropriately. The procedures
described below assume that we will dilute a 1\% solution of Gd-LS; if
another option is chosen these procedures will be appropriately
modified. We also assume:
\begin{enumerate}
 \item Detector modules will be treated as matched pairs, so
that they are known to be as identical as can be prepared. Comparison
of any differences in the operating characteristics of these identical
detectors will provide crucial information about the control of
systematic uncertainties. Thus, at a minimum, 40~tons of 0.1\% Gd-LS, well
mixed and equilibrated, will have to be prepared at one time, to be
able to fill two of the antineutrino detector modules at the same time.
 \item To reduce the volume of the required space underground and of the
costs of excavations, the `stock solutions' of LAB and of 1\% Gd
in LAB will be stored on the surface. 
 \item Some procedures, such as stirring or mixing of a liquid in a 
tank on surface, will be required to ensure uniformity within that 
liquid. With that capability in hand, it becomes feasible to dissolve 
the fluors in the LAB that is stored on the surface.
 \item Preparation of the 0.1\% Gd in LAB will be done
underground prior to filling the inner detector modules.
\end{enumerate}
The following points represent the preferred conceptual option for
handling large amounts of organic liquids at the Daya Bay site. The
first three subsections deal with conditions on the surface, during
transport, and underground, while the fourth section deals with
general issues that are relevant to all of the subsections.

\subsubsubsection{Surface LS Handling}
\label{ssssec:det_LS_surface}

\begin{itemize}
 \item  LAB (undoped) storage tank (if possible located not too far from
tunnel entrance, but this is not an absolute necessity).
 \begin{enumerate}
  \item Minimum capacity is 140~tons, the total amount of undoped LAB
           needed to prepare the solution of 0.1\% Gd in LAB that will
           fill all eight antineutrino detector modules.
  \item If desired, this tank can be large enough to store ALL of the LAB
           required to fill all of the $\gamma$-catcher regions as well
           as the antineutrino detector modules. Maximum capacity needed is 400~tons.
  \item This tank will not be for passive storage. It will have to
           contain some mixing apparatus (e.g., mechanical stirrer or
           gas bubbling) to ensure uniformity of the contained liquid
           (assuming that more than one shipment of LAB is received
           from the manufacturer).
  \item This tank on surface will also be used to dissolve and mix the fluors in LAB.
  \item This tank will require a purification column (containing
           Al$_2$O$_3$) in its inlet line, since the LAB will have to
           undergo final purification prior to addition of the fluors.
 \end{enumerate}
 \item Storage tank for 1\% Gd in LAB that has been synthesized elsewhere. 
 \begin{enumerate}
   \item Capacity of 20~tons.
   \item This tank will also need mixing apparatus, to ensure uniformity of
           the total volume of liquid.
 \end{enumerate}
 \item Mineral Oil will not need a storage tank on surface. If it is
delivered in a tanker truck, it can be moved directly underground. If delivered
in drums, the mineral oil will be transferred on the surface to an ISO tank for
shipment underground.
\end{itemize}

\subsubsubsection{Transportation of LS Underground}
\label{ssssec:det_LS_transport}

\begin{itemize}
 \item As discussed in Section~\ref{ssssec:det_LS_points} below,
           transport of batches of liquids from the tanks of LAB and 1\% Gd in LAB
           on the surface to the underground hall will be done in dedicated ISO
           containers.  
 \item Dilution of the 1\% Gd LAB mixture with
           LAB will be done in the underground filling hall.
\end{itemize}

\subsubsubsection{Underground LS Filling}
\label{ssssec:det_LS_underground}

\begin{itemize}
 \item There will be three tanks, each with 40-ton capacity, and each devoted
    to use with only one of the liquids that will go into the
    detector module.
 \item Each of these three tanks will serve as a filling station, by being
    outfitted with exit ports that will connect to a centralized,
    instrumented system with plumbing designed to allow the filling of
    the three zones of each detector module simultaneously: the
    inner zone with 0.1\% Gd in LAB, the intermediate zone with
    undoped LAB, the outer zone with mineral oil.
 \item Two of these tanks will be for intermediate passive storage, i.e.,
    one for undoped LAB and one for mineral oil.
 \item The third tank will serve as a chemical processing tank, to
    prepare the 0.1\% Gd in LAB, by diluting and mixing the 1\% Gd-LAB
    with the undoped LAB. This tank will contain a mixing apparatus.
\end{itemize}
		 
\subsubsubsection{General Points Regarding LS Handling}
\label{ssssec:det_LS_points}

\begin{itemize}
 \item All tanks containing LAB-based liquids, on surface and
    underground, will need Teflon-like inner liners. The tanks
    themselves may be stainless steel or other material.
We want to avoid rusting as well as contact of the
    metal with the organic liquids. The choice of materials may depend on
    the cost.
 \item To remove air, all of the tanks will have the capability of gas
    purging, by bubbling inert gas (such as nitrogen) through the
    organic liquids in the tanks or by establishing a blanket of inert
    gas to cover the liquid.
 \item ISO containers, with thermal insulation or possibly active
    temperature control, will either be leased or purchased. These
    have capacities of 5000--9000 gallons. They will be
    carefully cleaned prior to use. Individual ISO containers will be
    dedicated to use with particular liquids, one for LAB, one for 1\%
    Gd in LAB, and if needed, one for mineral oil. Their main use will
    be for transport of the liquids from the surface to underground.
 \item Procedures will have to be developed for clean transfers of the
    liquids, both on surface and underground, when hooking up to
    plumbing, etc., to minimize contamination by dust, dirt, and by
    air-borne radon.
\end{itemize}

\subsection{Antineutrino Detector Photomultiplier Tubes}
\label{ssec:det_PMT}

Optical photons produced by charged particles or $\gamma$ rays in the
antineutrino detector are detected with 224 PMTs submerged in the
buffer oil inside the steel vessel.  The PMTs are arranged in seven
horizontal rings, each with 32 PMTs.  The rings are staggered in such
a way that the PMTs on two adjacent rings are offset by half the PMT
spacing in a ring.  Simulation studies indicate the adopted number of
PMTs and configuration can provide good energy resolution, about 16\%
at 1~MeV.  From the experience of the other experiments, the failure
of PMT in the detector is about 1\%. We thus have sufficient number of
PMTs in each detector module to ensure reliable performance.

We require the PMT to have a spectral response that matches the
emission spectrum of the liquid scintillator and good quantum
efficiency for detecting single optical photon at a nominal gain of
about $10^7$. It is desirable to have good charge response, i.e. the
peak-to-valley ratio, for identifying the single photo-electron
spectrum from the noise distribution. Since the energy of an event is
directly related to the number of optical photons collected, the PMTs
operating at the nominal gain must have excellent linearity over a
reasonable broad dynamic range.  In addition, the dark current,
pre-pulse and after pulse should be low to minimize the noise
contribution to the energy measurement.  Furthermore, the natural
radioactivity of the materials of the PMT must be kept low so that the
$\gamma$-ray background in the detector module is as small as possible.
These specifications will be quantified with simulations and by
detailed studies of a small number of PMTs purchased from the
manufacturers prior to the production order.

Taking the photo-cathode coverage of the detector module, number of PMTs to
be used, and cost into account, we plan to use 20-cm-diameter PMTs for
the antineutrino detector modules.

\subsubsection{PMT Selection Options}
\label{sssec:det_PMT_candidates}

There are currently two candidate photomultiplier tubes for use in
the antineutrino detector modules, the Hamamatsu R5912~\cite{HPK} and the Electron Tubes
9354KB~\cite{EMI}. Both are 2$\pi$ PMTs with a 190~mm-wide photocathode and
peak wavelength sensitivity near 400~nm.  They are similar in design
and construction. However, the R5912 has 10 dynodes while the 9354KB
has 12. The Hamamatsu R5912 is an improved version over the R1408,
which was used by SNO~\cite{NIM449}. The R5912 is used by MILAGRO and AMANDA. 
Both PMTs will be extensively tested.

The manufacturers will be asked to integrate potted bases and
oil-resistant high-voltage and signal cables into the construction of
the deliverable products.  Also we plan to specify the type of the
voltage divider optimized by us for Daya Bay operation, which the
manufacturer will build from high radiopurity components and seal in
the PMT/base assembly.  The final decision on the selection of a
specific manufacturer will be made after verifying the compliance with
the required level of radiopurity, detailed performance comparisons,
and price.

\subsubsection{PMT Testing}
\label{sssec:det_PMT_testing}

Uniform performance, stable, reliable and lasting operation of the
PMT system are essential to the successful execution of the experiment.
These requirements demand a comprehensive program of testing and
validation conducted prior to installation and commissioning of the PMTs.

We will ask the selected manufacturer to provide certificates of
acceptance for the PMTs. The certificates document measured results,
compliant to our specifications, that typically include: cathode and anode 
luminous sensitivity, cathode blue sensitivity, anode dark current and 
dark counting rate, operating voltage for a gain of  10$^{7}$, charge 
response and transit time spread. 

Testing and validation of the received PMTs will be conducted using a
custom test-stand.  An LED will be pulsed to simulate scintillation
light.  The light will be collected within optical fibers and
transported to the PMT.  This setup allows us to adjust the intensity
and position (on the photocathode) of the light reaching the
phototube.  The purpose of this exercise is to gather a set of
physical parameters for each PMT, such as gain vs. high-voltage,
operating voltage at the nominal gain, quantum efficiency, dark rate,
transition time spread, and linearity.  In addition, test of
radio-purity will be made. A couple randomly selected tubes from each
batch will be radio-assayed non-destructively.  If the K, Th, or U
content exceeds the specified level of contaminations, additional
randomly selected tubes from the same batch will be radio-assayed.  If
more PMTs exceed the specified contamination level, the whole batch
will be rejected. The collected data will be used in simulation and
analysis.

\subsubsection{PMT Support Structure}
\label{sssec:det_PMT_support}

\begin{figure}[!htb]
\begin{center}
\includegraphics[width=10cm]{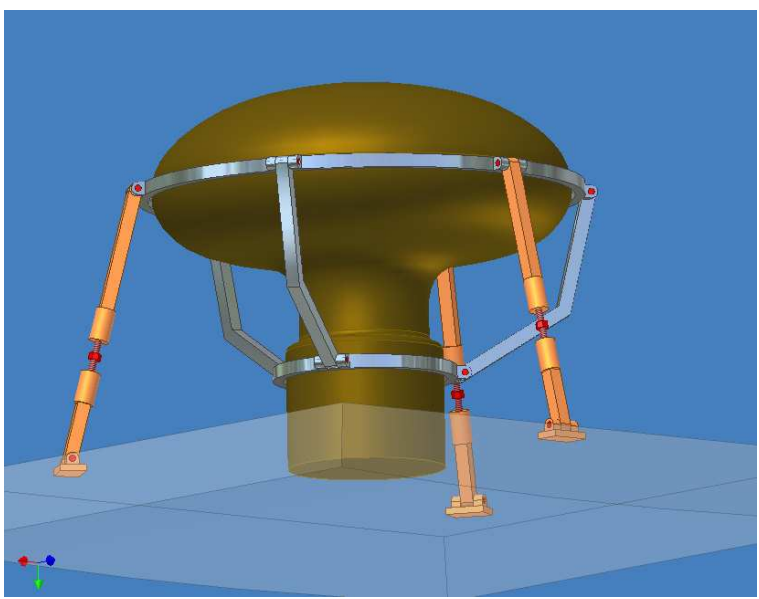}
\caption{PMT mechanical support structure.
\label{fig:fig7_PMTsupport}}
\end{center}
\end{figure}

The mechanical support of each PMT is a tripod structure which is
mounted on a frame attached to the inner wall of the steel vessel.  A
tripod is stable and convenient for adjusting the orientation of the
PMT. Figure~\ref{fig:fig7_PMTsupport} shows the support structure.
This structure is light and will be made of radio-pure materials.  The
orientation of the PMT can be adjusted by varying the lengths of the
three legs.  The circular grips provide reliable support of the PMT in
all possible positions relative to the direction of the buoyant force.

\subsubsection{High-Voltage System}
\label{ssec:det_HV}

The high voltage and its control system is designed around an Analog
Devices' variant of the 8052 microcontroller.  It functions for both
HV control and monitoring.  The system is capable of 0.5~V
resolution up to 2048~V.  An EMCO DC-to-DC HV generator~\cite{EMCO} will be
mounted near the phototube and will act as our voltage source. The
EMCO chips have low ripple, good regulation, and are economical. The
base will be a simple tapered divider design mounted directly to the
phototube, having a footprint smaller than the standard socket
provided by each manufacturer.

Control software is currently written using LabView. The design of
this unit is a modified version of the STAR EMC (4800 towers) code,
which allowed for 4096 units to be controlled from a single serial
port (plus a RS485 converter) in a multi-drop, master-slave network
using an RS485 bus. RS485 repeaters will be used to connect various
branches of the HV control network to the main control bus.  The
firmware on the microcontroller and the LabView based control
program are both available.

\subsection{Front End Electronics}
\label{ssec:det_fee}

The antineutrino detector readout system is designed to process
the PMT output signals. The essential functions
are as follows:
\begin{itemize}
 \item Determine the charge of each PMT signal to measure the energy
  deposit in the liquid scintillator.  This will enable us to select
  neutrino events, reject backgrounds and deduce the neutrino energy
  spectrum.  
  \item Determine the event time by measuring arrival time
  of the signals to the PMTs in order to build the time correlation
  between prompt and delayed sub-event.  The timing information can
  also help us to reconstruct the location of the antineutrino
  interaction in the detector, and to study and reject potential
  background events.  
  \item Provide fast information to the trigger system.
\end{itemize}

\subsubsection{Front-End Electronics Specifications}
\label{sssec:det_fee_spec}

When a reactor antineutrino interacts in the target, its energy is
converted into ultraviolet or visible light, some of which will
ultimately be transformed into photoelectrons (p.e.) at the
photocathodes of the PMTs. For a given PMT, the minimal number of
p.e.'s is one and, based on Monte Carlo simulation, the maximum number
is 50 when an antineutrino interaction occurs in the vicinity of the
interface between the Gd-doped liquid scintillator and the
$\gamma$-catcher.  Since typically 500~p.e.'s will be recorded by PMTs
for a cosmic-ray muon passing through the detector, the dynamic range
of the PMTs is required to be up to about 500~p.e.'s. The intrinsic
energy resolution for a single p.e.  is typically about 40\% with some
variation from PMT to PMT, while the energy threshold of a PMT is
constrained by the dark noise, typically at the level of about
1/3--1/4 of a p.e. The peak-to-peak noise and the charge resolution of
the PMT readout electronics is thus required to be less than 1/10 of a
p.e. in both cases.  The total charge measurement determined by the
center-of-gravity method will produce the total energy deposited by an
antineutrino signal or a background event.

The arrival time of the signal from the PMT will be measured relative
to a common stop signal, for example, the trigger signal. The time
jitter of a PMT for a single p.e.  is about 1--2~ns, caused by the PMT
transit time jitter, the PMT rise time, and the time walk effect of
the signal, etc.  The design goal for the time resolution of the
readout electronics is thus determined to be less than 0.5~ns.

Since an antineutrino event is a coincidence of the prompt and delayed
sub-event, their time interval is a crucial parameter for physics
analysis. The precision of this interval is dominated by the trigger
signal which is synchronized to the 100~MHz system clock. Hence a
10~ns precision is expected, which is sufficient given the fact that
the coincidence window of sub-events is 200~$\mu$s and the resultant
uncertainty in efficiency is less than 0.03\%. If a TDC counter is
employed at the trigger board to measure the actual trigger arrival
time with respect to the system clock, a better precision can be
achieved.

The dynamic range of the time measurement depends on the uncertainty
of the trigger latency and the maximum time difference between the
earliest and the latest arrival time of light to PMTs. The
range is chosen to be from 0 to 500~ns. 

The time measurement of the individual PMT time can also be used to
determine the event vertex. Although such a method is particularly
suitable for large detectors similar to KamLAND, it provides an
independent measurement which complements the charge-gravity method
for small detectors with diameters of several meters.  Hence it offers
a cross-check of systematic uncertainties and an additional handle for
studying backgrounds. The readout electronics specifications are
summarized in Table~\ref{tab:readout}.
\begin{table}[!htb]
\begin{center}
\begin{tabular}[c]{|l||c|} \hline
 Quantity               & Specification     \\ \hline\hline
Dynamic range           & 0--500~p.e.       \\ \hline
Charge resolution       & $<$ 10\% @ 1~p.e.      \\
                        & 0.025\% @ 400~p.e. \\ \hline
noise                   & $<$ 10\% @ 1~p.e.      \\ \hline
Digitization resolution & 14 bits           \\ \hline
Time range              & 0--500~ns         \\ \hline
Time resolution         & $<$ 500~ps          \\ \hline
Sampling rate           & 40~MHz            \\ \hline
Channels/module         & 16                \\ \hline
VME standard            & VME64xp-340~mm     \\ \hline
\end{tabular}
\caption{Readout Electronics Specifications.
\label{tab:readout}}
\end{center}
\end{table}

\subsubsection{Front End Boards}
\label{sssec:det_readout}

Each detector module is designed to have a readout system
without any relationship to the other modules except receiving a
common clock signal and GPS information. The positron and neutron
triggers are both recorded with time stamps and the their matching
in time will be done offline by software.  The readout electronics for
each detector module is housed in a 9U VME crate, each can handle up
to 16 readout modules, one trigger module, and one or two fan-out
modules. In such an arrangement, moveable modules can be easily
realized, and correlations among modules can be minimized.

Each readout module receives 16 channels of PMT signals and completes
the time and charge measurement. The sum of hit numbers and the total
energy of this module is fed into the trigger system for a fast
decision. After collecting information from all readout modules, a
trigger signal may send to all readout modules for data readout upon a
positive decision.

\begin{figure}[!htb]
\begin{center}
\includegraphics[height=9cm, width=0.85\textwidth]{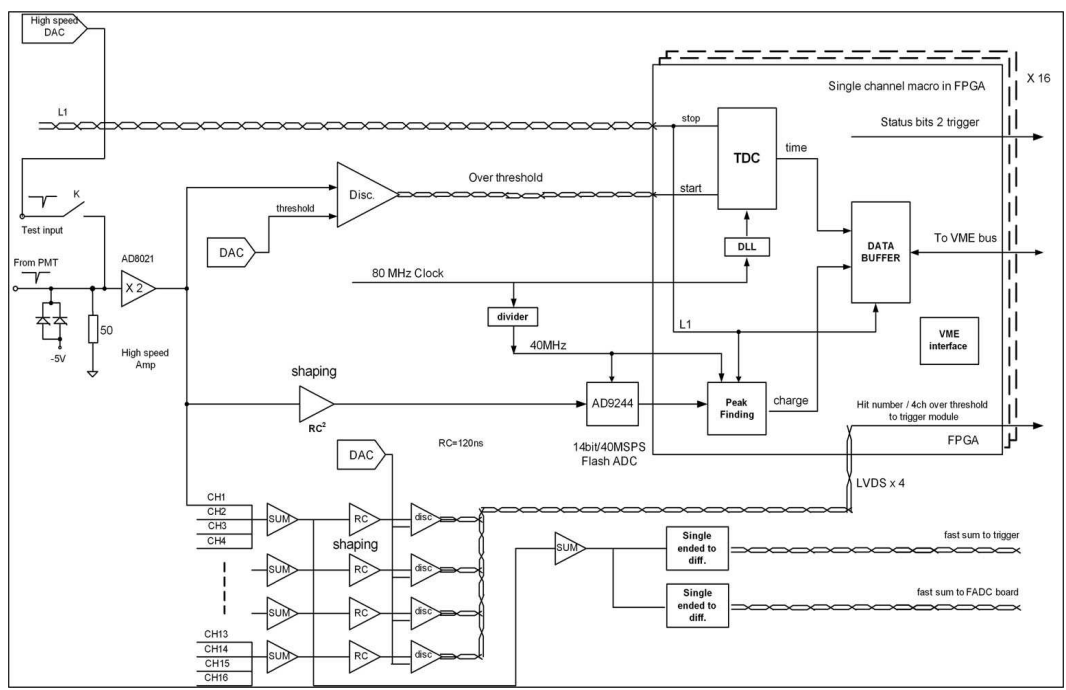}
\caption{Block diagram of front-end electronics module for the antineutrino
detector. \label{fig:fig7_feen}}
\end{center}
\end{figure}

A simplified circuit diagram of the electronic readout system,
showing its main functions, is given in Fig.~\ref{fig:fig7_feen}.

The analog signal from a PMT is amplified with a fast, low noise (FET
input stage) amplifier. The output of the amplifier is split into two
branches, one for time measurement, and the other for charge
measurement.

The signal for time measurement is first sent to a fast discriminator
with a programmable threshold to generate a timing pulse, whose
leading edge defines the arrival time of the signal. A stable
threshold set by a 14-bit DAC (AD7247)~\cite{AD} via the VME controller is
needed for the discrimination in order to achieve the required time
resolution.

The timing pulse is sent to a TDC as the start signal, while the
trigger signal is used as the common stop.  The TDC is realized by
using internal resources of a high-performance FPGA with key
components of two ultra high speed Gray-code counters. The first
counter changes at the rising edge of the 320~MHz clock, while the
second one changes at the falling edge.  Thus, the time bin is
1.5625~ns and the RMS of the time resolution is less than 0.5~ns.

To measure the charge of a PMT signal, an ultra low-noise FET input
amplifier (AD8066) is selected for the charge integrator. A passive RC
differentiator is used after the integrator to narrow the signal.
Since the signal rate of a typical PMT is about 5~kHz including noise,
a 300~ns shaping time is chosen, corresponding to an output signal
width of less than 1~$\mu$s. The analog signal is accurately digitized
by a 14-bit Flash ADC with 40~MHz sampling rate after a baseline
recovering. The digitized result goes directly into FPGA, in which all
data-processing like data pipelining, pedestal subtraction,
nonlinearity correction, and data buffering are implemented.

The readout module has a standard VME A24:D32 interface. Both ADC and
TDC data of the triggered event are sent to a buffer, which can store
a maximum of 256 events. The data will be readout through the VME
backplane by the DAQ system within a reasonable time span.

\subsection{Antineutrino Detector Prototype}
\label{ssec:det_prototype}

Valuable data on the performance of the
antineutrino detector has been obtained from a scaled down prototype at
at the Institute of High Energy Physics, Beijing, China.  The
goal of this R\&D work is multifold:
1) to verify the detector design principles such as reflectors 
at the top and the bottom, uniformity of the response in a 
cylinder, energy and position resolution of the detector, etc.; 
2) to study the structure of the antineutrino detector; 
3) to investigate the long term stability of the liquid scintillator; 
4) to practice detector calibration; 
5) to provide necessary information for the Monte Carlo simulation.

\subsubsection{Prototype Detector Design}
\label{sssec:det_configurations}

\begin{figure}[!htb]
\begin{center}
\includegraphics[width=14cm]{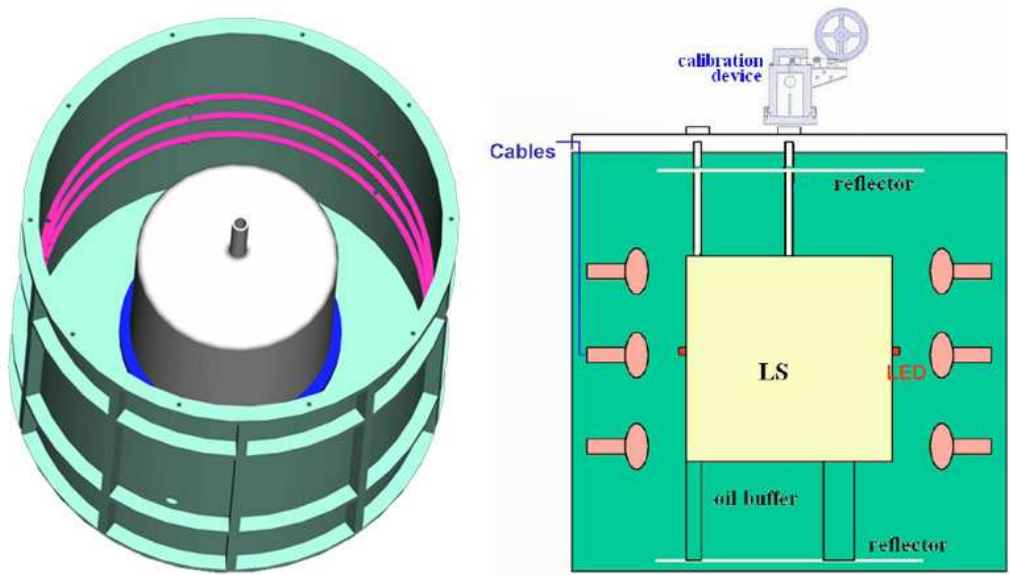}
\caption{Sketch of the antineutrino detector prototype (Left) Top view, (Right) Side
view.
 \label{fig:fig7_proto_sketch}}
\end{center}
\end{figure}

As shown in Fig.~\ref{fig:fig7_proto_sketch}, the prototype consists
of two cylinders: the inner cylinder is a transparent acrylic vessel
0.9~m in diameter and 1~m in height with 1~cm 
wall thickness. The outer cylinder is 2~m in diameter and 2~m in height
made of stainless steel.  Currently, the acrylic vessel is filled with
0.54~tons of normal liquid scintillator, while Gd-loaded liquid scintillator is
planned for the near future.  The liquid scintillator consists of 30\%
mesitylene, 70\% mineral oil with 5~g/l PPO and 10~mg/l bis-MSB.  The
space between the inner and outer vessel is filled with 4.8~tons of
mineral oil.  A total of 45 8" EMI 9350 and D642 PMTs, arranged in three
rings and mounted in a circular supporting structure, are immersed in
the mineral oil. The attenuation length of the LS is measured to be 10~m and that
of the mineral oil is 13~m.

An optical reflector of Al film is placed at the top and 
bottom of the cylinder to increase the effective coverage area
from 10\% to 14\%. The scintillator light yield is about 
10000~photons/MeV, and the expected detector energy response is
about 200~p.e./MeV.

\begin{figure}[!htb]
\begin{center}
\includegraphics[width=14cm]{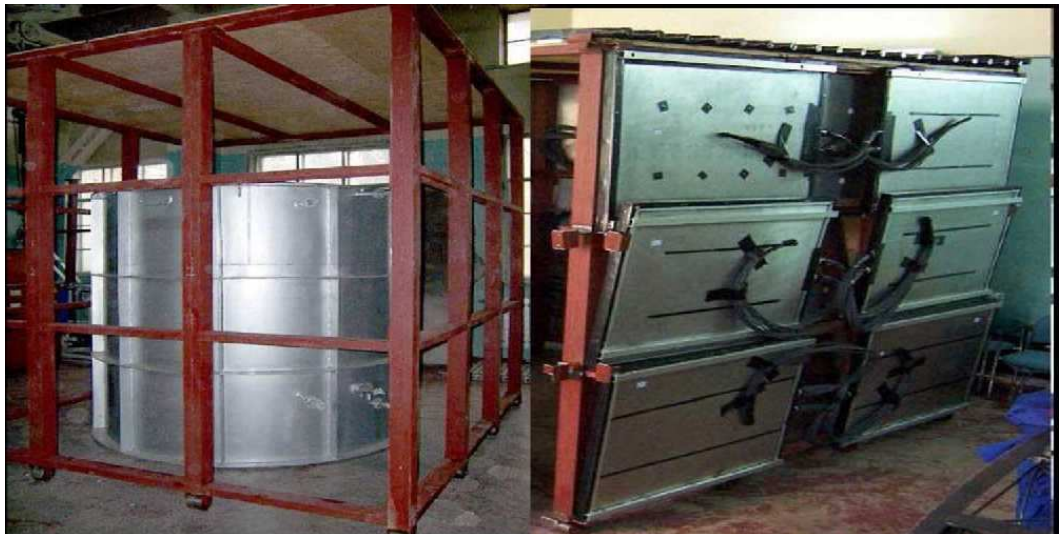}
\caption{The antineutrino detector: before (left) and after
(right) the muon detectors were mounted.
\label{fig:fig7_test_setup}}
\end{center}
\end{figure}

The prototype is placed inside a cosmic ray shield with dimensions of
3~m$\times$3~m$\times$3~m. It fully covers five sides (except the
bottom).  The top is covered by 20 plastic scintillator counters (from
the BES Time-of-Flight system), each 15~cm wide $\times$ 3~m long. The
four side walls are covered by 36 1.2~m$\times$1.2~m square
scintillation counters from the L3C experiment.
Figure~\ref{fig:fig7_test_setup} shows a photograph of the prototype
test setup, before and after the muon counters were mounted.

The readout electronics were designed as prototypes for the antineutrino
detector, according to the requirements
discussed in Section~\ref{ssec:det_fee}. The trigger system, 
DAQ system and online software are all assembled as prototypes for
the experiment (see Chapter~\ref{sec:trig}).

\subsubsection{Prototype Detector Test Results}
\label{sssec:det_result}

Several radioactive sources including $^{133}$Ba (0.356~MeV),
${^{137}}$Cs (0.662~MeV), ${^{60}}$Co (1.17+1.33~MeV) and
$^{22}$Na (1.022+1.275~MeV) are placed at different locations
through a central calibration tube inside the liquid scintillator
to study the energy response of the prototype.  The gain of all
PMTs are calibrated by using LED light sources, and the trigger
threshold is set at 30~p.e., corresponding to about 110~keV.

\begin{figure}[!htb]
\begin{minipage}[t]{0.48\textwidth}
\includegraphics[width=0.8\textwidth,angle=270]{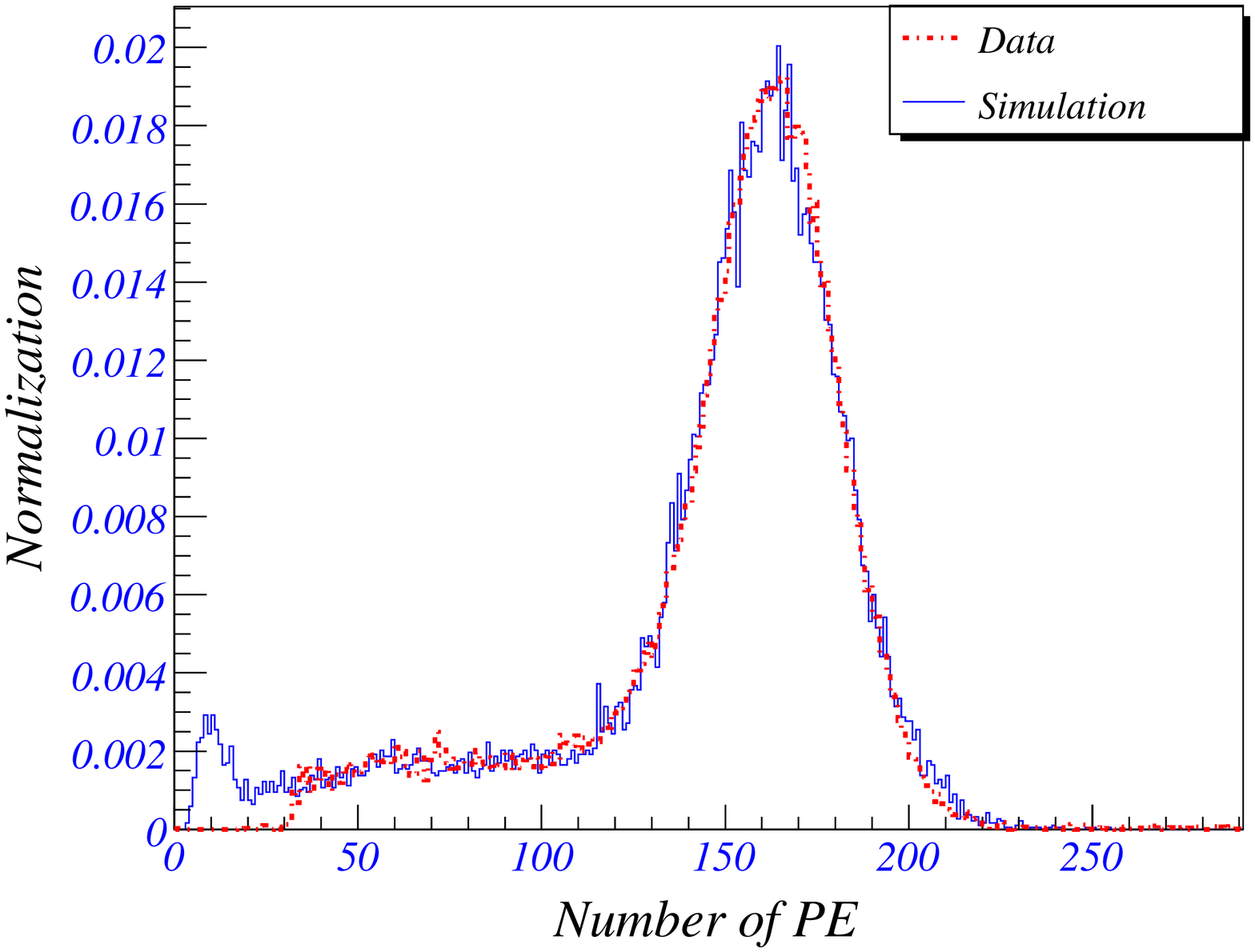}
\end{minipage}
 \hfill
\begin{minipage}[t]{0.48\textwidth}
\includegraphics[width=0.8\textwidth,angle=270]{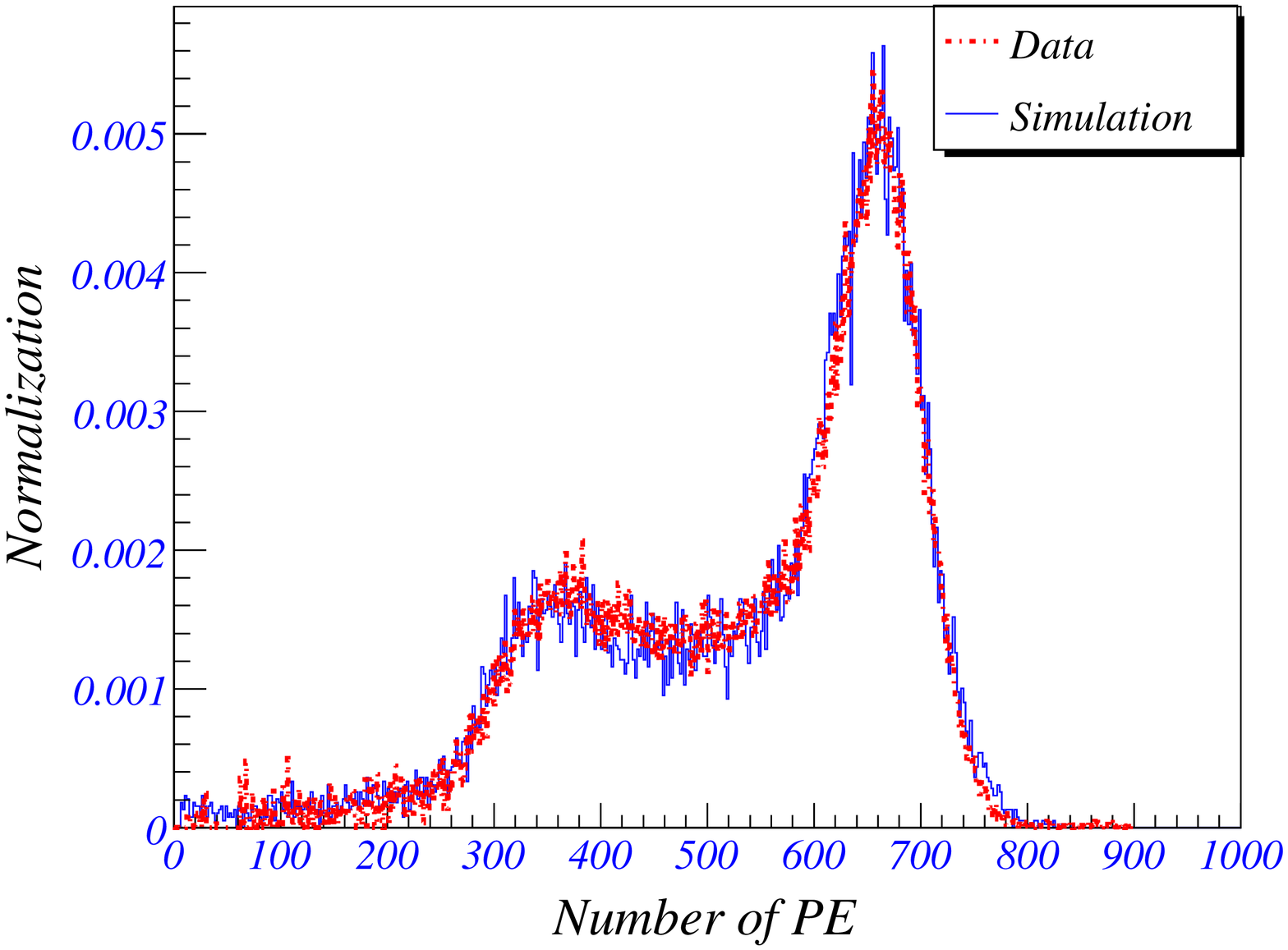}
\end{minipage}
\caption{Energy response of the prototype to ${^{137}}$Cs (left)
and $^{60}$Co (right) sources at the center of the detector with a
comparison to Monte Carlo simulation.} \label{fig:fig7_energyspec}
\end{figure}

Figure~\ref{fig:fig7_energyspec} shows the energy spectrum after
summing up all PMT response for the $^{137}$Cs and $^{60}$Co
sources located at the center of the detector. A total of about
160~p.e. for $^{137}$Cs is observed, corresponding to an energy
response of 240~p.e./MeV, higher than naive expectations.  The
energy resolution can be obtained from a fit to the spectra, resulting in
a value of about 10\%. A detailed Monte Carlo simulation is
performed to compare the experimental results with
expectations, as shown in Fig.~\ref{fig:fig7_energyspec}. Very good
agreement is achieved, showing that the detector behavior is well
understood.

\begin{figure}[!htb]
\begin{minipage}[t]{0.48\textwidth}
\includegraphics[width=\textwidth]{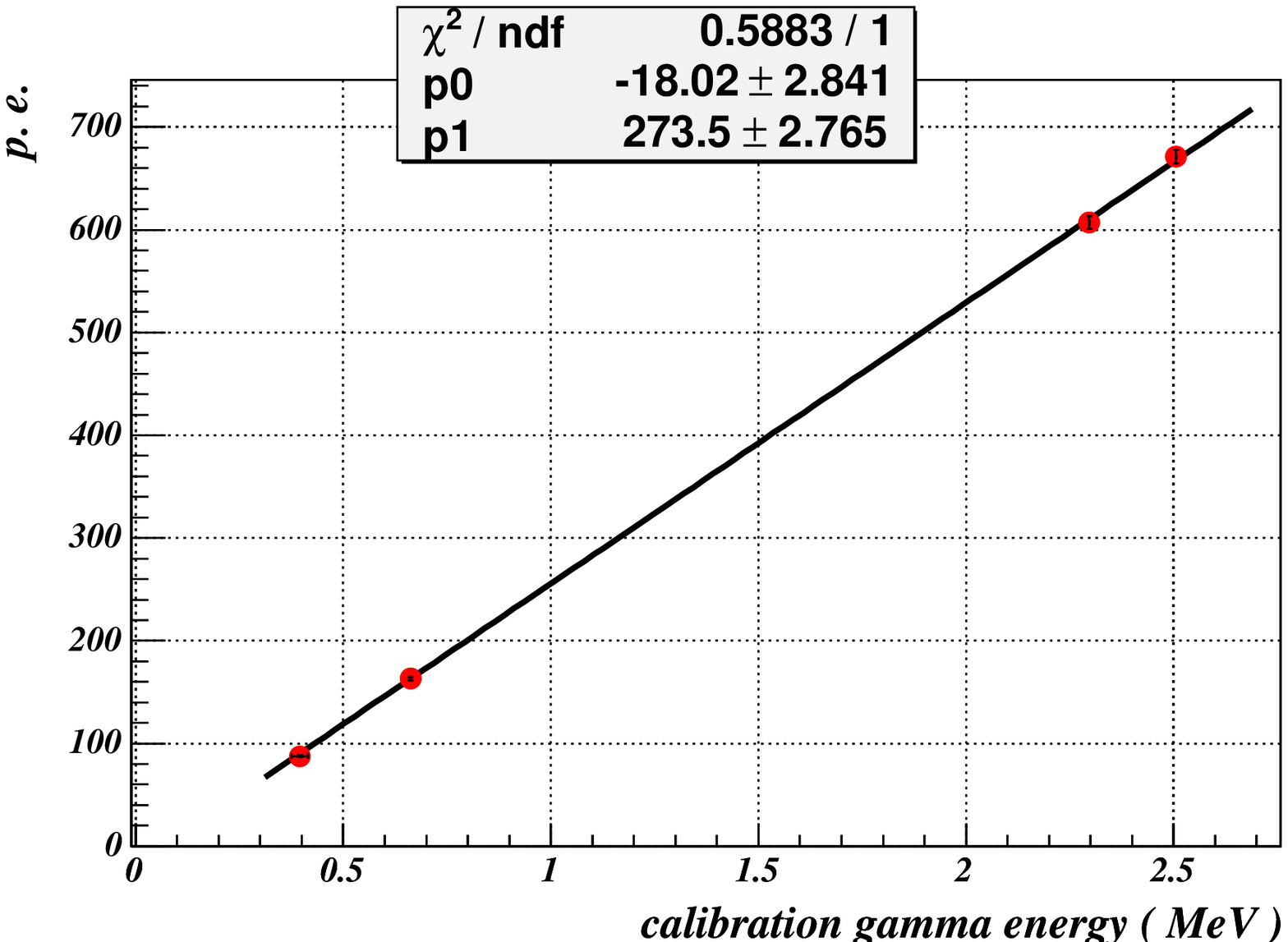}
\end{minipage}
 \hfill
\begin{minipage}[t]{0.48\textwidth}
\includegraphics[width=\textwidth]{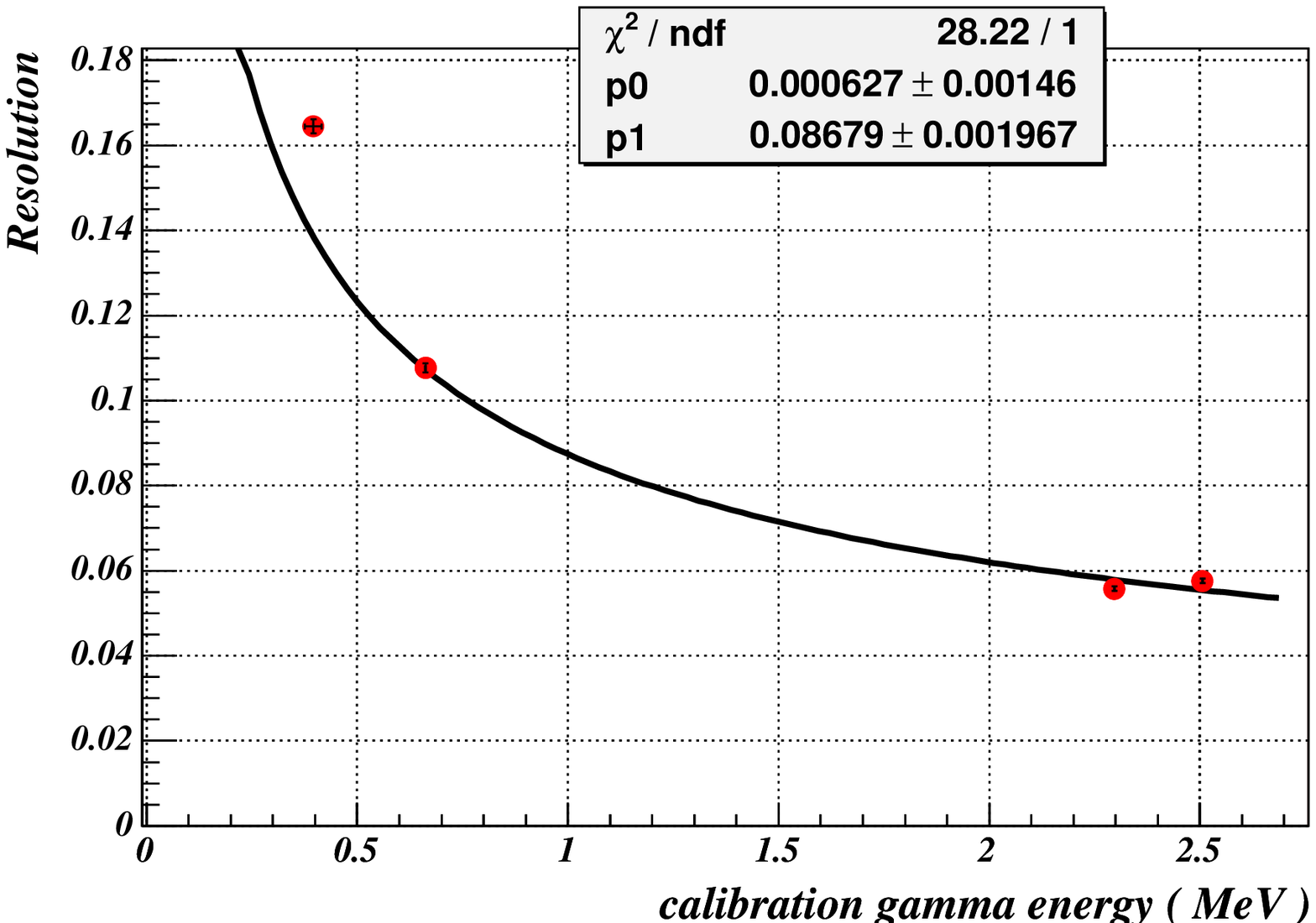}
\end{minipage}
\caption{Linearity of energy response of the prototype to various sources at
the center of the detector (left), and the energy
resolution (right).} \label{fig:fig7_resolinear}
\end{figure}

All of the sources were inserted into the center of the detector; the
energy response is shown in Fig.~\ref{fig:fig7_resolinear}
(left). Good linearity is observed, although at low energies
non-linear effects are observed which are likely due to light
quenching and Cherenkov light emission.  The energy resolution at
different energies is also shown in Fig.~\ref{fig:fig7_resolinear}
(right), following a simple expression of $\sim 9\%/\sqrt{E}$, in good
agreement with Monte Carlo simulation as shown in
Fig.~\ref{fig:fig7_energyspec}.

\begin{figure}[!htb]
\begin{center}
\includegraphics[height=7cm]{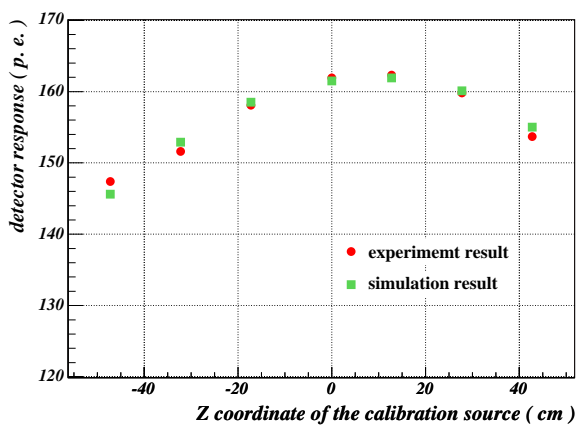}
\caption{Energy response of the prototype to a ${^{137}}$Cs source
as a function of z position, and a Monte Carlo
simulation.} \label{fig:fig7_zaxis}
\end{center}
\end{figure}

The energy response as a function of vertical depth along the
z-axis is shown in Fig.~\ref{fig:fig7_zaxis}. Very good uniformity
(better than 10\%) over the entire volume of the liquid
scintillator shows that the transparency of the liquid is good,
and the light reflector at the top and the bottom of the cylinder
works well as expected. The fact that the data and Monte Carlo
expectation are in good agreement, as shown in
Fig.~\ref{fig:fig7_zaxis}, demonstrates that the prototype,
including its light yield, light transport, liquid scintillator,
PMT response, and the readout electronics is largely understood.

\newpage
\renewcommand{\thesection}{\arabic{section}}
\setcounter{figure}{0}
\setcounter{table}{0}
\setcounter{footnote}{0}

\renewcommand{\labelitemi}{$\bullet$}

\section{Calibration and Monitoring Systems}
\label{sec:cal}

The measurement of $\sin^2 2 \theta_{13}$ to a precision of 0.01 in
the Daya Bay experiment will require extreme care
in the characterization of the detector properties as well as
frequent monitoring of the detector performance and condition. The
physics measurement requires that the neutrino flux be measured with
{\it relative} precision that is substantially better than 1\%. This
is accomplished by taking ratios of observed event rates in the
detectors at near and far sites to separate the oscillation effect.
This will require that differences between detector modules be
studied and understood at the level of $\sim$0.1\% and that
changes in a particular detector module (over time or after
relocation at another site) be studied and understood at $\sim$0.1\%.
Achieving these goals will be accomplished through a
comprehensive program of detector calibration and monitoring.

We anticipate a program with three different classes of procedures:
\begin{enumerate}
\item  ``complete'' characterization of a detector module,
\item  ``partial' characterization, and
\item routine monitoring.
\end{enumerate}
We envision that the complete characterization (procedure \#1.) will generally be performed once
during initial commissioning of a detector module before taking
physics data. Procedure \#2 would be employed after relocation of a
detector module or after some other change that requires a careful
investigation of the detector properties and will involve a subset
of the activities in procedure \#1. If substantial changes are
detected during procedure \#2, then we would likely opt for
reverting to procedure \#1. Finally, procedure \#3 will involve both
continuous monitoring of some detector parameters as well as
frequent (i.e., daily or weekly) automated procedures to acquire
data from LED light sources and radioactive sources deployed into
the detector volume.
The requirements and proposed solutions for procedure \#1. are listed
in Table~\ref{tab:proc1}.
\begin{table}[!hbt]
\begin{center}
\begin{tabular}[c]{|c||c|c|}\hline
\bf{Requirement} & \bf{Description} & \bf{Proposed Solution(s)} \\ \hline\hline
Optical Integrity & Spatial uniformity of response, light attenuation& LED, $\gamma$ sources\\ \hline
PMT gains& Match gains of all PMTs& LED - single p.e. matching \\ \hline
PMT timing & $\sim 1$~ns timing calibration for each PMT & Pulsed LED \\ \hline
Energy scale & Set scale of energy deposition & Gamma sources \\ \hline
H/Gd ratio & Measure relative Gd fraction & ${}^{252}$Cf neutron source \\ \hline
\end{tabular}
\caption{Requirements for the full manual calibration procedure \#1.} \label{tab:proc1}
\end{center}
\end{table}
These will be manually operated procedures using equipment and systems
to be described below, and will likely entail several weeks activity.

Procedure \#2 will be a subset (to be determined) of the activities
in procedure \#1 These will be also be manually operated procedures
using equipment and systems to be described below, and will likely
entail several days activity.

The requirements and proposed solutions for procedure \#3 are listed
in Table~\ref{tab:proc3}. 
\begin{table}[!hbt]
\begin{center}
\begin{tabular}[c]{|c||c|c|} \hline
\bf{Requirement} & \bf{Description} & \bf{Proposed Solution(s)} \\ \hline \hline
 Mechanical/thermal & Verify these properties are stable  & Load sensors, thermometers, etc.\\ \hline
Optical stability& Track variations in light yield& Gamma sources, spallation products \\ \hline 
Uniformity, light attenuation & Monitor spatial distribution of light & Gamma sources, spallation products \\ \hline 
Detection efficiency & Monitor $\epsilon$ for neutrons and
positrons & Gamma sources, neutron sources\\ \hline
PMT gains & Monitor 1~p.e. peaks & LED source\\ \hline
\end{tabular}
\caption{Requirements for automated calibration procedure \#3.} \label{tab:proc3}
\end{center}
\end{table}
Procedure \#3 will entail continuous in-situ monitoring
(Sec.~\ref{ssec:cal_in-situ}), monitoring of continuously produced
spallation-induced activity (Sec.~\ref{ssec:cal_data}), and regularly
scheduled automated deployment of sources
(Sec.~\ref{ssec:cal_automated}).

 \subsection{Radioactive Sources}
 \label{ssec:cal_sources}

The main goal of the source calibration is to reach the maximum
sensitivity to neutrino oscillations by comparing the energy spectra
measured by near and far detectors. The response of the detectors
at the far and near sites may have small differences, these minute
differences can lead to slight distortion in the measured energy
spectra of the antineutrinos. Therefore, it is necessary to
characterize the detector properties carefully before data taking
and monitor the stability of the detectors during the whole
experiment. The calibration sources must be deployed regularly
throughout the active volume of the detectors to simulate and
monitor the detector response to positrons, neutron capture gammas and
gammas from the environment.

The sources will be used in the calibration are listed in
Table~\ref{tab:source}.
\begin{table}[!hbt]
\begin{center}
\begin{tabular}[c]{|c||c|}\hline
\bf{Sources} & \bf{Calibrations}  \\ \hline\hline
 Neutron sources:           & Neutron response, relative and \\
 Am-Be and $^{252}$Cf       & absolute efficiency, capture time \\ \hline
 Positron sources:          & Positron response, energy scale \\
 $^{22}$Na, $^{68}$Ge       & trigger threshold \\ \hline
 Gamma sources:             & Energy linearity, stability, resolution \\
                            & spatial and temporal variations, quenching effect \\
 $^{137}$Cs                 & 0.662 MeV \\
 $^{54}$Mn                  & 0.835 MeV \\
 $^{65}$Zn                  & 1.351 MeV \\
 $^{40}$K                   & 1.461 MeV \\
 H neutron capture          & 2.223 MeV \\
 $^{22}$Na                  & annih + 1.275 MeV \\
 $^{60}$Co                  & 1.173 + 1.333 MeV \\
 $^{208}$Tl                 & 2.615 MeV \\
 Am-Be                      & 4.43 MeV \\
 $^{238}$Pu-$^{13}$C        & 6.13 MeV \\
 Gd neutron capture         & $\sim$ 8 MeV \\ \hline
\end{tabular}
\caption{Radioactive sources to be used for calibrations.}
\label{tab:source}
\end{center}
\end{table}
These sources cover the energy range from
about 0.5~MeV to 10~MeV and thus can be used for a thorough energy
calibration.

The Am-Be source can be used to calibrate the neutron capture
detection efficiency by detecting the 4.43~MeV gamma in coincidence
with the neutron. The absolute neutron detection efficiency can be
determined with a $^{252}$Cf source, because the neutron
multiplicity is known with an accuracy of about 0.3$\%$. In order to
absolutely determine the neutron detection efficiency, a small fission
chamber will be used to tag neutron events by detecting the
fission products. In addition, neutron sources allow us to determine
the appropriate thresholds of neutron detection and to measure the
neutron capture time for the detectors.

The positron detection can be simulated by a $^{22}$Na source. When a
$^{22}$Na source emits a 1.275~MeV gamma, a low energy positron will
be emitted along  with the gamma and then annihilate. The primary gamma
and the following annihilation gammas mimic the antineutrino
event inside the detector.

The sources must be encapsulated in a small containers to prevent any
possible contamination of the ultra-pure liquid scintillator. They
can be regularly deployed to the whole active volume of the
detectors and the $\gamma$-catcher.

 \subsection{LED Calibration System}
 \label{ssec:cal_LED}

\begin{figure}[htb!]
  \begin{center}
\includegraphics[height=10cm]{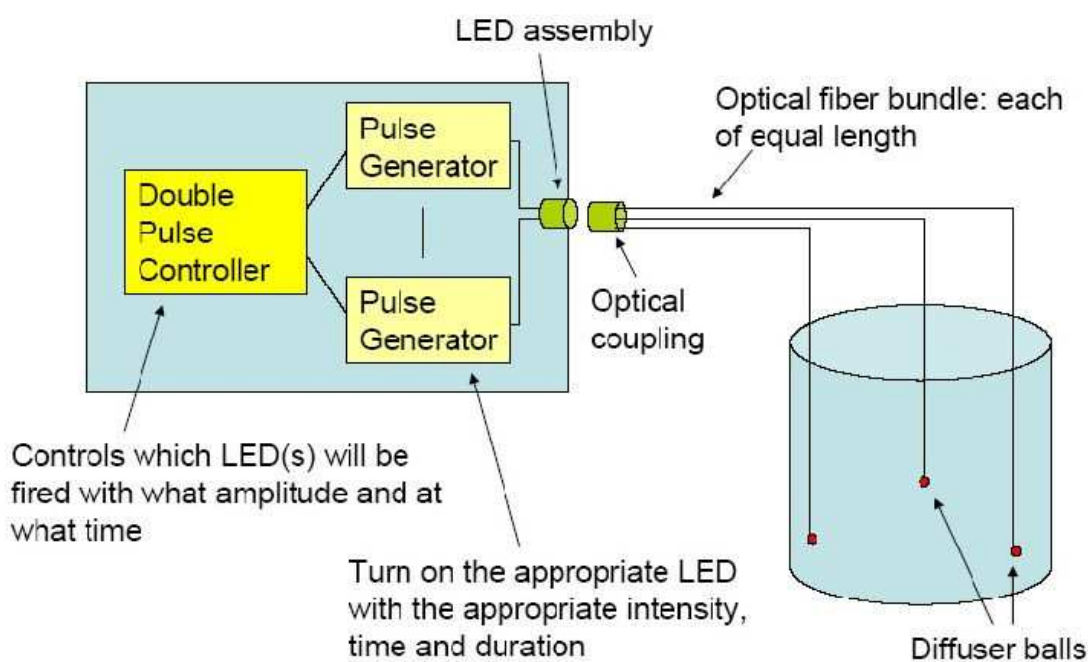}
  \caption{Schematic diagram of the LED calibration system.}
  \label{fig:LED}
  \end{center}
\end{figure}

LEDs have proven to be reliable and stable light sources that can
generate fast pulses down to ns widths.  They are therefore ideal light
sources for checking the optical properties of the liquid
scintillator, the performance of the PMTs and the timing
characteristics of the data acquisition systems.  A schematic diagram
of the LED calibration system is shown in Fig.~\ref{fig:LED}.

The LED controller box controls the pulsing of the individual LEDs
which are coupled through optical fibers to diffuser balls installed
inside the detector module.  To ensure consistency among detector
modules, a single controller box will be used which can be coupled
to the optical fiber bundle of each detector module when needed.
Some of the features of the controller box are as follows:
\begin{itemize}
\item   The pulse heights of each of the double pulses and their
separation are fully programmable to simulate the scintillation light
produced in an inverse beta-decay interaction. The first pulse will
simulate the annihilation photons and direct energy deposition from the positron and the second
pulse will simulate the gamma burst resulting from neutron capture
on Gd.
\item    The pulse separation can be generated randomly or
stepped gradually.
\item    The gamma burst can be simulated by
simultaneously flashing a number of diffuser balls at various
locations inside the detector.
\item    The controller can be
triggered by the muon system to test the detector response
following muon events.
\end{itemize}
The performance of the LEDs will be checked regularly against
scintillation produced by a gamma source in a solid scintillator
viewed by a PMT.  This could be done by coupling the controller box
to an optical fiber bundle that is viewed by the same PMT. The
number of diffuser balls and their locations inside the central
acrylic vessel and the $\gamma$-catcher will be determined through
computer simulation.  Most of the diffuser balls will be fixed while
a few can be moved in the vertical direction by using the same
deployment system for radioactive source calibration.  The diffuser
balls and optical fibers will have to be fully compatible with
liquid scintillator.

\subsection{In-situ Detector Monitoring}
\label{ssec:cal_in-situ}

 Each detector module will be equipped with
a suite of devices to monitor in-situ some of the critical detector
properties during all phases of the experiment. The in-situ
monitoring includes load and liquid sensors for the detector mass,
attenuation length measurements of the Gd-loaded LS target and the
LS $\gamma$-catcher, a laser-based monitoring system for the position
of the acrylic vessel, accelerometers, temperature sensors, and
pressure sensors for the cover gas system. A sampler for routine
extraction of a LS sample complements this multi-purpose suite of
monitoring tools.

The purpose of these tools is to provide close monitoring of the
experiment during three critical phases of the experiment:

\begin{enumerate}
  \item detector filling
  \item data taking
  \item detector transport and swapping
\end{enumerate}

During filling of the modules the changing loads and buoyancy forces
on the acrylic vessels and the detector support structure are
carefully monitored with load and level sensors to ensure that this
dynamic process does not exceed any of the specifications for the
acrylic vessels.

Most of the time during the duration of the experiment the detectors
will be stationary and taking data. Experience from past experiments
has shown that the optical properties of detectors will change over
time due to changes in the attenuation lengths of the liquid
scintillator or changes in the optical properties of the acrylic
vessel. It is important to track these characteristics to be able to
explain any possibly changes in the overall detector response as
determined in the regular, automated calibration. In-situ monitoring
of the LS attenuation length and regular extractions of LS samples
from the detector modules will help monitor some of the basic
detector properties.

The transport of the filled detectors to their location and the
swapping of detectors over a distance of up to $\sim$1.5~km is a
complex and risky task that will require close monitoring of the
structural health of the detectors modules during the move. The
proposed swapping scheme of the detectors is a novel method without
proof-of-principle yet. While conceptually very powerful, extreme care
has to be taken in the calibration and characterization of the
detectors before and after the move to be able to correct for all
changes in the detector response or efficiencies. The accelerometers,
pressure sensors, and the monitoring of the acrylic vessel positions
will provide critical real-time information during this procedure to
ensure that the detectors -- and in particular the acrylic vessels and
PMTs -- are not put at risk. Recording any changes in the detector
modules will also help us understand possible differences in the
detector response before and after the move. The acrylic vessel
position monitoring system will use a laser beam and reflective target
on the acrylic vessel surfaces. By measuring the angular deflection of
the laser beam over the length of the detector, transverse
displacements of the acrylic vessel can be monitored quite precisely.
\begin{figure}[htbp]
\begin{center}
\includegraphics[width=6in]{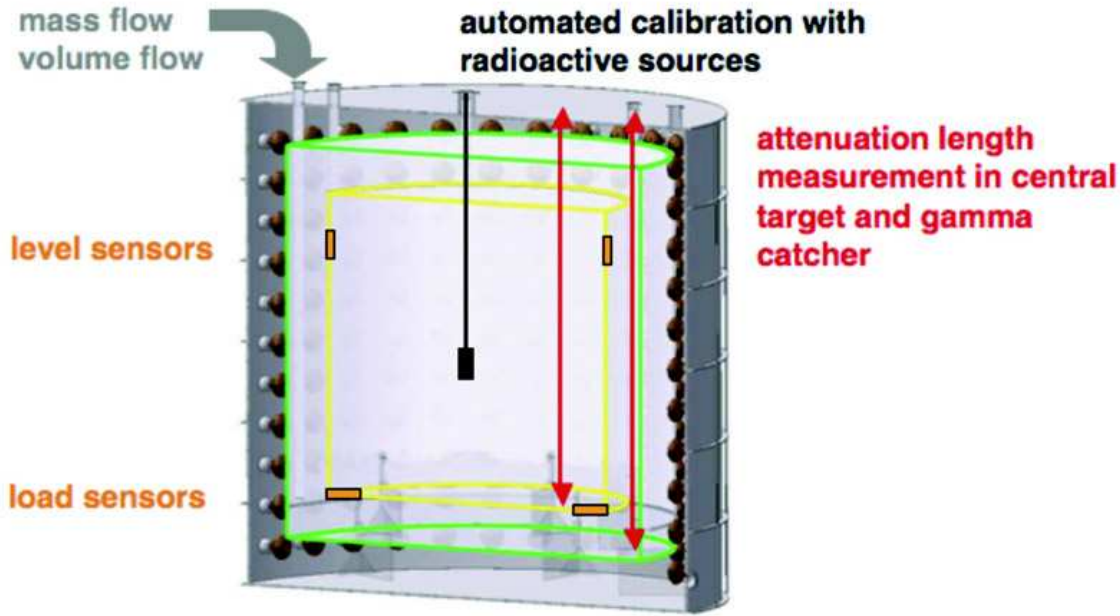}
  \caption{Diagram illustrating the variety of monitoring tools to be
integrated into the design of the antineutrino detector modules.}
  \label{fig:monitor}
\end{center}
\end{figure}

\subsection{Detector monitoring with data}
 \label{ssec:cal_data}

Cosmic muons passing through the detector modules will produce
useful short-lived radioactive isotopes and spallation neutrons.
These events will follow the muon signal (detected in the muon system as
well as the detector) and will be uniformly distributed throughout
the detector volume. Therefore, these provide very useful
information on the full detector volume which is complementary to
the information obtained by deploying point sources (Sec.~\ref{ssec:cal_automated} and
\ref{ssec:cal_manual}). For example, such events are used by KamLAND to study the
energy and position reconstruction as well as to determine the
fiducial volume. As with KamLAND, the Daya Bay experiment will use
primarily spallation neutron capture and ${}^{12}$B decay ($\tau=
29.1$~ms and $Q= 13.4$~MeV). The rates of these events for Daya Bay
are given in Table~\ref{tab:spallrate}.
\begin{table}[!hbt]
\begin{center}
\begin{tabular}[c]{|c||c|c|} \hline
\bf{Event type} & \bf{Near Site Rate} & \bf{Far Site Rate} \\ \hline \hline
Neutrons & 9000/day & 400/day\\ \hline 
${}^{12}$B & 300/day & 28/day \\ \hline
\end{tabular}
\caption{Estimated production rates (per 20~T detector module) for
spallation neutron and ${}^{12}$B events in the Daya Bay
experiment.} \label{tab:spallrate}
\end{center}
\end{table}

Regular monitoring of the full-volume response for these events,
compared with the regular automated source deployments, will provide
precise information on the stability (particularly of optical
properties of the detector, but also general spatial uniformity of
response) of the detector modules. With the addition of Monte
Carlo simulations, this comparison can be used to accurately assess
the relative efficiency of different detector modules as well as the
stability of the efficiency of each module.

 \subsection{Automated Deployment System}
 \label{ssec:cal_automated}

\begin{figure}[htb!]
  \begin{center}
\includegraphics[height=5in, angle=90]{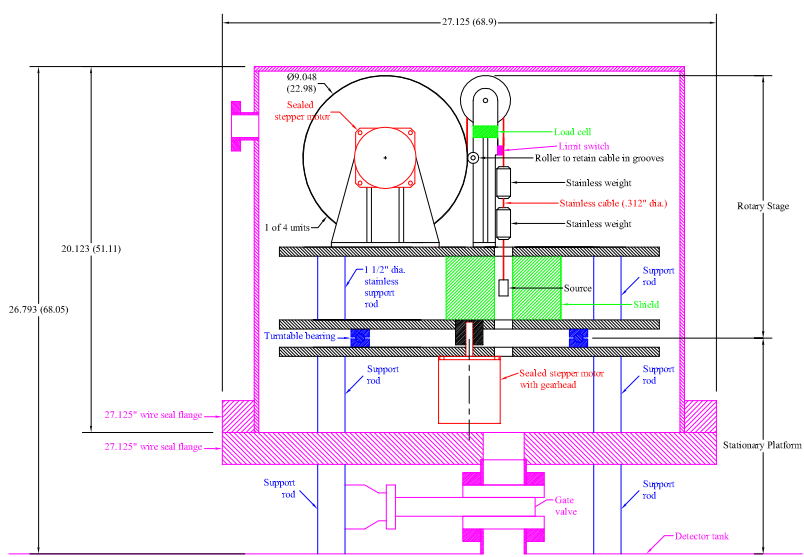}
  \caption{Schematic diagram of the automated deployment system concept.}
  \label{fig:autodeploy}
  \end{center}
\end{figure}

Automated deployment systems will be used to monitor all detector
modules on a routine (perhaps daily) basis. Each detector module
will be instrumented with three (or possibly four) identical
automated deployment systems. Each system will be located above a
single port on the top of the detector module, and will be capable
of deploying four different sources into the detector volume (see
Fig.~\ref{fig:autodeploy}).
This will be facilitated by four independent stepping-motor driven source
deployment units all mounted on a common turntable. The turntable and
deployment units will all be enclosed in a sealed stainless steel
vessel to maintain the isolated detector module environment from the
outside. All internal components must be certified to be compatible
with liquid scintillator. The deployment systems will be operated
under computer-automated control in coordination with the data
acquisition system (to facilitate separation of source monitoring data
from physics data). Each source can be withdrawn into a shielded
enclosure on the turntable for storage. The deployed source position
will be known to about 2~mm.

At present, we anticipate including three radioactive sources on each
deployment system:
\begin{itemize}
\item  ${^{68}}$Ge source providing two coincident 0.511~MeV $\gamma$'s to
simulate the threshold positron signal,
\item  ${^{60}}$Co source
providing a $\gamma$ signal at 2.506~MeV
\item  ${^{252}}$Cf fission
source to provide neutrons that simulate the neutron capture signal.
\end{itemize}
These sources can be deployed in sequence by each of the systems on
each detector module. During automated calibration/monitoring
periods, only one source would be deployed in each detector module
at a time. Simulation studies are in progress to determine the
minimal number of locations necessary to sufficiently characterize
the detector (in combination with spallation product data as
discussed in section~\ref{ssec:cal_data}). At present we anticipate that three or
four radial locations will be sufficient with at least three as
follows:
\begin{itemize}
  \item  Central axis
  \item  A radial location in the central Gd-loaded volume near (just
  inside) the inner cylindrical acrylic vessel wall
  \item  A radial location in the $\gamma$-catcher region.
\end{itemize}
An additional radial location may be instrumented if it is
demonstrated to be necessary by the ongoing simulation studies.

Simulation studies indicate that we can use these regular automated
source deployments to track and compensate for changes in:
\begin{itemize}
\item average gain of the detector (photoelectron yield per MeV)
\item number of PMTs operational
\item scintillation light attenuation length
\end{itemize}
as well as other optical properties of the detector system.

\begin{figure}[htb!]
  \begin{center}
\includegraphics[width=0.85\textwidth]{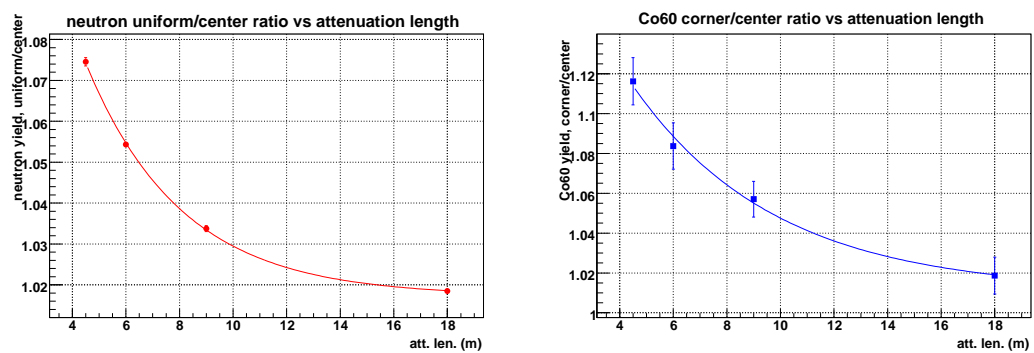}
  \caption{The left panel shows how the scintillator attenuation length can be determined from the ratio of
  the neutron capture peak from spallation neutrons (uniform distribution) to that from a source at the center of the
  detector. The right panel shows a similar measurement using the ratio of the ${}^{60}$Co peak for a source at the
  center to one at the corner ($r=1.4$~m, $z=1.4$~m) of the central volume (1000 events each).}
  \label{fig:source_atten}
  \end{center}
\end{figure}

As an example of how the system can be utilized to monitor the
attenuation length of the scintillator,
Figure~\ref{fig:source_atten} shows simulations of neutron captures
and ${}^{60}$Co source deployments. 
Both methods can clearly be used to measure the attenuation length of
the Gd-loaded liquid scintillator. Thus these methods will provide
frequent monitoring of the condition of the scintillator, and will
allow us to track changes and differences between detector modules.

 \subsection{Manual deployment system}
 \label{ssec:cal_manual}

\begin{figure}[htb!]
  \begin{center}
\includegraphics[width=0.6\textwidth]{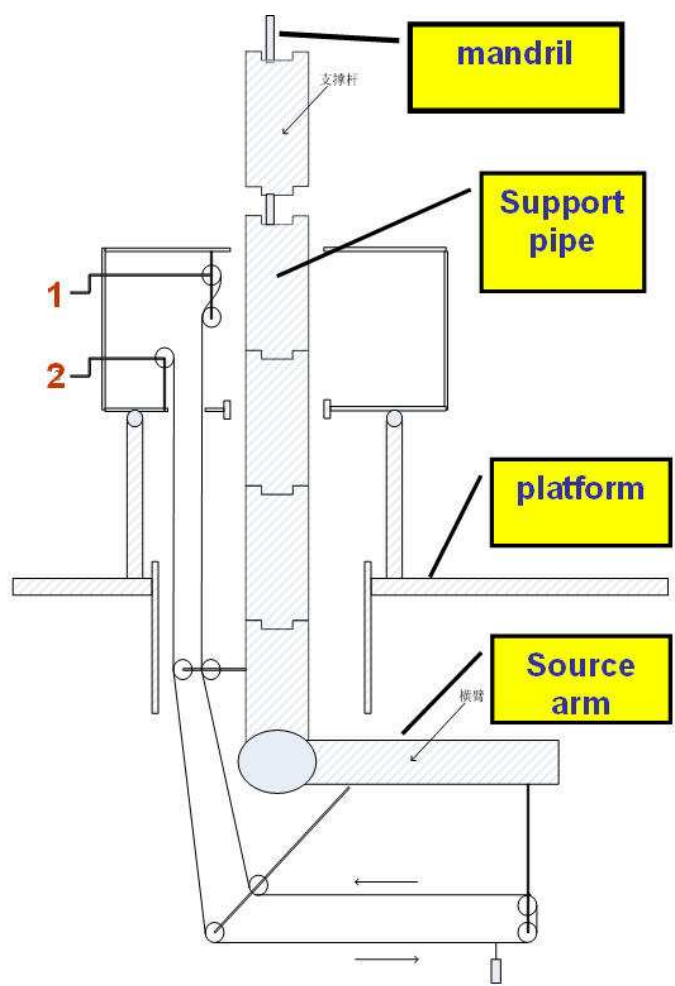}
  \caption{Schematic diagram of the manual source deployment system.}
  \label{fig:mandeploy}
  \end{center}
\end{figure}

A mechanical system will be designed to deploy sources throughout the
active volume of the detectors. The source inside the detector can be
well controlled and the position can be repeated at a level less than
5~cm. The whole deployment system must be treated carefully to prevent
any contamination to the liquid scintillator. The system must be easy
to setup and operate, tolerate frequent use and must have a reliable
method to put sources into the detectors and to take the sources out
as well. The space for operation should not be too large.
Figure~\ref{fig:mandeploy} shows a schematic view of the manual source
deployment system.
The philosophy of such a system is taken from the oil drilling
system. The support pipe is separated into several segments. They can
be connected one by one to make a long support pipe. This design would
reduce the requirement for large space for operations.

The operation procedure will be the following: first, the support
pipe and the source arm will be installed in line (vertical). Then,
it will be put into the desired position inside the detector, by
adjusting the number of segments. When it reach the measurement
position, the source arm is turned to the horizontal. After this, the source
position can be adjusted by the rope system. The rope system must be
designed to insert and remove the sources easily and the position
of the source must be accurately controlled. The whole system can be
rotated around the axis of the pipe on the platform, thus it can
deploy the sources to any position inside the detector.

\renewcommand{\labelitemi}{$\m@th\circ$}

\newpage
\renewcommand{\thesection}{\arabic{section}}
\setcounter{figure}{0}
\setcounter{table}{0}
\setcounter{footnote}{0}

\section{Muon System}
\label{sec:muon}

The main backgrounds to the Daya Bay Experiment are induced by
cosmic-ray muons. These backgrounds are minimized by locating the
detectors underground with maximum possible overburden. Background due
to muon spallation products at the depths of the experimental halls as
well as ambient gamma background due to the radioactivity of the rock
surrounding the experimental halls is minimized by shielding the
antineutrino detectors with 2.5~meters of water. Gammas in the range
of 1--2~MeV are attenuated by a factor $\sim$10 in 50~cm of
water~\cite{mu_att}.
Thus the 2.5~meters of water provides a reduction in the rock gamma
flux of approximately five orders of magnitude. This ``water buffer'' also
attenuates the flux of neutrons produced outside the water pool.

Events associated with fast neutrons produced in the water itself
remain a major potential background. A system of tracking detectors will be
deployed to tag muons that traverse the water buffer.
Events with a muon that passes through the water less than 200$\mu$s
before the prompt signal, which have a small but finite
probability of creating a fake signal event, can be removed from the
data sample without incurring excessive deadtime. By measuring the
energy spectrum of tagged background events and having precise
knowledge of the tagging efficiency of the tracking system, the
background from untagged events (due to tagging inefficiency) can be
estimated and subtracted statistically with small uncertainty. Our
goal is to keep the uncertainty of this background below 0.1\%.

The tracking system will also tag events that have a high likelihood to
produce other cosmogenic backgrounds, ${}^9$Li being the most important
one. While tagging muon showering events may help to suppress the ${}^9$Li
background, the working assumption is that no extra requirements are
imposed on the tracker in order to reduce the ${}^9$Li background.

\begin{figure}[tb]
\centerline{\includegraphics[width=0.96\textwidth]{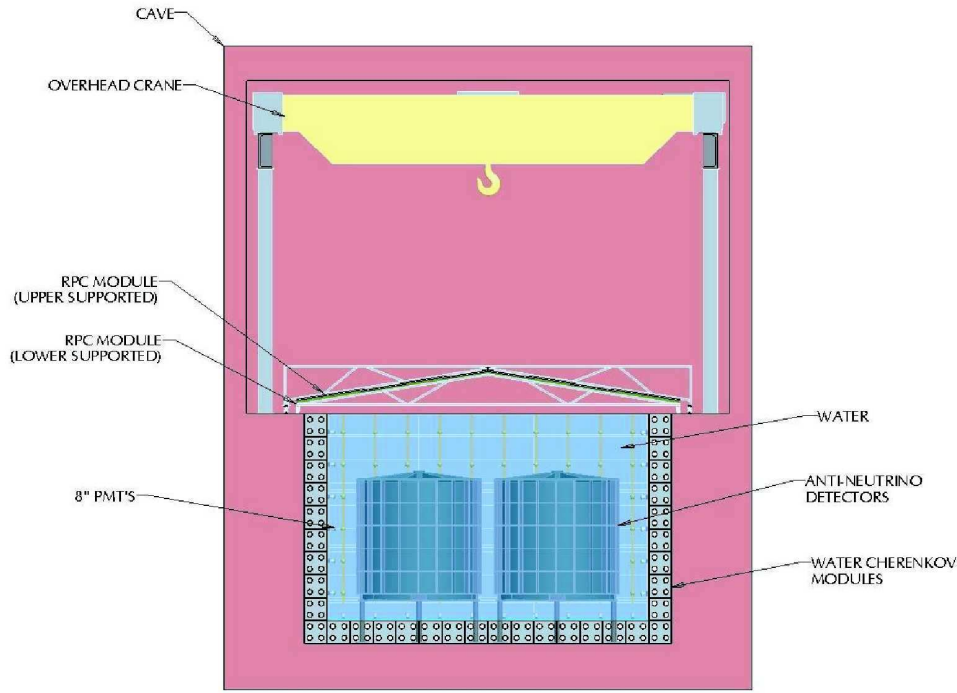}}
\caption{\label{fig:overall} Elevation view of an experimental hall.}
\end{figure}

The current baseline configuration for meeting these challenges is
shown schematically in Fig.~\ref{fig:overall}.  
The antineutrino detectors are
separated by 1~m from each other and immersed in a large pool of
highly-purified water.  The pool is rectangular in the case of the
near halls and square in the case of the far hall.  The minimum
distance between the detectors and the walls of the pool is 2.5~m.
The water shield constitutes the inner section of the pool and is
instrumented with phototubes to detect Cherenkov photons from
muons impinging on the water.  The sides and bottom of the pool are 
lined with 1~m $\times$ 1~m cross-section Water Cherenkov Modules (WCMs)
read out by phototubes at 
either end.  The muon tracker is completed by layers of Resistive Plate
Chambers (RPCs) above the pool.  The top layers extend 1~m beyond the edge
of the pool in all directions, both to minimize the gaps in coverage and
to allow studies of background caused by muon interactions in the rocks
surrounding the pool.

Expected rates of cosmic ray muons in the components of the muon 
system can be found in Table~\ref{tab:DAQ_TAB1}.

\subsection{Muon System Specifications}
\label{ssec:muon_intro}

Note that it is not envisioned that this system will act as an online
veto.  This will allow ample opportunity for careful offline studies
to optimize the performance of the system.

Requirements of the muon system are summarized in more
detail in the following subsections.

\subsubsection{Muon Detection Efficiency}
\label{sssec:muon_eff}

The combined efficiency of the muon tracker and the water shield has
to exceed 99.5\%, with an uncertainty $<$0.25\%.  This is driven by
the need to reject the fast neutron background from muon interactions
in the water and to measure its residual level. As can be seen in
Table~\ref{tab_fastn}, without suppression, this background would 
otherwise be $>$30 
times that of the fast neutron background from muon interactions 
in the surrounding rock, {\it i.e.} at a level roughly 2\% 
of that of the
signal.  A factor 200 reduction in this rate brings the fast neutron
background from the water safely below that from the rock, and the
total residual fast neutron background down to the 0.1\% level.  The
requirement on the uncertainty in the efficiency brings the systematic
due to the uncertainty on the fast neutron background from the water to
a level where it is small compared with other systematics.

\subsubsection{Muon System Redundancy}
\label{sssec:muon_redundancy}

It is difficult to achieve the requisite efficiency with only one
tracking system.  Moreover it is necessary to have a method of measuring
the residual level of background after the imposition of the muon rejection cuts.
Therefore it is desirable to have two complementary tracking systems to
cross check the efficiency of each system. 

As discussed below, the current baseline design is to instrument the
water shield as a Cherenkov tracker by deploying 8'' PMTs in the
water with 1.6\% 
coverage.  Such systems are expected to have $>$95\%
efficiency. A second tracking system, in our baseline a combination of
RPCs~\cite{mu_bes_rpc}~\cite{mu_belle_rpc} above and Water Cherenkov
Modules~\cite{mu_wcc} at the sides and bottom of the water shield, can
give an independent efficiency of $>$90\%. The two systems compliment
each other, with the probability of a muon being missed by both systems
below 0.5\%.

\subsubsection{Spatial Resolution}
\label{sssec:muon_spatial}

The fast neutron background due to muons interacting in the water
shield falls rapidly with distance from the muon track. The spatial
resolution of the muon tracker should be sufficient to measure this
falloff.  Measurements from previous experiments show
that the falloff is about 1~meter~\cite{li9_falloff}. A spatial
resolution of 50--100~cm in the projected position in the region of the
antineutrino detectors is necessary in order to study this radial
dependence. All the technologies we are considering are capable of achieving
sufficient resolution in each coordinate.

\subsubsection{Timing Resolution}
\label{sssec:muon_timing}
There are several constraints on the timing resolution.  The least
restrictive is on the time registration of the muon signal with respect
to that of the candidate event.  To avoid compromising the veto rejection to a
significant extent, this resolution need only be in the range of fractions
of a microsecond.  More stringent requirements are imposed by other,
technology-dependent, considerations.  The water shield PMTs need $\sim$2~ns
resolution to minimize the effect of accidentals and assure event
integrity.   The Water Cherenkov Modules require $\sim$3~ns resolution
to match the position resolution given in the transverse direction by the
1~m granularity of the system (see Sect.~\ref{sssec:wcm}).
If scintillator strips are used, 1ns time resolution will allow the
random veto deadtime from false coincidences in that system to be held
to the order of 1\%.  RPCs will need $\sim$ 25~ns resolution to
limit random veto deadtime from false coincidences in that system to a
similar level.

\subsubsection{Water Shield Thickness}
\label{sssec:muon_thickness}

As mentioned above the shield must attenuate $\gamma$'s and neutrons from the 
rock walls of the cavern by large factors to reduce the accidental background
in the antineutrino detectors. A minimum thickness of 2~m of water is required; 
2.5~m gives an extra margin of safety.

\subsubsection{Summary of Requirements}
\label{sssec:muon_requirements}

The requirements discussed above are summarized in
Table~\ref{table:muon_req}
\begin{table}[h]
\begin{tabular}{l} \hline
Product of inefficiencies for the $\mu$ tracker
\& water shield for cosmic rays should be $\le$0.5\%.\\
 The uncertainty on this
 quantity should be no greater than $\pm$0.25\%.\\
The uncertainty on the random veto  deadtime should be no greater than $\pm 0.05\%$ \\
The position of the muon in the region of the antineutrino detectors should be
determinable  to 0.5-1~m \\
Timing resolution of $\pm$ 1, 2, 3, \ 25~ns for scintillator, water shield, WCMs, and RPCs respectively\\
Thickness of the water buffer of at least 2~m \\
\hline
\end{tabular}
\caption{Muon system requirements
\label{table:muon_req}}
\end{table}


\subsection{Water Buffer}
\label{subsect:water_shield}

The neutrino detectors will be surrounded by a buffer of water with a
thickness of at least 2.5~meters in all directions. Several important
purposes are served by the water.  First, fast-neutron background
originating from the cosmic muons interacting with the surrounding rocks 
 will be significantly reduced by the water. Simulation shows that the
fast-neutron background rate is reduced by a factor of $\sim$2 for every 50-cm 
of water. Second, the water will insulate 
the neutrino detectors from the air, reducing background from the radon
in the air as well as gamma rays from surrounding rocks and dust in the
air. With the low-energy gamma ray flux reduced by a factor of $>$10 
per 50-cm of water, the water can very effectively reduce the 
accidental background rate associated with the gamma rays.
Third, the inner portion of the water buffer, the ``water shield'' 
can be instrumented with PMTs for observing
the passage of cosmic muons via the detection of the Cherenkov light.

The active water shield, together with the RPC and the Water Cherenkov
Module detectors, form an efficient muon tagging system with an
expected overall efficiency greater than 99.5\%. The ability to tag
muons with high efficiency is crucial for vetoing the bulk of the
fast-neutron background.   Finally, the large mass of water
can readily provide a constant operating temperature for the
antineutrino detectors at the near and far sites, eliminating one
potential source of systematic uncertainty.

\subsubsection{Water Buffer Design}
\label{subsubsect:ws_config}

The schematics of the water shield is illustrated in Fig.~\ref{fig:overall}
for the water pool configuration. The cylindrical neutrino detector
modules are placed inside a rectangular cavity filled with purified water, 
{\it i.e.} a water pool configuration. The dimensions of
the water pool are 16~m$\times$16~m$\times$10~m (high) for the far site, and
16~m$\times$10~m$\times$10~m (high) for the near sites. The four detector modules
in the far site will be immersed in the water pool forming a 2 by 2 
array. As shown in Fig. \ref{fig:overall}, the adjacent 
detector modules are separated by 1~meter and each module is
shielded by at least 2.5~meters of water in all directions. For the
near sites, the two neutrino detector modules are separated by 1~meter. 
Again, any neutrons or gamma rays from the rock must penetrate at least 2.5~m  
of water in order to reach the neutrino detector modules. The weight of
water is 2170~tons and 1400~tons, respectively, for the
far site and for each of the two near sites.

As discussed in Sect.~\ref{sssec:wcm}, Cherenkov water module
detectors of 1~m$\times$1~m cross sectional area and lengths of up to 16~m
will be laid against the four sides and the bottom of the water pool,
shown in Fig. \ref{fig:overall}. Therefore, the water buffer is
effectively divided into two independent sections, each 
capable of cross-checking the performance of the other.

\subsubsection{Water Shield PMT layout}
\label{subsubsect:ws_layout}

In the baseline design
the water shield will be instrumented with arrays of 8'' PMT as shown
in Fig. \ref{fig:overall}. Inward-viewing PMT arrays will be mounted
on frames placed at the sides and on the bottom of the pool, abutting the 
inner surfaces of the Water Cherenkov Modules (which will be covered
with Tyvek). The PMTs will be evenly distributed forming a
rectangular grid with a density of 1 PMT per 2~m$^2$. 
This corresponds to a 1.6\%  
areal coverage. The total number of PMTs for the
far site and the two near sites is 844, 
as detailed in Table~\ref{tab:veto_PMT}. 
\begin{table}[tbp]
\begin{center}
\begin{tabular}{|c||c|c|c|} \hline 
Site & bottom & sides & total\\ \hline\hline
DB Near & 60 & 192 & 252 \\ \hline
LA Near & 60 & 192 & 252 \\ \hline
Far & 100 & 240 & 340 \\ \hline
All three &220  &624  & 844 \\ \hline
\end{tabular}
\caption{\label{tab:veto_PMT} Number of PMTs for the water shield.}
\end{center}
\end{table}
%
The HV system will be very similar to that described in
Sect.~\ref{ssec:det_HV} for the antineutrino detector PMTs.

Optimized efficiency, position resolution, energy resolution, timing
resolution, etc. are to be determined from Monte-Carlo simulations now
in progress.  The baseline and a number of other possible arrangements
of PMTs have been studied so far.

\subsubsection{Water Shield Simulation Studies}
\label{subsubsect:ws_perform}

\begin{figure}
\noindent
\begin{minipage}[thb]{0.46\linewidth}
\centering{\includegraphics[width=\linewidth]{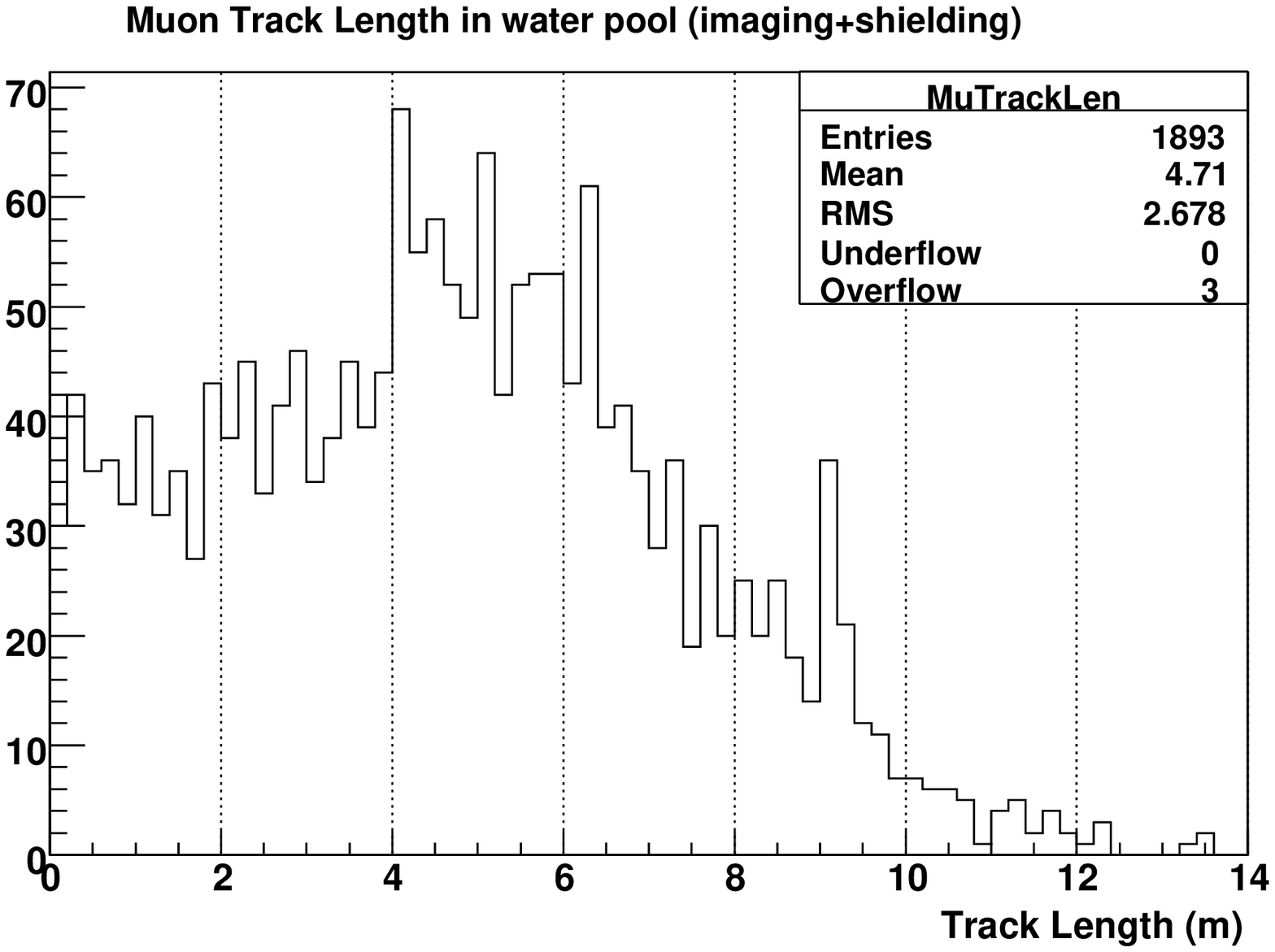}}
\caption{\label{fig:mu_length} Track length of muons in the water shield for
the Daya Bay Near Hall.}
\end{minipage}\hfill
\begin{minipage}[thb]{0.46\linewidth}
\centering{\includegraphics[width=\linewidth]{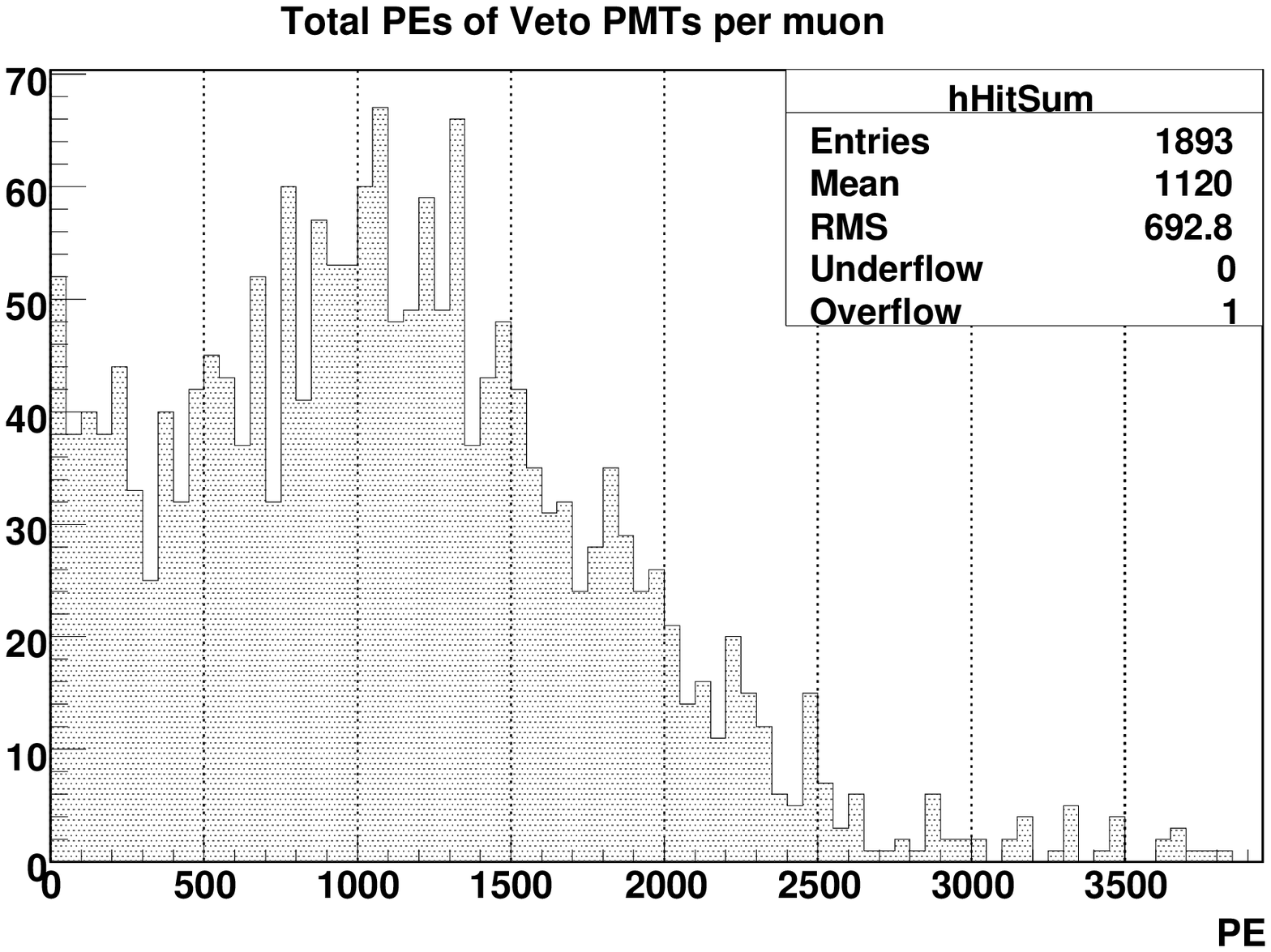}}
\caption{\label{fig:total_pe} Total number of photoelectrons observed
in baseline configuration in the Daya Bay Hall.}
\end{minipage}
\end{figure}

\begin{figure}
\noindent
\begin{minipage}[thb]{0.46\linewidth}
\centering{\includegraphics[width=\linewidth]{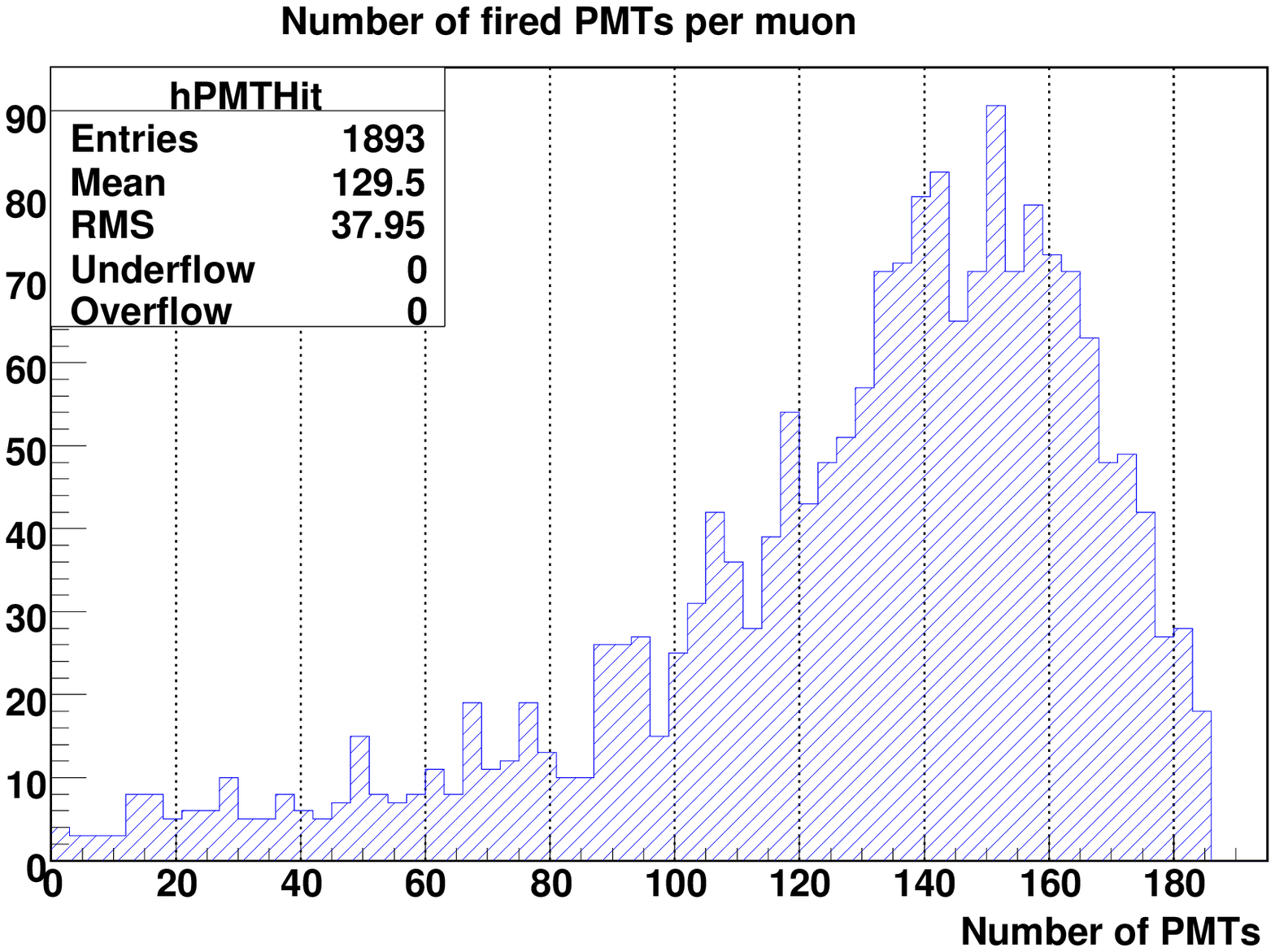}}
\caption{\label{fig:total_pmt} Number of phototubes hit in baseline
configuration of the Day Bay Near Hall.}
\end{minipage}\hfill
\begin{minipage}[thb]{0.46\linewidth}
\centering{\includegraphics[width=\linewidth]{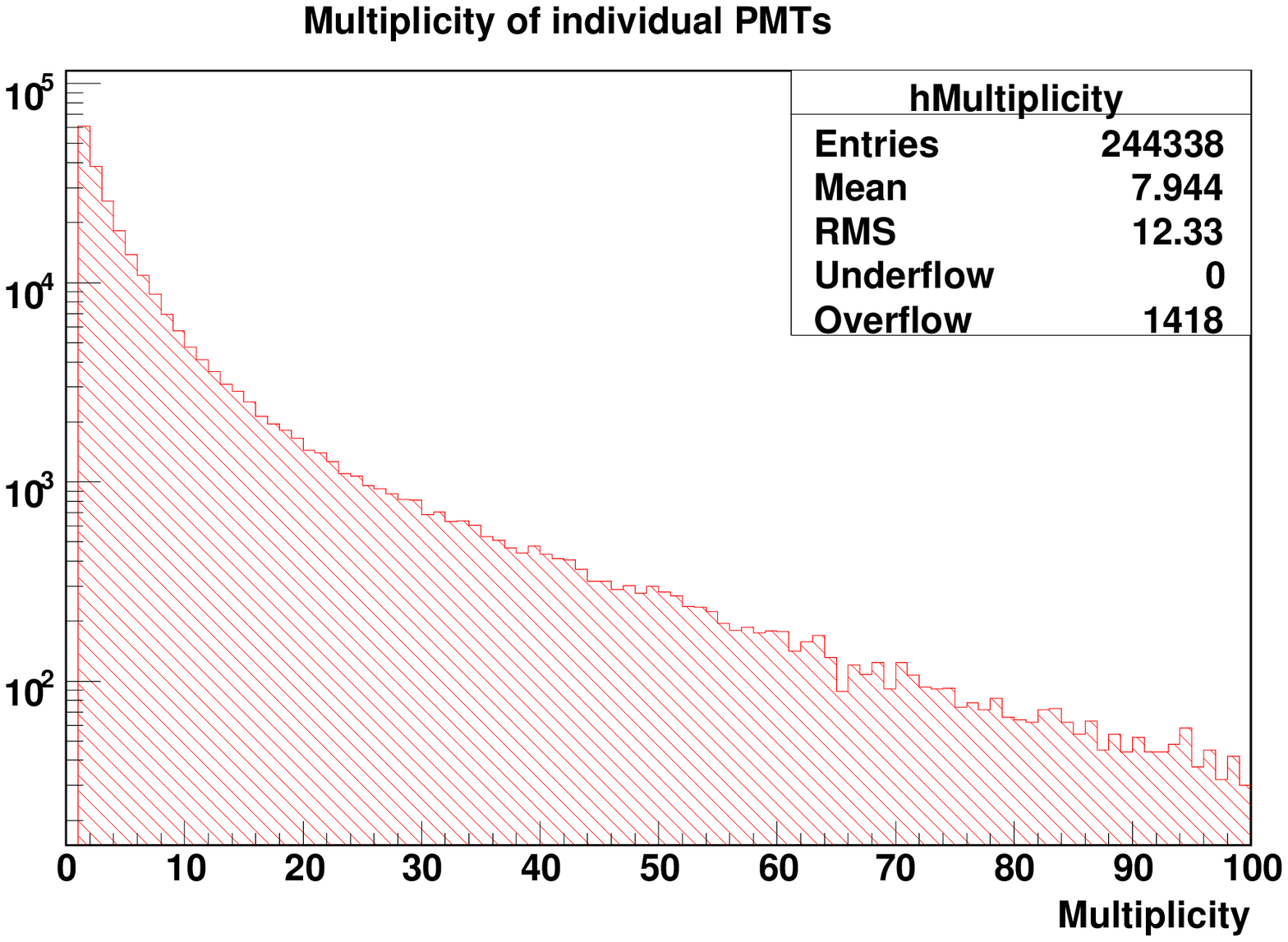}}
\caption{\label{fig:module_pe} Photoelectrons observed per PMT
in baseline configuration of the Daya Bay Near Hall.}
\end{minipage}
\end{figure}

Fig.~\ref{fig:mu_length} shows the simulated distribution of track length
of cosmic ray muons in the water shield of the Daya Bay Near Hall.  The
mean distance traveled through the water is about 5~m.  For full geometric 
coverage with a typical bialkali photo-cathode, one would expect about 
15,000 photoelectrons from a track of this length.
Taking into account the 1.6\% 
PMT geometric coverage, these muons would produce $\sim$240 
photoelectrons in the PMTs from photons collected directly.
As can be seen in Fig.~\ref{fig:total_pe}, 
the average is actually $>$1000 for the baseline configuration.  Our simulation
verifies that this is due to reflected photons.
These photoelectrons are spread over an average of $\sim$130 PMTs as
seen in Fig.~\ref{fig:total_pmt}, and the resulting distribution of
photoelectrons in a single PMT is shown in Fig.~\ref{fig:module_pe}.
As expected the average is about 8 photoelectrons, although there is a 
long tail exponential tail with a slope of $\sim$26.  

\begin{figure}[tb]
\noindent
\begin{minipage}[thb]{0.46\linewidth}
\centering{\includegraphics[width=\linewidth]{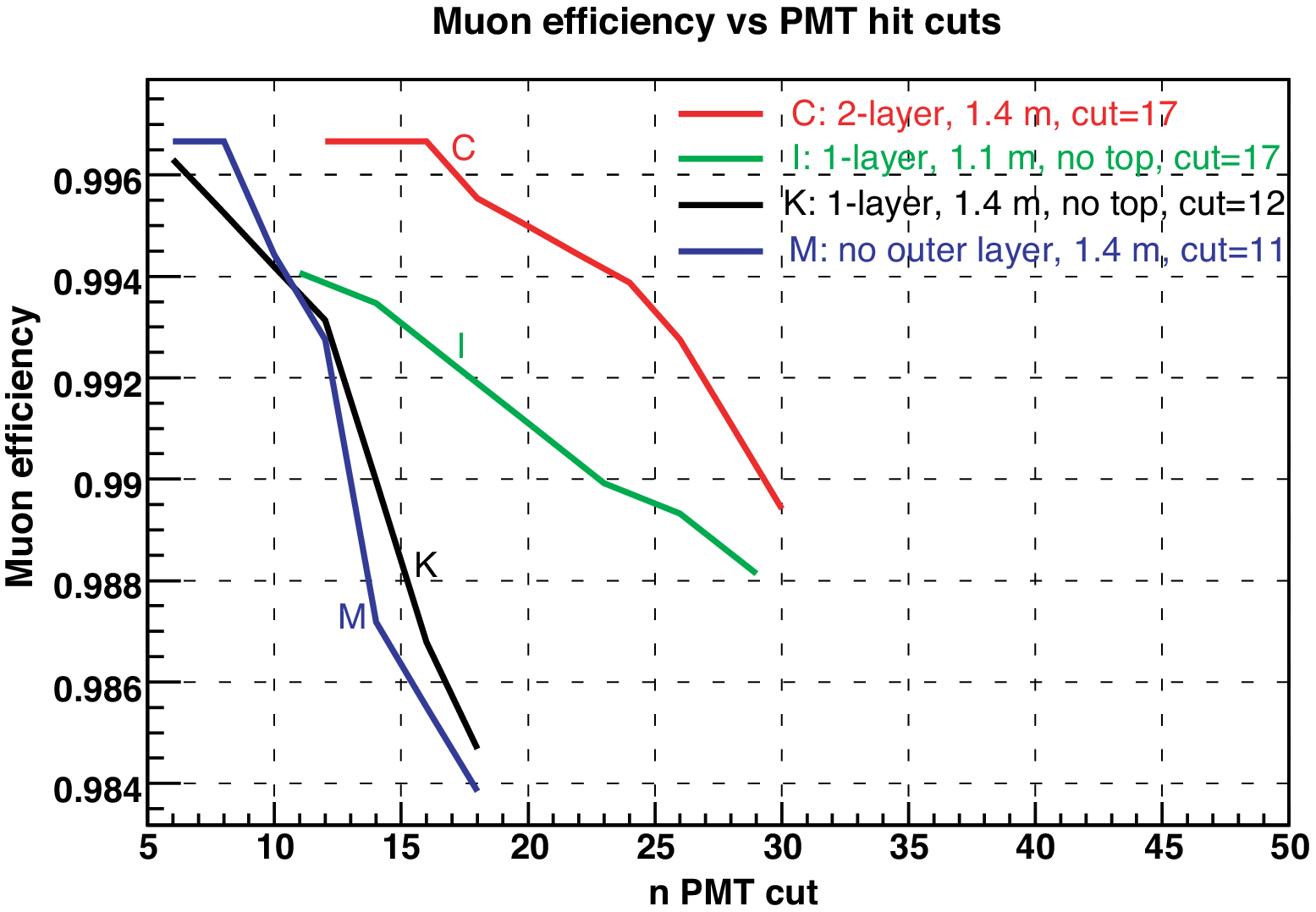}}
\caption{\label{fig:mu_eff} Muon efficiency of the water shield as a function
of threshold (in number of PMTs hit) for four different configurations
of PMTs.  The black curve represents the performance of the current baseline.}
\end{minipage}\hfill
\begin{minipage}[thb]{0.46\linewidth}
\centering{\includegraphics[width=\linewidth]{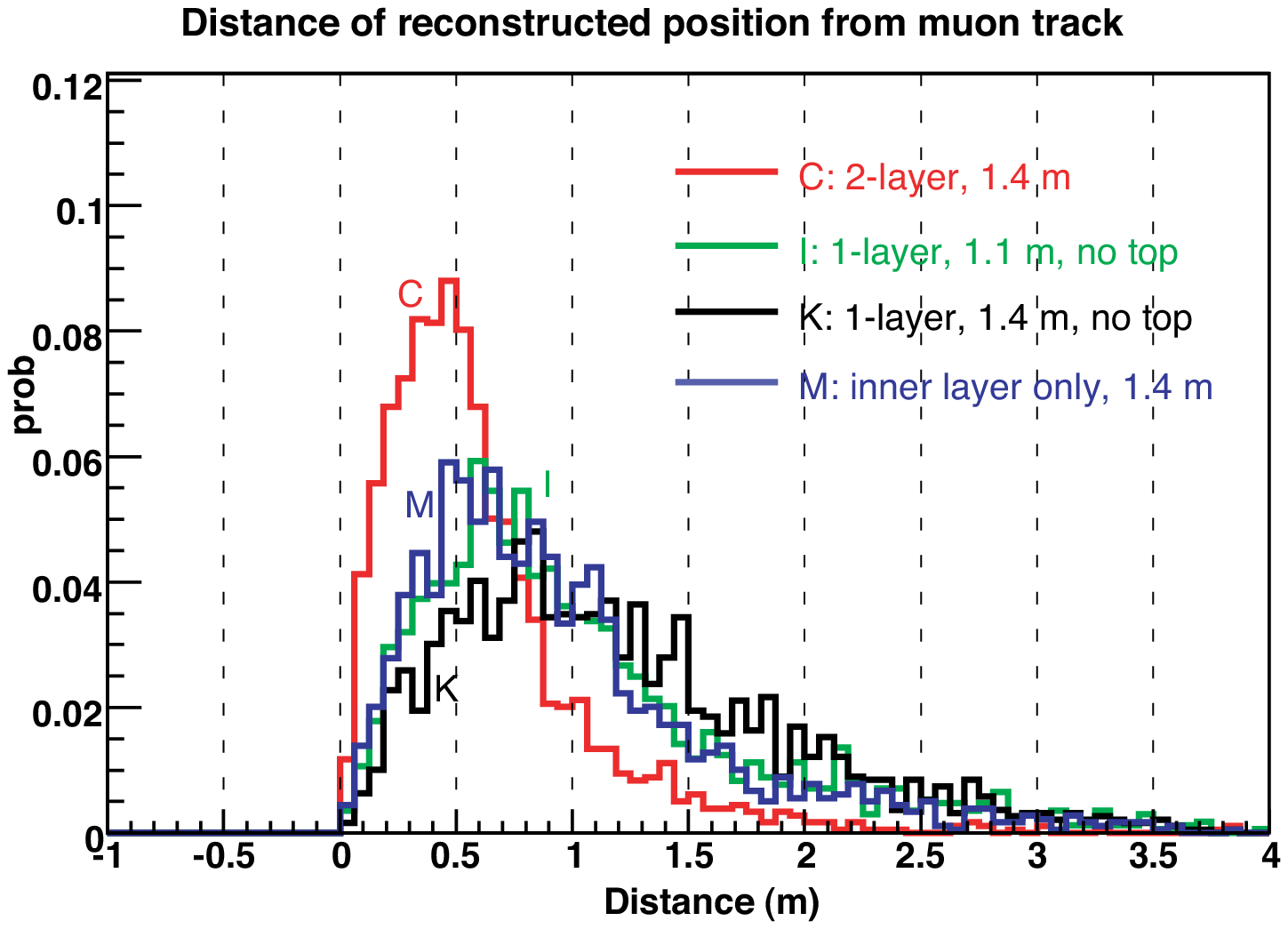}}
\caption{\label{fig:mu_time} Difference between the reconstructed position
and the nearest point on the actual muon trajectory for four different
configurations.  The black curve is for the current baseline.}
\end{minipage}
\end{figure}

Initial results on efficiency of the water shield as
a function of the number of PMTs demanded are shown in
Fig.~\ref{fig:mu_eff}. In each case a threshold number of PMTs is
determined by the requirement that the deadtime due to random coincidences
be 
$<$1\%.  Conservatively an effective singles rate (dark current plus
radioactivity) of 50~kHz/PMT was assumed for this calculation.  For
the baseline configuration described above, this level was reached
at a threshold of 12~PMTs, yielding an efficiency of $>$99.3\% as
can be seen from the black curve, labeled ``K'' in Fig.~\ref{fig:mu_eff}.  

Results from three other configurations are also shown.  The green
curve, labeled ``I'' corresponds to a configuration similar to that of
the baseline, but with a higher density of PMTs (1.1~m spacing
instead of 1.4~m).  A threshold of 17~PMTs is needed to reduce the
random trigger deadtime to below 1\%.  At this point the efficiency is
a little above 99.2\%, {\it i.e.} statistically identical to that of the
baseline.  However it's clear that if a higher threshold should be
needed, the efficiency of this configuration holds up better than that
of the baseline.  The red curve, labeled ``C'', corresponds to a case
when PMTs are deployed on many other surfaces of the pool (at the
same 1.4~m spacing).  In addition to the sides and bottom of the pool,
for this configuration, there is a plane of PMTs at the top of the
pool looking down into it, a plane parallel to this one just above the
antineutrino detectors looking up, and vertical planes abutting the
antineutrino detectors (forming a ``box'' around the detectors) with
the PMTs looking out into the pool.  For this configuration, the  
random coincidence rate reaches 30Hz, which results in a deadtime of
0.6\%, at a threshold of 17~PMTs. At this point the efficiency
is about 99.6\%.  This configuration has clearly better performance
than that of the baseline, although at the cost of twice as many PMTs
and much added complication in the PMT mounting scheme.  A
configuration more efficient in the number of PMTs than the baseline
is shown in blue, labeled ``M''.  This configuration has no
inward-looking PMTs except near the corners of bottom of the pool.
There are upward-facing PMTs above the antineutrino detectors and
outward-facing PMTs on the sides of the ``box'' surrounding those
detectors.  The PMTs are spaced at 1.4~m.  The ``M'' configuration has
very similar efficiency performance to that of the baseline, but
requires 25\% fewer PMTs (albeit at the cost of a more complicated mounting
system).

An initial attempt was made to develop a position-determining
algorithm.  The positions of the five PMTs with earliest light are
averaged to obtain an estimator of the muon position.  This was tested in
the same GEANT runs used to calculate efficiency. In
Fig.~\ref{fig:mu_time} the difference between the calculated position
and actual trajectory of the muon is plotted for the four
configurations discussed above.  For the baseline, the resolution
averages $\sim$0.9~m.  Doubling the numbers of PMTs in the same
configuration doesn't help much here.  However the ``M'' configuration
performs a little better than the baseline, and the ``C''
configuration with both more PMTs and more surfaces has substantially
better performance.  

These results on efficiency and position determination, although preliminary 
are extremely encouraging.

\subsubsection{Water Buffer Front-End Electronics}
\label{sssec:ws_PMT_elect}

Extrapolation of the curve in 
Fig.~\ref{fig:module_pe} indicates that although the number of photoelectrons
per PMT has a long tail, only about 0.01\% of the PMTs see more
than 100 photoelectrons.  Thus the performance of the antineutrino
detector electronics should certainly be adequate for the water buffer
readout. However the reduced dynamic range requirement may indicate that less
expensive options should be considered for the pulse height measurement.
In addition to the pulse height information, timing
information will also be provided by the readout electronics. With
0.5~ns/bin TDCs, a timing resolution of 2~ns 
is readily achievable for a single PMT channel. The energy
sum of the PMTs as well as the multiplicity of the struck PMTs will be
used for defining the muon trigger (see Sect.~\ref{ssec:mu_trig}).

\subsubsection{Calibration of the Water Buffer PMTs}
\label{sssec:ws_PMT_calib}

The gain stability and the timing of the PMTs will be monitored by a LED
system identical to that for the neutrino detectors. No radioactive sources
will be required.

\subsubsection{Purification of the Water Buffer}
\label{subsubsect:ws_pure}

We must purify the water to maintain constant water transparency and
to prevent microbial growth.

We must also recirculate the water to maintain a constant relatively low
temperature (15$^{\circ}$C) to inhibit microbial growth and maintain the
antineutrino detectors at constant temperature.  Also, we must remove 
impurities that have leached into the water from the detector materials and
wall during recirculation.

The level of purity needed to prevent growth will reduce radioactive
backgrounds to well below the level where they would make detectable
background in the antineutrino detectors.


\subsection{Muon Tracker}
\label{ssec:muon_tracker}

The muon tracker has the job of tagging the entering muons and
determining their path through the region of the antineutrino detectors.
In addition, it must measure the efficiency of the water shield for
muons. 

Three technologies are being considered for the muon tracker.
RPCs can be used on top of the water
shield, but to use them in the water would require a large program of
R\&D on techniques of encapsulation.  Water Cherenkov modules are
cheap and practical to operate in the water, but would be difficult to
remove from the top of the water shield when the antineutrino detectors need
maintenance or have to be moved.  Plastic strip scintillators can be
operated either on top of the water or in it, although the latter
requires developing an encapsulation scheme.

\subsubsection{Resistive Plate Chambers (RPC)}
\label{sssec:muon_RPC}

The RPC is an attractive candidate 
tracking detector since it is economical for instrumenting 
large areas.  Furthermore, RPCs are simple to fabricate.  The
manufacturing technique for both Bakelite (developed by IHEP for the
BES-III detector~\cite{mu_bes_rpc}) and glass RPCs (developed for
Belle~\cite{mu_belle_rpc}) are well established.

An RPC is composed of two resistive  plates  with  gas  flowing  between
them. 
High  voltage is applied
on the plates to produce a strong electric field in the  gas.  When a charged
particle passes through the gas, an avalanche or a
streamer is produced. The electrical signal is then registered by a pickup
strip and sent to the data acquisition system.  In our case,  the  RPCs  will
operate in the streamer mode.


The RPCs for the BES-III spectrometer were constructed using  a  new
type of phenolic paper laminate developed at IHEP. The surface  quality  of
these plates is markedly improved compared to the Bakelite plates  previously
used  to  construct  RPCs.  
IHEP has developed a technique to control the resistivity of the
laminates to any value within a range of $10^9\!-\!10^{13}\ \Omega \cdot$m.  About 1000 bare
chambers ($\sim$1500~m$^2$) without linseed oil coating have been produced for BES-III.  Tests show that the
performance of this type of RPC is comparable in performance
to RPCs made with linseed oil-treated Bakelite and to glass RPCs
operated in streamer mode.  Applying linseed oil to Bakelite is a 
time-consuming step in the productions, and presents a major risk
factor in the long-term operation of the chambers.

\begin{figure}[tb]
\noindent
\begin{minipage}[thb]{0.46\linewidth}
\centering{\includegraphics[width=\linewidth]{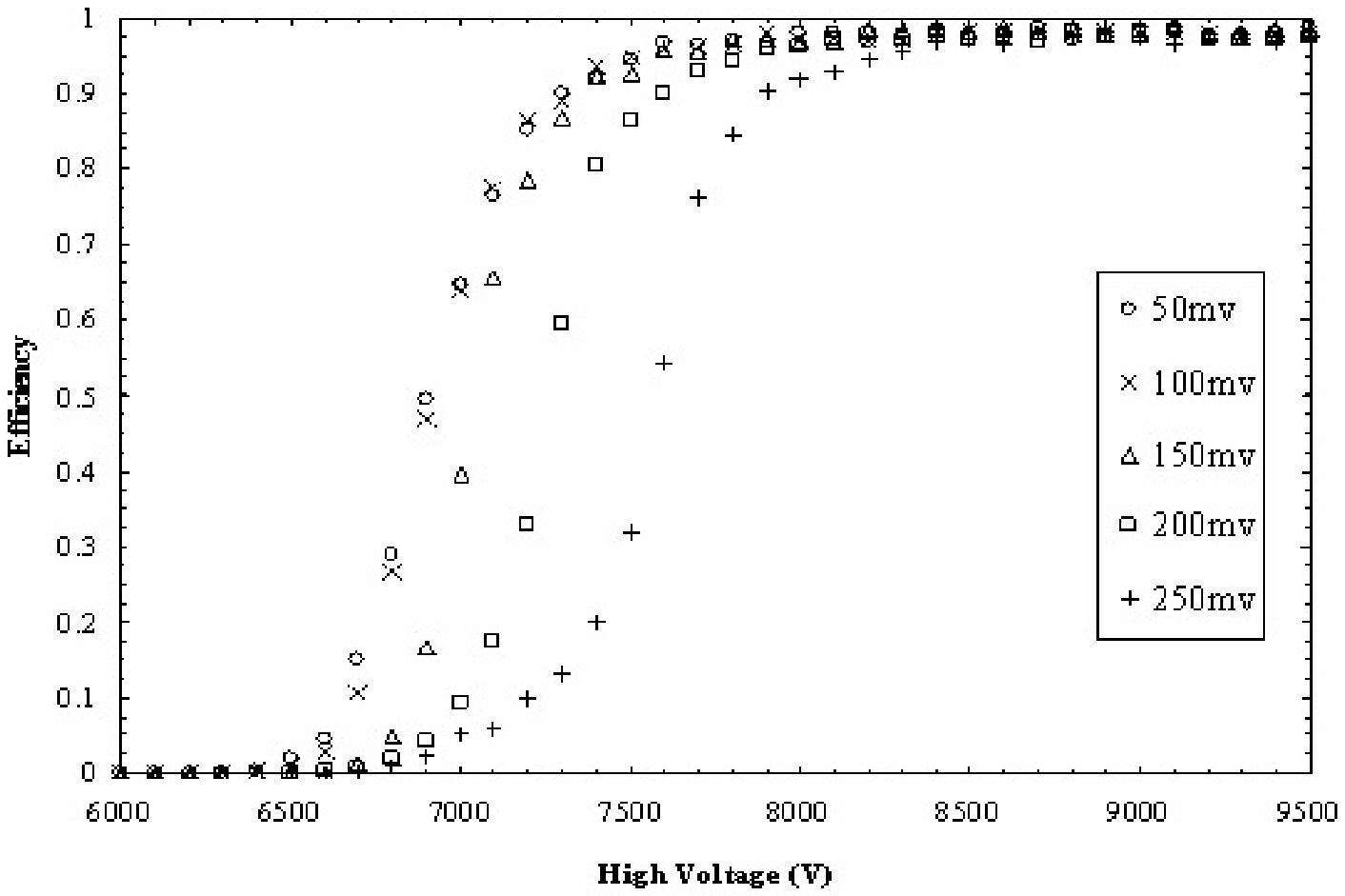}}
\caption{\label{fig:rpc_eff_test} Efficiency of the BES-III RPC versus high voltage
for different thresholds.}
\end{minipage}\hfill
\begin{minipage}[thb]{0.46\linewidth}
\centering{\includegraphics[width=\linewidth]{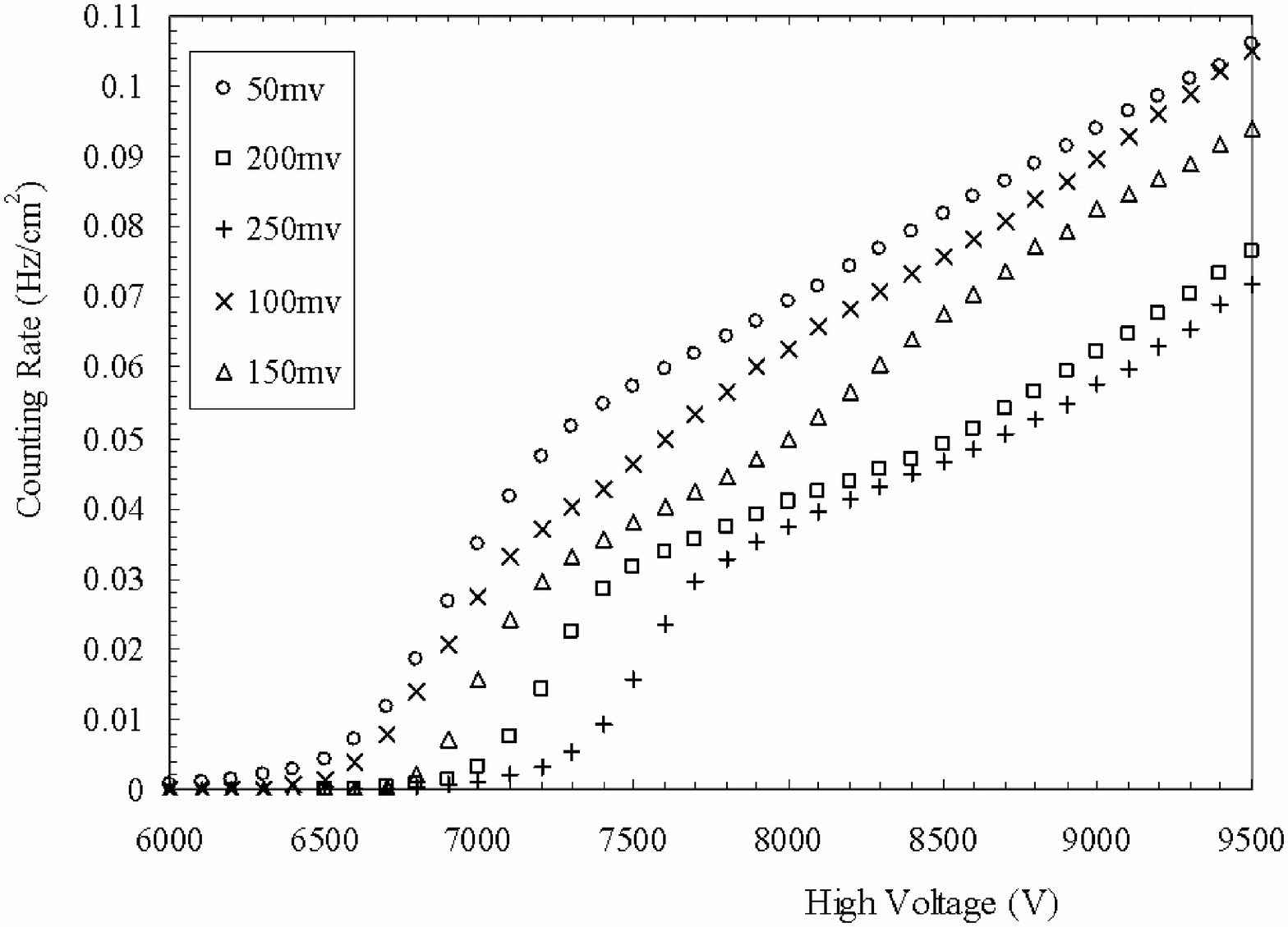}}
\caption{\label{fig:rpc_rate} Noise rate of the BES-III RPC versus high voltage for different thresholds.}
\end{minipage}
\end{figure}

The efficiency and noise rate of the BES-III RPCs have been measured.  In
Fig.~\ref{fig:rpc_eff_test}, the efficiencies versus high voltage are
shown for threshold settings between 50 and 250~mV. The efficiency as
plotted does not include the dead area along the edge of the detector,
but does include the dead region caused by the insulation gasket. This
kind of dead area covers 1.25\% of the total detection area. The
efficiency of the RPC reaches plateau at 7.6~KV and rises slightly to
98\% at 8.0~KV for a threshold of 150~mV.  There is no obvious difference 
in efficiency above
8.0~KV for thresholds below 200~mV.  The singles rate of the RPC 
shortly after production is
shown in Fig.~\ref{fig:rpc_rate}. 

The typical singles rate at thresholds above 150~mV is $<$0.1~Hz/cm$^2$ 
after training.  The noise rate
increases significantly when the high voltage is higher than 8~kV.

\begin{figure}[bt]
\centerline{\includegraphics[width=0.9\textwidth]{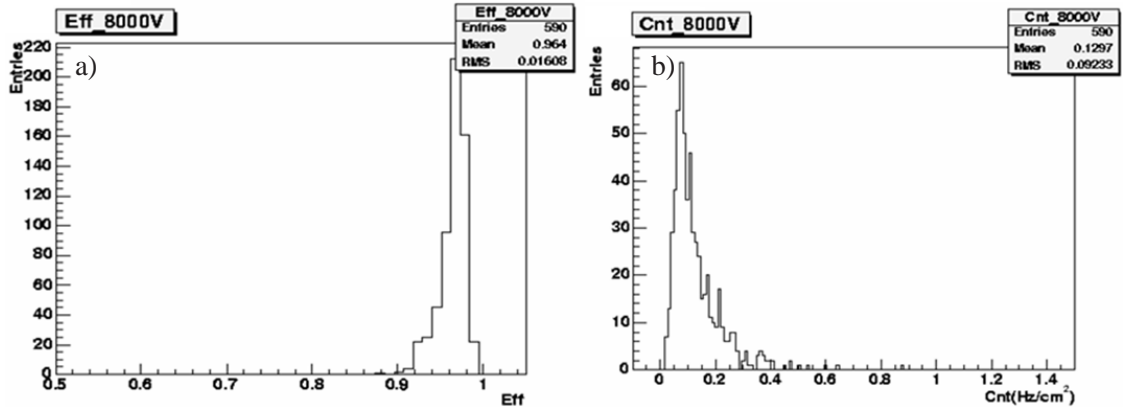}}
\begin{picture}(0,0)
\put (50,140) {a)}
\put (268,140) {b)}
\end{picture}
\caption{\label{fig:rpc_test} Distribution of tested RPC a) efficiencies and b)
singles rates.}
\end{figure}

In cosmic ray tests of a large sample of BES chambers, the average
efficiency was 97\%, and only 2 had efficiency less than 92\%.
Figure~\ref{fig:rpc_test}a shows the efficiency distribution.  This
efficiency was obtained with no corresponding excessive chamber noise.
Figure~\ref{fig:rpc_test}b shows the RPCs singles rate.  
The most
probable value was $\sim$0.08~Hz/cm$^2$ and the average was 0.13~Hz,
with only 1.5\% higher than 0.3~Hz/cm$^2$.

\subsubsection{RPC Design}
\label{sssec:muon_design}

\begin{figure}[tb]
\centerline{\includegraphics[width=0.6\textwidth]{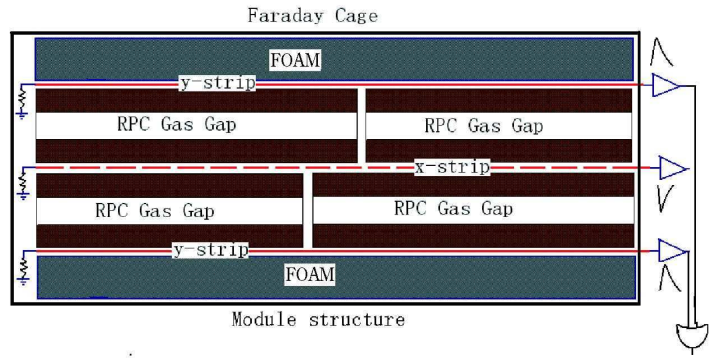}}
\caption{\label{fig:rpc_mod} Structure for a double-gap RPC module. Three
such layers are envisioned.}
\end{figure}

The above measurements were made with one dimensional readout RPCs.  For the
Daya Bay experiment, we are planning to use RPCs with readout in two dimensions in order
to get both x and y-coordinates of the cosmic muons
(similar in design to the BELLE RPCs~\cite{mu_belle_rpc}).  Three double gap
layers would be combined to form a module.  
The layers are electrically shielded from one another
to avoid cross talk.  The structure of a single such layer
is shown in Fig.~\ref{fig:rpc_mod}.


Plastic spacers will be used periodically to
precisely maintain the gap width.  These spacers are potentially a
source of dead space.  Therefore, within each module, spacers in
different layers will be offset, resulting in no aligned dead space.
Also, modules will overlap at the edges, so there will be no
inter-module dead space.  The double-gap design improves robustness
and efficiency, at the cost of doubling the noise rate\footnote{Note that
this is the case only for the true noise rate.  Cosmic rays provide
$\sim$0.018~Hz/cm$^2$ of the singles at sea level}.

Bakelite modules as large as 1~m$\times$2~m are straightforward to
manufacture.  Two of these will be bonded together to make a single
2~m$\times$2~m unit.  The chambers will be read out by strips of $\sim$14~cm 
dimension.  Thus each unit will have 28 
readout channels.  With adequate module overlap and the
peaked roof shown in Fig.~\ref{fig:overall} extending an extra 1m
on all sides of the pool, it will be necessary to
cover an area of 20~m$\times$18~m 
at the Far Hall and 20~m$\times$12~m 
at each of the near halls.  This will require a total of 630 
units for three layers, and a total of 17,640 
readout strips.

Alternatively glass modules as large as 1.5~m$\times$1.5~m are easy to 
make and convenient to handle.  They would have $\sim$19~cm-wide readout 
strips. These would require a total of 1092 
units and 17,472 
readout strips.


\subsubsection{RPC Mounting}
\label{sssec:muon_mounting}

\begin{figure}[tb]
\centerline{\includegraphics[width=0.8\textwidth]{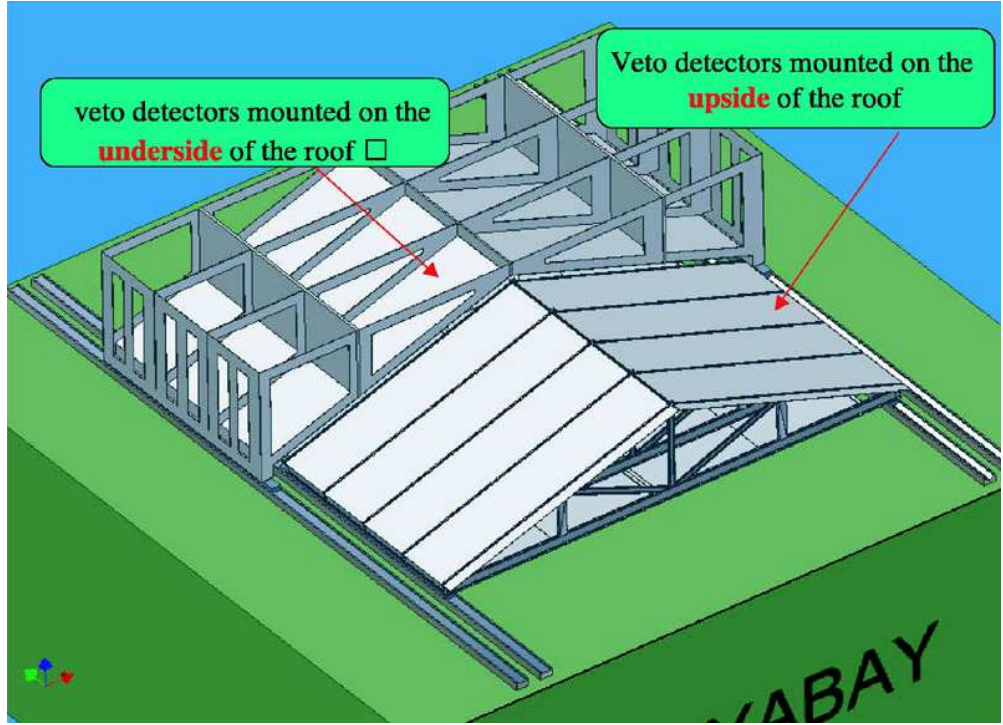}}
\caption{\label{fig:roof} Sliding roof mount for muon tracker modules above
the water pool.}
\end{figure}

Figure~\ref{fig:roof} shows a candidate scheme for mounting the RPCs on
a peaked roof over the water pool.  
The roof will be divided into two sections of different heights and
the RPCs mounted in a way that allows one section to slide over or
under the other.

\subsubsection{RPC Performance}
\label{sssec:muon_performance}

\begin{figure}[tb]
\centerline{\includegraphics[angle=270, width=1.\textwidth]{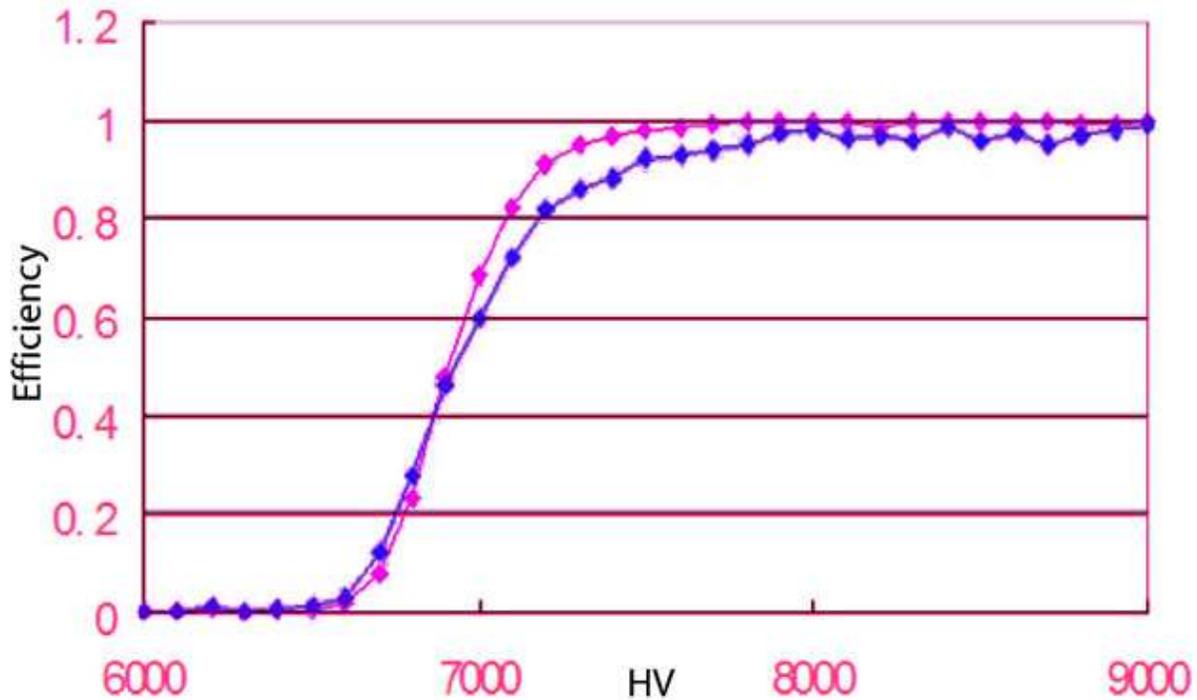}}
\caption{\label{fig:rpc_2/3} Efficiency as a function of gap voltage for the
individual modules of the Daya Bay prototype RPCs (blue) and for the system 
when two out of three hit modules are required (red).}
\end{figure}

Taking into account inefficiencies due to dead-spaces, we expect the
overall efficiency of a single layer to be at least $\varepsilon\sim$96\%. 
If we adopt the definition of a track as hits in at least two out of three layers then 
the coincidence efficiency is
$\varepsilon^3 + C^2_3\varepsilon^2(1-\varepsilon) = 0.96^3+3\times0.96^2\times(1-0.96)=99.5$\%, 
where $C^2_3=3$ is the binomial coefficient.
Assuming a bare chamber noise rate, $r$, of 1.6~kHz/m$^2$ 
is achieved (consistent with twice the BES chamber measurements), a signal 
overlap
width $\tau$, of 20~ns, 
and a coincidence area, $A$, of 0.25~m$^2$, 
the noise rate would be $C^2_3 A^2r^2\tau
(1$~m$^2/A^2)= 3\times 1600^2\times 0.2 \cdot 10^{-7}=0.154$~Hz/m$^2$. 
For the Far Hall, this gives a total accidental rate of 55~Hz 
and a corresponding contribution to the deadtime of 1.1\% 
in the case that a muon signal is defined
by a hit in the RPCs alone.  A test of the 3 layer scenario with
prototypes of the Daya Bay chambers, using a track definition of two
out of three hits, found a coincidence efficiency of 99.5$\pm$0.25\%,
which is consistent with the calculated efficiency.  The efficiency
curves are shown in Fig.~\ref{fig:rpc_2/3}.

Initial simulation results based on measurements of radioactivity in
the Aberdeen Tunnel predict singles rates from radioactivity of 
$\sim$650~Hz/m$^2$ 
and coincidence rate of 1~Hz/m$^2$, 
mainly from double Comptons.  This corresponds to a trigger of 
$\sim$300~Hz 
in the Far Hall and a contribution to the veto deadtime of $\sim$6\%. 



\subsubsection{RPC Front-End Electronics}
\label{sssec:muon_electronics}

\begin{figure}[tb]
\centerline{\includegraphics[width=0.75\textwidth]{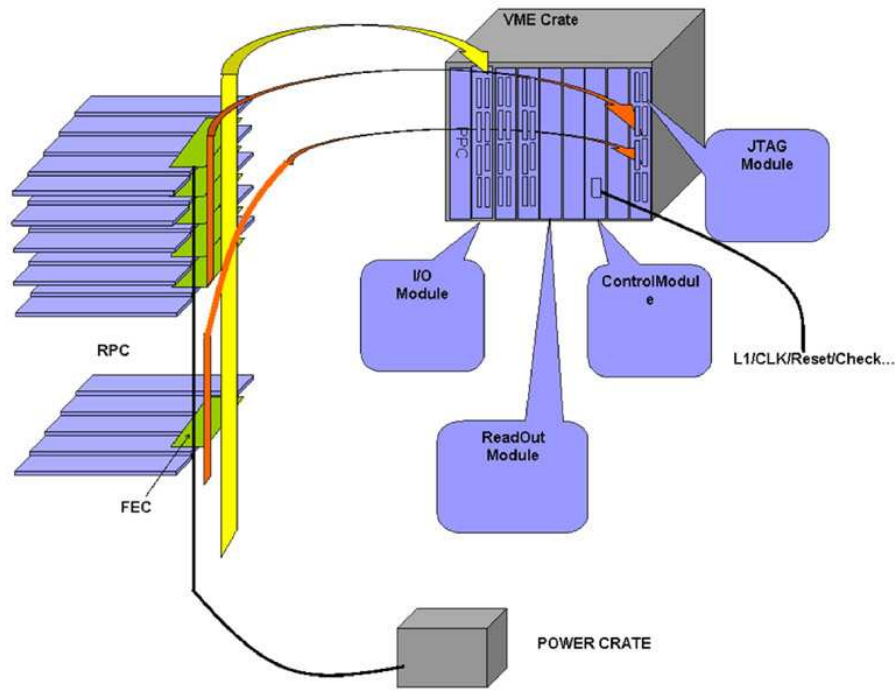}}
\caption{\label{fig:rpc_readout} Configuration of the electronics \& readout system.}
\end{figure}

The readout system consists of a readout subsystem, a threshold
control subsystem, and a test subsystem.  The readout system, shown in
Fig.~\ref{fig:rpc_readout}, contains a 9U VME crate located above the
detector, which holds a system control module, a readout module, an
I/O module, and a JTAG control module.  The system clock will operate
at 100~MHz.

\noindent \textbf{1) Control Module}

The control module receives the trigger signals (L1, Clock, Check, and
Reset) from the trigger system and transmits them to the Front End
Card (FECs) through the I/O modules. It also receives commands (such
as setting thresholds, testing, etc.) and transmits them to the FECs.
The control module is also a transceiver which transfers the FULL
signal between the readout module and FECs.

\noindent \textbf{2) I/O Modules}

The VME crate contains several I/O modules, each of which consists of
12 I/O sockets connected by a data chain.  The I/O module drives and
transmits the signals of the clock and trigger to all the FEC's, and
transmits control signals between the readout module and the FECs.

\noindent \textbf{3) VME Readout Modules}

The readout module is responsible for all the operations relative to
data readout. It not only reads and sparsifies the data from all the
data chains (it can read 40 of them in parallel), constructs the
sub-event data to save into the buffer, and requests the interrupt to
the DAQ system to process the sub-event data, but also communicates
the Full signals to the FEC's to control the data transmission. The
readout module checks and resets
control signals to the trigger system.  It also controls the reading
and sparcifying of the FEC data, the requesting of a DAQ interrupt,
and the counting and resetting of the trigger number.

\noindent \textbf{4) Front-End Cards}

The FECs are located on the RPC detector.  
Their task is to transform signals from the strips into a bit map,
store the data in a buffer and wait for a trigger signal.  Events with
a trigger will be transmitted in a chain event buffer in the VME
readout modules.  Events without a trigger are cleared.  Analog
signals from groups of 16 strips are discriminated and the output read
and stored in parallel into a 16-bit shift register, which is
connected to a 16-shift daisy chain.  A total of 16 FECs compose one
FEC Daisy-Chain, which covers 256 strips.  The data from each chain,
as position information, are transferred bit-by-bit to the readout
module in the VME crate through the I/O modules using differential
LVDS signals. Each datum of the chain will be stored temporarily in
the relative data chain buffer of the readout module.  After the data
sparsification, the whole data chain will be stored into the sub-event
data buffer awaiting DAQ processing.

In each FEC there is also a DAC chip, which is used to generate test
signals.  When a test command goes to the test signal generator which
resides in the system control module in the VME crate, the generator
sends timing pulses to the FEC's DAC chip through an I/O module.  This
chip then delivers a test signal to each channel's comparator.

The principle of the threshold setting circuit is the same as the test
circuit. The timing pulses are generated by the threshold controller
in the system control module, and sent to the DAC to generate the
threshold level at each of the input ports of the discriminators in
the FEC.

\noindent \textbf{5) JTAG Module}

The JTAG module gets the FPGA setting command from the VME BUS,
transforms the command into the JTAG control timing, and sends it to
the FECs.  Each of the JTAG modules has 12 slots on the panel of the
module, enough to satisfy the requirements of the whole readout
system.

\noindent \textbf{6) Test System}

The test system for the readout system consists of the test control
module in the VME crate, and a test function generator in the FEC.

\noindent \textbf{7) Threshold-Setting System}

The threshold-setting system for the readout system consists of the
threshold-setting control module in the VME crate, and a
threshold-setting generator in the FEC.

\subsubsection{Water Cherenkov Modules}
\label{sssec:wcm}

As a part of the preliminary R\&D for a long baseline neutrino
oscillation experiment, the novel idea of a water Cherenkov
calorimeter made of water tanks was investigated~\cite{mu_wcc}. A
water tank prototype made of PVC with dimensions
1$\times$1$\times$13~m$^3$ was built.
The inner wall of the tank is covered by Tyvek film 1070D from
DuPont. At each end of the tank is a Winston cone that can collect
parallel light at its focal point, where an 8-in photomultiplier is
situated. The Winston cone is again made of PVC, covered by aluminum
film with a protective coating. Cherenkov light produced by
through-going charged particles is reflected by the Tyvek and the Al
film and collected by the photomultiplier.

\begin{figure}[htbp]
\begin{center}
\includegraphics[height=9cm,width=0.85\textwidth]{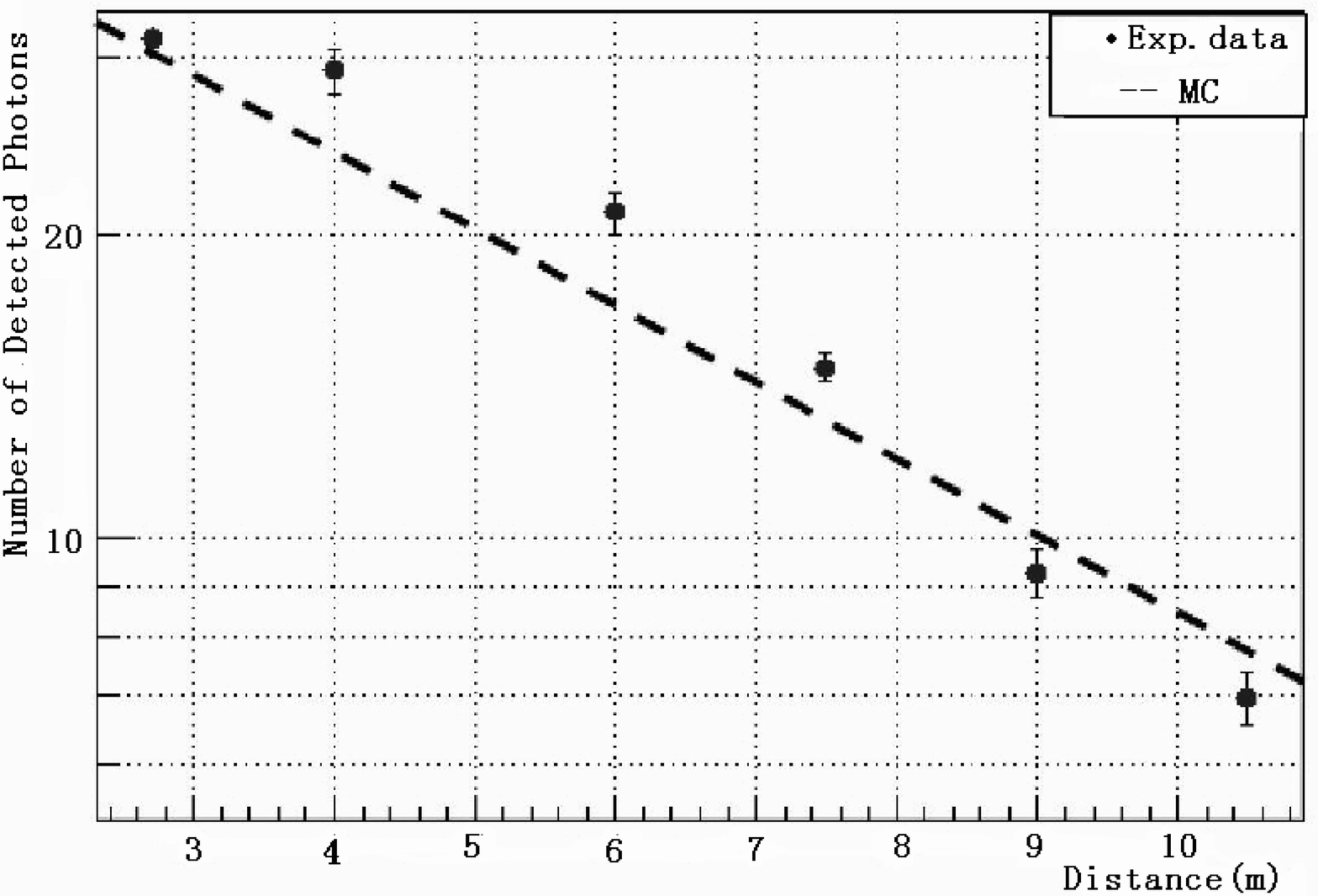}\\
\begin{minipage}[c]{10cm}
\caption{Position dependent response of the water tank to
cosmic-muons. $X$ is the distance from trigger counters to the PMT at
right. The line represents the Monte Carlo prediction with an
effective attenuation length of 5.79~m. The measured effective
attenuation length of the water tank is (5.74$\pm$0.29)~m.
\label{fig:wabs}}
\end{minipage}
\end{center}
\end{figure}

The light collected due to cosmic-muons is a function of the distance
from the point of incidence of the muon to the phototube. Such a
position dependent response of the tank is critical to its energy
resolution and pattern recognition capability. Typically it is
characterized by an exponential behavior of $e^{-x/\lambda}$, where
$x$ is the distance of the muon event to the phototube and $\lambda$
is the characteristic parameter, often called the ``effective attenuation
length''.  The characteristic parameter $\lambda$ depends on the water
transparency, the reflectivity of the Tyvek film, and the geometry of
the tank. Using trigger scintillation counters to define the muon
incident location, keeping the $y$ coordinate constant, 
the total light collected as a function of $x$ at
several locations was obtained as shown in Fig.~\ref{fig:wabs}.  It
can also be seen from Fig.~\ref{fig:wabs} that, for a through-going
muon entering the center of the tank, $\sim$20 photoelectrons are
collected by each PMT, corresponding to a statistical determination of
about 7\%/$\sqrt{E(GeV)}$.

As discussed in Section~\ref{subsect:water_shield} the Daya Bay
antineutrino detector modules are to be shielded from external
radiation (such as gamma-ray and cosmic-ray induced neutrons) by a
$\sim$2.5-m thick water buffer. One could use modules similar to the
above-mentioned Cherenkov units as the muon tracker
(Fig.~\ref{fig:overall}).
\begin{figure}[htbp]
\begin{center}
\includegraphics[width=13cm]{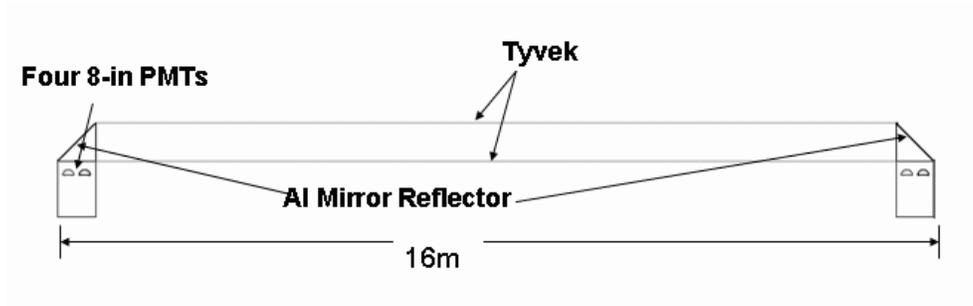}
\begin{minipage}[c]{10cm}
\caption{The geometry of a water Cherenkov module unit. Four 8-in PMTs are
installed in each end, with an Al mirror.}
\label{fig:modules}
\end{minipage}
\end{center}
\end{figure}
Such modules have many advantages: (1) very good cosmic-ray detection
efficiency (theoretically, the efficiency is close to 100\%); (2)
insensitivity to the natural radioactivity of Daya Bay's rock; (3)
very modest requirements on the support system because the whole
detector is immersed in the water; (4) low cost because water is used
as the medium.

The geometry of the water Cherenkov units located at the four sides of
water pool is shown in Fig~\ref{fig:modules}. We use four PMTs at each
end to decrease the risk of failure of a unit if one PMT dies. The
unique geometry of two ends reduces the optically dead region.
Similar modules would be placed at the bottom of the pool.  

It is possible that instead of a Tyvek lining, we may be able to rely
on the reflectivity of TiO2-loaded PVC, of the sort being studied for
the NOvA experiment at Fermilab.  R\&D on this possibility is in
progress.

A modified version of the LED system discussed in
Sec.~\ref{ssec:cal_LED} would be used for gain calibration of the
WCMs.

To cover the Daya Bay and Ling Ao Near Sites and the Far Site will
require 108 16~m-long units and 36 10~m-long units, with a 
total of 1152 8'' PMTs.  The WCMs will not be water-tight and will
share the purification system of the water shield.

\subsubsubsection{Water Cherenkov Module Performance}
\label{ssssec:muon_wm_performance}

A GEANT4 simulation tuned to match the performance of the prototype
described above was adapted to the longer modules proposed for Daya
Bay.  
The
optical parameters come from our previous MC simulation program of
the prototype.  Preliminary MC results are shown in
Table~\ref{table:wc}.  These show that adequate signal can be obtained
from both ends of these modules for muons at any point along them.
\begin{table}[htbp]
\begin{center}
\begin{minipage}[c]{10cm}
\caption{Number of photoelectrons detected in the 16~m-long water Cherenkov 
module with different incident positions of (vertical) muons from the Monte 
Carlo simulation.  These modules have four PMTs on each end.
\label{table:wc}}
\end{minipage}
\begin{tabular}{|l||l|l|l|l|l|l|l|} \hline
{}&-7m&-5m&-2m&0&2m&5m&7m   \\ \hline\hline
Left&9.8&23.1.&52.8&86.1&149.5&343.8&1044.9 \\ \hline
Right&985.3&340.8&152,7&88.6&51.3&23.0&10.0 \\ \hline
\end{tabular}
\end{center}\
\end{table}

Position resolution in the direction transverse to the axis of the module
will be given by its size: $\sigma_x = 100{\rm cm}/\sqrt{12} =$ 29~cm.  In
the other dimension, 2~ns 
resolution on the end-to-end timing will contribute
$\sim$32~cm 
to the position uncertainty.  We can also use the comparison of pulse
height from the two ends for determining the position.  Including both
methods we expect a resolution comparable or better than that in the
coordinate determined by the module cross-section.  For muons that
traverse the water pool and hit modules on either side,
these resolutions yield uncertainties of $\sim$21~cm 
on the position at the center of the trajectory.

Accidental rates in this system, which is essentially blind to rock
radioactivity, are expected to be negligibly small.

From Table~\ref{table:wc}, we see it is desirable to measure signals
that are $\sim$1 
photoelectron to those that are several hundred
photoelectrons.  Thus the same electronics discussed in
Sect.~\ref{sssec:ws_PMT_elect} are appropriate for use here.

\subsubsection{Plastic Scintillator Strips}
\label{ssssec:muon_strips}

Plastic scintillator strips serve as a backup option for both the top
and in-water tracking systems.  For both purposes we propose to use
the extruded plastic scintillator strip technology developed by MINOS,
OPERA and other previous experiments.  The parameters of this system
are shown in Table ~\ref{table:strips}.
\begin{table}[h]
\begin{center}
\begin{tabular}{lll}\hline
Name & value & unit \\ \hline
number of strips & 5304 & \\
length of strip & 5.25 & m \\
width of strip & 0.2 & m \\
thickness of strip & 1 & cm \\
fibers/strip & 5 & \\
length of fiber & 7.25 & m\\
diameter of fiber & 1 & mm \\
strips/module & 6 & \\
modules (full/top only)& 884/295 & \\
phototubes (full/top only)& 1768/530 & \\ \hline
\end{tabular}
\caption{Parameters of scintillator strip detectors
\label{table:strips}}
\end{center}
\end{table}

If the scintillators are used for the entire muon tracker system they
will be set back from the walls and floor of the pool by 50~cm to allow
attenuation of the gammas from rock radioactivity.  For a similar
reason they will be mounted 50~cm below the top of the pool.  There
will be two orthogonal layers on each wall, the floor and the top.  In
the case where only the top is to be covered by scintillator strips,
they would be arrayed in the manner described above for the RPCs, {\it
i.e.} there would be three layers mounted on the sliding roof.  In
this case a triple coincidence could be demanded if made necessary by
the random rates.

\subsubsubsection{Scintillator Strip Design}
\label{ssssec:muon_scint}

\begin{figure}[tb]
\centerline{\includegraphics[width=0.8\textwidth]{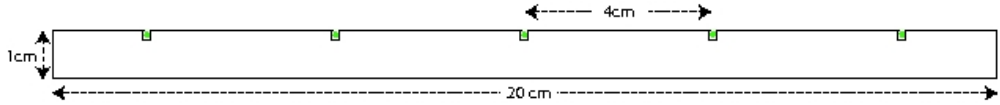}}
\caption{\label{fig:cntr} Cross-section of a single scintillator strip.}
\end{figure}

Almost all the scintillators will be of the same type:
5.25~m$\times$0.2~m$\times$0.01~m extruded polystyrene, co-extruded
with a coating of TiO2-doped PVC.  Five 1~mm Kuraray Y-11(200) S-type
wavelength-shifting fibers will be glued into 2~mm deep $\times$
1.6~mm wide grooves in the plastic using optical glue~\cite{glue}.
Six such scintillators will be placed in a single frame and read out
as one 1.2~m-wide unit. Figure~\ref{fig:cntr} shows the cross section of
one scintillator.

\subsubsubsection{Scintillator Strip Photoreadout}
\label{ssssec:muon_PMT}

A 1$\frac{1}{8}$-inch photomultiplier tube such as a Hamamatsu R6095 or
Electron Tubes 9128B will be used to read out 30 fibers on each end of
the six-scintillator module.  The PMTs will be run at positive HV,
via a system similar to that discussed in Sect.~\ref{ssec:det_HV}.
Calibration will be via thin-film ${}^{241}$Am sources placed near the
ends of the scintillators.  The sources provide $\sim$400~Hz of
$\sim$0.5~MeV signals.

\subsubsubsection{Counter Housing and Support}
\label{ssssec:muon_housing}

\begin{figure}
\noindent
\begin{minipage}[thb]{0.46\linewidth}
\centering\includegraphics[width=\linewidth]{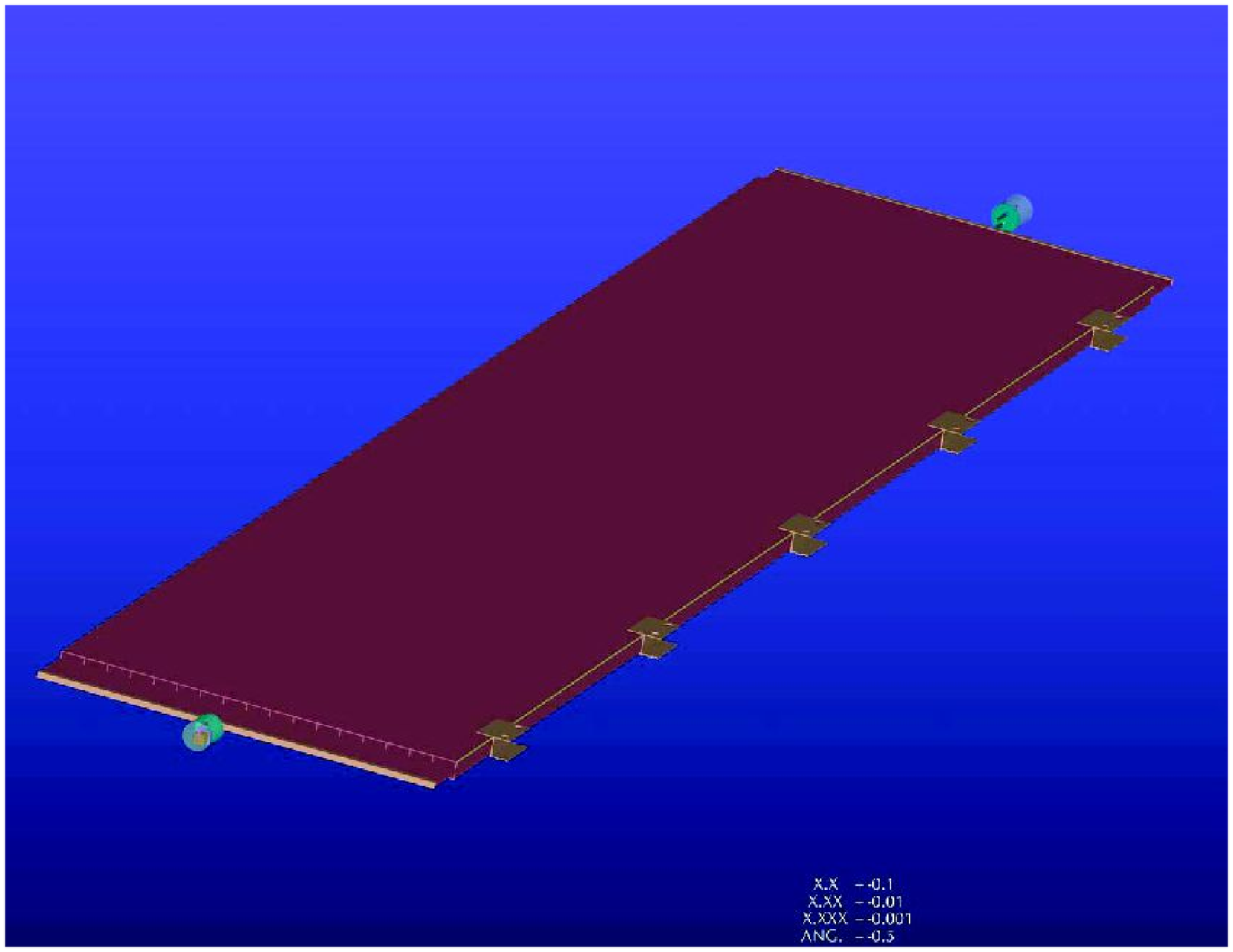}
  \caption{Extruded container for six-scintillator module}.
    \label{fig:housing}
\end{minipage}\hfill
\begin{minipage}[thb]{0.46\linewidth}
 \centering\includegraphics[width=\linewidth]{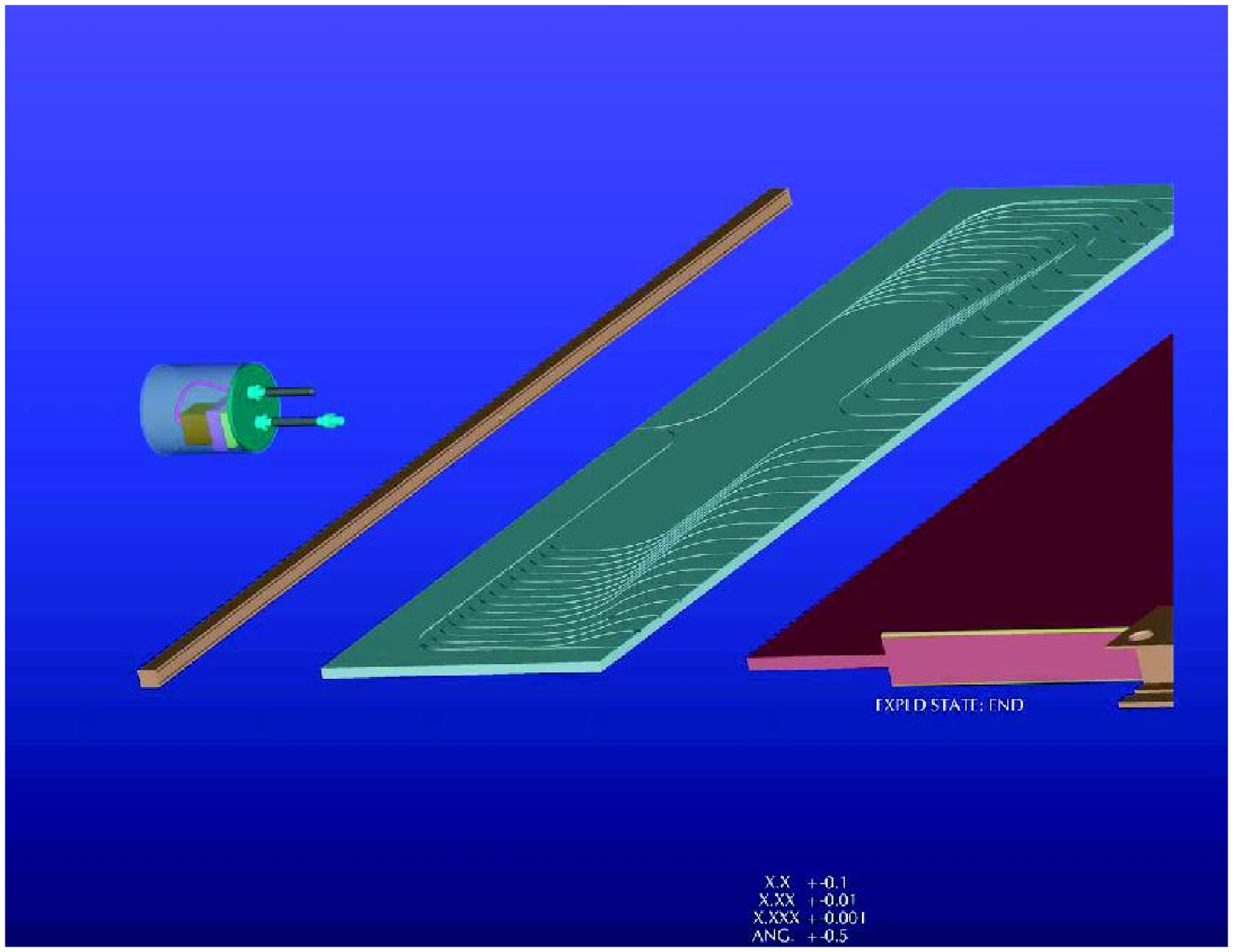}
  \caption{Exploded view of the end of the scintillator housing module showing routing of 
fibers, PMT containment, and other details.}
    \label{fig:routing}
\end{minipage}
\end{figure}

Above the water, the counters will be mounted on a simple system of
strongbacks supported by the sliding roof.  In the water, the
requirements for deployment are much more demanding.  The six
scintillator strips will be housed in an RPVC extruded box, shown in
Fig.~\ref{fig:housing}.  The box ends are closed by custom manifolds
that contain the fiber ends which are dressed to have an equal length
of $\sim$90~cm.  The fibers will be routed through a molded cookie,
gathered into single bundle and conducted in a PVC pipe through the
water into a separate small enclosure containing the PMT/base
assembly.  Figure~\ref{fig:routing} shows the module end, routing
cookie and PMT containment.

\begin{figure}
\noindent
\begin{minipage}[thb]{0.46\linewidth}
\centering\includegraphics[width=\linewidth]{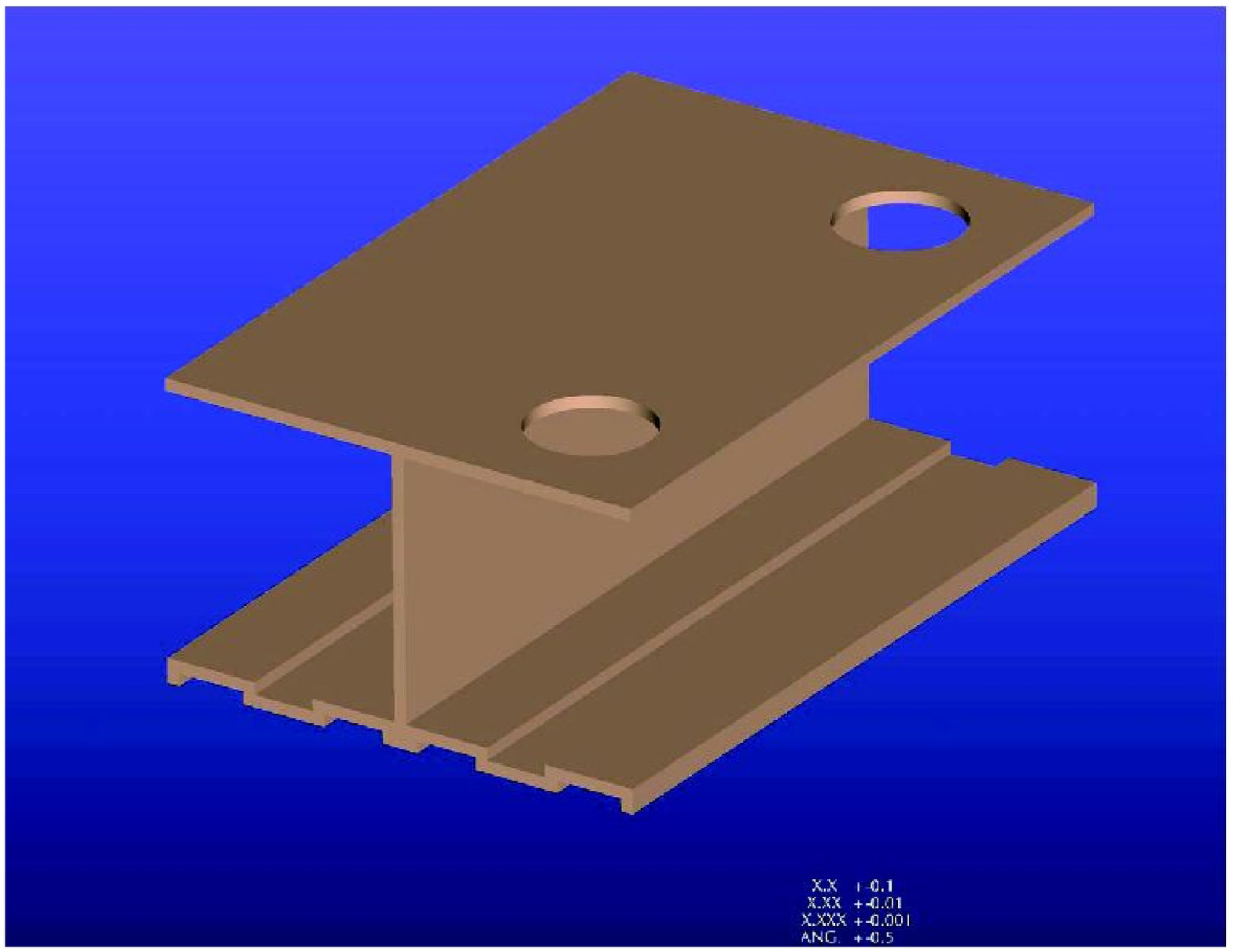}
  \caption{H-clip to hold the scintillator housing.}
    \label{fig:hclip}
\end{minipage}\hfill
\begin{minipage}[thb]{0.46\linewidth}
 \centering\includegraphics[width=\linewidth]{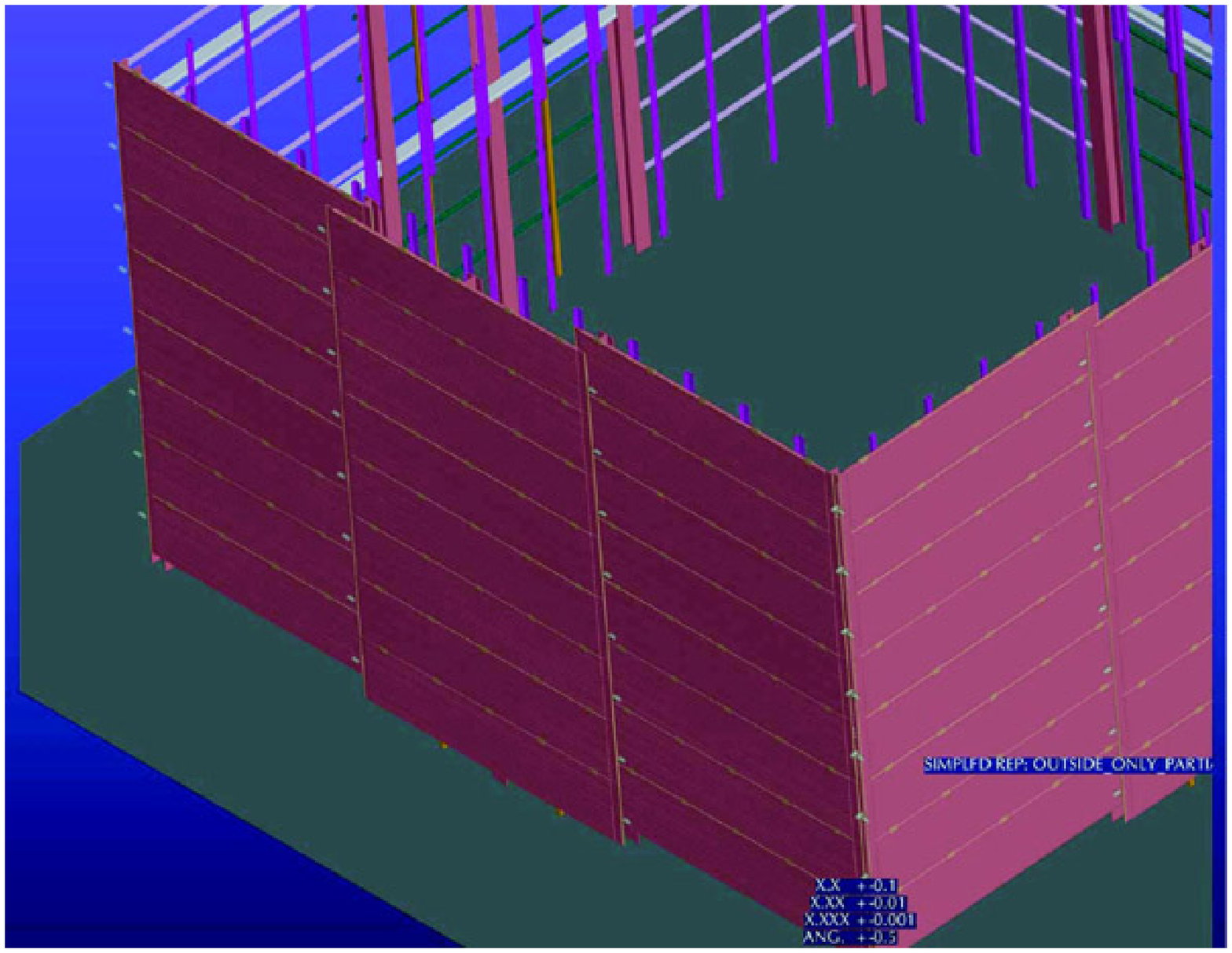}
  \caption{Side walls of the scintillator strip system partially assembled.}
    \label{fig:partial}
\end{minipage}
\end{figure}

The scintillator housings will be supported by a steel frame in a manner similar
to the H-clip technique used by MINOS~\cite{minos_scint}, although our version,
shown in Fig.~\ref{fig:hclip}, will be made of RPVC.  It will be glued onto the
module housing and fixed to the frame with two hole-drilling screws or blind rivets.
Figure~\ref{fig:partial} shows the support scheme for the side walls of the
scintillator system.

\subsubsection{Scintillator Strip Performance}
\label{ssssec:muon_strip_performance}

\begin{figure}[h]
 \includegraphics[height=.49\linewidth, width=0.85\textwidth]{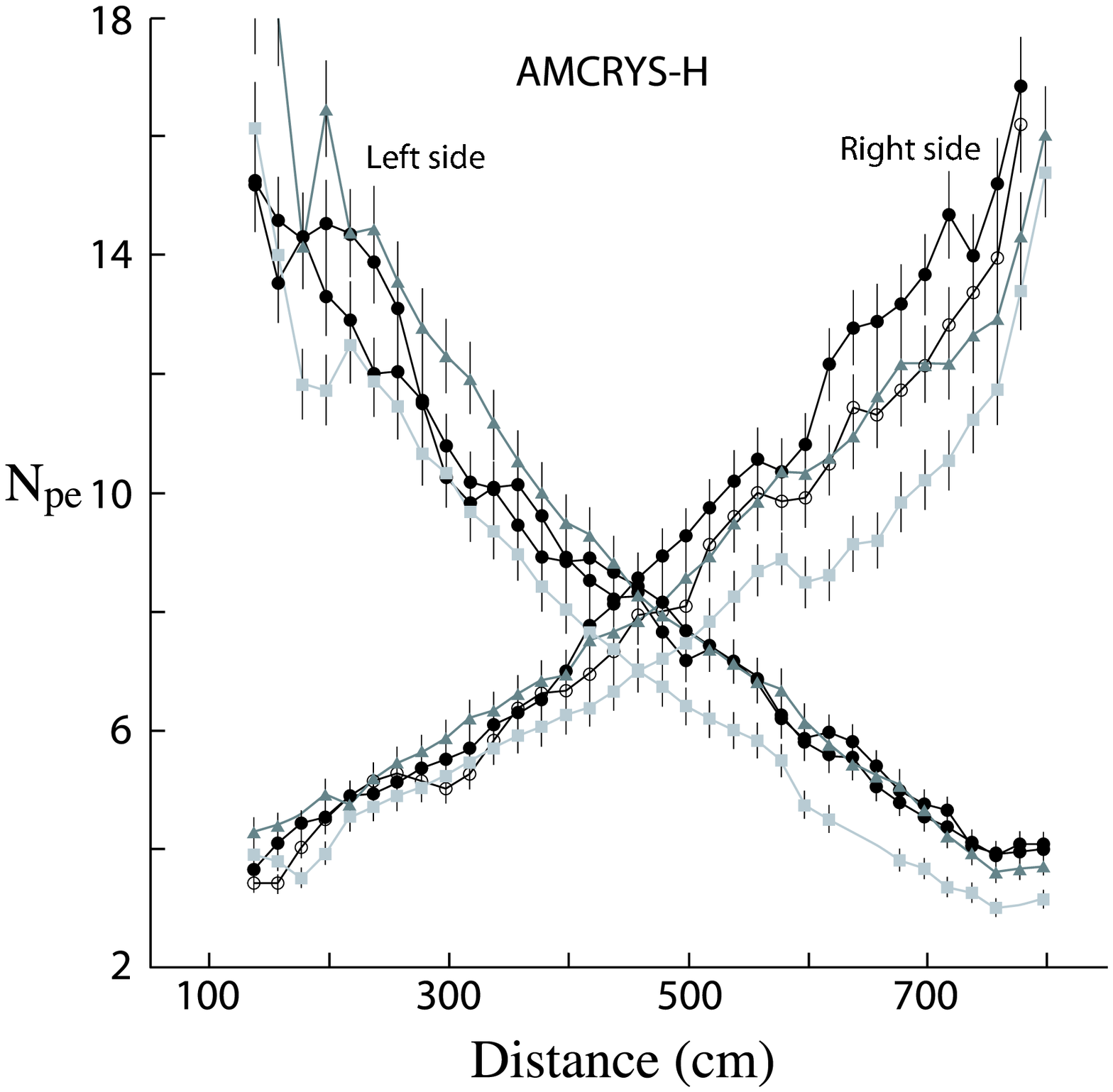}
  \caption{Number of photoelectrons detected on each side of several
AMCRYS-H plastic scintillator strips versus the distance to the photomultipliers
(from Dracos {\it et al.}).}
    \label{fig:OPERA}
\end{figure}

We base our expectation of performance on that of the prototype OPERA
target tracker scintillators~\cite{mu_OPERA}.  Figure~\ref{fig:OPERA} shows
the yield of photoelectrons versus distance to the photomultiplier
tubes.  
Note that our counters are only a little longer than 5~m, a
point at which the OPERA strips yield about 6~p.e.

The OPERA strips are 26~mm wide by 10.6~mm thick.  Our strips are 200~mm
wide by 10~mm thick.  The MINOS GEANT3 Monte Carlo was adapted to compare
the two cases.  For collection into the wavelength-shifting fibers,
the fraction of OPERA performance for 4, 5, and 6 fibers per 20~cm is
0.74, 0.89, and 1.02 respectively.  OPERA uses Hamamatsu H7546-M64 PMTs, which have a
photocathode efficiency about 80\% as high as either of the single-anode tube we
are planning to use.  Thus any of the 4--6 fiber cases should achieve
performance similar to that of OPERA.  For estimation purposes we
choose 5 fibers, which nominally should give 1.15 times OPERA
performance in our system.  The single photoelectron pulse height
distribution will reduce the effective number of photoelectrons by
a factor of (1 + the variance of the distribution).  With PMTs of
the type discussed, this will result in an inefficiency of $\sim$0.6\%
in the worst case (hit at one end of the counter).  An upper limit
on the position resolution is given by the granularity of the counters:
$\sigma_x = 120 {\rm cm}/\sqrt{12} \approx 35\, {\rm cm}$.  For a muon that
hits two sides of the pool, the resolution on the position at the center
of its trajectory through the pool will be $\sim$ 25~cm.  End-to-end
timing and pulse height are expected to improve this.  A timing resolution
of 1~ns will contribute $\sim$15~cm to the resolution along the counter
and $\sim$11~cm to the resolution at the center of the trajectory
for through-going muons.

Plastic scintillators are sensitive to the ambient radioactivity from
rock.  Tests of these rates were carried out with a scintillator
telescope in the Aberdeen Tunnel in Hong Kong~\cite{Aberdeen}, which
has similar granite to that of Daya Bay.  These indicate that the
true coincidence rate of two 1cm layers at a threshold of 0.5~MeV
would be $\sim$7~Hz/m$^2$.  For the relevant active area of the top of the Far
Hall water pool, this gives an overall rate of $\sim$1800~Hz. 
If the top scintillator array alone
were used as a 200~$\mu$s veto, it would give an unacceptable
random veto deadtime.  However, for through-going muons that could
be required to register as well on the side or bottom of the pool, the
random veto rate would be reduced to a negligible level
(with the shielding from 50~cm of water, the coincidence
rate in scintillators in the pool would be expected to be only 0.7~Hz/m$^2$, 
so 180~Hz 
on the bottom which is the worst case).
Since the background from stopping muons is extremely small, scintillators
seem acceptable in either or both roles.

\subsubsection{Scintillator Strip Front-End Electronics} 
\label{sssect:muon_strip_readout}

Once again, the electronics and readout discussed in
Sect.~\ref{sssec:ws_PMT_elect} would be adequate for this system.
However since it is not necessary to measure energies above a few MeV,
a smaller dynamic range would be acceptable.  Whether it is worth it
to develop separate electronics for this case is under study.  In any
case the readout would be similar to that discussed in
Sect.~\ref{ssec:det_fee}.

\newpage
\renewcommand{\thesection}{\arabic{section}}
\setcounter{figure}{0}
\setcounter{table}{0}
\setcounter{footnote}{0}

\section{Trigger and Data Acquisition System}
\label{sec:trig}

The trigger event selection and estimated rates are presented, along
with the timing synchronization between all of the electronics
elements. The processing of the trigger data from the front-end
modules through data storage is discussed, along with the detector
control system.

\subsection{The Trigger System}
\label{sssec:trig_trig}

The trigger system of the Daya Bay experiment makes trigger decisions
for the antineutrino and muon detectors to select neutrino-like
events, muon-related events, periodic trigger events and calibration
trigger events.  The following sections will describe the requirements
and technical baseline for the trigger system.

\subsubsection{Requirements}
\label{sssec:trig_req}

The signature of a neutrino interaction in the Daya Bay antineutrino
detectors is a prompt positron with a minimum energy of 1.022~MeV plus
a delayed neutron.  About 90\% of the neutrons are captured on
Gadolinium, giving rise to an 8~MeV gamma cascade with a capture time
of 28~$\mu$s. The main backgrounds to the signal in the antineutrino
detectors are fast neutrons produced by cosmic muon interactions in the
rock, $^{8}$He/$^{9}$Li, which are also produced by cosmic muons and
accidental coincidences between natural radioactivity and neutrons produced
by cosmic muons. All three major backgrounds are related to
cosmic muons. The following are the main trigger requirements imposed
by the physics goals of the Daya Bay experiment:

\begin{enumerate} 
\item 
{\bf Energy threshold:} The trigger is required to independently
 trigger on both the prompt positron signal of 1.022~MeV and the
 delayed neutron capture event with a photon cascade of $\approx$8~MeV 
with very high efficiency. The threshold level of the trigger is
 set at 0.7~MeV. This level corresponds to the minimum visible
 positron energy adjusted for a $3\sigma$ energy resolution effect.
 This low threshold requirement fulfils two trigger goals. For the
 neutrino signal, it allows the DAQ to record all prompt positron
 signals produced from the neutrino interactions, enabling a complete
 energy spectrum analysis that increases the sensitivity to
 $\theta_{13}$. For background, it allows the DAQ to register
 enough uncorrelated background events due to either PMT dark noise or
 low energy natural radioactivity to enable a detailed analysis of
 backgrounds offline.

\item
{\bf Trigger efficiency:} In the early stages of the experiment, the
trigger efficiency is required to be as high as possible for signal
and background, provided that the event rate is still acceptable and
will not introduce any dead time. After an accurate characterization
of all the backgrounds present has been achieved, the trigger system
can then be modified to have more powerful background rejection
without any efficiency loss for the signal. To measure the efficiency
variation, the system should provide a random periodic trigger with no
requirement on the energy threshold at trigger level. A precise
spectrum analysis also requires an energy-independent trigger
efficiency for the whole signal energy region.

\item 
{\bf Time stamp:} Since neutrino events are constructed offline from
the time correlation between the prompt positron signal and the
neutron capture signal, each front-end (FEE), DAQ and trigger unit
must be able to independently time-stamp events with an accuracy
better than 1~$\mu$s. The trigger boards should provide an independent
local system clock and a global time-stamp to all the DAQ and FEE
readout boards in the same crate. The trigger boards in each DAQ crate
will receive timing signals from a global GPS based master clock
system as described in Section~\ref{ssec:timing}. Events recorded by
the antineutrino detectors and muon systems can thus be accurately
associated in time offline using the time-stamp.

\item 
{\bf Flexibility:} The system must be able to easily implement 
various trigger algorithms using the same basic trigger board design 
for different purposes such as
\begin{enumerate}

\item Using different energy thresholds to adapt to the possible aging 
effect of liquid scintillator, or for triggering on calibration source
events which have lower energy signatures.

\item Using different hit multiplicities to increase the rejection power 
due to the uncorrelated low energy background and for special
calibration triggers.

\item Implementing different pattern recognition for
triggering on muon signals in the different muon systems.

\item Using an OR of the trigger decision of different trigger algorithms to
provide a cross-check and cross-calibration of the different
algorithms as well as a redundancy to achieve a high trigger
efficiency.

\end{enumerate}
\item
{\bf Independence:} Separate trigger system modules should be used for
each of the antineutrino detectors, and the muon systems. This is to
reduce the possibility of introducing correlations between triggers
from different detector systems caused by a common hardware failure.

\end{enumerate}

\subsubsection {The Antineutrino Detector Trigger System}
Neutrino interactions inside a detector module deposit an energy
signature that is converted to optical photons which are then detected by a
number of the PMTs mounted on the inside of the detector module. Two
different types of triggers can be devised to observe this
interaction:

\begin{enumerate}
\item An energy sum trigger.
\item A multiplicity trigger. 
\end{enumerate}

In addition to neutrino interaction triggers, the antineutrino detector trigger
system needs to implement several other types of triggers for calibration
and monitoring: \\

\begin{enumerate}
\setcounter{enumi}{2}
\item Calibration triggers of which there are several types:

\begin{enumerate}
\item Triggers generated by the LED pulsing system that
        routinely monitors PMT gains and timing.
\item Triggers generated by the light sources periodically
        lowered into the detector volume to monitor spatial uniformity
        of the detector response and the light attenuation.
\item Specialty energy and multiplicity triggers used to test
        detector response using radioactive sources 
\end{enumerate}

\item A periodic trigger to monitor detector stability and random backgrounds. 
\item An energy sum and/or multiplicity trigger (with looser threshold
   and multiplicity requirements) generated in individual
   antineutrino detector modules which is initiated by a delay trigger from
   the muon system. This trigger records events to study muon induced
   backgrounds. This trigger should be able to operate in both tag
   and veto modes. 
\end{enumerate}

\begin{figure}[!hbt]
\begin{center}
 \includegraphics[width=0.6\textwidth]{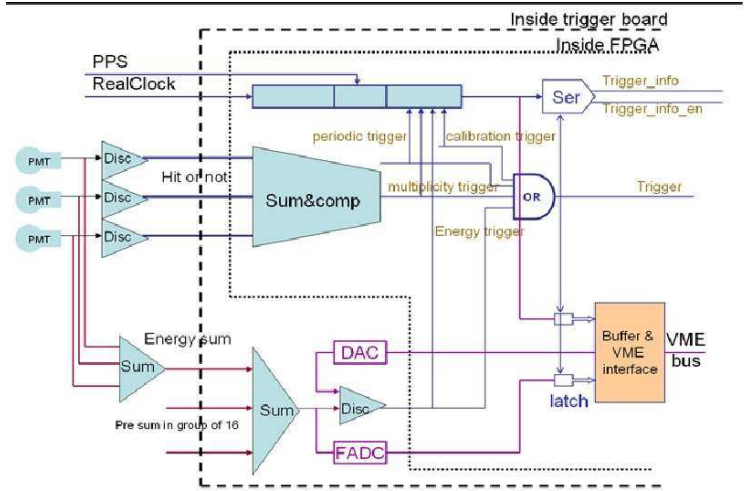}
\caption{A simplified trigger scheme.}
\label{fig:TRG_FIG1}
\end{center}
\end{figure}

A VME module with on-board Field Programmable Gate Arrays (FPGA)s is
used to implement the antineutrino detector trigger scheme outlined in
Fig.~\ref{fig:TRG_FIG1} based on experiences gained at the Palo
Verde~\cite{TRG_REF1} and KamLAND experiments.
We use an OR of both an energy sum and a multiplicity trigger to
signal the presence of neutrino interactions in the antineutrino
detector. These two triggers provide a cross-check and
cross-calibration of each other.

The multiplicity trigger is implemented with FPGAs which can perform
complicated pattern recognition in a very short time. FPGAs are
flexible and can be easily reprogrammed should trigger conditions
change. In addition, different pattern recognition software can be
downloaded remotely during special calibration runs, such as might be
needed for detector calibration with sources.  The signal from
different PMTs is compared with the threshold on on-board
discriminators in the front-end readout cards as described in
Section~\ref{ssec:det_fee}. The output of the PMT discriminators are
input into the trigger module FPGA which performs clustering and
pattern recognition and generates the multiplicity trigger
decision. The dark current rate of the low activity antineutrino
detector PMTs is typically around 5~kHz at 15$^\circ$~C. For a
detector with $N$ total PMTs, a dark current rate of $f$~Hz, and an
integration time of $\tau$~ns, the trigger rate $R$ given a
multiplicity threshold $m$ is
\begin{equation}
 R = \frac{1}{\tau} \sum_{i=m}^N i C_N^i (f\tau)^i(1-f\tau)^{N-i}, \ \ \
f\tau << 1
\label{eqn:TRG_EQN1}
\end{equation}
where $C_N^i$ are the binomial coefficients.

\begin{figure}[!hbt]
\begin{center}
 \includegraphics[width=0.6\textwidth]{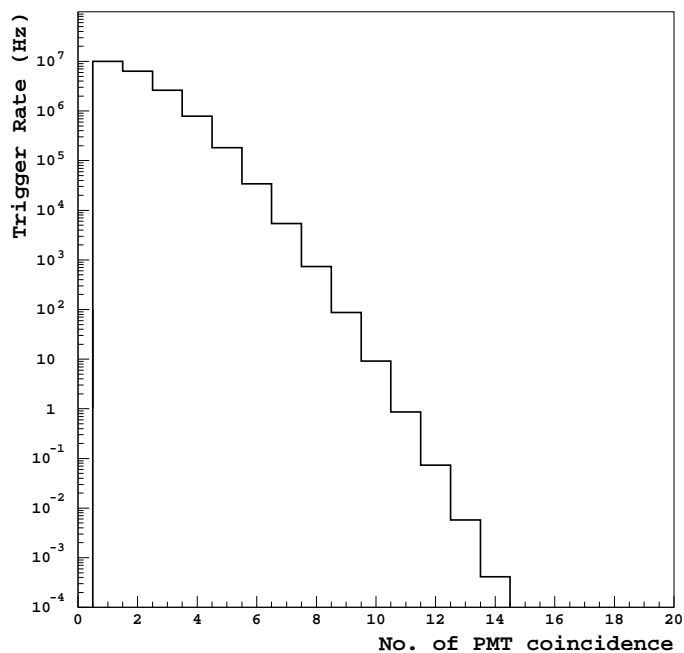}
\caption{Calculated trigger rates caused by PMT dark current as a
function of the multiplicity threshold. The maximum number of PMTs is
200, the PMT dark current rate used is 50k with a 100~ns integration
window.
\label{fig:TRG_FIG4}}
\end{center}
\end{figure}

To be conservative, we assume a PMT dark current rate of 50~kHz when
estimating the dark current event rate from the multiplicity
trigger. For the multiplicity trigger, an integration window of 100~ns
will be used for the central detector PMTs.  The dark current
rate calculated using Eq.~\ref{eqn:TRG_EQN1} as a function of
the number of PMT coincidences is shown in Fig.~\ref{fig:TRG_FIG4}.
At a multiplicity of 10 PMTs, the total trigger rate would be of order
1~Hz with a 100~ns integration window.

The energy sum trigger is the sum of charges from all PMTs obtained
from the front-end readout boards with a 100~ns integrator and
discriminator. The threshold of the discriminator is generated with a
programmable DAC which can be set via the VME backplane bus. The
energy sum is digitized using a 200~MHz flash ADC (FADC) on the
trigger module. We plan to have an energy trigger threshold of 0.7~MeV
or less to be compatible with the positron energy of 1.022~MeV within
3$\sigma$ of the energy resolution. At such low energy thresholds, the
trigger will be dominated by two types of background: One is natural
radioactivity originating in the surrounding environment which is less
than 50~Hz as shown in a Monte Carlo simulation in
Section~\ref{sssec:sys_radioactivity}, and the other is from cosmic
muons (negligible at the far site). At this threshold, the energy sum
trigger rate from the PMT dark current with a 100~ns integration
window is negligible.

Tagging antineutrino interactions in the detector requires measuring
the time-correlation between different trigger events.  The
time-correlation will be performed offline, therefore each triggered
event needs to be individually timestamped with an accuracy of order
of microseconds or better. It may become necessary to have a
correlated event trigger in the case the background rate is too high.

A periodic trigger to monitor the PMT dark-current, the cosmic ray
background, and detector stability will be included. 

\subsubsection {The Muon Trigger System}
\label{ssec:mu_trig}
The muon system will utilize three separate trigger and DAQ VME
crates, one for each of the muon detector systems: The water
Cherenkov detector, the RPC system and the muon tracker
system (scintillators or water trackers).

The presence of a muon which goes through the water Cherenkov detector
can be tagged with energy sum and multiplicity triggers using a
similar scheme and hardware modules as used for the antineutrino
detector. In addition, a more complicated pattern recognition scheme
using localized energy and multiplicity information may be used. The
trigger rate in the water Cherenkov detector is dominated by the
cosmic muon rate which is $<$15~Hz in the far hall and $<$300~Hz in
the near halls (see Table~\ref{tab:DAQ_TAB1}).  In addition to the
water pool Cherenkov detector trigger, muons will be
tagged by a system of RPCs and either water tracker modules or 
double layers of scintillator strips.

The FPGA logic used for the RPC and scintillator strip detectors forms
muon ``stubs'' from coincident hits in two overlapping layers of
scintillator or two out of three layers of RPC. Although the readout
electronics of RPC is very different from that of the PMT, the trigger
board can still be similar to the other trigger boards. As we
discussed before, each FEC of RPC readout electronics can provide a
fast OR signal of 16 channels for the trigger.  All the fast OR
signals will be fed into the trigger board for further decision by
FPGA chips.  The principal logic is to choose those events with hits
in two out of three layers within a time window of 20~ns in a
localized region of typically 0.25~m$^2$. Since the noise rate of a
double gap RPC is estimated to be about 1.6~kHz/m$^2$ (consistent with
twice the BES single gap chamber rates), the false trigger rate from
noise can then be controlled to be less than 50~Hz in such a
scheme. The coincidence rate in the RPC system due to radioactivity is
estimated to be 1~Hz/m$^2$ obtained from simulation results based on
measurements in the Aberdeen tunnel (see Section~\ref{sssec:muon_performance}). This
corresponds to a radioactivity trigger rate of about 360~Hz in the far
hall.

For the water tracker modules, we need three types of triggers:
\begin{enumerate}
\item an AND of the two ends with a threshold of approximately 
3~p.e. on each end.
\item  A prescaled single ended trigger with a lower threshold.
\item The energy sum of the two ends with a threshold of $\sim$ 20~p.e.
\end{enumerate}

The antineutrino detector trigger board can be used to implement the
trigger schemes for the water tracker. The fake trigger rates from
radioactivity in the water tracker modules is expected to be
negligible.

An alternative to the water tracker modules discussed above, two layers
of scintillator strips in the water pool can be used as described in
Section~\ref{sec:muon}. The 0.5~m of water between the water pool
walls and the scintillator strips provides some shielding from
radioactivity in the rock which generates a rate of 180~Hz of
background in the largest plane (bottom of the far detector).  The
scintillator PMTs noise rate is $<$2~kHz at 15$^\circ$~C. Requiring a
coincidence of two hits in overlapping layers with a 100~ns integration
window reduces the fake trigger rate from the scintillator strip PMT
noise to a negligible level. In principal, the same trigger module
design can be used for both RPCs and scintillator strips with
different FPGA software to handle the stub formation in the different
geometries.

The global muon trigger decision is an OR of the three muon detector
trigger systems: RPC, water  Cherenkov and muon tracker. The muon
trigger decision may be used to launch a higher level delay trigger
looking for activity inside the antineutrino detector at lower thresholds
and/or multiplicities for background studies.

\subsection{The Timing System}
\label{ssec:timing}

The design of the trigger and DAQ system is such that each
antineutrino detector and muon detector system has independent DAQ and
trigger modules. In this design it is necessary to synchronize the
data from the individual DAQ and trigger systems offline. This is
particularly important for tagging and understanding the backgrounds
from cosmic muons. A single cosmic muon candidate will be
reconstructed offline from data originating in three independent
systems: the water Cherenkov pool, muon tracker and RPC
tracker. Cosmic muon candidates reconstructed in the muon detector
systems have to then be time correlated with activity in the
antineutrino detector to study muon induced backgrounds. To this end,
the Daya Bay timing system is required to provide a global time
reference to the entire experiment, including the trigger, DAQ, and
front-end boards for each module (LS, water Cherenkov, and tracker) at
each site. By providing accurate time-stamps to all components various
systematic problems can easily be diagnosed.  For instance, common
trigger bias, firmware failure, and dead time can all be tracked by
looking for time-stamp disagreements in the data output from each
component. Furthermore, by having multiple sites synchronized to the
same time reference, it will be possible to identify physical
phenomena such as supernova bursts or large cosmic-ray air showers.

The timing system can be conceptually divided into four subsystems: the
(central) master clock, the local (site) clock, the timing control
board, and the timing signal fanout.

\subsubsection{Timing Master Clock}

The global timing reference can easily be provided by a GPS (Global
Positioning System) receiver to provide a UTC (Universal Coordinated
Time) reference.  Commercially-available units are typically accurate
to better than 200~ns relative to UTC~\cite{TRG_REF2,TRG_REF3}.

\begin{figure}[!hbt]
 \begin{center}
 \includegraphics[width=12cm]{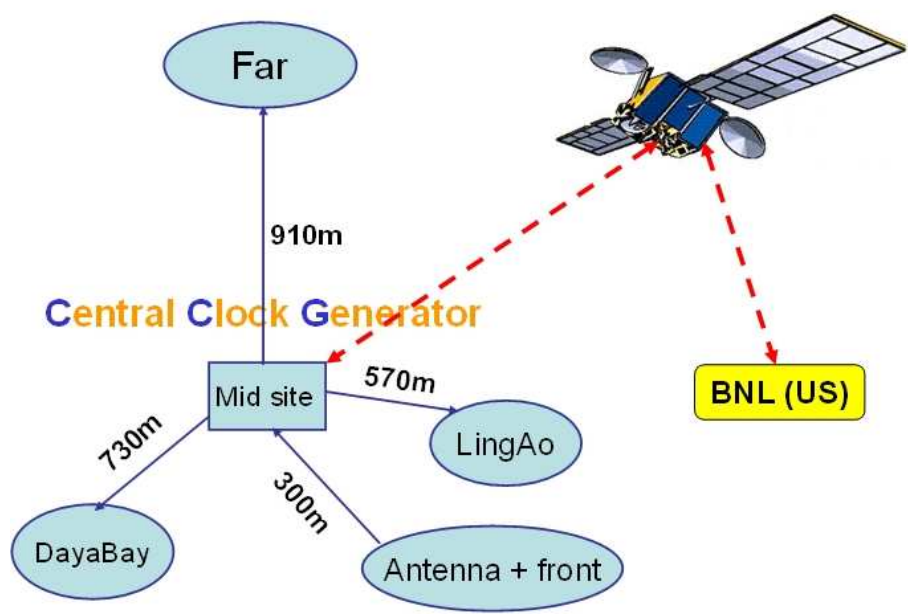}
 \caption{Schematic layout of the global clock.}
 \label{fig:TRG_FIG2}
 \end{center}
\end{figure}

This GPS receiver can be placed either at one of the detector sites
(most conveniently the mid hall) or in a surface control building. A
master clock generator will broadcast the time information to all
detector sites. If the master clock is located underground, the GPS
antenna may require an optical fiber connection to the surface, which
again is commercially available.  One such possibility is illustrated
in Fig.~\ref{fig:TRG_FIG2}

\begin{figure}[!hbt]
\begin{center}
\includegraphics[width=12cm]{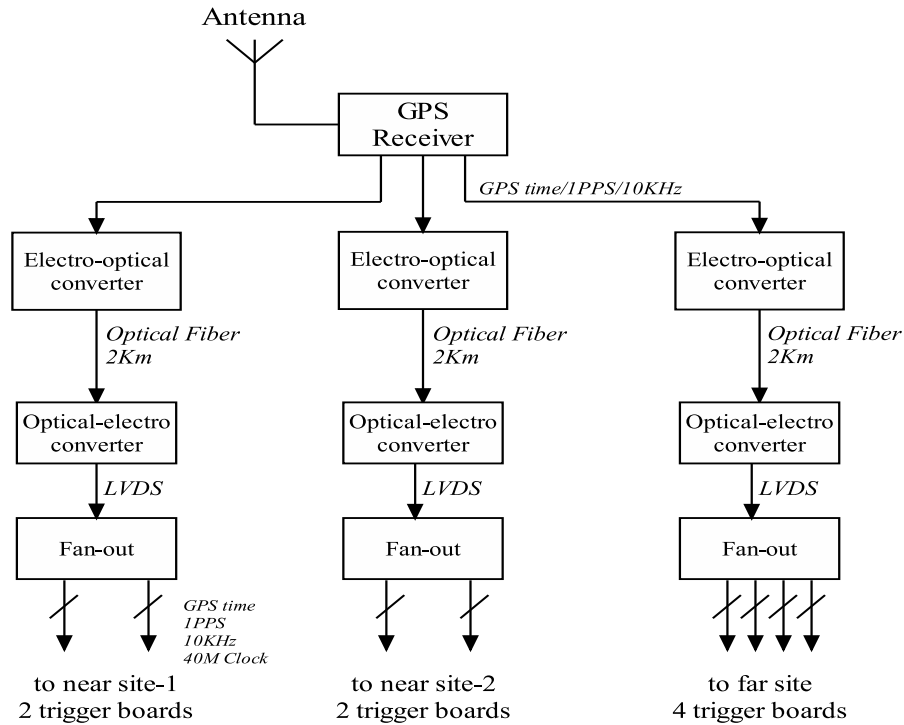}
\caption{Block diagram of the Daya Bay clock system.}
\label{fig:TRG_FIG3}
\end{center}
\end{figure}

The master clock will generate a time reference signal consisting of a
10~MHz clock signal, a PPS (Pulse Per Second) signal, and a date and
time.  These signals can be encoded onto a one-way fiber optic link to
be carried to each of the detector halls where they are then
fanned-out to individual trigger boards as shown in
Fig.~\ref{fig:TRG_FIG3}

Additionally, the GPS receiver will be used to synchronize a local
computer. This computer can then be used as a Tier-1 network time
protocol (NTP) peer for all experiment computers, in particular the
DAQ,

Each site will receive the signals from the master clock and use them
to synchronize a quartz crystal oscillator via a phase-locked
loop. This local clock can then be used as the time reference for that
site.

This method allows each site to operate independently of the master
clock during commissioning or in the case of hardware failure, but in
normal operation provides good time reference.  This clock could be
used to multiply the 10~MHz time reference to the 40~MHz and 100~MHz
required for the front ends. This clock will reproduce the PPS, and 10
(or 40/100)~MHz signals and supply them to the timing control board.

\subsubsection{Timing Control Board}

The timing control board will act to control the local clock
operations (i.e. to slave it to the master clock or let it run freely)
and to generate any timing signals required by the trigger, DAQ, or
front end that need to be synchronously delivered.  Typical examples
include buffer swap signals, run start/stop markers, and electronic
calibration triggers.  In addition, this board could be used to
generate pulses used by optical calibration sources.  This board would
be interfaced to the detector control computers.

\subsubsection{Timing Signal Fanout}

The signals from the timing control board need to be delivered to the
individual detector components: every FEE board, DAQ board, and trigger
unit.  This will allow each component to independently time-stamp
events at the level of 25~ns. 

This fanout system could work, for example, by encoding various
signals by encoding them on a serial bus, such as HOTLink. The trigger
board in each FEE and DAQ VME crate could then receive the serial
signal and distribute it via the crate backplane.  The crate
backplanes will then carry the 40~MHz clock (100~MHz clock for RPCs),
the PPS signal (to reset the clock counters), and the other timing
signals (run start/stop marker, calibration, etc).

Individual components of the trigger, DAQ, and front end can employ
counters and latches to count seconds since start of run and clock
ticks since start of second. These will provide sufficient data to
assemble events and debug the output data streams.

\subsection{The Data Acquisition System}
\label{ssec:trig_daq}

The data acquisition (DAQ) system is used to:
\begin{enumerate}
\item Read data from the
front-end electronics.
\item Concatenate data fragments from all FEE readout into a complete
  event.
\item Perform fast online processing and event reconstruction for
  online monitoring and final trigger decisions.
\item Record event data on archival storage.
\end{enumerate} 

A brief review of the DAQ design requirements is followed by
a discussion of the system architecture, DAQ software, and
detector control and monitoring system.

\subsubsection{Requirements}
\label{sssec:trig_daqreq}

The Daya Bay DAQ system requirements are listed in Table~\ref{tab:DAQ_TAB1}.
\begin{table}[!htbp]
 \centering
 \begin{tabular}[h]{|r|r|r|r|r|r|r|r|} \hline
 &  & \multicolumn{3}{c|}{Trigger Rates (Hz)} & & & data rate\\ \cline{3-5}
Detector & Description & DB & LA & Far & Occ & Ch size & (kB/s) \\ \hline \hline
$\bar\nu$ module & cosmic-$\mu$ & $36\times 2$ & $22\times 2$ & $1.2\times 4$ & 100\% & $228 \times 64$ bits & 217 \\ \cline{2-5} \cline {8-8}
      & Rad. & $50.0\times 2$ & $50.0\times 2$ & $50.0\times 4$ &  & & 730  \\ \hline \hline
  RPC & Rad. \& Noise & 260 & 260 & 415 & 10\% & $5040/7560 \times 1$ bit & 72 \\ \cline{2-5} \cline {8-8}
      & cosmic-$\mu$ & 186 & 117 & 10.5 & & & 20 \\ \hline \hline
 Pool & cosmic-$\mu$ & 250 & 160 & 13.6 & 50\% & $252/340$ $\times 64$ bits & 437  \\ \hline \hline
$\mu$-tracker & cosmic-$\mu$ & 1390 & 819 & 57.8 & 100\% & $8 \times 64$ bits &  145 \\ \hline \hline \hline
site totals & (kB/s) & 683 & 500 & 436 & \multicolumn{2}{c|}{} &  1620 \\ \hline
  \end{tabular}
  \caption{Summary of data rate estimations.  kB/s = 1000 bytes per second.  The total data throughput
  rate for all 3 sites is estimated to be 1620~kB/s.  The trigger rate for the central detector has
substantial components from natural radioactivity and from muons.  The trigger rate in the
RPCs has, in addition, some trigger rate from noise. The trigger rate in the water pool
and $\mu$-tracker comes predominantly from muons.}
  \label{tab:DAQ_TAB1}
\end{table}

\begin{enumerate}
\item{\bf Architecture requirements:} The architecture requires separate DAQ
  systems for the three detector sites. Each antineutrino detector module
  will have an independent VME readout crate that contains the trigger
  and DAQ modules. In addition, the water  Cherenkov detector and muon
  tracking detectors will also have their own VME readout crates.  The
  trigger and DAQ for the antineutrino and muon detector modules are
  kept separate to minimize correlations between them. The DAQ
  run-control is designed to be operated both locally in the detector
  hall during commissioning and remotely in the control room. In
  addition, run-control will enable independent operation of
  individual antineutrino and muon detector modules.

\item {\bf Event rates} 
The trigger event rates at the Daya Bay, Ling Ao and Far site from
various sources are summarized in Table~\ref{tab:DAQ_TAB1}. The rate
of cosmic muons coming through the top of a detector are calculated
using Table~\ref{tab.near-far}.  To turn this into a volumetric rate,
we use a MC simulation to calculate the ratio of muon rates entering
the top to all muons entering the detector's volume. The total rates
from cosmic muons in the different different systems are shown in
Table~\ref{tab:DAQ_TAB1}. At the far site, the trigger rates in the
central detectors are dominated by natural
radioactivity ($<$50~Hz/detector) and at the near sites both cosmic
and natural radioactivity contribute.

The trigger rate in the water Cerenkov pool is dominated by the cosmic
muon rate, the singles rate from PMT noise, gammas and fast neutron
backgrounds are negligible.

The RPC noise rates are taken from the BES chamber measurements in
Section~\ref{sssec:muon_performance}. We scale the BES noise rates by
a factor of 2 to account for the different geometry of the Daya Bay
RPC modules (3 double gap layers). This increases the coincidence rate
due to RPC noise by a factor of 4. The singles rates shown in
Table~\ref{tab:DAQ_TAB1} are the sum of the noise and natural
radioactivity rates in the RPC systems at the various sites.

For the purposes of calculating the overall data throughput for the
muon tracker modules, each module is treated as an independent
detector and the muon rate through each module is assumed to be its
``trigger rate''.  The occupancy is 100\% but only 8 channels are read
out.  In reality of course muons will tend to hit multiple modules in
one full-detector trigger but the entire detector will not be read
out.  In the end the two methods simply trade trigger rate for \# of
hit channels and the results should be the same.

While the trigger rate in the antineutrino detectors at each site is
of order a few 100~Hz, an OR of the three muon trigger systems could
produce a maximum trigger rate of $<$1~kHz. The Daya Bay trigger and
DAQ system will be designed to handle a maximum event rate of
1~kHz. In addition, to trigger on the correlated neutrino and fast
neutron signals in the antineutrino detector, the DAQ needs to be
able to acquire events that occur 1~$\mu$s or more apart.

\item{\bf  Bandwidth}

The maximum number of electronics channels for the antineutrino
detectors, water Cherenkov pool, and muon tracker PMTs at the far site
is estimated to be at most 2000 channels as shown in
Table~\ref{tab:DAQ_TAB2}.
\begin{table}[h!]
\begin{center}
\begin{tabular}{llc}
\hline
Detector Option	 &    Geometry	     & 	Approximate number of channels \\  \hline
\hline \hline
\multicolumn{3}{c}{PMT channels} \\ \hline \hline
     Scint tracker	 &    2 layers	     & \\
     strip module	 &    side/bottom    & \\
     1.2~m $\times$ 5.25~m		 &    water pool     & 	$\sim 530$\\			     
     OR		         & 		     & \\
     Water tracker       &    8 PMTs	     & 	$\sim 450$\\
     modules		 &      per module   & \\
     1~m $\times$ 16~m module	 &      side/bottom  & \\		    
                         & 		     & \\
     Water Cherenkov	 &    1 PMT/2~m$^2$ & \\
     pool	         &    4 sides/bottom & 	$\sim 350$ \\
                         & 		     & \\
     Antineutrino detector	 &    4 modules	     & 	896\\ 
                         & 		     & \\ \hline
     Total PMT channels  &	             & $\sim 1900$ \\
\hline \hline
\multicolumn{3}{c}{RPC channels} \\ \hline \hline
                         & 		     & \\
     RPC on top		 &     3 layers of   & \\
     of water pool	 &     double gap    & \\
     2~m $\times$ 2~m module	 &     modules	     & 	7560 \\
                         & 		     & \\ \hline
\end{tabular}
\end{center}
\caption{     Estimated number of readout channels from various detector
     systems  at the far site.}
\label{tab:DAQ_TAB2}
\end{table}
We assume that the largest data block
needed for each PMT channel is 64 bits or less, provided waveform
digitization is not used, the breakdown of the channel data block
could be as follows: \\
      Address : 12 bits \\ 
      Timing(TDC+local time): 32 bits \\ 
      FADC : 14 bits \\ 
For the RPC readout its 1bit/channel + header (12bits) + global
time-stamp (64bits) = 1~kBytes maximum.
     
Assuming zero suppression, and maximum occupancy numbers of 10\% for
the RPC system, 100\% for 1 out of the 4 antineutrino detectors, 10\%
for the water tracker and 50\% for the water Cherenkov (with
reflecting surfaces), we estimate the maximum event size at the far
will not exceed 10~kBytes/event including DAQ/Trigger header words and
global time-stamps. The event sizes at the near sites are smaller than
the far site due to a smaller number of channels.  The expected data
throughput from each site is estimated by combining the number of
readout channels with the trigger rates and occupancies as shown in
Table~\ref{tab:DAQ_TAB1}. The site totals in Table~\ref{tab:DAQ_TAB1}
do not include global header words, trigger words and time-stamps which
add a small overhead. Therefore, we estimate that the expected data
throughput rate is $<$1~MBytes/second/site. If waveform digitization
is used for the PMTs, this could increase the maximum desired data
throughput by an order of magnitude to $<$10~MBytes/second/site.

\item{\bf Dead-time:} The DAQ is required to have a negligible readout 
dead-time ($<$0.5\%). This requires fast online memory buffers that can 
hold multiple detector readout snapshots while the highest level DAQ CPUs
perform online processing and final trigger decisions and transfer to
permanent storage.  It may also require some low level pipelines at
the level of the PMT FADCs.

\end{enumerate}
\subsubsection{The DAQ System Architecture}
\label{sssec:trig_daqcons}

The main task of the DAQ system is to record antineutrino candidate
events observed in the antineutrino detectors. In order to understand
the background, other types of events are also recorded, such as
cosmic muon events, low energy radiative backgrounds... etc.
Therefore, the DAQ must record data from the antineutrino and muon
detectors (RPCs, water Cherenkov and Tracker), with precise timing
information. Offline analysis will use timing information between
continuous events in the antineutrino detector and in both the muon
and antineutrino detectors to select antineutrino events from
correlated signals or study the muon related background in the
antineutrino detectors.

\begin{figure}[h!]
 \begin{center}
 \includegraphics[width=12cm]{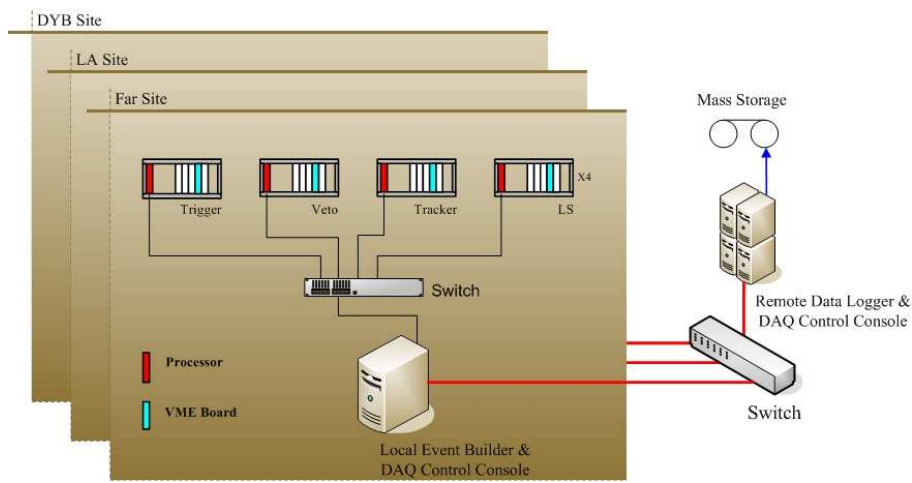}
 \caption{Block diagram of data acquisition system.}
 \label{fig:DAQ_FIG1}
 \end{center}
\end{figure}

The DAQ architecture design is a multi-level system
using advanced commercial computer and network technology as
shown in Fig.~\ref{fig:DAQ_FIG1}. 
There should be three sets of DAQ systems: one for each of the
three detector sites. The DAQ system levels shown in
Fig.~\ref{fig:DAQ_FIG1} are as follows:
\begin{enumerate}
\item {\bf VME front-ends:} The lowest level is the
VME based front-end readout system. Each VME crate is responsible for
one detector or muon system. Each module of the antineutrino detector
will have its own independent VME crate. Therefore, The lowest level
VME readout system of the far detector hall will consist of the
trigger boards for each system, the front-end readout boards from
three muon systems, and the four antineutrino detector readout
boards. All readout boards are expected to be 9U VME boards.

The Far and Near detector halls, will have the same DAQ architecture
but with different number of VME readout crates to accommodate the
different number of readout channels in the Far/Near halls.  Each VME
crate holds a VME system controller, some front-end readout (FEE)
modules and at least one trigger module which supplies the clock
signals via the VME backplane to the FEE modules.  The VME processor,
an embedded single board computer, is used to collect, preprocess, and
transfer data. The processor can read data from a FEE board via
D8/D16/D32/MBLT 64 transfer mode, allowing a transfer rate up to
80~MB/s per crate which is sufficient to meet the bandwidth
requirement. All readout crates of the entire DAQ system at a single
site are connected via a fast asynchronous Ethernet switch to a single
local event builder computer.

\item {\bf Event Builder and DAQ control:}
At each site an Event Builder computer collects the data from the
different VME crates for the different detectors and concatenates the
FEE readout to form single antineutrino or muon events.  The data
stream flow can work in two ways, depending on the requirements of
offline analysis. One scheme is to send muon events and antineutrino
events out into one data stream on the readout computer. Another
scheme is that each type of sub-event, muon events, or antineutrino
events, have a different data stream and will be recorded as separate
data files in permanent storage. The second scheme is simpler from a
DAQ design viewpoint and complies with the DAQ system design principal
of keeping each detector system completely independent for both
hardware and software. The Event Builder computer at each site also
allows for local operation and testing of the DAQ system.

\item {\bf Data Storage and Logging:}

Data from the Event Builder computer at each site are sent via fast
optical fiber link through a dedicated switch at a single surface
location where it is then transferred to local hard disk arrays. The
hard disk arrays act as a buffer to the remote data archival storage
or as a large data cache for possible further online processing.  Each
day will produce about 0.6~Terabyte of data that needs to be
archived. Although implementation of data logging has not yet been
finalized, there are two obvious options:
\begin{enumerate}

\item Set up a high bandwidth network link
between Daya Bay and the Chinese University of Hong Kong, China, and
distribute the data via the GRID (high bandwidth computing network and
data distribution applications for high energy physics
experiments). This is the preferable scheme.

\item Record the data locally on tape. This scheme
requires a higher level data filter to reduce data
throughput to a manageable level.
\end{enumerate}

Whichever option is realized, the local disk array should have the
capability to store a few days worth of data in the case of temporary
failures of the network link or the local tape storage.

\end{enumerate}

Since the DAQ system is required to be dead time free, each DAQ level
should have a data buffer capability to handle the random data
rate. In addition, both the VME bus and network switches should have
enough margin of data bandwidth to deal with the data throughput of
the experiment.

The DAQ control and monitoring systems should be able to run both
remotely from the surface control room computers and locally on the
Event Builder computer in each detector hall. The run control design
should be configurable allowing it to run remotely for data taking
from all systems and locally. Run control should allow both global
operation of all detector systems simultaneously, and local operation
of individual detector systems for debugging and commissioning.

\subsubsubsection{Buffer and VME Interface}
\label{ssssec:trig_buffer}

For each trigger, the event information (including the time stamp,
trigger type, trigger counter) and the snapshot of the FADC values
should be written into a buffer that will be read out via the VME
bus for crosscheck.

\par The global event information which includes absolute time-stamps
and trigger decision words will be read out from the trigger board,
while individual channel data are read out from the FEE boards. In
this case the event synchronization between the DAQ boards and the
trigger board is critical, and an independent event counter should be
implemented in both the DAQ boards and the trigger boards. The trigger
board in each crate provides the clock and synchronization signals for
the local counters on each FEE board. The global timing system is
designed to enable continuous synchronization of the local clocks in
different crates and at different sites.

The event buffers are envisioned to be VME modules that are in the
same crates as the FEE boards. Data from the trigger and FEE boards is
transfered via the VME bus to the VME buffers. An alternative design
is to have the VME buffer modules in separate crates and have data
transfered from the FEE modules via fast optical GHz links (GLinks) to
the VME buffer modules. We envision VME buffers with enough capacity
to store up to 256 events.

\subsection{Detector Control and Monitoring}
\label{ssec:trig_control}

The detector control system (DCS) controls the various devices of
the experiment (e.g., high voltage systems, calibration system,
etc.), and monitors the environmental parameters and detector
conditions (e.g., power supply voltages, temperature/humidity, gas
mixtures, radiation, etc.). Some safety systems, such as rack
protection and fast interlocks are also included in the DCS.

The DCS will be based on a commercial software package implementing
the supervisory, control, and data acquisition (SCADA) standard in
order to minimize development costs, and to maximize its
maintainability.  LabVIEW with Data logging and Supervisory control
module is a cost effective choice for the DCS.

The endpoint sensors and read modules should be intelligent, have
digitalized output, and conform to industrial communication
standard. We will select the minimum number of necessary field bus
technologies to be used for communication among the SCADA system
and the readout modules.

\bibliographystyle{unsrt}

\newpage
\renewcommand{\thesection}{\arabic{section}}
\setcounter{figure}{0}
\setcounter{table}{0}
\setcounter{footnote}{0}

\section{Installation, System Testing, and Detector Deployment}
\label{sec:test}

The construction and installation of the Daya Bay experiment requires
the civil construction of the underground 
halls, the assembly and testing of the antineutrino detectors, and the
transport and deployment of the detectors in their appropriate
locations.  Well-coordinated activities underground and on surface are
essential for the timely start of the experiment. While the civil
construction of the underground tunnels and halls 
is being completed the Collaboration will start the assembly and
testing of the first detector modules above-ground so that they can be
filled and deployed as early as possible.

The Collaboration has a wide range of experience in the installation
and operation of large detector systems including underground
installations at SNO and KamLAND, the IceCube experiment at the South
Pole, and the STAR detector. Members of the Daya Bay Collaboration
have also been involved in the engineering and installation activities
at MINOS, and Chris Laughton is directly involved in the evaluation of
the tunnel design and specification for the civil construction at Daya
Bay.

All of the assembly work of the antineutrino detector modules except
for the filling will be performed above-ground in a Surface Assembly
Building (SAB). This will provide a facility for the assembly and
testing of two antineutrino detector modules at a time. The SAB
will also include storage and testing facilities for other
subsystems such as the muon system as well as some storage and mixing
facilities for the antineutrino detector liquids (see
Chapter~\ref{ssec:det_LS}.)  A separate Storage Building (SB) will
also be available for storage of arriving equipment. Some elements,
such as the mineral oil storage tanks, will arrive ready for
installation on the surface or in the tunnel. Other elements, such as
the muon tracker will require brief retesting to ensure that no damage
occurred during transport. However, elements such as the antineutrino
detector tanks will require assembly under cleanroom conditions and
system testing prior to the transport underground and filling with
liquid scintillator.

Careful logistical coordination will be essential for the receiving,
assembly, installation, and testing of all detector components and
subsystems. This chapter discusses some of the basic considerations in
the installation process and outlines a plan for the assembly,
filling, and deployment of the antineutrino detector modules.

The logistics of assembly and installation of the antineutrino
detector include:
\begin{enumerate}
 \item All detector modules will be fully assembled and tested with
 inert gas under cleanroom conditions in the surface assembly building.
 \item Only empty detector modules will be moved down the ramp of the
access tunnel to the underground laboratory.
During transport down the access tunnel the
antineutrino detectors are unfilled (and are therefore only about 20~T
or 20\% their final weight).
 \item All detector liquids will be transported underground in special
 ISO liquid containers to ensure clean and safe handling of all
 liquids (see Section~\ref{ssec:det_LS}).
 \item The detector modules will be filled in pairs in the underground
filling hall to ensure ``identical'' target mass and composition
 for each pair.
 \item Once a pair of detector modules has been filled, the detectors
 are ready for deployment in one of the experimental halls.
 \item Once a module is filled with liquid it will only be moved in
 the horizontal tunnels ($<$0.5\% grade) between the experimental halls.
 \item The filling hall is designed to allow for the draining of
 all detector modules at the end of the experiment.
\end{enumerate}

\subsection{Receiving and Storage of Detector Components}
\label{ssec:test_receiving}

\begin{figure}[!htb]
\begin{center}
\includegraphics[height=7cm]{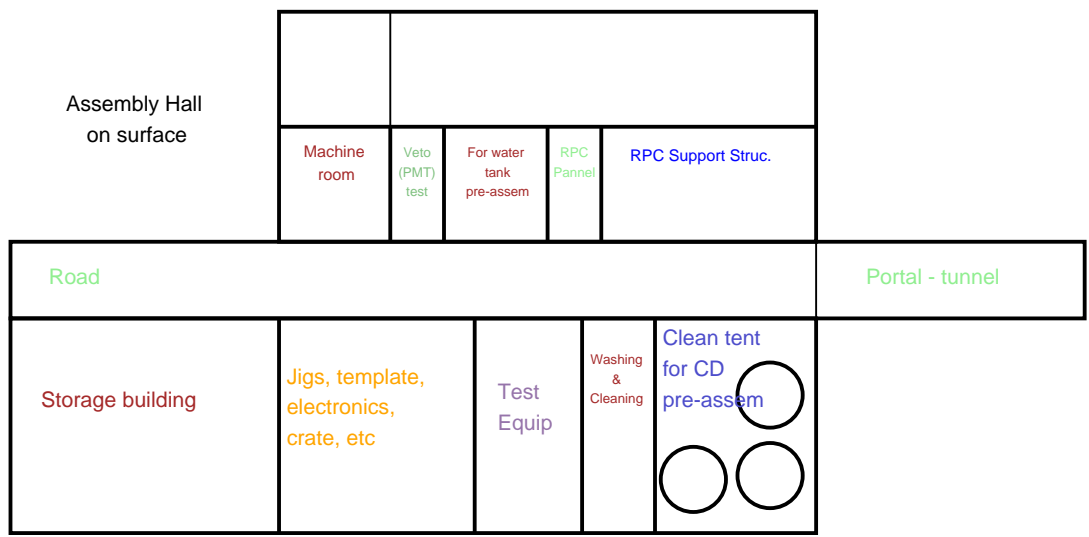}
\caption{Layout of the surface assembly building and storage facilities 
along the road to the tunnel portal.}
\label{fig:SAB2}
\end{center}
\end{figure}

The logistics of receiving, storing, assembling, and testing
components for the Daya Bay experiment requires the construction of
suitable surface facilities including the SAB and SB. As detector
subsystem elements arrive at the Daya Bay site, they will be delivered
to one of these buildings. Special arrangements will be made for the
handling of the detector liquids: the unloaded liquid scintillator,
Gd-loaded liquid scintillator, and mineral oil.  The major
elements of the detector (antineutrino detector tanks, acrylic
vessels, calibration systems, muon detectors, PMTs, liquid storage
tanks) will arrive on a well coordinated timeline to avoid space
problems and to allow the assembly of two detector modules at a time
in the SAB.  The space required for the storage of components would
otherwise quickly overrun the available storage space.  The storage
building will be of sufficient size to hold the large elements of the
antineutrino detector and muon systems, but only for a few of these
elements and for short periods of time before they are moved into the
surface assembly building.  Space for two steel outer vessels plus two
sets of nested acrylic vessels as well as several large muon detector
panels and boxes of PMTs will be sufficient. This requires a building
of roughly 200~m$^2$ area and a crane with two hooks (20~T and 5~T). A
possible layout of the on-site storage and assembly facilities is
shown in Fig.~\ref{fig:SAB2}.

The logistics underground also requires special consideration: the
storage tanks for Gd-loaded liquid scintillator, pure liquid
scintillator and mineral oil will need to be in place prior to the
arrival of antineutrino detector elements or completed antineutrino
detectors will stack up down in the tunnel waiting to be filled.

\subsection{Surface Assembly Building}
\label{ssec:test_SAB}

A surface assembly building of the scale of 25~m$\times$50~m
(1250~m$^2$) is required to assemble, survey, and test two
antineutrino detectors at once. This building will be large enough to
house two detector tanks and their associated inner acrylic vessels.
The rest will be stored in the storage building.  It will also have a
crane of sufficient capacity to assemble the nested vessels and to
lift the completed (but dry) antineutrino detectors onto their
transporter.  The surface assembly building will require clean
assembly space for working on the open vessels to maintain the
appropriate surface cleanliness. Once the detector modules are
assembled and tested as required, they will be moved underground for
filling and subsequent installation in the experimental halls.

In parallel with the assembly of the antineutrino detectors, the muon
detectors will be inspected and tested.  A building of this size will
allow us to set up several inspection and testing stations and have a
station for survey and alignment.  It is sized to handle the assembly
of two antineutrino detectors in parallel plus a short incoming RPC
panel test station.  If the building is arranged in a long (50~m),
orientation, a single 30~T bridge crane with rails along the building
and a smaller 5--10~T crane utilizing the same rails are sufficient.
This allows for the manipulation of partially or fully completed (dry)
antineutrino detectors while moving muon detector panels or staging
other structures in parallel.

To accomplish these multiple testing, assembly, and QA tasks,
appropriate test stations will be assembled. We may need to provide
appropriate gas mixes and high and low voltage power as well as a
low-noise test environment.

The surface assembly building will be designed to ensure several
levels of cleanliness control. Detector components arriving on site
will be stored under sealed conditions in the surface assembly
building. During the assembly of the detector modules more stringent
cleanliness requirements apply. Both the level of particulates and the
environmental air will have to be monitored. Cleanrooms of class
1000--10000 inside the surface assembly building or movable clean tents
with HEPA filters to cover the detector modules can be used to provide
the appropriate environment for the detector assembly.

\subsection{Assembly of the Antineutrino Detectors}
\label{ssec:test_CD}

The major components of the antineutrino detectors will be fabricated
at different places worldwide and shipped to the Daya Bay site for
assembly and testing. The tasks involved in the assembly of the
detector modules include:
\begin{enumerate}
\item Cleaning and inspection of stainless steel tank
\item Installation of the PMTs and cabling inside the detector tank
\item Installation of monitoring equipment in tank
\item Lifting the acrylic vessels into the detector tank
\item Connecting all fill lines, calibration, and instrumentation ports
\item Precision survey of tank and acrylic vessel geometry
\item Final cleaning throughout the entire assembly process.
\item Pressure/leak testing of acrylic vessels and detector tank after assembly
\item Test installation of automated calibration systems (to be removed before 
transport underground)
\end{enumerate}

The entire assembly of the detector modules will be performed in class
1000--10000 cleanroom conditions. This complex assembly and integration
task will require close coordination of several working groups
(detector design, engineering, calibration, monitoring) and the
on-site presence of key scientific and technical personnel.

\subsection{Precision Survey of Detector Modules}
\label{ssec:test_fiducial}

Before transporting the detector modules underground the geometry of
the detector modules is surveyed to high precision using modern laser
surveying techniques. The precision commonly achieved in modern
equipment over the scale of the antineutrino detector ($\sim5$~m) is
of the order of $<$25~$\mu$m in both the radial and the longitudinal
direction. This will serve as a baseline reference for the as-built
detector geometry. In-situ monitoring equipment inside the detector
modules will then be used to track any changes during the transport or
filling of the modules.

Similarly, the muon chambers, PMT support structure and other detector
subsystems will be surveyed in the surface assembly building prior to
the transport underground. Relating internal system geometries to
external fiducial points in the experimental halls will ultimately
allow a precise relative understanding of detector geometry to the
experimental hall and the outside world.

\subsection{Subsystem Testing}
\label{ssec:test_subsystem}

Following the assembly of detector modules and subsystems, testing
becomes a critical task to ensure a smooth turn on and commissioning
the detectors underground. The collaboration's QA and QC experience,
such as from IceCube, will be invaluable in preparing subsystems, getting them
ready, and finally installing them underground with a high success
rate.

All incoming equipment will be inspected for obvious damage.  System
elements that are completely assembled and tested to meet
specifications at far away sites (the US and Beijing for example) will
require a limited retest to ensure no internal damage occurred during
shipment.  Testing for broken channels or shorts in RPC chambers, PMT
function, calibration system function, etc., will all be required.  To
accomplish these tasks, appropriate test stations will be assembled
and utilized in the surface assembly building.  The test stations will
be manned by technicians, grad students, post-docs and physicists and
will utilize a small set of simple electrical tests performed to a
written test specification.  It is not likely we will repeat all the
original performance tests performed at the originating institutions.
However, we may need the capability to provide appropriate gas mixes
and high and low voltage power as well as a low-noise test
environment.

\vskip 0.1in

\noindent
Once the antineutrino detectors have been assembled in the surface
assembly building we plan to perform a suite of tests of their
mechanical integrity and functionality including

\begin{enumerate}
\item pressure and leak tests of the detector tank and acrylic vessels
\item running the PMTs and all cabling with a gas fill inside the detector zones
\item testing the functionality of all ports, calibration, and monitoring equipment
\end{enumerate}

\noindent
Once a detector module passes these tests it is ready for transport
underground. It will be moved down the access tunnel into the
underground filling hall  on  transporters at very low speed.

\subsection{Filling the Detector Modules}

The underground filling station is designed to accommodate two
detector modules during the filling process. The three components:
Gd-loaded liquid scintillator, liquid scintillator for the
$\gamma$-catcher, and mineral oil will be filled simultaneously into
each detector module. The goal is to fill two detector modules from
the same batch of liquids to ensure the same target mass and
composition in pairs of detectors. They can then be deployed either
both at the same near site for a check of the relative detection
efficiency, or one at the near and the other one at the far site for a
relative measurement of the antineutrino flux. An alternate plan under
consideration would be to fill the two detectors in their respective
experimental halls. This would require very different logistics and
transport arrangements.

The underground filling hall houses a 40-t storage/mixing tank for the
0.1\% Gd-liquid scintillator and two more storage tanks for the
$\gamma$-catcher scintillator and the mineral oil for the buffer
region.  All three regions of the detector modules will be filled
simultaneously while maintaining equal liquid levels in each vessel to
minimize stress and loads on the acrylic vessels.  The underground
storage tanks are sufficiently large to hold the full liquid volumes
for two detector modules.

Dedicated fill lines for the Gd-liquid scintillator, the $\gamma$-catcher
liquid scintillator, and the mineral oil connect the storage tank to
the detector module during the filling process.  The filling station
will be equipped with a variety of instrumentation on the storage
tanks and the fill lines for a precise and redundant measurement of
the target mass and composition. Each fill line may use multiple
flowmeters in series for additional systematic control. The
instrumentation we envision using during the filling procedure
includes (see chapter~\ref{ssec:cal_in-situ}):
\vskip 0.15in
\noindent
{\bf On the storage tanks}
\begin{enumerate}
\item liquid level sensors
\item load sensors
\item  temperature sensors
\item access ports for extracting liquid samples
\end{enumerate}
\noindent
{\bf  In the fill lines}
\begin{enumerate}
\item Coriolis mass flowmeters + density measurement
\item conventional volume flow meters
\item  temperature sensors
\end{enumerate}
\noindent
{\bf  In each detector module}
\begin{enumerate}
\item load sensors in the support of each acrylic vessel
\item liquid level sensors in each zone or volume
\item CCD imaging of inside of detector modules
\end{enumerate}
In addition, all fill lines will be equipped with the necessary
filtration and liquid handling systems.
The filling of the different detectors will be performed
sequentially. This ensures that the same set of instrumentation and
flowmeters is used in determining the target mass in each detector. In
this scenario the systematic uncertainty on the relative target mass
between detector modules comes from the repeatability of the mass flow
measurements of one set of instrumentation while the uncertainty on
the relative masses between different detector zones (which is less
critical) comes from the absolute difference between different sets of
instrumentation. After filling, the two detector modules will
be deployed in the appropriate experimental halls.

\subsection{Transport to Experimental Halls}
\label{ssec:test_transport}

Detector modules and related systems will be transported to the filling
hall and experimental halls from the surface assembly building using
flat-bed trailers or self-propelled transporters.  There are several
issues associated with this task that make it somewhat more difficult
than simply using conventional transportation equipment:

\begin{enumerate} 
 \item Due to cost and civil construction constraints the tunnel itself is not very large.  
 \item Entrance to the underground laboratory is through an access tunnel with an incline of up to 10\%.
 \item Transport systems have to be compliant with ventilation and underground safety requirements. 
\end{enumerate}

\vskip 0.1in
\noindent
Because of this, the transport mechanism should:
\begin{enumerate}
 \item have a low bed height ($\leq$0.5~m). 
 \item be powered by an electric drive or by some very clean burning fuel such as propane gas. 
 \item be capable of accurate tracking along an electric or mechanical guide.
 \item utilize an active horizontal level mechanism while transporting the antineutrino detector down the 10\% entrance tunnel into underground laboratory. This will maintain the detector in an upright and vertical position and minimize any stress on the inner acrylic vessels.

\end{enumerate}

Note that the detector modules will be filled in the underground
filling station and they will only be transported in the horizontal
tunnels between experimental halls after they have been filled.
During the transport to the experimental halls the antineutrino
detectors are filled and have a total mass of $\sim$100~T. We are
currently investigating several transportation systems.  Custom-made
(short and wide) flatbed, `lowboy' trailers pulled by an electric
airport pushback tugs are one option.  A second option is 
self-propelled, remote-controlled, small wheel diameter transporters
which often have hydraulic lifting capability.  A third option is an
electric power train-like transporter which runs on rails.  All of
these systems have the benefit of having a bed height of around 0.5~m.
All of these can be powered by electric motors for clean, safe
operation in confined spaces.

\subsection{Final Integration in the Experimental Halls}
\label{ssec:test_EH}

After the antineutrino detectors have been filled they will be slowly
transported through the tunnels to the appropriate experimental
halls. They will be deployed by crane into the drained water pool and
onto their stands. All cabling and electronics will be connected
and their calibration systems installed so they can be calibrated
and checked.

The muon-detector elements (RPC chambers, structures and PMTs) will be
delivered to the experimental halls after test and checkout on the
surface.  These elements will be installed in the pool (PMTs and PMT
supports) and over the roof of the pool (RPCs).

\subsection{Early Occupancy of the Experimental Halls at the Near and Mid Sites}

The civil construction of the underground laboratory including the
experimental halls and tunnels will take about 24~months. The time
scale is set by the excavation of the tunnels between the experimental
halls. The near and mid sites will be completed first before the
tunnel to the far site is finished. Completion of experimental halls
at the Daya Bay near site and the mid site suggests the implementation
of an early experiment utilizing the halls at these two sites. Early
occupancy of these sites would provide the opportunity to commission
the detector modules at the near site and to make a first, ``fast''
measurement with a sensitivity of sin$^22\theta_{13}<$0.035 (see
Chapter~\ref{sec:sys}).

The use of these experimental halls during the ongoing excavation and
construction of the tunnel to the far site poses significant but not
insurmountable logistical challenges for the work underground. While
shared underground occupancy is largely to be avoided because of
issues of safety (traffic, blasting, explosives, fumes etc..) and
interference with mining work other experimental facilities such as
KamLAND in the Kamioka mine have demonstrated that data taking with a
sensitive neutrino experiment is possible while a new underground hall
is excavated some few hundred feet away. In the case of KamLAND, a new
underground hall for a liquid scintillator purification system was
built in 2006. The experiment continued data taking and access to the
experimental facilities for scientists was arranged on a specific
schedule together with the mining and construction crews. A similar
situation can be found at SNOLab in the Creighton mine in Canada which
is being constructed during the active phase of the Sudbury Neutrino
Observatory.

The possibility of commissioning the detector modules at the near site
and making an early measurement of sin$^22\theta_{13}$ with detector
modules at the near and mid sites may be worth the additional
logistical challenge of coordinating the underground construction work
and the installation of the first detector modules. In this case, the
experiment may start commissioning detector modules 18 months after
beginning of civil construction and taking first neutrino data at two
different experimental halls and distances about two years after
breaking ground.

\vskip 0.1in
\noindent
The planning for this scenario requires that

\begin{enumerate}
\item the necessary infrastructure for the operation of the detector modules (power, air, etc) can be installed at the near and mid sites while the construction of the tunnel to the far site is ongoing. 
\item a plan for the installation of the detector modules will be developed that does not impact the day-to-day mining operation
\item safety issues with respect to escape routes and personnel underground are addressed
\end{enumerate}

This possibility requires further discussion and negotiations with the contractors for the underground construction of the tunnel and experimental halls. 

\subsection{Precision Placement and Alignment}
\label{ssec:test_location}

Precise knowledge of the `global' location of each hall with respect
to the reactor cores is important for the accurate determination of
the distances between the reactor cores and each neutrino
detector. Permanent survey markers in each experiment hall will serve
as reference marks for the positioning of the detectors.  These survey
markers will be placed and known to a precision of better than tens of
centimeters, with respect to the outside world, even though the halls
are hundreds of meters inside underground tunnels. This precision is
commonly achieved in the construction of tunnels and in mining.

Within the experimental halls the position of the detectors can be
determined quite precisely. The antineutrino detectors will be
surveyed into approximate but precisely known location on their stands
at the bottom of each water pool.  The knowledge of the location of
each antineutrino detector, with respect to the fiducial markers in
the halls, will be at the sub-mm level. The location of the muon
system elements also can be surveyed and understood at the same sub-mm
level. This is both with respect to the antineutrino detectors and the
experimental hall.

With this information the distance between the detector modules and
the reactor cores will be known to the required precision of better
than 30~cm.
\begin{table}[htdp]
\begin{center}
\begin{tabular}{|l|l|}
\hline
{ Element of Experiment }& { Positioning Accuracy} \\
\hline
\hline
experimental hall & $\mathcal{O}$(10~cm) \\
detector position in experimental hall & $<$~mm \\
acrylic vessels within detector tank & $<25$~$\mu$m \\
\hline
\end{tabular}
\caption{Positioning accuracy of the principal elements of the Daya Bay experiment.}
\end{center}
\label{default}
\end{table}

\newpage
\setcounter{figure}{0}
\setcounter{table}{0}
\setcounter{footnote}{0}

\section{Operations of the Daya Bay Experiment}
\label{sec:ops}

Operation of the Daya Bay Experiment requires a
variety of tasks including:

\begin{enumerate}
\item Underground transport of detector modules and deployment into the experimental halls. 
\item Data taking with the detector modules in the experimental halls.
\item Frequent automated calibration of each detector module.
\item Full-volume calibration of the detector modules as needed.
\item Monitoring of the state of the detector modules and the underground lab conditions.
\item Monitoring and maintenance of the muon system.
\item Maintenance and repair of the calibration system.
\item Monitoring, maintenance, and repair of electronics and data acquisition system.
\item Monitoring the target liquid and and performing regular chemical assays on the liquid scintillator samples.
\end{enumerate}

The routine monitoring of the experiment will be performed by members
of the collaboration and special technical personnel trained in
emergency procedures of the underground lab facility. A surface
control room will be set up in the vicinity of the access portal
for monitoring and data taking. Daily walk-around checks in the
underground facility will ensure the safe operation of all
underground systems.

The operations of the detector will consist of monitoring the detector
performance and data quality, routine calibration, and online data
analysis. The calibration procedure will include automated calibration
runs to be performed by shift members operating the detector. Special
manual calibration runs are to be performed by expert personnel.

All shift duties related to data taking and monitoring will be shared
between the members of the Day Bay Collaboration. On-site shifts as
well as remote, off-site shifts will be part of running the Daya Bay
experiment. Groups responsible for specific subsystems will make
arrangements for the maintenance of detector subsystems. 

The scientific and technical team of the Daya Bay experiment will have
support from the China Guangdong Nuclear Power Group (CGNPG) which
operates the Daya Bay reactor complex, and which is a collaborator on
the experiment.

Operation of the Daya Bay experiment includes data taking with the
detector modules in different configurations. The default
configuration, as well as other optional configurations, are outlined in
the following section. A variety of alternative operations plans are
currently being evaluated with varying frequency of detector swaps.

\subsection{Configurations of Detector Modules}
 \label{ssec:ops_config}

This section describes the different possible detector configurations
of the Day Bay experiment and their possible use during different
phases of the experiment. The different deployment options and run
scenarios are currently being evaluated from the point of view of
logistics, cost and physics reach:
\begin{enumerate}

\item {\bf Initial Detector Deployment:} The eight detector modules 
are built, assembled, and filled in pairs to ensure that their
characteristics and target mass and composition are as identical as
possible. Once a detector pair has been filled underground there are
two options:

\begin{enumerate}
\item the detectors can either be deployed together at the Daya Bay 
near site for a commissioning run and check of their relative 
detector efficiencies, or 
\item one of them can be installed at the near site and the other one 
at the mid or far site to immediately start data taking with two 
detectors of the same pair at different distances
\end{enumerate}
 
A commissioning run of both detectors at the near hall is a unique
opportunity to test the operation of each detector module before one
of them is deployed at the mid or far site. The intrinsic detector
background, the cosmogenic background at the near site, and the
relative detection efficiency of the detector modules can be checked
during this commissioning phase. This step is trivial for the first
detector pair, as the default plan for the Daya Bay near site allows
for the installation of two detector modules. Commissioning of the
other detector pairs at the near site requires deploying the detectors
to the other experimental halls immediately after the commissioning
run.
A drawing of the detector configuration is shown in
Fig.~\ref{fig:DYB_nearcommissioning}.
\begin{figure}[h]
\begin{center}
\includegraphics[height=7.5cm]{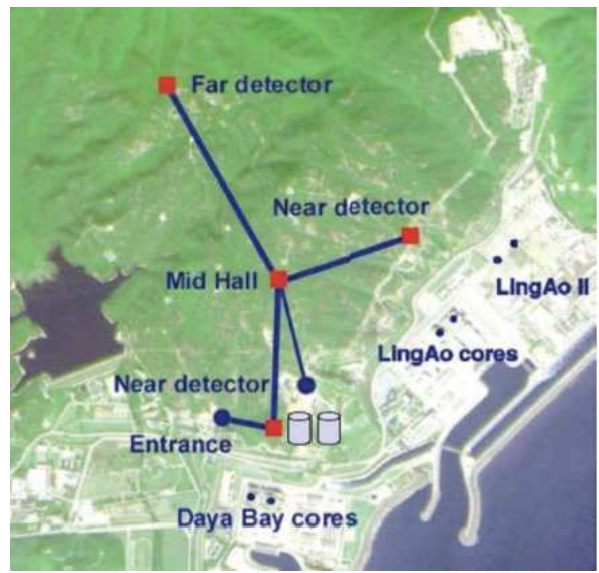}
\caption{Optional commissioning runs of pairs of detectors at the 
Daya Bay near site. With sufficient runtime of a few months 
for systematic checks of the detectors a relative comparison of 
the detector response is possible.}
\label{fig:DYB_nearcommissioning}
\end{center}
\end{figure}

Including the time for detector installation and start-up, the total
time for such a commissioning run is likely to be $\sim$6 months. The
collaboration may decide to skip this initial commissioning step and
immediately deploy the two detectors from each pair at the near and
far (or mid) sites to expedite the overall experiment. In
this second scenario two detectors are assembled and filled at the
same time and then one of them is deployed at a near site and the
other one is immediately moved to the far site. Data taking and a
relative measurement of the neutrino flux between these two detectors
can then commence immediately.

\item{\bf Using the Mid Site:} The default plan for the construction of 
the underground laboratory at Daya Bay includes a mid-site at a
distance of about 1156~m from the Daya Bay cores and 873~m from the
center of the Ling-Ao cores (as discussed in
Section~\ref{ssec:exp_layout}). Civil construction of this site will
finish earlier than the excavation of the tunnel to the far site. By
deploying two or four 20-ton detector modules at the mid-site along
with two 20-ton detectors at the Daya Bay near it may be possible to
make a first, ``fast'' measurement of sin$^22\theta_{13}$ at this
intermediate distance. See chapter~\ref{sec:test} for a discussion of
the logistical and construction issues relating to early occupancy
of the near and mid sites. A drawing of this detector configuration is
shown in Fig.~\ref{fig:DYB_early}.
\begin{figure}[h]
\begin{center}
\includegraphics[height=7.5cm]{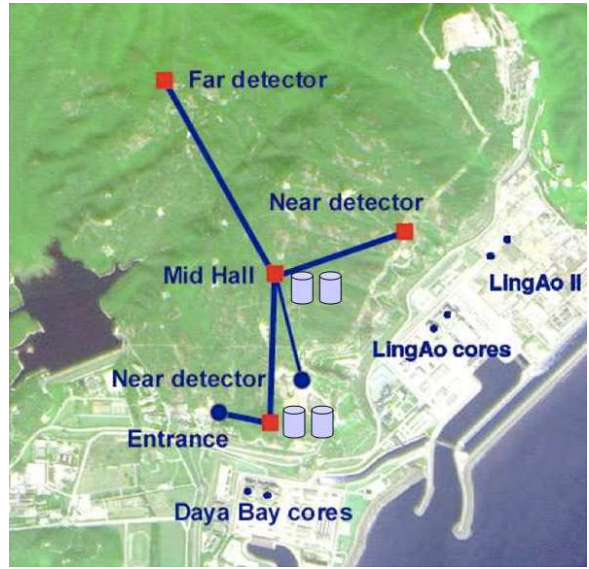}
\caption{Optional near-mid configuration of the Daya Bay experiment 
for an early physics run. With two 20-ton detectors at the near and 
mid site a sensitivity of sin$^22\theta_{13}\sim$0.035 can be achieved 
in $\sim$1 year of data taking. }
\label{fig:DYB_early}
\end{center}
\end{figure}

One can also envision using the mid-site for a systematic cross
check. By running the experiment in the mid-far configuration it is
possible to probe the $\theta_{13}$ oscillation with a different
combination of distances. The ultimate sensitivity of the experiment
is somewhat reduced but the ratio of the energy spectra from the mid
and far site provide a different oscillation signature as a function
of energy.

\item {\bf Default Configuration of Full Experiment:} To achieve the best 
sensitivity in the Daya Bay experiment two 20-ton detector modules are
deployed at each one of the Daya Bay and Ling-Ao near sites along with
four 20-ton detector modules at the far site. The total active target
mass at the far site is 80~tons. In the default scenario, the mid-site
is unused. It is possible to operate in this configuration either with
or without the swapping of pairs of detectors. A drawing of this
detector configuration is shown in Fig.~\ref{fig:DYB_default}.
\begin{figure}[!htb]
\begin{center}
\includegraphics[height=7.5cm]{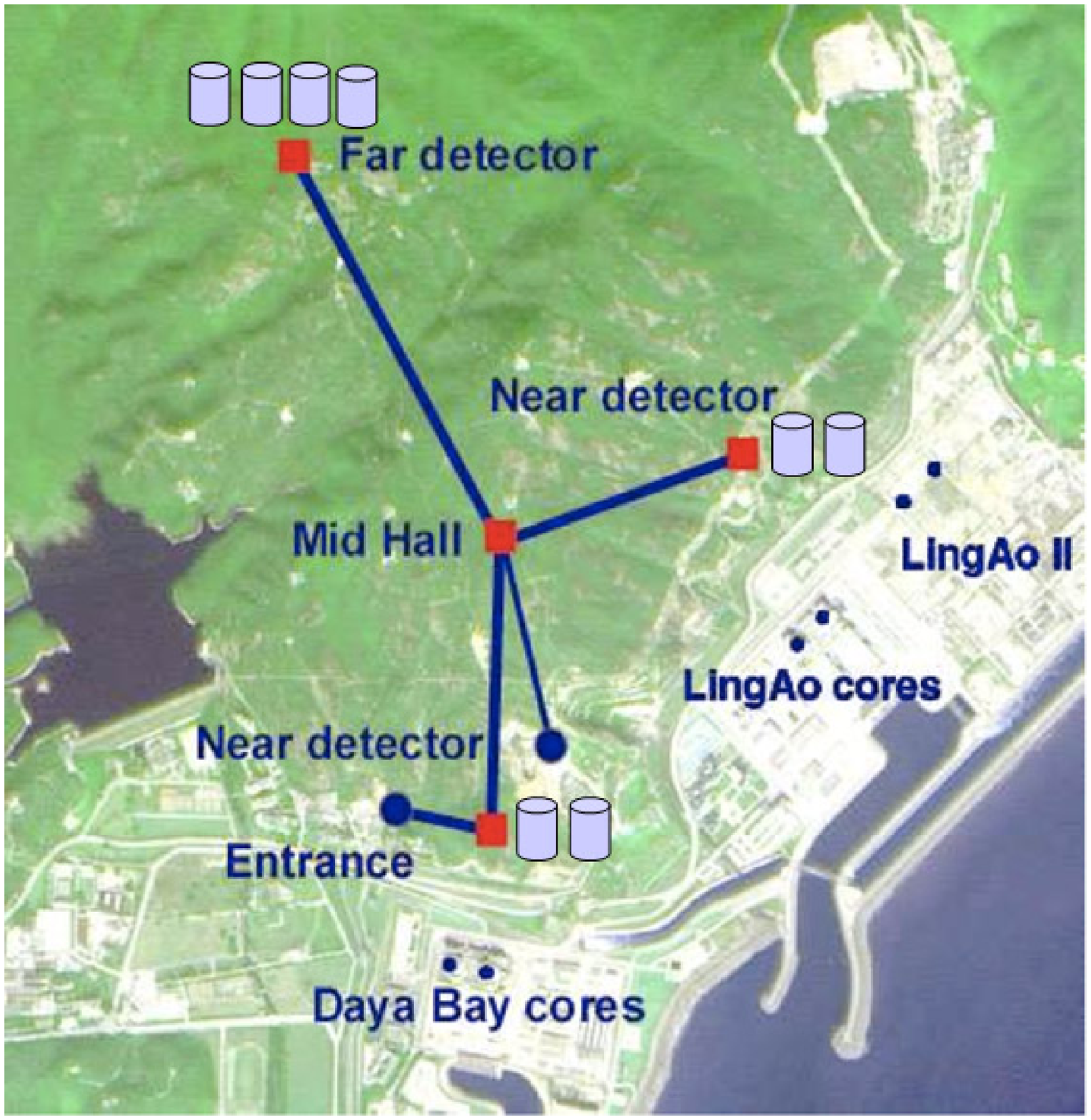}
\caption{Default configuration of the Daya Bay experiment, optimized 
for best sensitivity in sin$^22\theta_{13}$. Data taking can occur in 
a static configuration or with swapping of detectors.}
\label{fig:DYB_default}
\end{center}
\end{figure}

 \item {\bf Optional Swapping in the Daya Bay Run Plan:} Swapping of
 detector modules is an option but not a necessity in the Daya Bay
 experiment. The target sensitivity of sin$^22\theta_{13} < 0.01$ at
 90\% C.L.  can be achieved without swapping detectors. The design of
 the Daya Bay experiment provides the option of swapping detectors for
 systematic checks and to ultimately increase the sensitivity of the
 experiment to about sin$^22\theta_{13} < 0.006$ (see
 Table~\ref{tab:sens}). After all detectors are commissioned and
 located at their initial sites swapping of detectors can occur
 either:
\begin{enumerate} 
  \item throughout the experiment in regular 6-months intervals for the 
optimal cancellation of the experimental systematics (as described in 
Table~\ref{tab:swap}), or
 \item after an initial static experiment with data taking for 2-3 years 
that reaches the design goal of sin$^22\theta_{13} < 0.01$
\end{enumerate}
 
 The collaboration has not decided yet which approach to choose. It
 will depend on the outcome of the design studies of the antineutrino
 detectors, their transportation system, and R\&D on the calibration
 and monitoring of the detector response. In addition, the timeliness
 and potential impact of a first measurement of $\theta_{13}$ at Daya
 Bay will drive the detector deployment and run plan.
 
 \end{enumerate}
 
   \subsection{Detector Swapping}
   \label{sssec:ops_swap}

The purpose of swapping detectors has been described in
Sections~\ref{sec:over} and~\ref{sec:sys}.  An overview of the steps
involved in the swapping procedure is given below.  Detector swapping
will utilize the standard transportation methods developed for the
underground movement of the detectors. As such, detector swapping uses
all of the same techniques and procedures that are developed for the
initial deployment of the detector modules and the installation of the
experiment. Even the initial deployment of the detector modules at the
far site requires filled modules to be transported from the filling
station to the far experimental hall. Swapping is different from the
initial deployment of the detector modules in that it is critical to
characterize any change in the detector response during the
swap. Without a complete characterization of the detector response the
performance of a module cannot be compared before and after the
swap. This poses a unique challenge and sets the criteria for the
development of a comprehensive calibration and monitoring program.

 \subsection{Logistics of Detector Swapping}

The total estimated time for detector swapping in the baseline water
pool configuration is several days. We anticipate that
the transport of each detector module in the tunnel can be performed
in less than a day. With a transportation speed of $\sim$5~m/minute a distance
of 1500~m can be covered in less than 7~hrs.

Detector swapping includes the following sequence of steps:
\begin{enumerate} 
 \item Perform final detector calibration to establish detector response 
 immediately prior to the move.
 \item Shut down HV and DAQ.
 \item Disconnect large area RPC roof system as necessary to ready this 
portion of muon system for sliding back and out of the way for antineutrino 
detector lift operation. 
 \item Drain water pool to a level below the antineutrino detector module 
($\sim$1000--1500~m$^3$).  (Replace with fresh, filtered water when refilling.)
 \item Install a personnel man bridge over the open pit to allow safe access
to the top of antineutrino detector. 
 \item Disconnect PMT HV and signal
cables, LS overflow plumbing, etc. as required to prepare for move.
 \item Remove calibration system \& piping as required from top
of antineutrino detector. 
 \item Attach the lifting device to the antineutrino detector. 
 \item Using a 150~T crane, lift the antineutrino detector vertically
out of pool and translate it horizontally onto a transporter. 
 \item Transport the antineutrino detector to the new location.
 \item Reverse the operation at the previously prepared new location.
\item Calibrate the detector in the new location to establish the detector response immediately after the move.
\end{enumerate}

\newpage
\renewcommand{\thesection}{\arabic{section}}
\setcounter{figure}{0}
\setcounter{table}{0}
\setcounter{footnote}{0}

\section{Schedule and Scope}
\label{sec:scope}

In this chapter, the overall project plan will be described.  This
will include an overview of the project schedule, and the concept for
the international division of scope.  Later, the planned U.S.
scope range will be outlined.  This is a joint project with
an international collaboration.

\subsection{Project Schedule}
\label{ssec:scope_schedule}

Briefly, the first significant construction event of the Daya Bay
experiment schedule begins with the initiation of civil construction
on the tunnels in the spring of 2007. The Project's goal is to
complete the civil construction of the tunnels, experimental halls and
utility infrastructure before the middle of 2009.

There is an additional goal to complete the Daya Bay Near Hall (and
Filling Hall) as early as possible --- approximately 12 months earlier
than the final (far) hall.  The schedule for the detector elements is
therefore driven by the completion of the first two antineutrino
detectors and one third of the muon system hardware by the fall of
2008 in order to deploy these in this first experimental hall. This
Daya Bay Near Hall will be used as an early opportunity to install,
test and begin partial experiment operations --- a chance to debug and
gain insight into detector operations. This would occur in the early
summer of 2009.  The next hall to follow will be the Mid Hall. This
hall and its detectors will most likely be available for installation
tasks 3--4 months after the Daya Bay Near Hall (early in calendar
2009). This would then allow us an opportunity to install and begin
measurements of $\theta_{13}$ by late summer of 2009.

The remainder of the detectors will be installed and commissioned in
the Ling Ao Near and Far Halls by early summer of calendar 2010 so
that the full complement of near and far detectors can begin data
taking. A more complete view of the project schedule is shown in
Fig.~\ref{fig:proj_sched} below.
\begin{figure}[!h]
\includegraphics*[width=0.9\textwidth, height=0.8\textheight,angle=0]{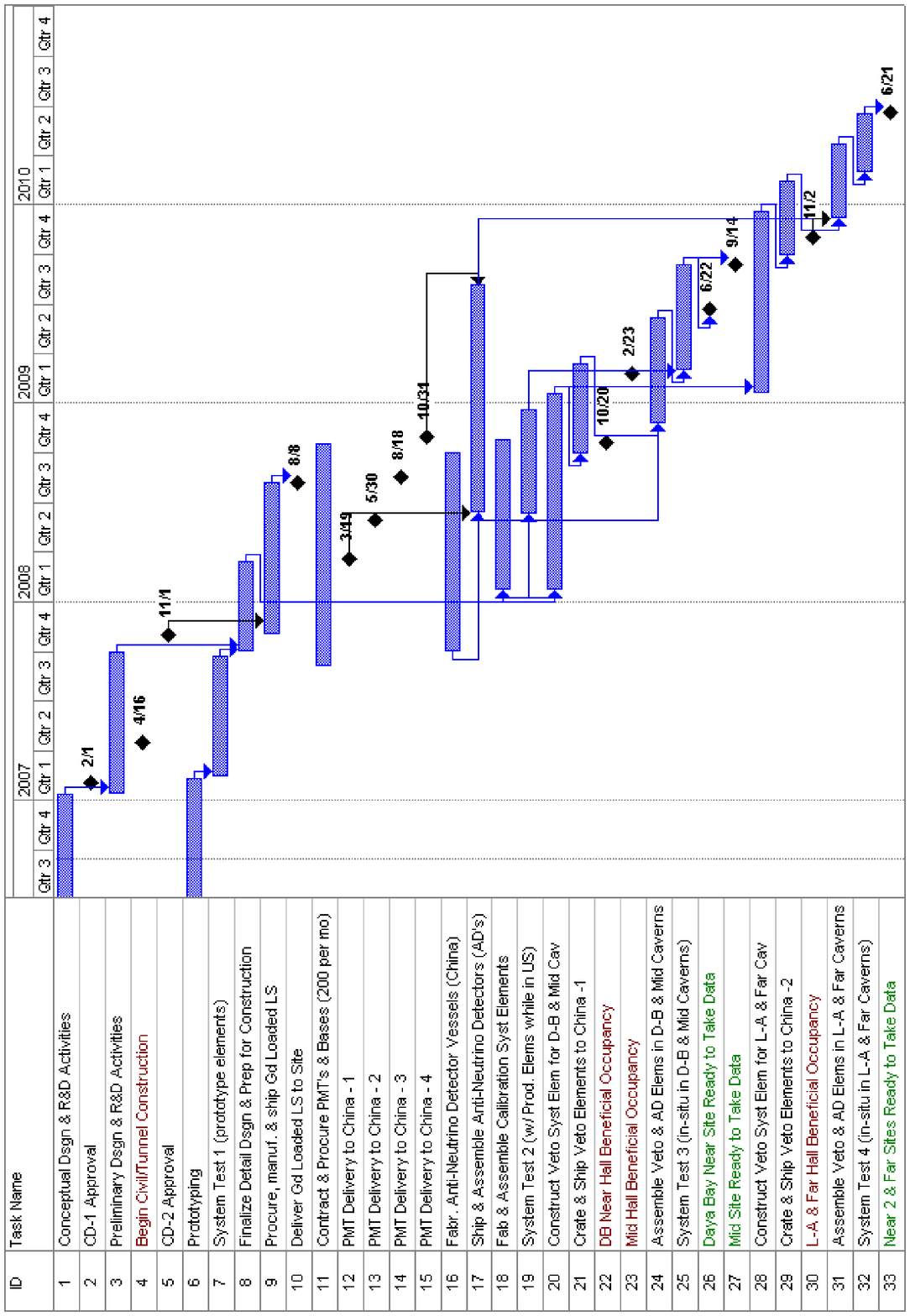}
\caption{Daya Bay Project Summary Schedule.}
\label{fig:proj_sched}
\end{figure}

\subsection{Project Scope}
\label{ssec:scope_scope}

The project's entire technical scope has been described in the
previous chapters.  The total Daya Bay project includes the civil
construction of the experimental facility at the Daya Bay nuclear
reactor complex as well as the construction of the detector elements
(antineutrino detectors, muon system, calibration system,
DAQ/Trigger/Online and offline).  Crucial to all of these activities
are the project integration elements: Installation and System Test,
System Integration and Project Management.

The division of the Project scope will not be finalized until a formal
MOU is developed and signed between the U.S., China and other
countries.  Therefore, the scope shown in Table~\ref{tab:scope} is
based on the current status of discussions within the Collaboration.
\begin{table}[tbp]
\begin{center}
\begin{tabular}{|c||l|c|c|} \hline 
WBS & Description & China & U.S.  \\ 
    &             & lead  & lead  \\ \hline \hline
1  & {\bf Antineutrino Detector}                     & {\bf X} &  \\ \hline
   & System design, steel vessels, LS, mineral oil   &  *       &         \\ \hline
   & FEE co-design and manufacture, racks            &  *       &         \\ \hline
   & safety systems, assem. and test                 &  *       &         \\ \hline
   & Gd-LS,  LS purification/mixing/filling          & $\circ$  & $\circ$ \\ \hline
   & acrylic vessels, PMTs and support,              &          &  *      \\ \hline
   & transporter, FEE co-design, cables, crates      &          &  *      \\ \hline
2  & {\bf Muon System}                               &          & {\bf X} \\ \hline
   & System design, muon tracker, water Cherenkov    &          & *       \\ \hline
   & PMTs and support, assem. and test               &          & *       \\ \hline
   & FEE, safety systems                             &  *       &         \\ \hline
3  & {\bf Calibration and Monitoring}                &          & {\bf X} \\ \hline
   & automated system, glove box                     &          &  *      \\ \hline
   & monitoring system and system test               &          &  *      \\ \hline
   & manual system, LED, radioactive sources         &  *       &         \\ \hline
   & low background counting system                  &  *       &         \\ \hline
4  & {\bf Trigger/DAQ/Online}                        & {\bf X}  &	  \\ \hline
   & Trig/DAQ board co-design and manufacture        &  *       &         \\ \hline
   & monitoring/controls hardware and software       &  *       &         \\ \hline
   & racks                                           &  *       &         \\ \hline
   & Online hardware and software                    & $\circ$  &  $\circ$\\ \hline
   & Trig/DAQ board co-design, crates, cables        &          &  *      \\ \hline
   & system test platform                            &          &  *      \\ \hline
5  & {\bf Offline}                                   & {\bf $\circ$} & {\bf $\circ$}\\ \hline
   & offline architecture and data archiving in U.S. &          &  *      \\ \hline
   & offline hardware and software and simulations   & $\circ$  & $\circ$ \\ \hline
6  & {\bf Conventional Construction}                 & {\bf X}  &         \\ \hline
   & tunnels, halls,  underground utilities          &  *       &         \\ \hline
   & safety systems, surface facilities              &  *       &         \\ \hline
7  & {\bf Installation and Test}                     & {\bf $\circ$} & {\bf $\circ$} \\ \hline
   & Onsite installation and system testing          & $\circ$  &  $\circ$ \\ \hline
   & planning, execution                             & $\circ$  &  $\circ$ \\ \hline
8  & {\bf System Integration}                        & {\bf $\circ$} & {\bf $\circ$} \\ \hline
   & System level mechanical  engineering            & $\circ$  &  $\circ$ \\ \hline
   & System level electronics engineering            & $\circ$  &  $\circ$ \\ \hline
   & Common Fund                                     &          &          \\ \hline
9  & {\bf Project Management}                        & {\bf $\circ$} & {\bf $\circ$} \\ \hline
   & Planning, communication, coordination           &  $\circ$ &  $\circ$ \\ \hline
   & reporting, reviews                              &  $\circ$ &  $\circ$ \\ \hline
\end{tabular}
\caption{\label{tab:scope} Daya Bay project scope. 
The {\bf X}'s refer to which country has the lead on a given task. The {\bf *}'s refer
to responsibility for scope deliverables. The {\bf $\circ$}'s refer to shared responsibility.}
\end{center}
\end{table}

The major elements of U.S. scope deliverables include parts of the
antineutrino detector:  Gadolinium
loaded Liquid Scintillator, PMTs (w/bases and control boards), 
PMT support structure and the transporter system for moving the
assembled and filled antineutrino detectors.  The U.S. scope also includes the
majority of the Muon System: the PMT based water Cherenkov system
and the roof tracking system (possibly RPCs).  A significant portion
of the Calibration system will also be a U.S. deliverable: the
automated deployment system and the full monitoring system.
Additionally, many elements will be cooperatively developed: the
front-end and trigger electronics design for both the antineutrino
and muon systems and many of the infrastructure items (e.g.,
online and offline software).  System design integration, installation
and test and project management will be jointly planned, managed and
executed by the U.S. and China.

\appendix
\section{Acknowledgements}
\label{sec:acknowledgements}

This work was supported in part by the the Chinese Academy of
Sciences, the National Natural Science Foundation of China (Project
numbers 10225524, 10475086, 10535050 and 10575056), the Ministry of
Science and Technology of China, the Guangdong provincial goverment,
the Shenzhen Municipal government, the China Guangdong Nuclear Power
Group, the Research Grants Council of the Hong Kong Special
Administrative Region of China (Project numbers 400805 and 400606),
the United States Department of Energy (Contracts DE-AC02-98CH10886,
DE-AS02-98CH1-886, and DE-FG02-91ER40671 and Grant DE-FG02-88ER40397),
the U.S. National Science Foundation (Grants PHY-0555674 and
NSF03-54951), the University of Houston (GEAR Grant number 38991), the
University of Wisconsin and the Ministry of Education, Youth and
Sports of the Czech Republic (Project numbers MSM0021620859 and
LC527).

\newpage

\section{Acronyms}
\label{sec:acronyms}

\bigskip
\begin{tabular}{ll}
AC    & alternating current \\
Access& database program from Microsoft Corporation \\
ADC   & analog to digital converter \\
BES   & Beijing Spectrometer at the Beijing Electron Positron Collider \\
BINE  & Beijing Institute of Nuclear Energy \\
BNL   & Brookhaven National Laboratory \\
CAS   & Chinese Academy of Sciences \\
CC    & charged-current neutrino interactions \\
CCG   & central clock generator \\
CD-1  & Critical Decision \#1 --- Site Selection (CDR) \\
CD-2  & Critical Decision \#2 --- Cost/Schedule/Scope well defined (TDR) \\
CERN  & European Organization for Nuclear Research\\
CGNPC & China Guandong Nuclear Power Group (Daya Bay owner) \\
CL    & confidence level \\
$CP$  & charge, parity symmetry \\
$CPT$ & charge, parity, time reversal symmetry \\
CVS   & code versioning system \\
DAC   & digital to analog converter \\
DAQ   & data acquisition \\
DC    & direct current \\
DCS   & detector control system \\
DOE   & U.S. Department of Energy \\
ES    & elastic neutrino scattering \\
ES\&H   & environment, safety \& health \\
FADC  & flash ADC \\
FEC   & front-end card \\
FEE   & front-end electronics \\
FET   & field effect transistor \\
FPGA  & field programmable gate array \\
FY    & fiscal year \\
FWHM  & full width at half maximum \\
Gallex & Gallium Experiment\\
Gd-LS & Gd loaded liquid scintillator \\
GEANT & detector description and simulation tool \\
GNO   & Gallium Neutrino Observatory\\
GOC   & global operation clock \\
GPS   & Global Positioning System \\
H/C   & ratio of hydrogen to carbon \\
H/Gd  & ratio of hydrogen to gadolinium \\
HOTLink & bus for clock distribution \\
HV    & high voltage \\
HVPS  & high voltage power supplies \\
\end{tabular}

\begin{tabular}{ll}
IFC   & International Finance Committee \\
IGG   & Institute of Geology and Geophysics \\
IHEP  & Institute for High Energy Physics \\
ILL   & Institut Laue-Langevin \\
ISO   & International Standards Organization \\
JINR  & Joint Institutes for Nuclear Research \\
JTAG  & electronic standard for testing \& downloading FPGA's \\
KamLAND & Kamioka Liquid Scintillator Antineutrino Detector \\
K2K   & KEK to Kamiokanda neutrino oscillation experiment \\
KARMEN& Karlsruhe Rutherford Medium Energy Neutrino experiment\\
KEK   & High Energy Accelerator Research Organization in Japan\\
Kr2Det & Two Detector Reactor Neutrino Oscillation experiment at Krasnoyarsk\\
$L/E$ & distance divided by energy \\
L3C   & L3 cosmic ray experiment \\
LAB   & Linear Alkyl Benzene\\
LabVIEW & Laboratory Virtual Instrument Engineering Workbench \\
LBNL  & Lawrence Berkeley National Laboratory \\
LED   & light emitting diode \\
LENS  & Low Energy Solar Neutrino Spectrometer\\
LIGO  & Laser Interferometric Gravity Observatory \\
LMA   & Large Mixing Angle solution \\
Ln    & lanthanides \\
LS    & liquid scintillator\\
LSND  & Liquid Scintillator Neutrino Detector\\
LVDS  & low voltage differential \\
MBLT  & Multiplexed Block Transfer \\
m.w.e.& meters of water equivalent \\
MC    & Monte Carlo \\
MINOS & Main Injector Neutrino Oscillation experiment \\
MoST  & Ministry of Science and Technology of China \\
MOU   & Memorandum of Understanding \\
MSB   & 1,4-bis[2-methylstyrl]benzene\\
MSPS  & mega-sample per second \\
NC    & neutral current neutrino interactions \\
NSFC  & Natural Science Foundation of China \\
NPP   & nuclear power plant \\
NTP   & Network Time Protocol \\
ODH   & oxygen deficiency hazard \\
OPERA & Oscillation Project with Emulsion-tRacking Apparatus\\
p.e.  & photo-electrons \\
PC    & pseudocumene \\
PC    & personal computer \\
PMT   & photomultiplier tube \\
PPS   & Pulse Per Second \\
PRD   & Pearl River Delta (elevation above sea level) \\
PVC   & Poly Vinyl Chloride\\
PWR   & pressurized water reactors \\
\end{tabular}

\begin{tabular}{ll}
QA    & quality assurance \\
QC    & Quality control \\
QE    & quantum efficiency \\
REE   & rare earth elements \\
R\&D  & research and development\\
RS    & Richter scale \\
RPC   & resistive plate chamber \\
RPVC  & rigid polyvinyl chloride\\
SAGE  & Soviet American Gallium solar neutrino Experiment\\
SCADA & supervisory, control, and data acquisition \\
s.p.e.& single photo-electron \\
SNO   & Sudbury Neutrino Observatory \\
SNO+  & proposed solar and geo-neutrino experiment using liquid scintillator in the existing SNO detector \\
SS    & stainless steel \\
SAB   & surface assembly building \\
TDC   & time to digital converter \\
TSY   & Fourth Survey and Design Institute of China Railways\\
UTC   & Universal Coordinated Time \\
UV    & ultraviolet light\\
VME   & Versa Module Europa \\
WBS   & work-breakdown structure \\
YREC  & Yellow River Engineering Consulting Co. Ltd.\\
\end{tabular}


\end{document}